%% file: main.tex
\documentclass[11pt,letterpaper,oneside,hidelinks]{article}
\usepackage{titlesec}
\PassOptionsToPackage{nodisplayskipstretch}{setspace}
\renewcommand{\thefootnote}{\fnsymbol{footnote}}
\input{preamble}

\setlist[enumerate]{
    itemsep=2pt,
    topsep=2pt,
    parsep=0pt,
    partopsep=0pt
}
\makeatletter
\def\thm@space@setup{%
    \thm@preskip=0pt
    \thm@postskip=0pt
}
\makeatother
\hypersetup{colorlinks,citecolor=blue,linkcolor=blue}
\titlespacing{\section}
    {0pt}{\parskip}{-\parskip}
\titlespacing{\subsection}
    {0pt}{\parskip}{-\parskip}
\titlespacing{\subsubsection}
    {0pt}{\parskip}{-\parskip}

\makeatletter
\def\thm@space@setup{
    \thm@preskip=0pt
    \thm@postskip=0pt
}
\makeatother

\begin{document}
{
\title{Supervised Factor-Augmented Matrix Regression with Weak Factors}
\date{\today}
\author{
Wenyan Hou$^1$, 
Zhaoxing Gao$^{1,2,3}$\thanks{
Corresponding author: zhaoxing.gao@uestc.edu.cn (Z. Gao), 
School of Mathematical Sciences, University of Electronic Science and Technology of China (UESTC), 
Chengdu 611731, P.R. China.
}, 
and Ruey S. Tsay$^4$
\\[1ex]
{\normalsize $^1$School of Mathematical Sciences,  UESTC}
\\
{\normalsize $^2$The Advanced Institute of Humanities and Social Sciences, UESTC}
\\
{\normalsize $^3$School of Economics and Management, UESTC}
\\
{\normalsize $^4$Booth School of Business, The University of Chicago}
}
}

\begin{titlepage}
\maketitle
\thispagestyle{empty}

\begin{abstract}
We develop a supervised factor-augmented matrix regression framework for high-dimensional analysis with weak factors. When factor loadings are concentrated on unknown subsets of rows and columns, conventional full-matrix factor extraction may suffer substantial signal dilution, with weakness in one matrix dimension propagating to estimation in the other. We characterize this two-way weak-loading phenomenon for both \(\alpha\)-PCA and projected estimators and show that their loading and factor estimates may deteriorate under weak signals. To address this problem, we propose a sequential response-guided screening procedure that identifies informative rows and columns prior to factor extraction. We establish exact support recovery and show that post-screening factor re-estimation converts the weak-loading problem into a pervasive one relative to the retained dimensions, yielding sharper convergence rates under substantially weaker dimensional growth conditions and consistent factor-number determination. We further establish asymptotic distributions for both unrestricted and bilinear regression estimators based on the estimated matrix factors. Simulations demonstrate substantial improvements in support recovery and factor estimation under weak signals. We illustrate the gains obtained by the response-guided factor construction with 
two empirical applications.
\end{abstract}							\noindent\textbf{Keywords:}  
Factor-augmented matrix regression; Weak factors; Response-guided screening; Matrix factor model; Supervised factor learning

\end{titlepage}

\setlength{\belowdisplayskip}{1pt}
\setlength{\belowdisplayshortskip}{1pt}
\setlength{\abovedisplayskip}{1pt}
\setlength{\abovedisplayshortskip}{1pt}
\setlength{\intextsep}{4pt plus 2pt minus 2pt}
\setlength{\textfloatsep}{3pt plus 2pt minus 2pt}
\setlength{\floatsep}{3pt plus 2pt minus 2pt}

\newpage
\section{Introduction}\label{sec:introduction}

Forecasting with a large number of predictors is often made feasible by summarizing the information in the predictors by a small number of latent factors and using the estimated factors in a predictive regression. This diffusion-index, or factor-augmented, approach has become a widely used framework for high-dimensional economic and financial forecasting; see, among others, \citet{StockWatson_JASA_2002,StockWatson_JBES_2002,BaiNg_Econometrica_2006,HuangJiangLiTongZhou_MS_2022,CaiKongWuZhao_JBES_2025,GaoTsay_JASA_2025}. The effectiveness of this forecasting approach, however, depends not only on the existence of a low-dimensional factor structure but also on how accurately the latent factors can be recovered from the observed predictors. Because the forecasting regression is constructed using estimated rather than observed factors, deterioration in factor recovery directly affects the quality of the latent regressors used in prediction. Accurate factor extraction is therefore a critical requirement for the success of factor-augmented regression with high-dimensional predictors.

In many applications, the predictors are naturally matrix-valued. For example, macroeconomic variables may be observed across countries or regions, and  financial variables may be indexed simultaneously by firms and characteristics. To preserve such two-way structure, following \citet{WangLiuChen_JoE_2019}, we consider the matrix factor model
\begin{equation}\label{equa:factor model}
    \bX_t=\bR\bF_t\bC^\top+\bE_t,
\end{equation}
where \(\bX_t\in\mathbb R^{p_1\times p_2}\), \(\bR\in\mathbb R^{p_1\times k_1}\) and \(\bC\in\mathbb R^{p_2\times k_2}\) are row and column loading matrices, \(\bF_t\in\mathbb R^{k_1\times k_2}\) is a latent factor matrix, and \(\bE_t\) is an idiosyncratic error matrix. Its vectorized representation is
\[
\vect(\bX_t)=(\bC\otimes\bR)\vect(\bF_t)+\vect(\bE_t),
\]
so the matrix factor model can be viewed as a vector factor model with a Kronecker-structured loading matrix. This matrix formulation retains the row--column structure and reduces the number of loading parameters from \(p_1p_2k_1k_2\) to \(p_1k_1+p_2k_2\).

A key premise behind conventional factor extraction is that the common signal is sufficiently pervasive across the observed predictor array. We instead study a setting in which the common components are  concentrated on unknown subsets of rows and columns 
{of the predictor matrix}. Let
\[
\mathcal I_R
=
\{i\in[p_1]:\bR_{i\cdot}\neq\bzero\},
\qquad
\mathcal I_C
=
\{j\in[p_2]:\bC_{j\cdot}\neq\bzero\},
\]
where $[n] = \{1,2,\ldots,n\}$, and 
$\bR_{i\cdot}^\top$ and $\bC_{j\cdot}^\top$ denote the $i$th and $j$th rows of $\bR$ and \(\bC\), respectively. Suppose that
\(
|\mathcal I_R|\sim p_1^{\delta_1},
\;
|\mathcal I_C|\sim p_2^{\delta_2},
\; \mbox{with}\; 
\delta_1,\delta_2\in(0,1].
\)
Under the usual loading-strength normalization, the conventional pervasive
regime corresponds to $\delta_1=\delta_2=1$. When either exponent is less than one, the common-factor signal is concentrated on an effective
\(p_1^{\delta_1}\times p_2^{\delta_2}\) submatrix, while conventional factor extraction continues to use the entire \(p_1\times p_2\) array. Increasing the ambient dimensions may then add some inactive coordinates without a proportional increase in common-factor signal.

Two-way weak loadings affect factor estimation via two distinct but
related channels. First, the loading signal in each target mode is itself weak: the effective row and column signal strengths grow only at the rates \(p_1^{\delta_1}\) and \(p_2^{\delta_2}\), rather than the ambient dimensions \(p_1\) and \(p_2\). Second, the matrix structure couples estimation across the two modes, so that estimation in one loading space inherits weakness from the other. In particular, weak column loadings reduce the
effective signal available for estimating the row loading space, and vice versa. We refer to this second, matrix-specific phenomenon as \emph{cross-mode signal dilution}. Our analysis shows that the combination of target-mode weakness and cross-mode signal dilution can introduce error terms in full-matrix factor estimation
that {may not} vanish under asymptotic regimes compatible with weak loadings. Consequently, increasing an ambient matrix dimension may not improve factor recovery when the additional coordinates carry little or no common-factor signal. Moreover, the resulting loss of effective signal can weaken the eigenvalue separation required for {the determination of the numbers of row and column common factors.}

For factor-augmented {matrix} regression, these estimation difficulties arise upstream of the regression stage. This observation suggests the need to screen the predictor array before factor extraction. Importantly, the role of screening here is not merely to discard predictors with little {predictive power}, but to remove inactive dimensions that dilute the common signal and thereby improve estimation of the latent factor structure. This motivates the key question studied in this paper: \emph{can information in the response be used to locate and remove inactive matrix dimensions before factor extraction?}

To address this question, we augment model \eqref{equa:factor model} with the predictive regression
\begin{equation}\label{equa:factor-augmented-regression}
    y_{t+h}
=\balpha^\top\bF_t\bbeta+\epsilon_{t+h},
\end{equation}
where the forecast horizon $h\geq 1$ is fixed. Equations (\ref{equa:factor model}) and (\ref{equa:factor-augmented-regression}) define jointly a factor-augmented matrix regression model. 
{The key observation of our study is that the response 
should contain information about the active loading supports which could be helpful in estimating the latent factors.} Under the predictive-signal conditions developed below, appropriately
aggregated population associations between $y_{t+h}$ and the entries of
$\bX_t$ separate active rows and columns from their inactive counterparts.
The response can therefore be used to identify where the common factor
signal relevant to the response is concentrated. 

A distinctive feature of the matrix-valued setting is that the screening should exploit the two-way structure. Consider, for example, a row-first procedure. The initial row statistics must be constructed from the full matrix and therefore face dilution from inactive columns. Once informative rows have been retained, however, the column statistics are recomputed using only the selected rows, so that one source of signal dilution has already been reduced. The column-first procedure is defined symmetrically. This sequential structure is important because screening one mode changes the signal-to-noise {ratio in recovering the other mode.}

More importantly, support recovery changes the statistical regime for the subsequent factor re-estimation. Once \(\mathcal I_R\) and \(\mathcal I_C\) are recovered, the row and column dimensions of the retained submatrix are of order $p_1^{\delta_1}$ and $p_2^{\delta_2}$, respectively, and the loading matrices are pervasive relative to these effective dimensions. Response-guided screening therefore transforms a weak-loading problem of an ambient matrix into a pervasive-loading problem of the informative submatrix. This transformation mitigates the dilution of the common signal by inactive rows and columns, providing a more favorable setting for both factor extraction and factor-number determination.

Our analysis is related to several strands of the factor-model literature. A growing literature develops factor methods that explicitly exploit matrix and multilinear structures. For matrix-valued time series, \citet{WangLiuChen_JoE_2019} use auto-cross-covariance structures to estimate loading spaces, \citet{ChenTsayChen_JASA_2020} incorporate structural restrictions through constrained matrix-factor models, \citet{ChenFan_JASA_2023} propose the \(\alpha\)-PCA estimator, and \citet{YuHeKongZhang_JoE_2022} develop a projected estimator that uses the estimated loading space in one direction to improve estimation in the other. Related work extends multilinear factor modeling to tensor-valued time series; see, among others, \citet{ChenYangZhang_JASA_2022,HanChenZhang_EJS_2022,ChenLam_AoS_2024,HanChenYangZhang_AoS_2024}. More recently, \citet{CaiKongWuZhao_JBES_2025} and \citet{ChenFanZhu_JASA_2026} study
factor-augmented regression with matrix-valued predictors. 

Another strand of the factor-model literature examines
weak factors and their implications for estimation. In vector factor models,
\citet{LamYaoBathia_Biometrika_2011,lamyao2012,
Onatski_JoE_2012,BaiNg_JoE_2023}
study weak or heterogeneous factor strengths, while
\citet{UematsuYamagata_JBES_2023_Estimation}
investigate sparsity-induced weak factors and loading selection. More recently, \citet{ChenLam_arXiv_2024}
consider factor-strength estimation in vector and matrix
time series. Our setting instead focuses on the rotation-invariant
row and column loading supports of a matrix factor model, which we recover using response-guided screening.

The response-guided nature of our procedure connects the paper to supervised and targeted dimension reduction. \citet{BOIVINNG_JOE_2006} show that enlarging the information set does not necessarily improve factor-based forecasts. \citet{BairHastiePaulTibshirani_JASA_2006} propose supervised principal components, in which predictors are screened according to their association with the response before principal-component extraction, while \citet{BaiNg_JoE_2008} estimate factors from predictors targeted to a particular forecasting variable. Related procedures use predictive information to screen, weight, or transform the predictor space; see, for example, \citet{FanXueYao_JoE_2017,HuangJiangLiTongZhou_MS_2022,GaoTsay_JASA_2025}. These methods establish the broader principle that predictive information can be useful in constructing latent predictors.

Our screening step is also related to the sure independence screening literature initiated by \citet{FanLv_JRSSB_2008} \citep[see also][]{%
FanSamworthWu_JMLR_2009,
FanSong_AoS_2010,
FanFengSong_JASA_2011}. The statistical role of screening is different here. Conventional sure screening uses marginal response--predictor associations to reduce an ultrahigh-dimensional predictor set before further variable selection. In our framework, the screening step is not itself the final dimension-reduction step. Instead, marginal predictive information is used to recover the support of a latent loading structure and thereby improve estimation of the low-dimensional predictive state \(\bF_t\). The subsequent factor extraction, rather than the selected coordinates themselves, provides the latent regressors entering the factor-augmented regression. 

The contributions of this paper are threefold. First, we derive convergence rates for $\alpha$-PCA and projected estimation under two-way weak loadings, quantifying the effects of target-mode weakness and \emph{cross-mode signal dilution}. Second, we develop sequential response-guided screening and establish exact recovery of the informative row and column supports. Building on this recovery result, we show that post-screening projected estimation eliminates the cross-mode signal dilution and extends factor estimation consistency to weaker loading regimes. We also establish consistency of factor-number determination on the screened matrix. Third, we establish asymptotic distributions for factor-augmented regression based on the estimated factors, covering both the unrestricted estimator of $\btheta=\bbeta\otimes\balpha$ and the identified bilinear coefficient estimators $(\widehat{\balpha},\widehat{\bbeta})$.
{Finally, it is worth mentioning that 
the proposed analysis can be extended to supervised factor-augmented regression with higher-order tensor-valued predictors.}

The remainder of the paper is organized as follows. Section~\ref{sec:methodology} presents the estimation and screening procedures, and Section~\ref{sec:theory} develops their asymptotic theory. Sections~\ref{sec:simulation} and~\ref{sec:empirical studies} report simulations and empirical applications, respectively. Section~\ref{sec:conclusion} concludes.

Throughout the paper, for $\bx\in\mathbb R^p$, let $\|\bx\|=(\sum_{i=1}^p x_i^2)^{1/2}$.
For a matrix $\bA=(a_{ij})$, let
$\|\bA\|_{\max}=\max_{i,j}|a_{ij}|$,
$\|\bA\|_1=\max_j\sum_i|a_{ij}|$,
$\|\bA\|_F=(\sum_{i,j}a_{ij}^2)^{1/2}$, and
$\|\bA\|=\sqrt{\lambda_{\max}(\bA^\top\bA)}$
denote the elementwise maximum, matrix $\ell_1$,
Frobenius, and operator norms, respectively.
Here, $\lambda_{\max}(\cdot)$ and $\lambda_{\min}(\cdot)$
denote the largest and smallest eigenvalues.
For matrices $\bA$ and $\bB$ of the same dimensions,
$\langle\bA,\bB\rangle=\Tr(\bA^\top\bB)$ denotes their Frobenius inner product.
For index sets $I$ and $J$, $\bA(I,J)$ denotes the
submatrix with rows in $I$ and columns in $J$,
with $\bA(I,:)$ and $\bA(:,J)$ selecting all columns and all rows, respectively.
For random sequences $X_n,Y_n$,
$X_n\lesssim Y_n$ means $X_n=O_p(Y_n)$,
and $X_n\gtrsim Y_n$ means $Y_n=O_p(X_n)$.
For deterministic sequences $a_n,b_n$,
$a_n\sim b_n$ means $a_n/b_n\to1$, whereas
$a_n\asymp b_n$ means $a_n=O(b_n)$ and
$b_n=O(a_n)$.

\section{Methodology}\label{sec:methodology}
This section develops a complete screen–extract-regress procedure.
We begin with full-matrix factor estimation under two-way weak loadings and then introduce sequential response-guided screening.
We next describe post-screening re-estimation and factor-number determination. Finally, we present factor-augmented regression using the resulting factor estimates. 
Throughout the paper, $\bX_t$ and $y_{t+h}$ denote the standardized predictor matrix and response, respectively, and are assumed to satisfy
the factor-augmented matrix regression model.

\subsection{Factor Estimation under Two-Way Weak Loadings}
Throughout this subsection, let \(T,p_1,p_2\to\infty\) with fixed
\((k_1,k_2)\), and suppose that the process is centered so that
\(
\E\bX_t=0,\;
\E\bF_t=0,\;
\E\bE_t=0.
\)
Under the weak-loading identification,
\(
\frac{\bR^\top\bR}{p_1^{\delta_1}}
\to\bI_{k_1},
\;
\frac{\bC^\top\bC}{p_2^{\delta_2}}
\to\bI_{k_2}.
\)
Accordingly, estimated loading bases are represented on the corresponding effective-dimension scales. If
\(\widehat\bU_R\) and \(\widehat\bU_C\) denote orthonormal bases for
estimated row and column loading spaces, we use
\(
\widehat\bR
=
\sqrt{p_1^{\delta_1}}\,\widehat\bU_R,
\;
\widehat\bC
=
\sqrt{p_2^{\delta_2}}\,\widehat\bU_C.
\)
This scaling is an identification convention and does not affect the
estimated loading spaces.

\begin{remark}[Normalization]
The same estimated loading spaces could instead be represented using
the conventional ambient-dimension scales
\(
\widehat\bR^{\mathrm{amb}}
=
\sqrt{p_1}\,\widehat\bU_R,
\;
\widehat\bC^{\mathrm{amb}}
=
\sqrt{p_2}\,\widehat\bU_C.
\)
This alternative changes only the allocation of scale between the
loading matrices and the corresponding factor representative. It does
not alter the estimated loading spaces or the fitted common component.
In particular, under consistent loading-space estimation, the
corresponding factor representative satisfies, up to rotations,
\(
\widehat\bF_t^{\mathrm{amb}}
\approx
p_1^{(\delta_1-1)/2}
p_2^{(\delta_2-1)/2}
\bF_t.
\) Thus, moving between the ambient-dimension
and effective-dimension normalizations only requires a corresponding
rescaling of the factor estimates. Importantly, knowledge of $\delta_1$ and $\delta_2$ is not required to
estimate the loading spaces themselves. These exponents enter only
through the scale used to represent the loading matrices and factors.
After screening, the selected support sizes
$|\widehat{\mathcal I}_R|$ and
$|\widehat{\mathcal I}_C|$ provide direct data-based counterparts of the effective dimensions.
\end{remark}

\subsubsection{\texorpdfstring{$\alpha$}{alpha}--PCA Estimation}
\label{sec:alphaPCA estimation}

Following \citet{ChenFan_JASA_2023}, we use the \(\alpha\)-PCA
estimator with \(\alpha=0\). Since the process is centered, loading-space estimation is based on contemporary row- and column-side sample covariance matrices. For theoretical convenience and to align
their scale with the weak-loading identification, define
\begin{equation}\label{def:alpha-PCA-statistics}
\widehat \bM_R
=
\frac{1}{T p_1^{\delta_1}p_2^{\delta_2}}
\sum_{t=1}^T \bX_t\bX_t^\top,
\qquad
\widehat \bM_C
=
\frac{1}{T p_1^{\delta_1}p_2^{\delta_2}}
\sum_{t=1}^T \bX_t^\top \bX_t.
\end{equation}
Let \(\widehat\bU_R\) and \(\widehat\bU_C\) contain the leading \(k_1\) and \(k_2\) eigenvectors of \(\widehat\bM_R\) and \(\widehat\bM_C\), respectively, and define \(\widehat\bR\) and \(\widehat\bC\) as above. The corresponding factor estimator is
\begin{equation}\label{eq:alpha-pca-factor-estimator}
\widehat\bF_t
=
\frac{1}{p_1^{\delta_1}p_2^{\delta_2}}
\widehat\bR^\top \bX_t \widehat\bC,
\qquad t=1,\ldots,T.
\end{equation}

A short decomposition clarifies why the full-matrix estimator can be sensitive to weak loadings. Substituting \(\bX_t=\bR\bF_t\bC^\top+\bE_t\) into the row aggregation matrix gives
\[
\widehat\bM_R
=
\underbrace{\frac{1}{T p_1^{\delta_1}p_2^{\delta_2}}
\sum_{t=1}^T
\bR\bF_t\bC^\top\bC\bF_t^\top\bR^\top}_{\text{signal component}}
+
\underbrace{\frac{1}{T p_1^{\delta_1}p_2^{\delta_2}}
\sum_{t=1}^T\bE_t\bE_t^\top}_{\text{idiosyncratic component}}
+\text{cross terms}.
\]
The weak-loading normalization keeps the leading signal eigenvalues on a constant scale, but the idiosyncratic and cross terms continue to aggregate over the ambient matrix dimensions while being normalized by the smaller effective dimensions \(p_1^{\delta_1}\) and \(p_2^{\delta_2}\). They therefore may not be negligible relative to the signal when the effective loading strength in either mode is sufficiently weak. Section~\ref{sec:theory} makes these effects precise through the resulting loading and factor convergence rates.

\subsubsection{Projected Estimation}\label{sec:projected-estimation-in-main}

The preceding analysis raises a natural question: whether projection
can mitigate the weak-loading difficulties of direct covariance-based
factor extraction. The projected estimator of
\citet{YuHeKongZhang_JoE_2022} uses an estimated loading space in one
mode to project the data before estimating the loading space in the
other. We use the following calculation to show its potential benefit.

Suppose temporarily that \(\bC\) were known and normalized so that
\(
\frac{\bC^\top\bC}{p_2^{\delta_2}}
=
\bI_{k_2}.
\)
Then the row-side projected observation is
\(
\bY_t^{(\mathrm{row})}
=
\frac{1}{p_2^{\delta_2}}
\bX_t\bC
=
\bR\bF_t
+
\frac{1}{p_2^{\delta_2}}
\bE_t\bC.
\)
The original \(p_1\times p_2\) observation is therefore compressed to
a \(p_1\times k_2\) projected observation, while the projected
idiosyncratic component is a loading-weighted aggregation of the
original errors along the column mode. Under weak cross-sectional
dependence, this projection can substantially reduce the idiosyncratic
variation entering estimation of the row loading space. The corresponding column-side calculation is symmetric.

In practice, however, the loading spaces are unknown. Let
\(\check\bR\) and \(\check\bC\) denote initial loading estimators
obtained by \(\alpha\)-PCA and define
\(
\bY_t^{(\mathrm{row})}
=
\frac{1}{p_2^{\delta_2}}
\bX_t\check\bC,
\;
\bY_t^{(\mathrm{col})}
=
\frac{1}{p_1^{\delta_1}}
\bX_t^\top\check\bR.
\)
The corresponding projected covariance matrices are
\[
\widetilde\bM_R
=
\frac{1}{T p_1^{\delta_1}}
\sum_{t=1}^T
\bY_t^{(\mathrm{row})}
\{\bY_t^{(\mathrm{row})}\}^\top,
\qquad
\widetilde\bM_C
=
\frac{1}{T p_2^{\delta_2}}
\sum_{t=1}^T
\bY_t^{(\mathrm{col})}
\{\bY_t^{(\mathrm{col})}\}^\top.
\]

Let \(\widetilde\bU_R\) and \(\widetilde\bU_C\) contain their leading
\(k_1\) and \(k_2\) eigenvectors, respectively, and define
\(
\widetilde\bR
=
\sqrt{p_1^{\delta_1}}\,\widetilde\bU_R,
\;
\widetilde\bC
=
\sqrt{p_2^{\delta_2}}\,\widetilde\bU_C.
\)
The corresponding factor estimator is
\begin{equation}\label{eq:projectd-factor-estimator}
\widetilde\bF_t
=
\frac{1}{p_1^{\delta_1}p_2^{\delta_2}}
\widetilde\bR^\top
\bX_t
\widetilde\bC,
\qquad
t=1,\ldots,T.
\end{equation}

The calculation shows that projection can substantially reduce
the idiosyncratic variation entering loading-space estimation. Under
two-way weak loadings, however, two difficulties remain: \textup{(i)} Even
with an accurately estimated projection direction, the second-stage
problem retains the weak loading strength of the target mode. \textup{(ii)} The feasible projection directions \(\check\bR\) and \(\check\bC\) are themselves estimated from the full matrix. They therefore inherit the weak-loading errors of the full-matrix initial estimators. Projection consequently
attenuates idiosyncratic noise without generally eliminating the effects of two-way weak loadings.

These observations suggest changing the estimation domain rather than only
modifying the factor estimator. If the informative rows and columns can
first be located, factor extraction can instead be performed on a
submatrix whose dimensions are aligned with the effective signal
dimensions. This motivates the response-guided screening procedure
developed next.

\subsection{Sequential Response-Guided Screening}\label{subsec:screening}
To identify the unknown row and column supports $\mathcal I_R$ and $\mathcal I_C$, we exploit the predictive associations between $y_{t+h}$
and the entries of $\bX_t$. For each matrix entry, define the absolute sample correlation based on the normalized data as
\(
\widehat s_{ij}
=\left|
\frac{1}{T}\sum_{t=1}^T y_{t+h}X_{t,ij}
\right|
\).
Under the conditions in Section~\ref{sec:theory}, the idiosyncratic and regression-error terms do not contribute to the population association, so the predictive signal is generated by
\(
\E\bracket{y_{t+h}\paren{\bR_{i\cdot}'\bF_t\bC_{j\cdot}}}.
\)
Hence inactive rows or columns have zero common-component contribution, whereas the informative block is separated from zero under Assumption~\ref{ass:signal}. This makes the response informative about the unknown loading supports even though the factors themselves remain latent.

The key difficulty is that support recovery is two-way. If rows are screened on the full matrix, the natural row statistic is
\(
\widehat{s}_i^{(R)}
=
\frac{1}{p_2}\sum_{j=1}^{p_2}\widehat{s}_{ij}.
\)
Only \(|\mathcal I_C|\sim p_2^{\delta_2}\) columns contribute nonzero common-factor signal. Under the signal condition in Section~\ref{sec:theory}, an informative row has signal at least of order \(p_2^{\delta_2-1}\). Specifically,
\(
\widehat{s}_i^{(R)}
\gtrsim\,p_2^{\delta_2-1},
\; i\in\mathcal I_R,
\)
whereas an inactive row has zero population signal. Symmetrically, full-matrix column screening has population signal of order \(p_1^{\delta_1-1}\). Screening both modes simultaneously would therefore require both diluted signals to dominate sampling fluctuations.

To avoid these simultaneous requirements, we screen the two modes sequentially. For the row-first procedure:
\begin{enumerate}
 
    \item Compute \(\widehat s_{ij}\) for all \((i,j)\in[p_1]\times[p_2]\).
  
    \item Form the row statistics
    \(
    \widehat s_i^{(R)},
    \; i=1,\ldots,p_1,
    \)
    and select
    \(
    \widehat{\mathcal I}_R
    =
    \{i\in[p_1]:\widehat s_i^{(R)}>\tau_R\}.
    \)

    \item Recompute the column statistics using only the selected rows,
    \(
    \widehat s_j^{(C)}
    =
    \frac{1}{|\widehat{\mathcal I}_R|}
    \sum_{i\in\widehat{\mathcal I}_R}\widehat s_{ij},
    \; j=1,\ldots,p_2,
    \)
    and select
    \(
    \widehat{\mathcal I}_C
    =
    \{j\in[p_2]:\widehat s_j^{(C)}>\tau_C\}.
    \)

    \item Form the screened matrices
    \(
    \bX_t^{\mathrm{scr}}
    =
    \bX_t(\widehat{\mathcal I}_R,\widehat{\mathcal I}_C),
    \; t=1,\ldots,T.
    \)
\end{enumerate}

The benefit of sequential screening arises in the second stage. If the row-support is correctly recovered, the column statistic becomes
\(
\widehat{s}_{j\mid\mathcal I_R}^{(C)}
=
\frac{1}{|\mathcal I_R|}\sum_{i\in\mathcal I_R}\widehat{s}_{ij}.
\)
For \(j\in\mathcal I_C\), this quantity is bounded away from zero under the aggregate signal condition discussed in Section~\ref{sec:theory}. It no longer carries the factor \(p_1^{\delta_1-1}\) generated by averaging over inactive rows. Thus, row-first screening
requires full-matrix separation only in the first stage. The column-first procedure follows symmetrically. The preferable first stage is therefore the one with the more favorable full-matrix separation. Section~\ref{sec:theory} formalizes this comparison and gives the corresponding threshold requirements. In the numerical studies, we use theoretically motivated threshold orders in the simulations and select the thresholds by cross-validation in the empirical analysis. The Supplementary Material further develops a data-driven iterative screening procedure that exploits the
two-way structure, together with additional simulation evidence.

The proposed procedure uses average-type screening statistics, while alternative aggregation rules, such as maximum-type aggregation, are discussed in the Supplementary Material.

\paragraph{Spearman rank-based robust screening.}
The covariance-based entrywise score may be sensitive to heavy-tailed
observations and isolated outliers. As a robust alternative, we replace
the covariance-based score by the absolute sample Spearman correlation
between \(X_{t,ij}\) and \(y_{t+h}\):
\(
\widehat s_{ij}^{\,\mathrm{rank}}
=
\left|
\widehat\rho^{}
\bigl(X_{\cdot,ij},y_{\cdot+h}\bigr)
\right|.
\)
Spearman's correlation depends only on the ranks of the observations
and is therefore less sensitive to extreme magnitudes. The rank-based
procedure is obtained by using
\(\widehat s_{ij}^{\,\mathrm{rank}}\) in place of
\(\widehat s_{ij}\) throughout the sequential screening procedure.
All cross-sectional aggregation and sequential row--column updating
steps remain unchanged. The corresponding support-recovery result is
given in Section~\ref{sec:theory}.
\begin{remark}
A median-type cross-sectional aggregation may also provide robustness to
extreme entrywise screening scores. However, under the weak-loading regime
considered here, the informative entries in the opposite mode may comprise
less than one half of all coordinates. In that case, the population median
can be determined entirely by inactive entries and may therefore fail to
separate informative rows or columns. For this reason, we retain the
average aggregation and robustify the entrywise association through ranks.
\end{remark}

\subsection{Post-Screening Re-estimation and Factor-Number Determination}
Write
\(
\widehat p_1=|\widehat{\mathcal I}_R|,
\;
\widehat p_2=|\widehat{\mathcal I}_C|.
\)
On the event of exact support recovery,
\(
\widehat p_1=|\mathcal I_R|\sim p_1^{\delta_1},
\;
\widehat p_2=|\mathcal I_C|\sim p_2^{\delta_2},
\)
and the retained loading matrices satisfy
\[
\frac{1}{\widehat p_1}
\bR(\widehat{\mathcal I}_R,:)^\top
\bR(\widehat{\mathcal I}_R,:)
\to\bI_{k_1},
\qquad
\frac{1}{\widehat p_2}
\bC(\widehat{\mathcal I}_C,:)^\top
\bC(\widehat{\mathcal I}_C,:)
\to\bI_{k_2}.
\]

For fixed factor numbers \((k_1,k_2)\), we apply projected estimation to \(\{\bX_t^{\mathrm{scr}}\}_{t=1}^T\), replacing \(p_1^{\delta_1}\) and \(p_2^{\delta_2}\) by \(\widehat p_1\) and \(\widehat p_2\). Let \(\widetilde\bU_R^{\mathrm{scr}}\) and \(\widetilde\bU_C^{\mathrm{scr}}\) denote the resulting orthonormal loading-space estimators. We use
\(
\widetilde\bR^{\mathrm{scr}}
=
\sqrt{\widehat p_1}\,\widetilde\bU_R^{\mathrm{scr}},
\;
\widetilde\bC^{\mathrm{scr}}
=
\sqrt{\widehat p_2}\,\widetilde\bU_C^{\mathrm{scr}},
\)
and define
\[
\widetilde\bF_t^{\mathrm{scr}}
=
\frac{1}{\widehat p_1\widehat p_2}
(\widetilde\bR^{\mathrm{scr}})^\top
\bX_t^{\mathrm{scr}}
\widetilde\bC^{\mathrm{scr}},
\qquad t=1,\ldots,T.
\]

When the factor numbers are unknown, we estimate them from the screened data using the iterative eigenvalue-ratio procedure of \citet{YuHeKongZhang_JoE_2022}. The procedure iteratively updates the row and column factor numbers using ridge-adjusted eigenvalue ratios of the projected covariance matrices constructed from the screened data. The iteration alternates between the two modes and terminates when both estimated factor numbers stabilize. The complete algorithm is provided in the Supplementary Material.

After \((\widehat k_1,\widehat k_2)\) is obtained, the post-screening loading spaces and factors are re-estimated using these dimensions. The consistency of the resulting factor-number estimates is established jointly with the post-screening estimation theory in Section~\ref{sec:theory}. In contrast, as discussed in the Supplementary Material, the same eigenvalue-ratio separation {may} not be preserved on the original full matrix under sufficiently weak loadings.

\subsection{Factor-Augmented Matrix Regression}
\label{sec:matrix-factor-augmented-regression-methodology}

Let \(\balpha_0\) and \(\bbeta_0\) denote the true bilinear coefficients in \eqref{equa:factor-augmented-regression}. Because \(\bF_t\) is latent, the feasible unrestricted factor-augmented regression uses
\(\vect(\widetilde\bF_t^{\mathrm{scr}})\) as the regressor. The
ordinary least-squares estimator is
\begin{equation}
\label{eq:unrestricted-ols}
\widehat{\btheta}
=
\left[
\frac{1}{T}
\sum_{t=1}^{T}
\vect\left(\widetilde{\bF}_t^{\mathrm{scr}}\right)
\vect\left(\widetilde{\bF}_t^{\mathrm{scr}}\right)^\top
\right]^{-1}
\left[
\frac{1}{T}
\sum_{t=1}^{T}
\vect\left(\widetilde{\bF}_t^{\mathrm{scr}}\right)y_{t+h}
\right].
\end{equation}
Since matrix factors are identified up to nonsingular rotations,
\(\widehat{\btheta}\) estimates a rotation-adjusted version of
\(\btheta_0=\bbeta_0\otimes\balpha_0\).
Section~\ref{sec:theory} states the corresponding target explicitly
and derives its limiting distribution.

To retain the bilinear structure of the predictive coefficient, we also
estimate \((\balpha,\bbeta)\) from
\[
\min_{\balpha,\bbeta}
\sum_{t=1}^{T}
\left(
y_{t+h}
-
\balpha^\top
\widetilde\bF_t^{\mathrm{scr}}
\bbeta
\right)^2,
\]
subject to the scale and sign normalization described below. Following
\citet{ChenXiaoYang_JoE_2021} and
\citet{CaiKongWuZhao_JBES_2025}, we solve this problem by alternating
least squares. Given \(\widehat\bbeta^{(\ell)}\), update
\begin{equation}
\label{eq:update-alpha}
\widehat{\balpha}^{(\ell+1)}
=
\left[
\sum_{t=1}^{T}
\widetilde{\bF}_t^{\mathrm{scr}}
\widehat{\bbeta}^{(\ell)}
\widehat{\bbeta}^{(\ell)\top}
\left(
\widetilde{\bF}_t^{\mathrm{scr}}
\right)^\top
\right]^{-1}
\sum_{t=1}^{T}
y_{t+h}
\widetilde{\bF}_t^{\mathrm{scr}}
\widehat{\bbeta}^{(\ell)},
\end{equation}
and then update
\begin{equation}
\label{eq:update-beta}
\widehat{\bbeta}^{(\ell+1)}
=
\left[
\sum_{t=1}^{T}
\left(
\widetilde{\bF}_t^{\mathrm{scr}}
\right)^\top
\widehat{\balpha}^{(\ell+1)}
\widehat{\balpha}^{(\ell+1)\top}
\widetilde{\bF}_t^{\mathrm{scr}}
\right]^{-1}
\sum_{t=1}^{T}
y_{t+h}
\left(
\widetilde{\bF}_t^{\mathrm{scr}}
\right)^\top
\widehat{\balpha}^{(\ell+1)}.
\end{equation}
The iterations continue until a prespecified convergence criterion is
met.

Let \((\bar\balpha,\bar\bbeta)\) denote the converged pair. To resolve
the scale and sign indeterminacy, define
\(
c_\alpha
=
\operatorname{sign}(\bar\alpha_1)
\|\bar\balpha\|,
\;
(\widehat\balpha,\widehat\bbeta)
=
\left(
\frac{\bar\balpha}{c_\alpha},
c_\alpha\bar\bbeta
\right),
\)
where \(c_{\alpha}\) is used for sign normalization. Under this
normalization,
\(
\|\widehat\balpha\|=1,
\;
\widehat\alpha_1>0.
\)
Section~\ref{sec:theory} gives the corresponding identification
conditions and establishes the joint limiting distribution of
\((\widehat\balpha,\widehat\bbeta)\).

\section{Theoretical Results}\label{sec:theory}
The theoretical analysis follows the statistical mechanism underlying the {proposed} procedure in Section~\ref{sec:methodology}. Throughout this section, $T,p_1,p_2\to\infty$
with fixed factor numbers $(k_1,k_2)$.
We begin with the assumptions used in the
subsequent analysis.

\begin{assumption}[Weak-loading structure]
\label{ass:weak loading}
For some $\delta_1,\delta_2\in(0,1]$, assume that
\[
\frac{1}{p_1^{\delta_1}}\bR^\top\bR\to\bI_{k_1},
\;
\frac{1}{p_2^{\delta_2}}\bC^\top\bC\to\bI_{k_2},
\]
and
\(
\max_{i\le p_1}\|\bR_{i\cdot}\|=O(1),
\;
\max_{j\le p_2}\|\bC_{j\cdot}\|=O(1).
\)
In addition, the informative row and column sets $\mathcal I_R$ and $\mathcal I_C$, defined in
Section~\ref{sec:introduction}, satisfy
\(
|\mathcal I_R|\sim p_1^{\delta_1},
\;
|\mathcal I_C|\sim p_2^{\delta_2}.
\)
\end{assumption}

\begin{remark}
Assumption~\ref{ass:weak loading} extends the weak-loading framework of
\citet{BaiNg_JoE_2023} to the matrix setting while specializing the
source of weakness to support sparsity along both modes. The normalized Gram-matrix conditions and uniformly bounded loading-row norms do not themselves require sparsity. Indeed, the convergence results in Section~\ref{sec:Estimation under Two-Way Weak Loadings} for full-matrix $\alpha$-PCA and projected estimation can be established under these
loading conditions, without imposing the support-size restrictions. Nevertheless, we retain these restrictions throughout the paper to simplify the proofs and provide a unified framework for subsequent response-guided screening
and post-screening estimation.
\end{remark}
\begin{assumption}[\(\alpha\)-mixing and noise process]
\label{ass:mixing-noise}
Let
\(
\bz_t=
\left(
\vect(\bF_t)^\top,
\vect(\bE_t)^\top
\right)^\top.
\)
The process \(\{\bz_t:t\in\mathbb Z\}\) is strictly stationary and
\(\alpha\)-mixing with mixing coefficients \(\alpha(k)\) satisfying
\[
\sum_{k=1}^{\infty}\alpha(k)^{1-2/\gamma}<\infty
\]
for some \(\gamma>2\). Moreover, for some constant \(C>0\),
\(
\E e_{t,ij}=0,
\) \(
\E |e_{t,ij}|^{2\gamma}\le C,
\) and \(\E e_{t,ij}f_{t,lh}=0\)
uniformly over all admissible \(t,i,j,l,h\).
\end{assumption}
\begin{assumption}[Factor process]
\label{ass:factor}
The factor process \(\{\bF_t:t\in\mathbb Z\}\) is strictly stationary and
satisfies
\(
\E\bF_t=\bzero,
\;
\E\|\bF_t\|^4\le C.
\)
In addition,
\(
\frac{1}{T}\sum_{t=1}^T
\bF_t\bF_t^\top\xrightarrow[]{p}\bSigma_1
\)
and
\(
\frac{1}{T}\sum_{t=1}^T
\bF_t^\top\bF_t\xrightarrow{p}\bSigma_2,
\)
where, for $i=1,2$, $\bSigma_i$ is a $k_i\times k_i$ positive definite matrix with distinct eigenvalues and spectral decomposition $\bSigma_i=\bGamma_i\bLambda_i\bGamma_i^\top$.
\end{assumption}
\begin{assumption}[Idiosyncratic error dependence]
\label{ass:noise-dependence}
There exists a constant \(C>0\) such that the following conditions hold.
\begin{enumerate} 

\item\label{Ee_te_s}  For any \(t\in[T]\), \(i\in[p_1]\), and \(j\in[p_2]\),
\(
\sum_{s=1}^{T}\sum_{\ell=1}^{p_1}\sum_{h=1}^{p_2}
\left|
\E(e_{t,ij}e_{s,\ell h})
\right|
\le C.
\)

\item For any \(i,\ell\in[p_1]\), \(t\in[T]\), and \(j\in[p_2]\),
\[
\sum_{m=1}^{p_1}\sum_{s=1}^{T}\sum_{h=1}^{p_2}
\left|
\Cov\left(
e_{t,ij}e_{t,\ell j},
e_{s,ih}e_{s,mh}
\right)
\right|
\le C,
\]
and, symmetrically, for any \(j,h\in[p_2]\), \(t\in[T]\), and \(i\in[p_1]\),
\[
\sum_{m=1}^{p_2}\sum_{s=1}^{T}\sum_{\ell=1}^{p_1}
\left|
\Cov\left(
e_{t,ij}e_{t,ih},
e_{s,\ell j}e_{s,\ell m}
\right)
\right|
\le C.
\]
\end{enumerate}
\end{assumption}
Assumptions~\ref{ass:mixing-noise} and
\ref{ass:noise-dependence} impose standard moment and weak-dependence
conditions on the factor--error process and the idiosyncratic component.
Related conditions are common in vector and matrix factor models
\citep{
Bai_Econometrica_2003,
YuHeKongZhang_JoE_2022,
ChenFan_JASA_2023,
CaiKongWuZhao_JBES_2025}.
Assumption~\ref{ass:factor} imposes the usual factor moment and
eigenvalue conditions, with distinct eigenvalues used to identify the
corresponding eigenvectors.

\subsection{Estimation under Two-Way Weak Loadings}\label{sec:Estimation under Two-Way Weak Loadings}

We first establish convergence rates for
full-matrix $\alpha$-PCA estimation. For $\alpha$-PCA, we impose the following additional conditions.

\begin{assumption}[Idiosyncratic spatial dependence for \(\alpha\)-PCA estimation]
\label{ass:alpha_spatial_noise}
There exists a constant \(C>0\) such that the following conditions hold.
\begin{enumerate}

    \item \label{EE_tE_t'} Define
    \(
    \bU_E
    =
    \E\left[
    \frac{1}{p_2T}\sum_{t=1}^T \bE_t\bE_t^\top
    \right],
    \;
    \bV_E
    =
    \E\left[
    \frac{1}{p_1T}\sum_{t=1}^T \bE_t^\top\bE_t
    \right].
    \)
    Then
    \(
    \|\bU_E\|_1\le C,
    \;
    \|\bV_E\|_1\le C.
    \)

    \item For any \(s,t\in[T]\), \(i_1\in[p_1]\), and
    \(j_1,h_1,j_2\in[p_2]\),
    \[
    \sum_{i_2\in[p_1]}\sum_{h_2\in[p_2]}
    \left|
    \Cov\left(
    e_{s,i_1j_1}e_{t,i_1h_1},
    e_{s,i_2j_2}e_{t,i_2h_2}
    \right)
    \right|
    \le C.
    \]
    Analogously, for any \(s,t\in[T]\), \(j_1\in[p_2]\), and
    \(i_1,\ell_1,i_2\in[p_1]\),
    \[
    \sum_{j_2\in[p_2]}\sum_{\ell_2\in[p_1]}
    \left|
    \Cov\left(
    e_{s,i_1j_1}e_{t,\ell_1j_1},
    e_{s,i_2j_2}e_{t,\ell_2j_2}
    \right)
    \right|
    \le C.
    \]
\end{enumerate}
\end{assumption}
\begin{assumption}[Factor-noise interaction for $\alpha$-PCA estimation]
\label{ass:Factor-noise interaction for alpha-PCA}
There exist constants \(\gamma>2\) and \(1<a,b<\infty\), with
\(a^{-1}+b^{-1}=1\), such that, for some constant \(C>0\),
\begin{enumerate}

\item For any $t\in[T]$,
\(
\E\|\bF_t\|^{\gamma a}\le C.
\)

\item For any $t\in[T],\;i\in[p_1]$,
\(
\E\biggl\|
\frac{1}{\sqrt{p_2^{\delta_2}}}
\sum_{j=1}^{p_2}\bC_{j\cdot}^{\top}e_{t,ij}
\biggr\|^{\gamma b}\le C.\)

\item For any $t\in[T],\;j\in[p_2]$,
\(\E\biggl\|
\frac{1}{\sqrt{p_1^{\delta_1}}}
\sum_{i=1}^{p_1}\bR_{i\cdot}^{\top}e_{t,ij}
\biggr\|^{\gamma b}\le C.\)

\item For any $t\in[T]$,
\(\E\biggl\|
\frac{1}{\sqrt{p_1^{\delta_1}p_2^{\delta_2}}}
\sum_{i=1}^{p_1}\sum_{j=1}^{p_2}
\bR_{i\cdot}\bC_{j\cdot}^{\top}e_{t,ij}
\biggr\|_F^{\gamma b}\le C.
\)
\end{enumerate}
\end{assumption}
The first condition of Assumption~\ref{ass:alpha_spatial_noise} is implied by
Assumption~\ref{ass:noise-dependence}, but is stated explicitly in the
form used by the perturbation analysis. The factor--noise interaction
conditions are satisfied under standard benchmark settings, such
as cross-sectionally independent Gaussian idiosyncratic errors with
suitable factor moments. More generally, they allow weak spatial and
temporal dependence provided that the loading-weighted error sums retain
their square-root order.
The following theorem establishes the asymptotic properties of the \(\alpha\)-PCA estimator under weak loadings.
\begin{theorem}
[\(\alpha\)-PCA estimation]
\label{thm:alpha-pca-estimation}
Suppose Assumptions~\ref{ass:weak loading}--\ref{ass:Factor-noise interaction for alpha-PCA} hold.
Assume further that
\begin{equation}
\frac{p_1^{2-2\delta_1}}{p_2^{2\delta_2-1}}=o(1),\;
\frac{p_2^{2-2\delta_2}}{p_1^{2\delta_1-1}}=o(1).\tag{C1}
\label{cond:C1}
\end{equation} The following assertions hold.
\begin{enumerate}
\item[\textup{(I)}] There exist matrices
\(
\widehat{\bH}_R\in\mathbb R^{k_1\times k_1}\) and \(
\widehat{\bH}_C\in\mathbb R^{k_2\times k_2},
\)
such that
\(
\widehat{\bH}_R^\top\widehat{\bH}_R\xrightarrow{p} \bI_{k_1},
\;
\widehat{\bH}_C^\top\widehat{\bH}_C\xrightarrow{p} \bI_{k_2},
\)
and
\[
\frac{1}{p_1^{\delta_1}}
\left\|
\widehat{\bR}-\bR\widehat{\bH}_R
\right\|_F^2
=
O_p\left(
\frac{p_1^{2-2\delta_1}}{Tp_2^{2\delta_2-1}}
+
\frac{p_2^{2-2\delta_2}}{p_1^{2\delta_1-1}}
\right),
\]
\[
\frac{1}{p_2^{\delta_2}}
\left\|
\widehat{\bC}-\bC\widehat{\bH}_C
\right\|_F^2
=
O_p\left(
\frac{p_2^{2-2\delta_2}}{Tp_1^{2\delta_1-1}}
+
\frac{p_1^{2-2\delta_1}}{p_2^{2\delta_2-1}}
\right).
\]
\item[\textup{(II)}] Assume
further that
\begin{equation}
\frac{p_1^{1-\delta_1}p_2^{1-\delta_2}}{T}=o(1).\tag{C2}
\label{cond:C2}
\end{equation}
Then,
\[
\left\|
\widehat{\bF}_t
-
\widehat{\bH}_R^{-1}
\bF_t
\left(\widehat{\bH}_C^{-1}\right)^\top
\right\|
=
O_p\left(
\frac{p_1^{2-2\delta_1}}{p_2^{2\delta_2-1}}
+
\frac{p_2^{2-2\delta_2}}{p_1^{2\delta_1-1}}
+
\sqrt{
\frac{1}{
p_1^{5\delta_1-4}p_2^{5\delta_2-4}
}
}
\right).
\]
\end{enumerate}
\end{theorem}

We next turn to projected estimation. In addition to the common Assumptions~\ref{ass:weak loading}--\ref{ass:noise-dependence}, we impose the following additional conditions.
\begin{assumption}[Idiosyncratic spatial dependence for projected estimation]\label{ass:projected_spatial_noise}
There exists a constant \(C>0\) such that the following conditions hold.
\begin{enumerate}

\item For any $t\in[T],i\in[p_1],j\in[p_2]$,
\(
\sum_{l=1}^{p_1}\sum_{h=1}^{p_2}
\left|
\mathbb{E}e_{t,lj}e_{t,ih}
\right|
\le C.
\)

\item For any $i,l_1\in[p_1],j,h_1\in[p_2],t\in[T],$
\[
\sum_{s=1}^{T}\sum_{l_2=1}^{p_1}\sum_{h_2=1}^{p_2}
\left(
\left|
\cov
\left(
e_{t,ij}e_{t,l_1h_1},
e_{s,ij}e_{s,l_2h_2}
\right)
\right|
+
\left|
\cov
\left(
e_{t,l_1j}e_{t,ih_1},
e_{s,l_2j}e_{s,ih_2}
\right)
\right|
\right)
\le C.
\]
\end{enumerate}
\end{assumption}
\begin{assumption}[Factor-noise interaction for projected estimation]\label{ass:Factor-noise interaction for projected}
There exists a constant $C>0$ such that the following conditions hold.
\begin{enumerate}
\item For any deterministic vectors $\bv$ and $\bw$ satisfying
$\|\bv\|=1$ and $\|\bw\|=1$, 
\(
\mathbb{E}
\left\|
\frac{1}{\sqrt{T}}\sum_{t=1}^{T}
\bF_t (\bv^\top \bE_t \bw)
\right\|^2
\le C.
\)
\item For any $i,l_1\in[p_1]$ and $j,h_1\in[p_2]$,
\(
\sum_{h=1}^{p_2}
\left\|
\mathbb{E}\left(\bar\bzeta_{ij}\otimes \bar\bzeta_{ih}\right)
\right\|_{\max}
\le C,
\;
\sum_{l=1}^{p_1}
\left\|\mathbb{E}\left(\bar\bzeta_{ij}\otimes \bar\bzeta_{lj}\right)
\right\|_{\max}
\le C,
\)
and
\[
\sum_{l=1}^{p_1}\sum_{h_2=1}^{p_2}\left\|
\cov
\left(
\bar\bzeta_{ij}\otimes\bar\bzeta_{ih_1},
\bar\bzeta_{lj}\otimes\bar\bzeta_{lh_2}
\right)
\right\|_{\max}
\le C,
\]
\[
\sum_{l_2=1}^{p_1}\sum_{h=1}^{p_2}\left\|
\cov
\left(
\bar\bzeta_{ij}\otimes\bar\bzeta_{l_1j},
\bar\bzeta_{ih}\otimes\bar\bzeta_{l_2h}
\right)
\right\|_{\max}
\le C,
\]
where
\(
\bar\bzeta_{ij}
=
\vect
\left(
\frac{1}{\sqrt{T}}\sum_{t=1}^{T}\bF_t e_{t,ij}
\right).
\)
\end{enumerate}
\end{assumption}
These conditions are the projected-estimation counterparts of the
dependence and factor--noise controls imposed above. They restrict the
temporal accumulation of the factor--noise terms to the
standard square-root order and control the aggregate cross-sectional
dependence of their time-aggregated counterparts. Settings
with weak cross-sectional dependence, weak temporal dependence, and
independence between $\bF_t$ and $\bE_t$ satisfy these conditions. Related conditions are imposed by  \citet{YuHeKongZhang_JoE_2022}.
\begin{theorem}[Projected estimation]\label{thm:projected-estimation}
Suppose Assumptions~\ref{ass:weak loading}--\ref{ass:noise-dependence} and Assumptions~\ref{ass:projected_spatial_noise}--\ref{ass:Factor-noise interaction for projected} hold, together with Condition~\eqref{cond:C1}. The following assertions hold.
\begin{enumerate}

\item[\textup{(I)}] There exist $\widetilde {\bH}_R\in \mathbb{R}^{k_1\times k_1}$ and $\widetilde{\bH}_C\in\mathbb{R}^{k_2\times k_2}$  satisfying $\widetilde{\bH}_R^{\top}\widetilde{\bH}_R\xrightarrow[]{p}\bI_{k_1},\;\widetilde{\mathbf{H}}_C^\top\widetilde{\mathbf{H}}_C\xrightarrow[]{p}\mathbf{I}_{k_2}$ and 
\[
\begin{aligned}
\frac{1}{p_1^{\delta_1}}\norm{\widetilde{\mathbf{R}}-\mathbf{R}\widetilde{\mathbf{H}}_R}_F^2
&=O_p\left(
\widetilde{w}_1
\right),\;
\frac{1}{p_2^{\delta_2}}\norm{\widetilde{\mathbf{C}}-\mathbf{C}\widetilde{\mathbf{H}}_C}_F^2
=O_p\left(\widetilde{m}_1\right).
\end{aligned}
\]
where
\[
\widetilde{w}_1=
\frac{p_2^{2-2\delta_2}}{T^2p_1^{3\delta_1-1}}
+\frac{p_2^{4-4\delta_2}}{T^3p_1^{4\delta_1-2}}
+\frac{1}{T^2p_2^{4\delta_2-3}p_1^{4\delta_1-3}}+\frac{p_1^{3-3\delta_1}}{Tp_2^{4\delta_2-2}}+
\frac{p_1^{1-\delta_1}}{Tp_2^{\delta_2}}
+\frac{1}{p_2^{2\delta_2}p_1^{3\delta_1-1}}
+\frac{1}{p_2^{8\delta_2-4}p_1^{7\delta_1-5}},
\]
\[
\widetilde{m}_1=\frac{p_1^{2-2\delta_1}}{T^2p_2^{3\delta_2-1}}+\frac{p_1^{4-4\delta_1}}{T^3p_2^{4\delta_2-2}}+\frac{1}{T^2p_1^{4\delta_1-3}p_2^{4\delta_2-3}}+\frac{p_2^{3-3\delta_2}}{Tp_1^{4\delta_1-2}}+\frac{p_2^{1-\delta_2}}{Tp_1^{\delta_1}}+\frac{1}{p_1^{2\delta_1}p_2^{3\delta_2-1}}+\frac{1}{p_1^{8\delta_1-4}p_2^{7\delta_2-5}}.
\]

\item[\textup{(II)}] Assume further that
\begin{equation}
    \frac{\max\{p_1^{\delta_1},p_2^{\delta_2}\}}{T^2}=o(1)\;,\; \frac{p_1^{2-2\delta_1}}{p_2^{3\delta_2-2}}=o(1)\; , \;\frac{p_2^{2-2\delta_2}}{p_1^{3\delta_1-2}}=o(1).\label{cond:c3}\tag{C3}
\end{equation} 
Then, 
\[
\norm{\widetilde{\mathbf{F}}_t-\widetilde{\mathbf{H}}_R^{-1}\mathbf{F}_t\paren{\widetilde{\mathbf{H}}_C^{-1}}^\top}
=O_p\paren{\frac{1}{\sqrt{p_1^{\delta_1}p_2^{\delta_2}}}+\frac{p_1^{1-\delta_1}}{\sqrt{T}p_2^{\delta_2}}+\frac{p_2^{1-\delta_2}}{\sqrt{T}p_1^{\delta_1}}}
\]
\end{enumerate}
\end{theorem}
\begin{remark}
Some remarks on
Theorems~\ref{thm:alpha-pca-estimation}
and~\ref{thm:projected-estimation} are in order.
\begin{enumerate}[label=\textup{(\roman*)}]

\item
Setting $\delta_1=\delta_2=1$ recovers the
corresponding pervasive-loading rates for
$\alpha$-PCA and projected estimation established
by \citet{ChenFan_JASA_2023} and
\citet{YuHeKongZhang_JoE_2022}, respectively.

\item
For row loading estimation, the terms
\(
\frac{p_2^{2-2\delta_2}}{p_1^{2\delta_1-1}}
\)
and
\(
\frac{p_1^{2-2\delta_1}}{Tp_2^{2\delta_2-1}}
\)
in the $\alpha$-PCA bound reflect target-mode weakness and cross-mode signal dilution. In particular, the bound may not be \(o(1)\) when
\(
p_1^{2\delta_1-1}
=O(p_2^{2-2\delta_2}),
\)
with an symmetric conclusion for column loading estimation.
Projection attenuates idiosyncratic noise, although its loading-error bounds still reflect these weak-loading effects.
Under $p_1\asymp p_2\asymp p$ and
$\delta_1=\delta_2=\delta$,
Condition~\eqref{cond:C1} requires $\delta>3/4$.
The loading-error bounds for $\alpha$-PCA
and projected estimation are
\[
O_p\left(p^{3-4\delta}+\frac{p^{3-4\delta}}{T}\right)
\text{and}\;
O_p\left(
\frac{p^{6-8\delta}}{T^2}
+\frac{p^{5-7\delta}+p^{1-2\delta}}{T}
+p^{1-5\delta}+p^{9-15\delta}
\right),
\]
respectively. Projected estimation achieves a faster convergence rate for loading estimation than \(\alpha\)-PCA when \(3/4<\delta\le1\).

\item
For factor estimation, there are still weak-loading effects, including target-mode weakness and cross-mode signal dilution. For example, the $\alpha$-PCA bound contains the term
\(
\frac{1}{
\sqrt{p_1^{5\delta_1-4}p_2^{5\delta_2-4}}
},
\)
reflecting the joint loss of effective signal across the two modes. The second and third terms in the projected factor-error bound may not be \(o(1)\) when
\(
\sqrt{T}p_2^{\delta_2}
=O(p_1^{1-\delta_1})
\)
and
\(
\sqrt{T}p_1^{\delta_1}
=O(p_2^{1-\delta_2}),
\)
respectively. Under the same setting, sufficient conditions for factor consistency are $\delta>4/5$ and $p^{2-2\delta}/T=o(1)$ for $\alpha$-PCA, and $\delta>4/5$ and $p^\delta/T^2=o(1)$ for projected estimation. This common restriction on $\delta$ limits the degree of loading weakness allowed for consistent full-matrix factor estimation. The term $p^\delta/T^2$ for projected estimation is asymptotically smaller than $p^{2-2\delta}/T$ for $\alpha$-PCA when $p^{3\delta-2}/T=o(1)$, indicating less restrictive consistency requirements for projected estimation when $T$ grows sufficiently fast relative to $p^{3\delta-2}$.

The corresponding factor-error bounds are
\(
O_p(p^{4-5\delta})
\)
and
\(
O_p\left(
p^{-\delta}
+\frac{p^{1-2\delta}}{\sqrt{T}}
\right),
\)
respectively.
Under Condition~\eqref{cond:c3}, the latter reduces to $O_p(p^{-\delta})$, which is of smaller order for $4/5<\delta<1$,
while both bounds are of order $O_p(p^{-1})$ when $\delta=1$.

\end{enumerate}
\end{remark}

\subsection{Consistency of Sequential Support Recovery}
To establish exact recovery of the informative row and column supports, we impose the following additional
assumptions.

\begin{assumption}[Weak dependence and tail condition]\label{ass:screening}
For \(i\in[p_1]\) and \(j\in[p_2]\), define
\(
Z_{t,ij}:=
y_{t+h}X_{t,ij}-\E(y_{t+h}X_{t,ij}).
\)
For each \(i\in[p_1]\) and \(j\in[p_2]\), let
\[
\alpha_{ij}(n)
:=
\sup_{p\in\mathbb Z}
\alpha\left(
\sigma(Z_{t,ij}:t\le p),
\sigma(Z_{t,ij}:t\ge p+n)
\right).
\]
There exist constants \(C>0\), \(\gamma_1>0\), and \(\gamma_2>0\), such that
\[
\sup_{i\in[p_1]}\sup_{j\in[p_2]}\alpha_{ij}(n)
\le
\exp(-C n^{\gamma_1}),
\qquad n\ge1,
\]
and
\[
\sup_{t\in\mathbb Z}\sup_{i\in[p_1]}\sup_{j\in[p_2]}
\Pr\left(|Z_{t,ij}|>x\right)
\le
\exp(1-x^{\gamma_2}),
\qquad x>0.
\]
Let \(\gamma_s\) be defined by
\(
\frac1{\gamma_s}
=
\frac1{\gamma_1}
+
\frac1{\gamma_2},
\)
and assume that
\(
\gamma_s<1.
\)
\end{assumption}
\begin{assumption}\label{ass:signal}
There exists a constant \(C>0\), such that
\(
\left|
\E\left[
(\balpha^\top \bF_t\bbeta)
(\bR_{i\cdot}^\top \bF_t \bC_{j\cdot})
\right]
\right|
\ge C,
\)
uniformly over $i\in\mathcal I_R$,
$j\in\mathcal I_C$, and admissible $t$.
\end{assumption}
\begin{assumption}\label{ass:v-bound}
For any \(M>0\), define the truncation function
\(
\varphi_M(x):=(x\wedge M)\vee(-M).
\)
For each \(i\in[p_1]\) and \(j\in[p_2]\), define
\[
V_{ij}
:=
\sup_{M>0}\sup_{s\ge1}
\left\{
\var\left(\varphi_M(Z_{s,ij})\right)
+
2\sum_{u>s}
\left|
\cov\left(
\varphi_M(Z_{s,ij}),
\varphi_M(Z_{u,ij})
\right)
\right|
\right\}.
\]
Assume that
\(
V:=\sup_{i\in[p_1]}\sup_{j\in[p_2]}V_{ij}<\infty.
\)
\end{assumption}
\begin{assumption}\label{ass:epsilon-t}
The sequence \(\{\epsilon_t\}\) is i.i.d., has mean zero, and is
independent of \(\{\bE_s:s\in\mathbb Z\}\). Moreover, \(
\E(\epsilon_{t+h}\mid\mathcal F_t)=0
\)
for any integer \(h>0\), where \(\mathcal F_t\) is the \(\sigma\)-field generated by
\(\{y_s,\bF_s:s\le t\}\).
Furthermore,
\(
\sup_t\E(\epsilon_t^4)\le C.
\)
\end{assumption}
Assumption~\ref{ass:signal} guarantees a nonvanishing informative
signal, while the tail and dependence conditions, together with Assumption~\ref{ass:v-bound}, control the uniform sampling fluctuations. Assumption~\ref{ass:epsilon-t} imposes standard conditional mean-zero and moment conditions on the regression error, as in \citet{CaiKongWuZhao_JBES_2025}
and \citet{GaoTsay_JASA_2025}.
\begin{remark}\label{rem:aggregate-screening-signal}
Assumption~\ref{ass:signal} imposes an entrywise lower bound on the
marginal predictive signal over the informative block. This condition
is stronger than necessary for the average-type screening statistics
considered here, but provides a simple sufficient condition for the
required population separation. More generally, one may consider the weaker aggregate conditions
\(
\inf_{i\in\mathcal I_R}
\frac{1}{|\mathcal I_C|}
\sum_{j\in\mathcal I_C}|\mu_{ij}|
\ge c_R>0,
\;
\inf_{j\in\mathcal I_C}
\frac{1}{|\mathcal I_R|}
\sum_{i\in\mathcal I_R}|\mu_{ij}|
\ge c_C>0.
\)
These conditions allow some informative entries to have zero marginal
predictive signal, provided that the aggregate signal within each
informative row and column remains sufficiently strong. The entrywise
condition in Assumption~\ref{ass:signal} implies these aggregate conditions and is adopted here to streamline the theoretical
development.
\end{remark}
The preceding assumptions provide the population separation and
sample fluctuation control needed for screening consistency. For row-first screening, the first-stage population gap is of order
\(p_2^{\delta_2-1}\), while the uniform sampling fluctuation is of order
\(\sqrt{\log(p_1p_2)/T}\). After exact recovery of the row support, the
second-stage column gap is of constant order, with sampling fluctuation
of order \(\sqrt{\log(p_1^{\delta_1}p_2)/T}\).
The column-first case is symmetric. These comparisons motivate the
following dimensional growth condition.
\begin{equation}
\min\left\{
p_1^{2-2\delta_1},
p_2^{2-2\delta_2}
\right\}
\log(p_1p_2)=o\paren{T},\;
p_1p_2
\le
\exp\left(
C T^{\frac{\gamma_s}{2-\gamma_s}}
\right)\label{cond:C4}\tag{C4}
\end{equation}

Condition~\eqref{cond:C4} ensures that at least one
screening order admits sufficient first-stage
separation. For row-first screening, choose
thresholds $(\tau_R,\tau_C)$ satisfying
\begin{equation}\label{eq:threshold}
\begin{aligned}
&\sqrt{\frac{\log(p_1p_2)}{T}}
=o(\tau_R),
\qquad
\tau_R=o(p_2^{\delta_2-1}),\\
&\sqrt{\frac{\log(p_1^{\delta_1}p_2)}{T}}
=o(\tau_C),
\qquad
\tau_C=o(1).
\end{aligned}
\end{equation}
The column-first threshold conditions are symmetric.
\begin{theorem}\label{thm:screening consistency}
Suppose Assumptions~\ref{ass:weak loading}--\ref{ass:mixing-noise},
Assumptions~\ref{ass:screening}--\ref{ass:epsilon-t}, and
Condition~\eqref{cond:C4} hold. If
\(
p_2^{2-2\delta_2}
\le
p_1^{2-2\delta_1},
\)
then the row-first screening procedure, with thresholds satisfying
\eqref{eq:threshold}, satisfies
\[
\Pr\left(
\widehat{\mathcal I}_R=\mathcal I_R,\,
\widehat{\mathcal I}_C=\mathcal I_C
\right)\to1.
\]
If instead
\(
p_1^{2-2\delta_1}
\le
p_2^{2-2\delta_2},
\)
the corresponding column-first procedure is selection consistent.
If both mode-specific first-stage separation conditions hold, either
screening order may be used.
\end{theorem}
\vspace{-15pt}
\paragraph{Rank-based robustness extension.}
The preceding screening argument is not specific to the covariance score.
For the Spearman-based score introduced in Section~\ref{subsec:screening},
boundedness of the rank kernel yields a uniform concentration bound under
the rank-dependence conditions stated in Section~\ref{sec:screening-proof} of the Supplementary Material. Combining this bound with the same sequential separation
argument gives the following result.

\begin{corollary}[Rank-based screening consistency]
\label{cor:rank-screening}
Suppose Assumptions~\ref{ass:weak loading} and~\ref{ass:rank-beta}--\ref{ass:rank-signal} hold. For the row-first procedure, suppose further that
\(
\log\paren{p_1p_2}=O\paren{T^{1-\nu}},
\)
for some \(\nu\in(0,1)\) and \(p_2^{2-2\delta_2}\log(p_1p_2)=o(T)\).
If thresholds satisfy
\(
\sqrt{\frac{\log(p_1p_2)}{T}}
=o(\tau_R^{\mathrm{rank}}),
\;
\tau_R^{\mathrm{rank}}
=o(p_2^{\delta_2-1}),
\)
and
\(
\sqrt{
\frac{\log(p_1^{\delta_1}p_2)}{T}
}
=o(\tau_C^{\mathrm{rank}}),
\;
\tau_C^{\mathrm{rank}}=o(1).
\)
Then
\[
\Pr\left(
\widehat{\mathcal I}_R^{\,\mathrm{rank}}=\mathcal I_R,\,
\widehat{\mathcal I}_C^{\,\mathrm{rank}}=\mathcal I_C
\right)\to1.
\]
The corresponding result for the column-first procedure follows symmetrically.
\end{corollary}
\begin{remark}
Relative to covariance-based screening,
Corollary~\ref{cor:rank-screening} requires the additional dimensional
condition
\(
\log(p_1p_2)=O(T^{1-\nu})
\)
for some $\nu\in(0,1)$. The weak-loading separation requirement itself
is unchanged:
\(
\min\left\{
p_1^{2-2\delta_1},
p_2^{2-2\delta_2}
\right\}
\log(p_1p_2)=o(T).
\)
In fact, the dimensional envelope in Condition~\eqref{cond:C4},
\(
\log(p_1p_2)
=
O\!\left(T^{\gamma_s/(2-\gamma_s)}\right),
\)
already implies the additional rank-based condition whenever
\(
\frac{\gamma_s}{2-\gamma_s}\le 1-\nu.
\)
\end{remark}

\subsection{Post-Screening Estimation and Factor-Number Determination}
Under Theorem~\ref{thm:screening consistency}, the covariance-based
screening procedure recovers the informative row and column sets with
probability tending to one. On this exact-recovery event,
\(\bX_t^{\mathrm{scr}}\) equals the submatrix restricted to the true
informative rows and columns, and the retained loading matrices satisfy
\[
\frac{1}{|\mathcal I_R|}
(\bR^{\mathrm{scr}})^\top\bR^{\mathrm{scr}}
\to\bI_{k_1},
\qquad
\frac{1}{|\mathcal I_C|}
(\bC^{\mathrm{scr}})^\top\bC^{\mathrm{scr}}
\to\bI_{k_2}.
\] Thus the weak-loading structure in the ambient dimensions becomes pervasive relative to the retained dimensions. The regularity conditions used for projected estimation are inherited by the screened process. The factor process is unchanged, so its stationarity and moment conditions continue to hold. The temporal mixing and tail conditions are preserved under restriction to a subarray, while the spatial dependence bounds can only become weaker because the corresponding sums are taken over smaller row and column index sets. Likewise, a unit vector defined on a retained mode can be zero-padded to a unit vector in the ambient space, so the uniform factor--noise interaction bounds imposed for the full matrix continue to apply to the screened process. Consequently, conditional on exact support recovery, the screened model falls within the pervasive-loading projected-estimation regime with dimensions \(|\mathcal I_R|\) and \(|\mathcal I_C|\). This yields the following result for the post-screening loading and factor estimators and for Algorithm~\ref{alg:factor-number}.

\begin{corollary}[Post-screening estimation and factor-number recovery]\label{thm:screened estimators}
Suppose the conditions of
Theorem~\ref{thm:screening consistency} hold. In addition, suppose
Assumptions~\ref{ass:weak loading}--\ref{ass:noise-dependence} and
Assumptions~\ref{ass:projected_spatial_noise}--
\ref{ass:Factor-noise interaction for projected} hold. Then there exist matrices
\(\widetilde{\bH}_R^{\mathrm{scr}}\in\mathbb R^{k_1\times k_1}\) and
\(\widetilde{\bH}_C^{\mathrm{scr}}\in\mathbb R^{k_2\times k_2}\) such that \(\paren{\widetilde{\bH}_R^{\mathrm{scr}}}^\top\widetilde{\bH}_R^{\mathrm{scr}}\xrightarrow[]{p}\bI_{k_1},\;\paren{\widetilde{\bH}_C^{\mathrm{scr}}}^\top\widetilde{\bH}_C^{\mathrm{scr}}\xrightarrow[]{p}\bI_{k_2}\) and the
post-screening projected estimators satisfy
\begin{enumerate}

\item[\textup{(I)}]

    \[
    \frac{1}{p_1^{\delta_1}}\norm{\widetilde{\bR}^{\mathrm{scr}}-\bR^{\mathrm{scr}}\widetilde{\bH}_R^{\mathrm{scr}}}_F^2=O_p\paren{\frac{1}{Tp_2^{\delta_2}}+\frac{1}{p_1^{2\delta_1}p_2^{2\delta_2}}+\frac{1}{T^2p_1^{2\delta_1}}},
    \]
    \[
    \frac{1}{p_2^{\delta_2}}\norm{\widetilde{\bC}^{\mathrm{scr}}-\bC^{\mathrm{scr}}\widetilde{\bH}_C^{\mathrm{scr}}}_F^2=O_p\paren{\frac{1}{Tp_1^{\delta_1}}+\frac{1}{p_1^{2\delta_1}p_2^{2\delta_2}}+\frac{1}{T^2p_2^{2\delta_2}}},
    \]

    \item[\textup{(II)}]
    \[\norm{\widetilde{\bF}_t^{\mathrm{scr}}-\paren{\widetilde{\bH}_R^{\mathrm{scr}}}^{-1}\bF_t{\paren{\widetilde{\bH}_C^{\mathrm{scr}}}^{-1}}^\top}=O_p\paren{\frac{1}{\sqrt{T}\min\{p_1^{\delta_1},p_2^{\delta_2}\}}+\frac{1}{\sqrt{p_1^{\delta_1}p_2^{\delta_2}}}}
    \]

    \item[\textup{(III)}]
In addition, suppose that \(k_{\max}\) is a predetermined constant
satisfying
\(
k_{\max}\ge\max\{k_1,k_2\}.
\)
We initialize Algorithm~\ref{alg:factor-number} with
\(
\widehat k_1^{(0)}
=
\widehat k_2^{(0)}
=
k_{\max}.
\)
Then, for every fixed iteration \(s\ge1\),
\(
\Pr\left\{
\widehat k_1^{(s)}=k_1,\,
\widehat k_2^{(s)}=k_2
\right\}
\to1.
\)
Consequently, if the maximum number of iterations \(m\) is fixed, the factor-number estimates returned by
Algorithm~\ref{alg:factor-number} satisfy
\(
\Pr\left\{
\widehat k_1=k_1,\,
\widehat k_2=k_2
\right\}
\to1.
\)
\end{enumerate}
\end{corollary}
\begin{remark}Some remarks on Corollary~\ref{thm:screened estimators} and Theorem~\ref{thm:projected-estimation} are in order.
\begin{enumerate}[label=\textup{(\roman*)}] 
\item Although the corollary is stated for covariance-based screening, the same conclusions
hold for rank-based screening under
Corollary~\ref{cor:rank-screening},
since the post-screening argument relies on the exact-support-recovery event.
\item In contrast to full-matrix estimation, the post-screening estimation error bounds no longer contain the cross-mode signal dilution terms appearing in full-matrix estimation. Under $p_1\asymp p_2\asymp p$ and $\delta_1=\delta_2=\delta$, the
loading-error and factor-error bounds are
\(
O_p\left(
\frac{1}{Tp^\delta}
+\frac{1}{p^{4\delta}}
+\frac{1}{T^2p^{2\delta}}
\right)
\) and
\(O_p\left(
\frac{1}{p^\delta}
+\frac{1}{\sqrt{T}p^\delta}
\right),
\)
respectively. Compared with full-matrix projected estimation, post-screening estimation yields smaller-order loading-error bound for \(\delta<1\), while both loading-error bounds have the same order when \(\delta=1\). For factor estimation, post-screening estimation no longer contains the term \(\frac{p^{1-2\delta}}{\sqrt{T}}\), and both methods attain the same leading error-bound order \(O_p(p^{-\delta})\) under the full-matrix projected consistency conditions. 
\item For factor estimation consistency, full-matrix projected estimation requires $\delta>4/5$ and $p^\delta/T^2=o(1)$, while post-screening projected estimation requires $p^{2-2\delta}\log p/T=o(1)$. Notably, a substantial improvement is that the post-screening projected does not impose the full-matrix restriction $\delta>4/5$, allowing factor estimation consistency under weaker loadings. For example, when $\delta=0.6$ and $T\asymp p$, the post-screening growth condition still holds, while both full-matrix factor-consistency results may not apply.
Even when $\delta>4/5$, there exist growth regimes in which post-screening projected estimation can achieve factor consistency while the full-matrix projected estimation does not apply.  For example, when
$T\log p=o(p^{3\delta-2})$,
it is possible that
\(\frac{p^{2-2\delta}\log p}{T}=o(1)\)
while \(\liminf \frac{p^\delta}{T^2}\ge c\) for some $c>0$.

In particular, when $T\asymp p$, the sufficient lower bound on $\delta$ for factor estimation consistency decreases from $4/5$ for both full-matrix estimators to $1/2$ after screening.
\end{enumerate}
\end{remark}

\subsection{Regression Inference with Estimated Matrix Factors}
We next establish asymptotic distributions for factor-augmented regression using the post-screening factor estimates. We impose the following additional conditions.
\begin{assumption}\label{ass:reg-factor-moment}
The factor process satisfies 
\(
\frac{1}{T}\sum_t\vect\paren{\bF_t}\vect\paren{\bF_t}'\xrightarrow[]{p}\bSigma_F\succ0,
\)
where $\bSigma_F$ is a positive definite matrix.
\end{assumption}
\begin{assumption}\label{ass:factor-error-clt}
The factor process and predictive error satisfy 
    \(
    \frac{1}{\sqrt{T}}\sum_t\epsilon_{t+h}\vect\paren{\bF_t}\xrightarrow[]{d}\mathcal{N}\paren{0,\bSigma_{F\epsilon}},
    \)
    where $\bSigma_{F\epsilon}= {\lim} \frac{1}{T}\sum_t\E \paren{\epsilon_{t+h}^2\vect\paren{\bF_t}\vect\paren{\bF_t}'}$ is a positive definite matrix.
\end{assumption}
\begin{assumption}\label{ass:identification}
The coefficient vectors \(\balpha\) and \(\bbeta\) are nonzero.
To remove the scale and sign indeterminacies of the bilinear
parameterization, we impose
\(
\norm{\balpha}=1,
\;
\alpha_1>0,
\;
\norm{\bbeta}_{\max}\le \bar b,
\)
for some constant \(\bar b>0\), where \(\alpha_1\) denotes the first
element of \(\balpha'\bGamma_1\).
\end{assumption}
Assumptions~\ref{ass:reg-factor-moment} and ~\ref{ass:factor-error-clt} are standard moment and central-limit conditions for
factor-augmented regression. Related conditions are used in
\citet{BaiNg_Econometrica_2006}
and \citet{CaiKongWuZhao_JBES_2025}.
Assumption~\ref{ass:identification} imposes the usual normalization
needed to resolve the scale and sign indeterminacies of the bilinear
parameterization; see, for example,
\citet{ChenXiaoYang_JoE_2021}
and \citet{CaiKongWuZhao_JBES_2025}.

For the bilinear estimator, define
\(
\bK_t
=
\begin{pmatrix}
\bF_t\bbeta_0\\
\bF_t^\top\balpha_0
\end{pmatrix},
\;
\bG_0
=
\begin{pmatrix}
\bbeta_0^\top\otimes\bI_{k_1}\\
\bI_{k_2}\otimes\balpha_0^\top
\end{pmatrix}.
\)
Since
\(
\bK_t=\bG_0\vect(\bF_t),
\)
Assumptions~\ref{ass:reg-factor-moment} and
\ref{ass:factor-error-clt} imply
\begin{equation}\label{ass:bilinear normal condition}
\frac{1}{T}\sum_{t=1}^T\bK_t\bK_t^\top
\xrightarrow{p}
\bSigma_{kk},
\qquad
\frac{1}{\sqrt{T}}
\sum_{t=1}^T
\epsilon_{t+h}\bK_t
\xrightarrow{d}
\mathcal N(\bzero,\bSigma_{kk,\epsilon}),
\end{equation}
where
\(
\bSigma_{kk}
=
\bG_0\bSigma_F\bG_0^\top,
\;
\bSigma_{kk,\epsilon}
=
\bG_0\bSigma_{F\epsilon}\bG_0^\top.
\)

The following theorem establishes the asymptotic distributions of the unrestricted and bilinear
regression estimators defined in
Section~\ref{sec:matrix-factor-augmented-regression-methodology}.
\begin{theorem}\label{thm:regression coefficient estimator}
Suppose the conditions of Corollary~\ref{thm:screened estimators} and Assumptions~\ref{ass:reg-factor-moment}--\ref{ass:factor-error-clt} hold and \(\frac{\sqrt{T}}{p_1^{\delta_1}p_2^{\delta_2}}=o\paren{1}\). Then
\[
\sqrt{T}\bracee{\widehat\btheta-\bracket{{\paren{\widetilde \bH_C^{\mathrm{scr}}}}^\top\bbeta_0}\otimes\bracket{\paren{\widetilde {\bH}_R^{\mathrm{scr}}}^\top\balpha_0}}\xrightarrow{d}\mathcal{N}\paren{0,\paren{\bGamma_2\otimes\bGamma_1^{}}'\bSigma_F^{-1}\bSigma_{F\epsilon}(\bSigma_{F}^{-1})'\paren{\bGamma_2\otimes\bGamma_1^{}}}
\]
Assume further \({\bR'\bR}/\abs{\mathcal{I}_R}=\bI_{k_1},\;{\bC'\bC}/\abs{\mathcal{I}_C}=\bI_{k_2}\), where \(\abs{\mathcal{I}_R}\sim p_1^{\delta_1},\abs{\mathcal{I}_C}\sim p_2^{\delta_2}\), and Assumptions~\ref{ass:identification} holds. Denote \(\bgamma_0=\paren{\begin{matrix}\bGamma_1'\balpha_0\\\bzero\end{matrix}}\in\mathbb{R}^{\paren{k_1+k_2}\times1}\). Then 
\[
\sqrt{T}\paren{\begin{matrix}
    \widehat{\balpha}-\paren{\widetilde \bH_R^{\mathrm{scr}}}'\balpha_0\\
    \widehat{\bbeta}-\paren{\widetilde \bH_C^{\mathrm{scr}}}'\bbeta_0
\end{matrix}}\xrightarrow[]{d}\mathcal{N}\paren{0,\bTheta_0^{-1}\bPsi_0'\bSigma_{kk,\epsilon}\bPsi_0\bTheta_0^{-1}},
\]
where \(\bTheta_0=\bPsi_0'\bSigma_{kk}\bPsi_0+\bgamma_0\bgamma_0',\;\bPsi_0=\bracket{\begin{matrix}
    \bGamma_1&\\&\bGamma_2
\end{matrix}}\).
\end{theorem}

\section{Simulation Studies}\label{sec:simulation}
We evaluate the finite-sample performance of the proposed procedure in support recovery, factor estimation, factor-number determination, and regression inference. Unless otherwise stated, all experiments use the following data-generating process and 500 Monte Carlo
replications.

The observed data matrices are generated according to model~\eqref{equa:factor model}. Specifically, we set \(k_1 = k_2 = 3\). To allow for temporal dependence in the latent factor process, we generate \(\bF_t\) according to

\begin{equation}
\vect(\bF_t)
=
\phi\,\vect(\bF_{t-1})
+
\sqrt{1-\phi^2}\,\boldsymbol{\eta}_t,
\qquad
\boldsymbol{\eta}_t\overset{i.i.d.}{\sim}
\mathcal{N}(\mathbf 0,\bI_{k_1 k_2}).
\end{equation}
The idiosyncratic component is generated by 
\begin{equation}\label{equa:E_t generate}
\vect(\bE_t)
=
\psi\,\vect(\bE_{t-1})
+
\sqrt{1-\psi^2}\,\vect(\bU_t),
\qquad
\bU_t\overset{i.i.d.}{\sim}
\mathcal{MN}(\mathbf 0,\bU_E,\bV_E),
\end{equation}
where \(\bU_E\) and \(\bV_E\) describe the cross-sectional dependence across rows and columns, respectively. 
We set the diagonal entries of these covariance matrices to one and assign $1/p_1,1/p_2$ to the off-diagonal entries of $\bU_E,\bV_E$, respectively. Unless otherwise specified, we set \(\phi=0.1\) and \(\psi=0.1\). The response variables are generated by $y_{t+1}=\balpha^T\bF_t\bbeta+\epsilon_{t+1}$ with independent and identically distributed errors $\epsilon_t\sim \mathcal{N}\paren{0,1}$, that is, $h=1$ in model~\eqref{equa:factor-augmented-regression}. For simplicity, we set $\balpha=\bbeta=\paren{1,1,1}^\top$. For each active row of \(\bR_0\) and \(\bC_0\), the entries are independently generated in absolute value from \(U(0.5,1)\), with a common row-specific sign drawn independently to be positive or negative with equal probability. The inactive rows are set to zero. Let
\(p_{1,\mathrm{act}}=|\mathcal I_R|\) and
\(p_{2,\mathrm{act}}=|\mathcal I_C|\). We then normalize the loading
matrices by
\(
\bR=\bR_0\bS_R^{-1/2},
\;
\bC=\bC_0\bS_C^{-1/2},
\)
where
\(
\bS_R=p_{1,\mathrm{act}}^{-1}\bR_0^\top\bR_0,
\;
\bS_C=p_{2,\mathrm{act}}^{-1}\bC_0^\top\bC_0.
\)
Consequently,
\(
p_{1,\mathrm{act}}^{-1}\bR^\top\bR=\bI_{k_1},
\;
p_{2,\mathrm{act}}^{-1}\bC^\top\bC=\bI_{k_2}.
\) This construction ensures the signal-separation condition in Assumption~\ref{ass:signal} asymptotically. To make the simulation results reproducible, the random seed is fixed at \texttt{2026} in the \texttt{Python} implementation.

\subsection{Screening Consistency}\label{sec:screening consistency}
We first examine the finite-sample support recovery of covariance-based screening under the baseline design. We then compare covariance- and rank-based screening under heavy-tailed and contaminated settings.
\subsubsection{Support Recovery for Covariance-Based Screening}\label{sec:screening-consistency-by-covariance}
To assess whether the procedure can correctly identify the true active row and column sets, we report the empirical exact-selection probability
\(\Pr\!\left(
\widehat{\mathcal I}_R=\mathcal I_R,\;
\widehat{\mathcal I}_C=\mathcal I_C
\right)\) estimated from Monte Carlo replications, under different signal strengths and dimensional settings. For this experiment, we parameterize the sizes of the informative sets through the active proportion \(r\), setting
\(
|\mathcal I_R|
=
\lfloor r p_1\rfloor,
\;
|\mathcal I_C|
=
\lfloor r p_2\rfloor.
\)
This design uses the active proportion $r$ as an intuitive finite-sample measure of loading sparsity. We consider \(p_1=p_2\in\{100,200,300\}\)
with corresponding sample sizes
\(T\in\{100,200\}\), \(\{200,400\}\),
and \(\{300,500\}\), respectively.
The active proportion is
\(r\in\{1/5,1/3,1/2,2/3\}\), with
\(|\mathcal I_R|=|\mathcal I_C|=\lfloor rp_1\rfloor\).

Table~\ref{tab:screening-consistency} summarizes the screening performance in terms of the Monte Carlo frequency of exact selection. Motivated by the threshold conditions following Condition~\eqref{cond:C4}, we consider
\[
\tau_R(a)
=
\left\{
\sqrt{\frac{\log(p_1p_2)}{T}}
\right\}^{a}
\left(
p_2^{\delta_2-1}
\right)^{1-a},
\;
\tau_C(a)
=
\left\{
\sqrt{
\frac{\log(|\widehat{\mathcal I}_R|p_2)}{T}
}
\right\}^{a},
\; 0<a<1.
\] Given the symmetric design across the two matrix dimensions, we report row-first screening.
\begin{table}[!htb]
\centering
\caption{Exact-selection frequencies under alternative threshold exponents.}
\label{tab:screening-consistency}
\label{tab:generalized-a-screening}
\small
\setlength{\tabcolsep}{4pt}
\renewcommand{\arraystretch}{1.0}
\begin{threeparttable}
\begin{tabular}{c*{10}{r}}
\toprule
& \multicolumn{5}{c}{Smaller sample size} & \multicolumn{5}{c}{Larger sample size} \\
\cmidrule(lr){2-6}\cmidrule(l){7-11}
$r$ & $a=.20$ & $a=.35$ & $\boldsymbol{a=.50}$ & $a=.65$ & $a=.80$
    & $a=.20$ & $a=.35$ & $\boldsymbol{a=.50}$ & $a=.65$ & $a=.80$ \\
\midrule
\multicolumn{11}{l}{
$p_1=p_2=100$; $T=100$ (left), $T=200$ (right)} \\
$1/5$ & 0.000 & 0.002 & 0.014 & 0.050 & 0.142 & 0.198 & 0.250 & 0.306 & 0.370 & 0.404 \\
$1/3$ & 0.840 & 0.806 & 0.750 & 0.716 & 0.624 & 1.000 & 1.000 & 1.000 & 0.998 & 0.942 \\
$1/2$ & 0.996 & 0.996 & 1.000 & 0.996 & 0.922 & 0.998 & 0.998 & 1.000 & 1.000 & 1.000 \\
$2/3$ & 0.992 & 0.994 & 0.998 & 1.000 & 0.996 & 1.000 & 1.000 & 1.000 & 1.000 & 1.000 \\
\addlinespace[3pt]
\multicolumn{11}{l}{
$p_1=p_2=200$; $T=200$ (left), $T=400$ (right)} \\
$1/5$ & 0.522 & 0.670 & 0.802 & 0.884 & 0.948 & 1.000 & 1.000 & 1.000 & 1.000 & 1.000 \\
$1/3$ & 1.000 & 1.000 & 1.000 & 1.000 & 0.998 & 1.000 & 1.000 & 1.000 & 1.000 & 1.000 \\
$1/2$ & 1.000 & 1.000 & 1.000 & 1.000 & 1.000 & 1.000 & 1.000 & 1.000 & 1.000 & 1.000 \\
$2/3$ & 1.000 & 1.000 & 1.000 & 1.000 & 1.000 & 1.000 & 1.000 & 1.000 & 1.000 & 1.000 \\
\addlinespace[3pt]
\multicolumn{11}{l}{
$p_1=p_2=300$; $T=300$ (left), $T=500$ (right)} \\
$1/5$ & 1.000 & 1.000 & 1.000 & 1.000 & 1.000 & 1.000 & 1.000 & 1.000 & 1.000 & 1.000 \\
$1/3$ & 1.000 & 1.000 & 1.000 & 1.000 & 1.000 & 1.000 & 1.000 & 1.000 & 1.000 & 1.000 \\
$1/2$ & 0.998 & 1.000 & 1.000 & 1.000 & 1.000 & 1.000 & 1.000 & 1.000 & 1.000 & 1.000 \\
$2/3$ & 1.000 & 1.000 & 1.000 & 1.000 & 1.000 & 1.000 & 1.000 & 1.000 & 1.000 & 1.000 \\
\bottomrule
\end{tabular}
\begin{tablenotes}[flushleft]
\footnotesize
\item \textit{Notes:}
Each entry reports the frequency of exact recovery of both
the informative row set and the informative column set
over 500 Monte Carlo replications.
Here, $r=|\mathcal I_R|/p_1=|\mathcal I_C|/p_2$.
The bold heading $a=0.50$ denotes the baseline
geometric-mean threshold.
\end{tablenotes}
\end{threeparttable}
\end{table}

Table~\ref{tab:screening-consistency} shows that the exact-recovery
frequency increases substantially with the sample size and matrix
dimensions. For configurations with smaller active proportions $r$,
and hence weaker first-stage signal separation, exact recovery can be difficult in smaller samples, whereas the recovery frequency approaches one rapidly as the dimensional and sample-size conditions become more favorable. The results are also broadly stable over a range of threshold exponents. These patterns are consistent with Theorem~\ref{thm:screening consistency}. In the subsequent post-screening estimation experiments, we use the geometric-mean specification \(a=0.5\) as the baseline implementation. Additional results for the rank-based screening procedure under the Gaussian design are reported in the Supplement.

The reported results of 
exact-selection provide empirical support for the subsequent post-screening analysis. In configurations for which the exact-selection frequency is close to one, the post-screening estimators are, with high frequency, applied to the true informative submatrix, thereby removing the signal dilution and additional noise associated with inactive rows and
columns. 
\subsubsection{Screening Performance under Heavy Tails and Contamination}
\label{sec:robust-screening}
To assess the effects of heavy tails and contamination on screening performance, we modify only the distribution
of the idiosyncratic innovations in
\eqref{equa:E_t generate}. Specifically, we generate $\bU_t=\bL_U\bA_t\bL_V^\top$, where
$\bL_U\bL_U^\top=\bU_E$ and
$\bL_V\bL_V^\top=\bV_E$, and the entries of $\bA_t$ are independent standard normal variables under the baseline design. For the heavy-tailed design, these entries are replaced by standardized $t_3$ variables,
$t_3/\sqrt{3}$. For the contaminated designs, each entry is independently drawn from a standard Cauchy distribution with probability $\epsilon\in\{0.01,0.05,0.10\}$ and from the standard normal distribution otherwise. 

We retain the tuning specifications from the Gaussian-design experiments. For covariance-based screening, we use \(a=0.50\) as a representative choice. For rank-based screening, we use \(a=0.80\), together with the scaling constants \(\kappa_R=0.30\) and \(\kappa_C=0.275\), a relatively well-performing specification under the Gaussian design as reported in the Supplementary Material. Table~\ref{tab:robust-screening} reports the exact-selection frequencies of the covariance- and rank-based screening procedures under heavy-tailed and contaminated designs. 

\begin{table}[!htb]
\centering
\caption{Exact-selection frequencies under heavy tails and contamination.}
\label{tab:robust-screening}
\footnotesize
\setlength{\tabcolsep}{3pt}
\renewcommand{\arraystretch}{1.0}
\begin{threeparttable}
\begin{tabular}{@{}c
  @{\hspace{7pt}}
  *{3}{r@{\hspace{4pt}}r@{\hspace{5pt}}}
  r@{\hspace{4pt}}r
  @{\hspace{8pt}}
  *{3}{r@{\hspace{4pt}}r@{\hspace{5pt}}}
  r@{\hspace{5pt}}r@{}}
\toprule
& \multicolumn{8}{c}{Smaller sample size} & \multicolumn{8}{c}{Larger sample size} \\
\cmidrule(lr){2-9}\cmidrule(l){10-17}
& \multicolumn{2}{c}{$t_3$} & \multicolumn{2}{c}{$\epsilon=.01$} & \multicolumn{2}{c}{$\epsilon=.05$} & \multicolumn{2}{c}{$\epsilon=.10$}
& \multicolumn{2}{c}{$t_3$} & \multicolumn{2}{c}{$\epsilon=.01$} & \multicolumn{2}{c}{$\epsilon=.05$} & \multicolumn{2}{c}{$\epsilon=.10$} \\
$r$ & Cov. & Rank & Cov. & Rank & Cov. & Rank & Cov. & Rank & Cov. & Rank & Cov. & Rank & Cov. & Rank & Cov. & Rank \\
\midrule
\multicolumn{17}{l}{
$p_1=p_2=100$, $T=100$ (left), $T=200$ (right)} \\
$1/5$ & 0.016 & 0.000 & 0.000 & 0.000 & 0.000 & 0.000 & 0.000 & 0.000 & 0.396 & 0.002 & 0.000 & 0.000 & 0.000 & 0.000 & 0.000 & 0.000 \\
$1/3$ & 0.764 & 0.010 & 0.000 & 0.014 & 0.000 & 0.006 & 0.000 & 0.002 & 0.992 & 0.498 & 0.000 & 0.510 & 0.000 & 0.288 & 0.000 & 0.136 \\
$1/2$ & 0.994 & 0.436 & 0.000 & 0.464 & 0.000 & 0.300 & 0.000 & 0.136 & 1.000 & 0.948 & 0.000 & 0.948 & 0.000 & 0.812 & 0.000 & 0.584 \\
$2/3$ & 0.998 & 0.750 & 0.002 & 0.682 & 0.000 & 0.546 & 0.000 & 0.340 & 0.998 & 0.988 & 0.000 & 0.980 & 0.000 & 0.940 & 0.000 & 0.784 \\
\addlinespace[3pt]
\multicolumn{17}{l}{
$p_1=p_2=200$, $T=200$ (left), $T=400$ (right)} \\
$1/5$ & 0.794 & 0.692 & 0.000 & 0.660 & 0.000 & 0.186 & 0.000 & 0.020 & 0.992 & 0.988 & 0.000 & 0.984 & 0.000 & 0.802 & 0.000 & 0.338 \\
$1/3$ & 0.998 & 0.970 & 0.000 & 0.964 & 0.000 & 0.826 & 0.000 & 0.514 & 0.998 & 1.000 & 0.000 & 1.000 & 0.000 & 0.988 & 0.000 & 0.848 \\
$1/2$ & 1.000 & 0.960 & 0.000 & 0.960 & 0.000 & 0.930 & 0.000 & 0.786 & 1.000 & 1.000 & 0.000 & 1.000 & 0.000 & 0.996 & 0.000 & 0.962 \\
$2/3$ & 1.000 & 0.974 & 0.000 & 0.970 & 0.000 & 0.960 & 0.000 & 0.890 & 0.998 & 1.000 & 0.000 & 1.000 & 0.000 & 0.998 & 0.000 & 0.988 \\
\addlinespace[3pt]
\multicolumn{17}{l}{
$p_1=p_2=300$, $T=300$ (left), $T=500$ (right)} \\
$1/5$ & 0.998 & 0.988 & 0.000 & 0.994 & 0.000 & 0.762 & 0.000 & 0.316 & 1.000 & 1.000 & 0.000 & 1.000 & 0.000 & 0.946 & 0.000 & 0.638 \\
$1/3$ & 1.000 & 0.988 & 0.000 & 0.994 & 0.000 & 0.976 & 0.000 & 0.830 & 1.000 & 1.000 & 0.000 & 1.000 & 0.000 & 0.998 & 0.000 & 0.954 \\
$1/2$ & 1.000 & 0.994 & 0.000 & 0.998 & 0.000 & 0.998 & 0.000 & 0.942 & 1.000 & 1.000 & 0.000 & 1.000 & 0.000 & 1.000 & 0.000 & 0.984 \\
$2/3$ & 1.000 & 0.994 & 0.000 & 0.992 & 0.000 & 0.992 & 0.000 & 0.968 & 1.000 & 1.000 & 0.000 & 1.000 & 0.000 & 1.000 & 0.000 & 0.996 \\
\bottomrule
\end{tabular}
\begin{tablenotes}[flushleft]
\footnotesize
\item \textit{Notes:}
Each entry reports the frequency of exact recovery of both the informative
row set and the informative column set over 500 Monte Carlo replications.
For the covariance-based method, $a=0.50$ (the geometric-mean threshold).
For the rank-based method, $a=0.80$, $\kappa_R=0.30$, and $\kappa_C=0.275$.
\end{tablenotes}
\end{threeparttable}
\end{table}

Table~\ref{tab:robust-screening} shows that covariance-based screening remains effective under standardized $t_3$ innovations but deteriorates sharply under even mild Cauchy contamination.
In contrast, rank-based screening remains close to
its Gaussian benchmark under the $t_3$ design and is substantially more resilient to contamination. Although its exact-recovery frequency generally decreases as the contamination probability increases, recovery generally improves as $T$, the matrix dimensions, and the active proportion increase. Additional fixed-size ranking results in the Supplementary Material suggest that the deterioration of covariance-based screening under contamination is not merely due to threshold calibration, but reflects a loss of ranking information in the covariance scores themselves.

\subsection{Finite-Sample Performance of Factor Estimation}
\begin{figure}[!htb]
\caption{Factor estimation error under a simulation setting with sample size $T=100$, dimension $p_1=50$, and $p_2 \in \{50,100,150,200,250\}$. We consider symmetric loading strength $\delta_1=\delta_2 \in \{0.6,0.7,0.8,0.9\}$.}\label{fig:factor estimation}
\centering
\includegraphics[width=0.65\textwidth]{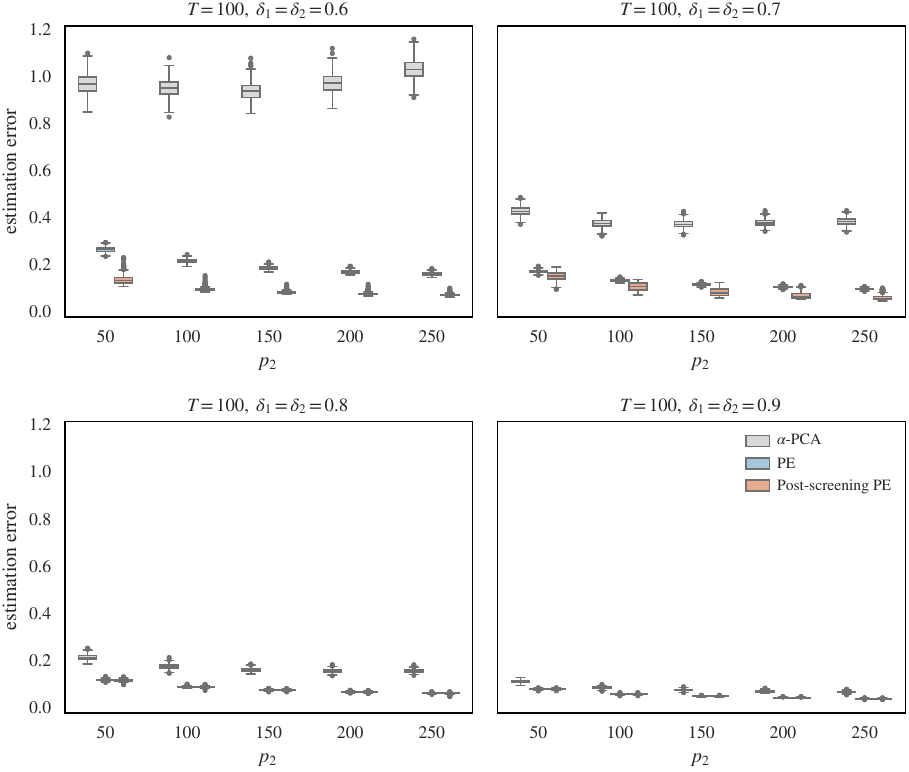}
\end{figure}

We next compare the finite-sample factor estimation accuracy of
$\alpha$-PCA, projected estimation (PE), and post-screening projected estimation (PS-PE) using the time-averaged error
\(
\mathcal{D}_F
=\frac{1}{T}\sum_{t=1}^T
\left\|\widehat{\bF}_t-
\widehat{\bH}_R^{-1}\bF_t
(\widehat{\bH}_C^{-1})^\top\right\|,
\) where $\widehat{\bF}_t$, $\widehat{\bH}_R$, and
$\widehat{\bH}_C$ denote the method-specific factor estimate and corresponding rotation matrices, respectively. For PS-PE, for example, they are $\widetilde{\bF}_t^{\mathrm{scr}}$,
$\widetilde{\bH}_R^{\mathrm{scr}}$, and
$\widetilde{\bH}_C^{\mathrm{scr}}$.
We fix $p_1=50$ and $T=100$, and vary
$p_2\in\{50,100,150,200,250\}$ and
$\delta_1=\delta_2=\delta\in\{0.6,0.7,0.8,0.9\}$. Table~\ref{tab:factor-estimation-error} reports the Monte Carlo means and standard deviations of \(
\mathcal{D}_F,
\) while Figure~\ref{fig:factor estimation} presents the corresponding boxplots.

\begin{table}[!htb]
\centering
\caption{Monte Carlo means and standard deviations (in parentheses) of the factor estimation error $\mathcal D_F$, with $p_1=50$ and $T=100$.}
\label{tab:factor-estimation-error}
\begin{threeparttable}
\fontsize{9.5}{11}\selectfont
\setlength{\tabcolsep}{3pt}
\renewcommand{\arraystretch}{1.10}
\begin{tabular}{@{}c c c c c @{\hspace{10pt}} c c c c@{}}
\toprule
 & \multicolumn{4}{c}{\(\delta=0.6,\ |\mathcal I_R|=10\)} & \multicolumn{4}{c}{\(\delta=0.7,\ |\mathcal I_R|=15\)} \\
$p_2$ & $|\mathcal I_C|$ & $\alpha$-PCA & PE & PS-PE & $|\mathcal I_C|$ & $\alpha$-PCA & PE & PS-PE \\
50 & 10 & 0.9672\hspace{1pt}(0.0436) & 0.2644\hspace{1pt}(0.0104) & \textbf{0.1378\hspace{1pt}(0.0203)} & 15 & 0.4283\hspace{1pt}(0.0197) & 0.1721\hspace{1pt}(0.0058) & \textbf{0.1507\hspace{1pt}(0.0171)} \\
100 & 15 & 0.9509\hspace{1pt}(0.0386) & 0.2152\hspace{1pt}(0.0087) & \textbf{0.0972\hspace{1pt}(0.0073)} & 25 & 0.3757\hspace{1pt}(0.0170) & 0.1330\hspace{1pt}(0.0044) & \textbf{0.1076\hspace{1pt}(0.0178)} \\
150 & 20 & 0.9384\hspace{1pt}(0.0383) & 0.1864\hspace{1pt}(0.0069) & \textbf{0.0833\hspace{1pt}(0.0050)} & 33 & 0.3723\hspace{1pt}(0.0159) & 0.1157\hspace{1pt}(0.0039) & \textbf{0.0843\hspace{1pt}(0.0168)} \\
200 & 24 & 0.9718\hspace{1pt}(0.0393) & 0.1712\hspace{1pt}(0.0067) & \textbf{0.0767\hspace{1pt}(0.0057)} & 40 & 0.3794\hspace{1pt}(0.0148) & 0.1054\hspace{1pt}(0.0034) & \textbf{0.0703\hspace{1pt}(0.0145)} \\
250 & 27 & 1.0305\hspace{1pt}(0.0434) & 0.1617\hspace{1pt}(0.0064) & \textbf{0.0718\hspace{1pt}(0.0046)} & 47 & 0.3838\hspace{1pt}(0.0148) & 0.0969\hspace{1pt}(0.0033) & \textbf{0.0616\hspace{1pt}(0.0122)} \\
\addlinespace[0.65em]
 & \multicolumn{4}{c}{\(\delta=0.8,\ |\mathcal I_R|=22\)} & \multicolumn{4}{c}{\(\delta=0.9,\ |\mathcal I_R|=33\)} \\
$p_2$ & $|\mathcal I_C|$ & $\alpha$-PCA & PE & PS-PE & $|\mathcal I_C|$ & $\alpha$-PCA & PE & PS-PE \\
50 & 22 & 0.2131\hspace{1pt}(0.0117) & 0.1160\hspace{1pt}(0.0038) & \textbf{0.1153\hspace{1pt}(0.0039)} & 33 & 0.1098\hspace{1pt}(0.0063) & 0.0771\hspace{1pt}(0.0023) & \textbf{0.0770\hspace{1pt}(0.0022)} \\
100 & 39 & 0.1734\hspace{1pt}(0.0091) & 0.0869\hspace{1pt}(0.0026) & \textbf{0.0865\hspace{1pt}(0.0027)} & 63 & 0.0828\hspace{1pt}(0.0046) & 0.0556\hspace{1pt}(0.0016) & \textbf{0.0555\hspace{1pt}(0.0016)} \\
150 & 55 & 0.1599\hspace{1pt}(0.0078) & 0.0733\hspace{1pt}(0.0023) & \textbf{0.0729\hspace{1pt}(0.0022)} & 90 & 0.0739\hspace{1pt}(0.0040) & 0.0466\hspace{1pt}(0.0014) & \textbf{0.0465\hspace{1pt}(0.0013)} \\
200 & 69 & 0.1557\hspace{1pt}(0.0073) & 0.0654\hspace{1pt}(0.0019) & \textbf{0.0651\hspace{1pt}(0.0019)} & 117 & 0.0688\hspace{1pt}(0.0038) & 0.0409\hspace{1pt}(0.0011) & \textbf{0.0408\hspace{1pt}(0.0011)} \\
250 & 82 & 0.1547\hspace{1pt}(0.0069) & 0.0599\hspace{1pt}(0.0017) & \textbf{0.0594\hspace{1pt}(0.0020)} & 143 & 0.0656\hspace{1pt}(0.0034) & \textbf{0.0369\hspace{1pt}(0.0010)} & \textbf{0.0369\hspace{1pt}(0.0010)} \\
\bottomrule
\end{tabular}
\begin{tablenotes}[flushleft]
\footnotesize
\item \textit{Notes:}
The active set sizes are
$|\mathcal I_R|=\lfloor p_1^{\delta_1}\rfloor$
and $|\mathcal I_C|=\lfloor p_2^{\delta_2}\rfloor$. PE and PS-PE denote the projected and post-screening projected estimators, respectively.
The $\alpha$-PCA and projected estimators are those of Chen and Fan (2023) and Yu et al. (2022), respectively. Boldface indicates the smallest mean error among the three methods for each simulation design.
\end{tablenotes}
\end{threeparttable}
\end{table}

The results in Table~\ref{tab:factor-estimation-error} and
Figure~\ref{fig:factor estimation} indicate that the post-screening projected estimator generally yields smaller factor estimation errors than the \(\alpha\)-PCA and the projected estimator. The improvement is most substantial under weak loading settings, where signal dilution is more severe. As the loading strength increases, all methods exhibit reduced estimation error, and the relative advantage of screening becomes less pronounced. 
In addition, under weaker loadings, particularly when
\(\delta_1=\delta_2\in\{0.6,0.7\}\), the improvement from screening
remains pronounced as \(p_2\) increases while \(T\) and \(p_1\) are
held fixed. This pattern is broadly consistent with the theoretical
analysis in Section~\ref{sec:theory}: the error bound for the
pre-screening estimators contains the dimension-dependent components (
\(\frac{p_2^{2-2\delta_2}}{p_1^{2\delta_1-1}}\) for \(\alpha\)-PCA and 
\(
\frac{p_2^{1-\delta_2}}{\sqrt{T}p_1^{\delta_1}}
\) for the projected method),
whereas the corresponding post-screening component \(\frac{1}{\sqrt{T}p_1^{\delta_1}}\) no longer carries
the factor \(p_2^{1-\delta_2}\).

\subsection{Finite Sample Performance of Factor-Number Determination}
\label{subsec:simulation-factor-number}

We study next the finite-sample performance of factor-number
determination under sparse loading support. To vary the idiosyncratic
noise level, we write
\(
\bE_t=\sigma_E\bE_t^{(0)},
\)
where $\bE_t^{(0)}$ follows the same process as in the baseline design.
We focus on $r\in\{1/5,1/3\}$ and compare a low-noise design,
$\sigma_E=1$, with a more challenging high-noise design,
$\sigma_E=5$. Factor numbers are estimated using $\alpha$-PCA and
projected estimation, both before and after screening.

For screening, we use the row-first procedure with $a=0.9$ and adjust
the thresholds according to the idiosyncratic noise scale,
\(
\tau_R^{(\sigma_E)}
    =\sigma_E\tau_R(0.9),
\;
\tau_C^{(\sigma_E)}
    =\sigma_E\tau_C(0.9).
\)
We also consider the ridge constants $c\in\{0,10^{-4},10^{-2}\}$. Since the joint factor-number selection frequencies are nearly unchanged across these values of \(c\), Table~\ref{tab:factor-number-joint-selection} reports the results for \(c=10^{-4}\), with the corresponding results for \(c=0\) and \(c=10^{-2}\) provided in the Supplementary Material.
\begin{table}[!htb]
\centering
\caption{Joint factor-number selection frequencies under sparse loading support.}
\label{tab:factor-number-joint-selection}
\begin{threeparttable}
\small
\setlength{\tabcolsep}{3pt}
\renewcommand{\arraystretch}{1.0}
\begin{tabular*}{\textwidth}{@{\extracolsep{\fill}}cl*{12}{c}@{}}
\toprule
& & \multicolumn{4}{c}{$p_1=p_2=100$} & \multicolumn{4}{c}{$p_1=p_2=200$} & \multicolumn{4}{c}{$p_1=p_2=300$} \\
\cmidrule(lr){3-6}\cmidrule(lr){7-10}\cmidrule(l){11-14}
& & \multicolumn{2}{c}{$T=100$} & \multicolumn{2}{c}{$T=200$} & \multicolumn{2}{c}{$T=200$} & \multicolumn{2}{c}{$T=400$} & \multicolumn{2}{c}{$T=300$} & \multicolumn{2}{c}{$T=500$} \\
\cmidrule(lr){3-4}\cmidrule(lr){5-6}\cmidrule(lr){7-8}\cmidrule(lr){9-10}\cmidrule(lr){11-12}\cmidrule(l){13-14}
$\sigma_E$ & Method & $1/5$ & $1/3$ & $1/5$ & $1/3$ & $1/5$ & $1/3$ & $1/5$ & $1/3$ & $1/5$ & $1/3$ & $1/5$ & $1/3$ \\
\midrule
$1$ & A & 0.000 & 0.000 & 0.000 & 0.000 & 0.000 & 0.000 & 0.000 & 0.000 & 0.000 & 0.002 & 0.000 & 0.002 \\
 & PS-A & 0.996 & 1.000 & 1.000 & 1.000 & 1.000 & 1.000 & 1.000 & 1.000 & 1.000 & 1.000 & 1.000 & 1.000 \\
 & P & 1.000 & 1.000 & 1.000 & 1.000 & 1.000 & 1.000 & 1.000 & 1.000 & 1.000 & 1.000 & 1.000 & 1.000 \\
 & PS-P & 1.000 & 1.000 & 1.000 & 1.000 & 1.000 & 1.000 & 1.000 & 1.000 & 1.000 & 1.000 & 1.000 & 1.000 \\
\addlinespace[5pt]
$5$ & A & 0.000 & 0.000 & 0.000 & 0.000 & 0.000 & 0.000 & 0.000 & 0.000 & 0.000 & 0.000 & 0.000 & 0.000 \\
 & PS-A & 0.000 & 0.000 & 0.008 & 0.000 & 0.056 & 0.000 & 0.050 & 0.000 & 0.588 & 0.002 & 0.518 & 0.002 \\
 & P & 0.000 & 0.000 & 0.000 & 0.000 & 0.002 & 1.000 & 0.000 & 1.000 & 1.000 & 1.000 & 1.000 & 1.000 \\
 & PS-P & 0.502 & 0.966 & 0.952 & 1.000 & 0.994 & 1.000 & 1.000 & 1.000 & 1.000 & 1.000 & 1.000 & 1.000 \\
\bottomrule
\end{tabular*}
\begin{tablenotes}[flushleft]
\footnotesize
\item[] \textit{Notes:}
Each entry reports the proportion of 500 Monte Carlo replications in which both factor numbers are correctly selected, $(\widehat{k}_1,\widehat{k}_2)=(3,3).$
A and P denote $\alpha$-PCA and projected estimators, respectively, while PS-A and PS-P denote their post-screening
counterparts. Screening is row-first with $a=0.9$, and
$r=|\mathcal I_R|/p_1=|\mathcal I_C|/p_2$.
\end{tablenotes}
\end{threeparttable}
\end{table}

Table~\ref{tab:factor-number-joint-selection} shows that projected
estimation is generally more reliable than $\alpha$-PCA for recovering
the two factor numbers, while screening provides an additional benefit
in the more challenging sparse and high-noise designs. Under $\sigma_E=1$, projected estimation already achieves nearly perfect joint recovery, and its post-screening version essentially preserves this performance. The gain from screening becomes pronounced when
$\sigma_E=5$. For example, when $(p_1,p_2)=(200,200)$, $T=200$, and
$r=1/5$, the joint recovery frequency increases from $0.002$ for
projected estimation to $0.994$ after screening. Post-screening also
improves $\alpha$-PCA, although it remains more sensitive to strong
idiosyncratic noise than the projected procedures. Overall, the results are consistent with screening improving
factor-number recovery when sparse loading support weakens effective
factor--noise separation.

\subsection{Finite-Sample Illustration of Asymptotic Normality}\label{sec:inference-simulation} 
Finally, we examine the finite-sample normal approximation
for the regression coefficient estimator \(\widehat{\btheta}\) in
Theorem~\ref{thm:regression coefficient estimator}. We consider the
configuration
\(
p_1=p_2=50,\;
|\mathcal I_R|=|\mathcal I_C|=20,\;
T\in\{150,200,250\}.
\)
We generate
\(
\vect(\bF_t)\overset{i.i.d.}{\sim}
\mathcal N\left(\bzero,\bSigma_F^{\mathrm{sim}}\right),
\)
where \(\bSigma_F^{\mathrm{sim}}\) is diagonal with diagonal entries
\((2,1.5,1,1.5,1.3,1.2,1,1.2,0.8)\).
The idiosyncratic error \(\bE_t\) is generated by setting \(\psi=0\)
in \eqref{equa:E_t generate}. The regression error is generated as
\(
\epsilon_{t+h}\overset{i.i.d.}{\sim}\mathcal N(0,1).
\)
The remaining data-generating mechanisms are
the same as those in Section~\ref{sec:screening-consistency-by-covariance}. The results are based on 1,000 Monte Carlo replications. 

\begin{figure}[htbp]
\caption{Histograms of the standardized regression coefficient estimator for different values of \(T\), with \(p_1=50\), \(p_2=50\) and \(\abs{\mathcal{I}_R}=20\), \(\abs{\mathcal{I}_C}=20\). The solid curve denotes the standard normal density.}
    \label{fig:regression-coefficient-normality}
    \centering

    \begin{subfigure}{0.31\textwidth}
    \centering
    \includegraphics[width=\linewidth]{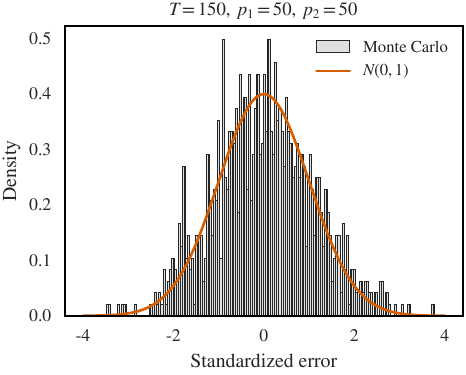}
    
\end{subfigure}
\hfill
\begin{subfigure}{0.31\textwidth}
    \centering
    \includegraphics[width=\linewidth]{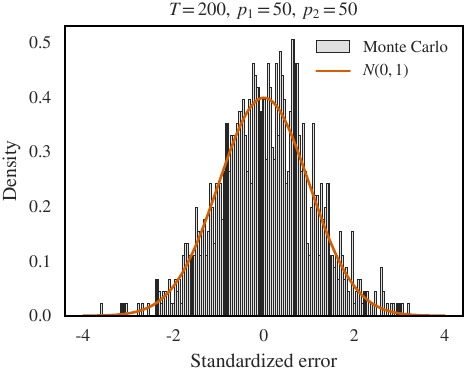}
    
\end{subfigure}
\hfill
\begin{subfigure}{0.31\textwidth}
    \centering
    \includegraphics[width=\linewidth]{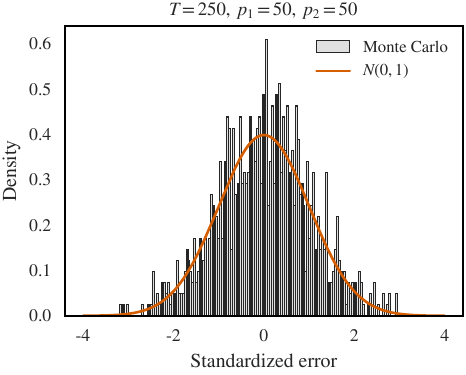}
    
\end{subfigure}
\end{figure}

Under this design, the asymptotic covariance matrix in
Theorem~\ref{thm:regression coefficient estimator} reduces to
\[
\bSigma^{\mathrm{sim}}
=\paren{\bGamma_2\otimes\bGamma_1^{}}'\bSigma_F^{-1}\bSigma_{F\epsilon}(\bSigma_{F}^{-1})'\paren{\bGamma_2\otimes\bGamma_1^{}}=
(\bSigma_F^{\mathrm{sim}})^{-1}
\bSigma_{F\epsilon}^{\mathrm{sim}}
(\bSigma_F^{\mathrm{sim}})^{-1}
=
(\bSigma_F^{\mathrm{sim}})^{-1}.
\]
Figure~\ref{fig:regression-coefficient-normality} displays the histogram of the first coordinate of
\(
\sqrt{T}\,
(\bSigma^{\mathrm{sim}})^{-1/2}
\left[
\widehat{\btheta}
-
\left\{
(\widetilde\bH_C^{\mathrm{scr}})^\top\bbeta_0
\otimes
(\widetilde\bH_R^{\mathrm{scr}})^\top\balpha_0
\right\}
\right],
\)
under different values of \(T\). The solid curve denotes the standard
normal density. The histograms show close agreement with the standard
normal distribution. Additional Q--Q plots in the Supplement provide further evidence supporting the asymptotic normal approximation in Theorem~\ref{thm:regression coefficient estimator}.

\section{Empirical Studies} \label{sec:empirical studies}

We evaluate the proposed procedure using two matrix-valued datasets: a quarterly multicountry macroeconomic panel and a monthly U.S. state--industry employment panel. The analysis examines the forecasting benefits of response-guided screening and post-screening factor-number determination.

\subsection{Empirical Design and Competing Methods}\label{sec:empirical-design-competing-methods}
Throughout the empirical analysis, we compare six one-step-ahead
forecasting procedures. Models 1--3 are matrix-factor regressions based on
\(\alpha\)-PCA \citep{ChenFan_JASA_2023}, projected estimation (PE)
\citep{YuHeKongZhang_JoE_2022}, and the proposed post-screening PE,
respectively. All three use
\begin{equation}
y_{t+1}
= \boldsymbol{\theta}^{\top}
\vect\left(\widehat{\bF}^{(m)}_t\right)
+\epsilon_{t+1}, \qquad m=1,2,3,
\label{eq:factor-augmented-regression}
\end{equation}
where
\(\widehat{\bF}^{(1)}_t=\widehat{\bF}_t\) (\(\alpha\)-PCA),
\(\widehat{\bF}^{(2)}_t=\widetilde{\bF}_t\) (PE), and
\(\widehat{\bF}^{(3)}_t=\widetilde{\bF}^{\,\mathrm{scr}}_t\)
(post-screening PE). At each forecast origin, \(\boldsymbol{\theta}\) is estimated by ordinary
least squares using the available training observations.

The remaining benchmarks are
\[
\begin{aligned}
\text{Model 4 (Vec-PCA):}\quad
& y_{t+1}=\boldsymbol{\xi}^{\top}
  \widehat{\mathbf f}^{\,\mathrm{VecPCA}}_t+\epsilon_{t+1},\\
\text{Model 5 (AR(2)):}\quad\;\quad
& y_{t+1}=\varphi_0+\varphi_1 y_t+\varphi_2 y_{t-1}+\epsilon_{t+1},\\
\text{Model 6 (Lasso):}\quad\;\;\quad
& y_{t+1}=\gamma_0+\bgamma^{\top}\vect(\bX_t)+\epsilon_{t+1}.
\end{aligned}
\]
For Vec-PCA, \(\widehat{\mathbf f}^{\,\mathrm{VecPCA}}_t\) is obtained by
applying conventional PCA to \(\vect(\bX_t)\), with the factor number
selected by the eigenvalue-ratio criterion of
\citet{AhnHorenstein_Econometrica_2013} on the initial training sample. \(\boldsymbol{\xi}\) is estimated by ordinary least squares. For Lasso,
\((\gamma_0,\bgamma)\) is fitted by least squares with an \(\ell_1\)
penalty \(\lambda\|\bgamma\|_1\), leaving the intercept unpenalized.

\subsection{Multicountry Macroeconomic Application}
\label{sec:oecd-application}
We consider a quarterly OECD macroeconomic panel with nine indicators for ten countries, organized as $\bX_t\in\mathbb R^{10\times9}$. Closely related matrix-valued macroeconomic datasets have been considered by \citet{
ChenTsayChen_JASA_2020,
YuHeKongZhang_JoE_2022,
ChenFan_JASA_2023}.
The countries included in the analysis are Australia, Canada, Finland,
France, Germany, the Netherlands, New Zealand, Norway, Sweden, and the
United States. The indicators cover four broad categories: consumer
prices, interest rates, production, and international trade. After applying the indicator-specific transformations, the resulting sample contains \(N=143\) quarterly
observations, yielding
\(
T=N-1=142
\)
usable predictor--response pairs
\(
\{(\bX_t,y_{t+1})\}_{t=1}^{N-1}
\) for one-step-ahead forecasting. Further details on the countries, variables, data sources, and transformations are provided
in the Supplementary Material.

We forecast the transformed United States--Total CPI,
$y_t=\Delta^2\log(\mathrm{CPI}^{\mathrm{US,Total}}_t)$,
over the final $L=40$ quarters using an
expanding-window scheme. Each newly observed data point is added to the estimation sample before the subsequent forecast is produced. The loading matrices, latent factors and regression coefficients are re-estimated using all information available at that time. 
For post-screening PE, the row and column
supports are reselected at each forecast origin. For Lasso, the regularization parameter \(\lambda\) is selected at each forecast origin by five-fold time-series cross-validation over the prespecified candidate grid. At forecast origin $t-1$, the loading spaces are estimated from $\{\bX_s\}_{s=1}^{t-1}$, while screening and regression estimation use the available pairs $\{(\bX_s,y_{s+1})\}_{s=1}^{t-2}$. 
Throughout the remainder of this section, United States--Total CPI refers
to the transformed series unless otherwise stated. All forecast accuracy
measures are evaluated on the transformed scale. 

At each forecast origin, the predictor series and response are standardized using the sample means and standard deviations computed from the current estimation window. The response forecasts are transformed back to the original transformed scale, whereas the reconstruction errors are evaluated on the standardized predictor scale. 

We assess matrix reconstruction accuracy for the matrix-factor methods using the retained-block and full-dimension MSEs, and forecasting accuracy for all competing methods using the response MSE and forecast $R^2$. For \(t=N-L+1,\ldots,N\), let
\(\widehat{\mathcal I}_{R,t}\) and
\(\widehat{\mathcal I}_{C,t}\) denote the row and column index sets selected at evaluation date \(t\), and let \(\widehat{\bX}_t^{\mathrm{retained}}\) denote the reconstruction on the selected submatrix. For the two unscreened matrix-factor methods, this is obtained by restricting the full-matrix reconstructions to the same selected block. The retained-block MSE is defined
as
\(
\frac{1}{L}
\sum_{t=N-L+1}^{N}
\frac{
\left\|
\widehat{\bX}_t^{\mathrm{retained}}
-
\bX_t\paren{
\widehat{\mathcal I}_{R,t},
\widehat{\mathcal I}_{C,t}}
\right\|_F^2
}{
|\widehat{\mathcal I}_{R,t}||\widehat{\mathcal I}_{C,t}|
}.
\)
For full-dimension reconstruction, let
$\widehat{\bX}_t^{\mathrm{full}}$ denote the $p_1\times p_2$ reconstructed matrix. For Post-screening PE,
it equals $\widehat{\bX}_t^{\mathrm{retained}}$ on the selected block and zero elsewhere.  For the unscreened methods, it is their original full-matrix reconstruction. The full-dimension MSE is defined as
\(
\frac{1}{L}
\sum_{t=N-L+1}^{N}
\frac{
\left\|
\widehat{\bX}_t^{\mathrm{full}}-\bX_t
\right\|_F^2
}{
p_1p_2
}.
\)

For forecasting, let \(\widehat y_t\) denote the one-step-ahead forecast of \(y_t\) based on the information available before \(t\). The response forecast MSE is
defined as
\(
\frac{1}{L}
\sum_{t=N-L+1}^{N}
\left(y_t-\widehat y_t\right)^2.
\)
We also report \(
R_{\mathrm{}}^2
=
1-
\frac{
\sum_{t=N-L+1}^{N}(y_t-\widehat y_t)^2
}{
\sum_{t=N-L+1}^{N}(y_t-\overline y_{\mathrm{}})^2
},
\)
where \(\overline y_{\mathrm{}}=
\frac{1}{L}\sum_{t=N-L+1}^{N}y_t\) is the sample mean of the evaluation
period. A negative \(R_{\mathrm{}}^2\) indicates that the factor-augmented model explains less variation in the evaluation sample than the constant-mean benchmark computed from the same evaluation period.

We consider both row-first and column-first screening. For each screening order, we consider two approaches to
factor-number determination. The screened specification estimates the row
and column factor numbers from the corresponding screened initial training
sample, whereas the unscreened specification estimates them from the full initial training sample. In either case, the estimated factor numbers are
held fixed throughout the forecast evaluation period.

Motivated by the empirical ranges of the screening statistics in the
initial training sample, we specify fixed candidate grids
\(
\mathcal T_R
=\mathcal T_C=
\left\{
0,0.005,0.01,
\ldots,0.30
\right\}
\).
For each combination of screening order and factor-number determination
scheme, the threshold pair is selected according to the response MSE. Table~\ref{tab:threshold-selection-summary} reports the threshold pair selected under each of the four implementations. Detailed threshold-grid
results are provided in the Supplementary Material.

\begin{table}[!htb]
\centering
\small
\renewcommand{\arraystretch}{1.0}
\caption{Threshold selection for United States--Total CPI.}
\label{tab:threshold-selection-summary}

\begin{tabular}{llrrrr}
\toprule
Screening order
& Factor-number source
& \(\tau_R\)
& \(\tau_C\)
& \(y\)-MSE \((\times 10^{3})\)
& \(R^2\) \\
\midrule
\textbf{Row-first}
& \textbf{Screened}
& \textbf{0.165}
& \textbf{0.215}
& \textbf{0.0310}
& \textbf{0.490} \\

Row-first
& Unscreened
& 0.110
& 0.145
& 0.0445
& 0.268 \\

Column-first
& Screened
& 0.195
& 0.190
& 0.0397
& 0.348 \\

Column-first
& Unscreened
& 0.110
& 0.145
& 0.0450
& 0.259 \\
\bottomrule
\end{tabular}

\begin{minipage}{0.94\textwidth}
\footnotesize
\textit{Notes:}
``Screened'' indicates that the factor numbers are estimated from the
screened initial training sample, whereas ``Unscreened'' indicates that
they are estimated from the full initial training sample.
\end{minipage}
\end{table}

The selection summary reveals two patterns. First, under either
factor-number determination scheme, row-first screening attains a lower
minimum response MSE and a higher corresponding \(R^2\) than column-first
screening. The difference is considerably larger when the factor numbers are estimated from the screened initial training sample. Second, forecast
accuracy is higher when factor numbers are estimated from the screened
rather than the unscreened initial training sample for either screening
order. The latter pattern is consistent with the weak-loading motivation for
post-screening factor-number determination: weak aggregate factor signals may obscure the eigenvalue separation underlying factor-number selection, whereas removing inactive rows and columns can produce a more pronounced eigenvalue structure. This interpretation is consistent with Corollary~\ref{thm:screened estimators}, which establishes consistent factor-number determination from the screened matrix. 

Among the four implementations, row-first screening with factor numbers
estimated from the screened initial training sample selects
\(
(\tau_R,\tau_C)=(0.165,0.215)
\)
and yields the smallest response MSE. We therefore use this specification
for the detailed empirical analysis below. The selected row threshold is consistent with the pattern of the ordered screening statistics shown in Figure~\ref{fig:sta-distribution} of the Supplementary Material. A detailed discussion of the screening statistics is also provided there. 
Table~\ref{tab:representative-method-comparison} compares post-screening PE with $\alpha$-PCA and PE in terms of reconstruction accuracy and with the
other five methods in terms of forecasting accuracy.

For Vec-PCA, the benchmark factor number is selected using the
eigenvalue-ratio criterion of
\citet{AhnHorenstein_Econometrica_2013} applied to the initial training
sample, yielding \(r=1\). Post-screening PE achieves the lowest response
MSE, RMSE, and MAE and the highest out-of-sample \(R^2\) among all
reported methods. The advantage of Post-screening PE is particularly substantial relative to the two unscreened matrix-factor procedures. In terms of reconstruction, screening
substantially reduces the \(\mathrm{Retained \ MSE}\), although its \(\mathrm{Full\text{-}dimension\ MSE}\) is higher than those of the unscreened matrix-factor methods. These findings are not contradictory. The \(\mathrm{Full\text{-}dimension\ MSE}\) penalizes the omission of
all screened-out entries, including variation that may be important for
reconstructing \(\bX_t\) but relatively less informative for predicting
\(y_{t+1}\).

\begin{table}[!htb]
\centering
\small
\renewcommand{\arraystretch}{1.0}
\caption{Reconstruction and one-step-ahead forecasting performance for
United States--Total CPI.}
\label{tab:representative-method-comparison}

\begin{adjustbox}{max width=\textwidth}
\begin{tabular}{lrrrrrr}
\toprule
& \multicolumn{2}{c}{Reconstruction accuracy}
& \multicolumn{4}{c}{Forecast accuracy} \\
\cmidrule(lr){2-3}\cmidrule(lr){4-7}
Method
& \(\mathrm{Retained\ MSE}\)
& \(\mathrm{Full\text{-}dimension\ MSE}\)
& \(y\)-MSE \((\times 10^{3})\)
& RMSE \((\times 10^{3})\)
& MAE \((\times 10^{3})\)
& \(R^{2}\) \\
\midrule

\(\alpha\)-PCA
& 0.9033 & \textbf{0.6066}
& 0.05731 & 7.571 & 6.240 & 0.058 \\

PE
& 0.9269 & 0.6256
& 0.05644 & 7.513 & 6.189 & 0.072 \\

Post-screening PE
& \textbf{0.3061} & 0.9062
& \textbf{0.03099}
& \textbf{5.567}
& \textbf{4.560}
& \textbf{0.490} \\
Vec-PCA (ER, \(r=1\))
& \textemdash & \textemdash
& 0.05577 & 7.468 & 6.260 & 0.083 \\

AR(2)
& \textemdash & \textemdash
& 0.04323 & 6.575 & 5.226 & 0.289 \\

Lasso
& \textemdash & \textemdash
& 0.04393 & 6.628 & 5.172 & 0.278 \\

\bottomrule
\end{tabular}
\end{adjustbox}

\begin{minipage}{0.97\textwidth}
\footnotesize
\textit{Notes:}
The table reports expanding-window one-step-ahead forecasting results over
40 evaluation periods. The response MSE, RMSE, and MAE are
multiplied by \(10^3\). Bold entries indicate the best-performing value
in each column.
\end{minipage}
\end{table}

We further assess the forecast improvements using
one-sided Diebold--Mariano tests
\citep{DieboldMariano_JBES_1995}.
Post-screening PE significantly outperforms
the three factor-based benchmarks at the 1\% level
and Lasso at the 5\% level, with the comparison
against AR(2) significant at the 10\% level.
Detailed statistics are reported in
Table~\ref{tab:representative-dm-tests}
of the Supplementary Material.

To illustrate the screening structure and provide an economic
interpretation of the retained predictors, we examine the loading
structures estimated by the three competing matrix-factor-based methods using the initial training sample. The proposed screening procedure retains Australia, Canada, France, Norway, and the United States, together with total-industry production, manufacturing production, and imports. The estimated row and column factor numbers change from \((1,5)\) for the full matrix to \((3,2)\) for the screened submatrix. After Varimax rotation, the two screened column-loading patterns are primarily associated with production activity and imports, respectively. The complete loading matrices for \(\alpha\)-PCA, PE, and Post-screening PE, together with further economic interpretation, are provided in the Supplementary Material.

\subsection{U.S. State--Industry Employment Application}
\label{sec:state-industry-application}

We next consider a monthly U.S. state--industry employment panel from January 2003 to February 2026, organized as $\bX_t\in\mathbb R^{41\times20}$,
with states as rows and industries as columns. The forecast target is the monthly growth rate of U.S.
manufacturing production. Both the employment series and manufacturing
production are transformed using monthly log differences. Further details on the states, industries, data sources, and transformations are provided in the Supplementary Material.

We use the same expanding-window forecasting framework and competing methods as in the macroeconomic application, evaluating the final
52 months from November 2021 to February 2026. The threshold search uses the same coarse grid, followed by local refinement. The selected post-screening specification uses
row-first screening with
$(\tau_R,\tau_C)=(0.070,0.144)$ and
$(\widehat k_1,\widehat k_2)=(1,1)$.

Table~\ref{tab:ip-method-comparison} reports the forecasting performance
of the competing methods. Post-screening PE achieves the smallest MSE,
RMSE, and MAE and the largest out-of-sample \(R^2\) among the reported
methods. The three unscreened factor procedures produce nearly identical
forecasts in this application.

\begin{table}[!htb]
\centering
\small
\renewcommand{\arraystretch}{1.0}
\caption{One-step-ahead forecasting performance for U.S. manufacturing
production growth.}
\label{tab:ip-method-comparison}
\begin{tabular}{lrrrr}
\toprule
Method
& \(y\)-MSE \((\times10^{3})\)
& RMSE \((\times10^{3})\)
& MAE \((\times10^{3})\)
& \(R^{2}\) \\
\midrule
\(\alpha\)-PCA
& 0.04182 & 6.467 & 4.810 & -0.004 \\

PE
& 0.04182 & 6.467 & 4.810 & -0.004 \\

Post-screening PE
& \textbf{0.03730}
& \textbf{6.107}
& \textbf{4.532}
& \textbf{0.105} \\
Vec-PCA (ER, \(r=1\))
& 0.04182 & 6.467 & 4.809 & -0.004 \\

AR(2)
& 0.04371 & 6.612 & 4.913 & -0.049 \\

Lasso
& 0.04200 & 6.480 & 4.815 & -0.008 \\
\bottomrule
\end{tabular}

\begin{minipage}{0.94\textwidth}
\footnotesize
\textit{Notes:}
The table reports expanding-window one-step-ahead forecasting results
over 52 evaluation periods from November 2021 through February 2026. The response MSE, RMSE, and MAE are multiplied by \(10^3\).
Bold entries indicate the best-performing value in each column.
\end{minipage}
\end{table}

The one-sided Diebold--Mariano tests favor
Post-screening PE over all reported benchmarks. The comparisons with the factor-based methods and Lasso are significant at the 10\% level, whereas that with AR(2) is significant at the 5\% level. Detailed results are provided in Table~\ref{tab:ip-dm-tests}
of the Supplementary Material.
\vspace{-15pt}
\section{Conclusion}\label{sec:conclusion}
This paper develops a supervised factor-augmented matrix regression framework under two-way sparsity-induced weak
loadings. We characterize how target-mode weakness and cross-mode signal dilution affect full-matrix factor
estimation. By recovering the informative row and column
supports, response-guided screening enables improved factor estimation under weaker growth conditions, consistent factor-number determination, and asymptotic inference for unrestricted and bilinear regression estimators.

Natural extensions include approximate sparsity with weak but nonzero loadings outside the informative block and formal theory for more adaptive screening and tuning procedures. Extensions to
higher-order tensor predictors and multivariate responses are also of
interest.

\section*{Supplementary Material}
The supplementary material provides implementation details for the proposed algorithms, proofs of the theoretical results, descriptions of the datasets used in the empirical analysis,
and additional simulation and empirical results.

\section*{Disclosure Statement}
The authors report there are no competing interests to declare.
\clearpage
\appendix

\setcounter{theorem}{4}
\setcounter{lemma}{0}
\setcounter{proposition}{0}
\setcounter{corollary}{2}
\setcounter{definition}{0}
\setcounter{remark}{8}
\setcounter{assumption}{15}
\setcounter{algorithm}{0}
\setcounter{page}{1}

\renewcommand*{\thefootnote}{\fnsymbol{footnote}}
\begin{center}
{\LARGE\bfseries
Supplementary Material for ``Supervised Factor-Augmented Matrix Regression with Weak Factors''}\\

\end{center}
\vspace{0.5cm}

\noindent
This supplementary material provides implementation details for the proposed algorithms, proofs of the theoretical results, descriptions of the datasets used in the empirical analysis,
and additional simulation and empirical results.

\vspace{0.5cm}
	
	\setcounter{section}{0}
	\setcounter{subsection}{0}
	
	\renewcommand{\thesection}{S.\Alph{section}}
	\renewcommand{\thesubsection}{\thesection.\arabic{subsection}}
	
    \numberwithin{equation}{subsection}
	
	\renewcommand{\theequation}{S.\arabic{equation}}%
	\renewcommand{\thefigure}{S.\arabic{figure}} \setcounter{figure}{0}
	\renewcommand{\thetable}{S.\Roman{table}} \setcounter{table}{0}
\section{Algorithms}
\subsection{Factor-Number Determination}
Let \(k_{\max}\) be a fixed upper bound on both factor numbers and let \(c\ge0\) be a small constant. We use the selected-dimension ridge adjustments
\(
a_R
=
\max\left\{
\frac{1}{\sqrt{T\widehat p_2}},
\frac{1}{\sqrt{T\widehat p_1}},
\frac{1}{\widehat p_1}
\right\},
\;
a_C
=
\max\left\{
\frac{1}{\sqrt{T\widehat p_1}},
\frac{1}{\sqrt{T\widehat p_2}},
\frac{1}{\widehat p_2}
\right\}.
\)
The complete factor-number determination procedure is given in Algorithm~\ref{alg:factor-number}.

\begin{algorithm}[htbp]
\caption{Iterative determination of factor numbers on the screened matrix}
\label{alg:factor-number}
\begin{algorithmic}[1]
\Require Screened matrices $\{\bX_t^{\mathrm{scr}}\}_{t=1}^T$, maximum candidate number $k_{\max}$, maximum number of iterations $m$, constant $c$, and ridge adjustments $a_R,a_C$.
\Ensure Estimated row and column factor numbers $\widehat k_1$ and $\widehat k_2$.
\State Initialize
$\widehat k_i^{(0)}
=\min\{k_{\max},\widehat p_i-1\}$,
$i=1,2$.
\For{$s=1,\ldots,m$}
    \State Estimate the column loading space from $\{\bX_t^{\mathrm{scr}}\}$ using $\widehat k_2^{(s-1)}$, and form the row-side projected matrix $\widetilde{\bM}_R^{(s)}$.
    \State Set
    \(
    \widehat k_1^{(s)}
    =
    \argmax_{j\le \min\{k_{\max},\widehat{p}_1-1\}}
    \frac{\lambda_j\paren{\widetilde{\bM}_R^{(s)}}}
    {\lambda_{j+1}\paren{\widetilde{\bM}_R^{(s)}}+c\mkern2mu a_R}.
    \)
    \State Estimate the row loading space from $\{\bX_t^{\mathrm{scr}}\}$ using $\widehat k_1^{(s)}$, and form the column-side projected matrix $\widetilde{\bM}_C^{(s)}$.
    \State Set
    \(
    \widehat k_2^{(s)}
    =
    \argmax_{j\le \min\{k_{\max},\widehat{p}_2-1\}}
    \frac{\lambda_j\paren{\widetilde{\bM}_C^{(s)}}}
    {\lambda_{j+1}\paren{\widetilde{\bM}_C^{(s)}}+c\mkern2mu a_C}.
    \)
    \If{$\widehat k_1^{(s)}=\widehat k_1^{(s-1)}$ and $\widehat k_2^{(s)}=\widehat k_2^{(s-1)}$}
        \State \textbf{break}
    \EndIf
\EndFor
\State Return $\widehat k_1=\widehat k_1^{(s)}$ and $\widehat k_2=\widehat k_2^{(s)}$.
\end{algorithmic}
\end{algorithm}
\subsection{Data-Driven Iterative Screening}
\label{sec:iterative-screening}

As noted in Section~\ref{subsec:screening}, the main procedure is motivated by
the separation between informative and inactive row and column statistics.
Here we provide a data-driven iterative implementation that determines the
retained dimensions directly from the empirical screening statistics rather
than from prespecified thresholds.

The procedure alternates between the two modes. Given the currently selected
columns, all row statistics are recomputed, and the updated row set is then
used to recompute all column statistics. This repeated refinement reduces
cross-mode signal dilution and can sharpen the separation between informative
and inactive coordinates. Because each update considers all indices in the
corresponding mode, previously excluded rows or columns are allowed to re-enter.
The retained dimension at each step is determined by the largest ridge-stabilized
break in the ordered screening statistics. Algorithm~\ref{alg:iterative-screening} summarizes the procedure for either the covariance- or rank-based entrywise screening score.

\begin{algorithm}[htbp]
\caption{Iterative Sequential Screening}
\label{alg:iterative-screening}
\begin{algorithmic}
\Require
Entrywise screening method
$\mathsf{M}\in\{\mathsf{Cov},\mathsf{Rank}\}$;
ridge adjustments $\nu_R,\nu_C> 0$; constant \(c\); upper bounds $p_{R,\max}$ and $p_{C,\max}$;
maximum number of iterations $L_{\max}$.
\Ensure
$\widehat{\mathcal I}_R$ and
$\widehat{\mathcal I}_C$.
\State Define the entrywise screening score
\(
\widehat q_{ij}
=
\begin{cases}
\widehat s_{ij},
& \mathsf{M}=\mathsf{Cov},\\
|\widehat\rho_{ij}|,
& \mathsf{M}=\mathsf{Rank},
\end{cases}
\;
i\in[p_1],\ j\in[p_2].
\)
\State Initialize
\(
\widehat{\mathcal I}_R^{(0)}=[p_1],
\;
\widehat{\mathcal I}_C^{(0)}=[p_2].
\)
\For{$\ell=1,\ldots,L_{\max}$}
    \State \textbf{Row update:} for each $i\in[p_1]$, compute
    \(
    \widehat s_i^{(R,\ell)}
    =
    \frac{1}
    {|\widehat{\mathcal I}_C^{(\ell-1)}|}
    \sum_{j\in\widehat{\mathcal I}_C^{(\ell-1)}}
    \widehat q_{ij}.
    \)
    \State Let
    \(
    \widehat s_{(1)}^{(R,\ell)}
    \ge \cdots \ge
    \widehat s_{(p_1)}^{(R,\ell)}
    \)
    denote the ordered row statistics, with corresponding indices
    $\pi_R^{(\ell)}(1),\ldots,\pi_R^{(\ell)}(p_1)$.
    \State Set
    \(
    \widehat r_\ell
    =
    \arg\max_{1\le r\le p_{R,\max}}
    \frac{
    \widehat s_{(r)}^{(R,\ell)}
    }{
    \widehat s_{(r+1)}^{(R,\ell)}+c\nu_R
    },
    \)
    and update
    \(
    \widehat{\mathcal I}_R^{(\ell)}
    =
    \left\{
    \pi_R^{(\ell)}(1),\ldots,
    \pi_R^{(\ell)}(\widehat r_\ell)
    \right\}.
    \)
    \State \textbf{Column update:} for each $j\in[p_2]$, compute
    \(
    \widehat s_j^{(C,\ell)}
    =
    \frac{1}
    {|\widehat{\mathcal I}_R^{(\ell)}|}
    \sum_{i\in\widehat{\mathcal I}_R^{(\ell)}}
    \widehat q_{ij}.
    \)
    \State Let
    \(
    \widehat s_{(1)}^{(C,\ell)}
    \ge \cdots \ge
    \widehat s_{(p_2)}^{(C,\ell)}
    \)
    denote the ordered column statistics, with corresponding indices
    $\pi_C^{(\ell)}(1),\ldots,\pi_C^{(\ell)}(p_2)$.
    \State Set
    \(
    \widehat c_\ell
    =
    \arg\max_{1\le q\le {p_{C,\max}}}
    \frac{
    \widehat s_{(q)}^{(C,\ell)}
    }{
    \widehat s_{(q+1)}^{(C,\ell)}+c\nu_C
    },
    \)
    and update
    \(
    \widehat{\mathcal I}_C^{(\ell)}
    =
    \left\{
    \pi_C^{(\ell)}(1),\ldots,
    \pi_C^{(\ell)}(\widehat c_\ell)
    \right\}.
    \)
    \If{
    $\widehat{\mathcal I}_R^{(\ell)}
    =
    \widehat{\mathcal I}_R^{(\ell-1)}$
    \textbf{and}
    $\widehat{\mathcal I}_C^{(\ell)}
    =
    \widehat{\mathcal I}_C^{(\ell-1)}$
    }
        \State \textbf{break}
    \EndIf
\EndFor
\State Set
\(
\widehat{\mathcal I}_R
=
\widehat{\mathcal I}_R^{(\ell)},
\;
\widehat{\mathcal I}_C
=
\widehat{\mathcal I}_C^{(\ell)}.
\)
\State \Return
$\widehat{\mathcal I}_R,\widehat{\mathcal I}_C$.
\end{algorithmic}
\end{algorithm}
The finite-sample performance of the data-driven iterative procedure is
examined in Section~\ref{sec:iterative-screening-simulation}.

\section{Proofs}\label{sec:appendix}
Throughout the appendix, Roman-numeral labels, with or without subscripts, are
used locally to denote intermediate terms in the decompositions. For instance,
the labels \(II_1\) and \(III_2\) are interpreted only within the decomposition
where they are introduced and may represent different quantities in different
proofs. We also use \(C\) to denote a generic positive constant whose value may
change from line to line.
\setcounter{subsection}{0}
\renewcommand{\theequation}{\thesubsection.\arabic{equation}}
\setcounter{subsubsection}{0}
\renewcommand{\thesubsubsection}{\thesubsection.\arabic{subsubsection}}
\subsection{\texorpdfstring{$\alpha$-PCA Estimation}{alpha-PCA}}
\setcounter{equation}{0}
In this section, the estimators $\widehat{\bR}$ and $\widehat{\bC}$ are defined as the matrices of 
$\sqrt{p_1^{\delta_1}}$ and $\sqrt{p_2^{\delta_2}}$ times the top $k_1$ and $k_2$ eigenvectors of
\begin{equation}
\widehat{\bM}_R = \frac{1}{p_1^{\delta_1} p_2^{\delta_2} T} \sum_{t=1}^T \bX_t \bX_t^\top,
\qquad
\widehat{\bM}_C = \frac{1}{p_1^{\delta_1} p_2^{\delta_2} T} \sum_{t=1}^T \bX_t^\top \bX_t,
\end{equation}
respectively, arranged in descending order of their corresponding eigenvalues.
\begin{equation}
\begin{aligned}
\widehat{\bM}_R
&=
\frac{1}{p_1^{\delta_1} p_2^{\delta_2} T}
\sum_{t=1}^T \bR \bF_t \bC' \bC \bF_t' \bR'+\frac{1}{p_1^{\delta_1} p_2^{\delta_2} T}
\sum_{t=1}^T \bR \bF_t \bC' \bE_t'\\
&\quad+
\frac{1}{p_1^{\delta_1} p_2^{\delta_2} T}
\sum_{t=1}^T \bE_t \bC \bF_t' \bR'
+
\frac{1}{p_1^{\delta_1} p_2^{\delta_2} T}
\sum_{t=1}^T \bE_t \bE_t'\\
&:=I+ II + III + IV.
\end{aligned}
\end{equation}
Let $\boldsymbol{\Lambda}_R \in \mathbb{R}^{k_1 \times k_1}$ and 
$\boldsymbol{\Lambda}_C \in \mathbb{R}^{k_2 \times k_2}$ denote the diagonal matrices 
containing the largest $k_1$ and $k_2$ eigenvalues of $\widehat{\bM}_R$ and 
$\widehat{\bM}_C$, respectively, in decreasing order. By the definition of 
eigenvalues and eigenvectors, we have
\begin{equation}
\frac{1}{p_1^{\delta_1} p_2^{\delta_2} T} \sum_{t=1}^T \bX_t \bX_t^\top \widehat{\bR}
=
\widehat{\bR} \boldsymbol{\Lambda}_R,
\qquad
\frac{1}{p_1^{\delta_1} p_2^{\delta_2} T} \sum_{t=1}^T \bX_t^\top \bX_t \widehat{\bC}
=
\widehat{\bC} \boldsymbol{\Lambda}_C.
\end{equation}
Define $\bH_R \in \mathbb{R}^{k_1 \times k_1}$ and 
$\bH_C \in \mathbb{R}^{k_2 \times k_2}$ as
\begin{equation}
\begin{aligned}
&\bH_R = \frac{1}{p_1^{\delta_1} p_2^{\delta_2} T} \sum_{t=1}^T 
\bF_t \bC^\top \bC \bF_t^\top \bR^\top \widehat{\bR}\, \boldsymbol{\Lambda}_R^{-1},\\
&\bH_C = \frac{1}{p_1^{\delta_1} p_2^{\delta_2} T} \sum_{t=1}^T 
\bF_t^\top \bR^\top \bR \bF_t \bC^\top \widehat{\bC}\, \boldsymbol{\Lambda}_C^{-1}.
\end{aligned}
\end{equation}
Then we obtain
\begin{equation}
\begin{aligned}
&\quad\;\widehat{\bR} - \bR \bH_R\\
&=
\frac{1}{p_1^{\delta_1} p_2^{\delta_2} T}
\left(
\sum_{t=1}^T \bX_t \bX_t^\top \widehat{\bR}
-
\sum_{t=1}^T \bF_t \bC^\top \bC \bF_t^\top \bR^\top \widehat{\bR}
\right)
\boldsymbol{\Lambda}_R^{-1}\\
&=
\left(
\frac{1}{p_1^{\delta_1} p_2^{\delta_2} T} \sum_{t=1}^T \bR \bF_t \bC^\top \bE_t^\top \widehat{\bR}
+
\frac{1}{p_1^{\delta_1} p_2^{\delta_2} T} \sum_{t=1}^T \bE_t \bC \bF_t^\top \bR^\top \widehat{\bR}
+
\frac{1}{p_1^{\delta_1} p_2^{\delta_2} T} \sum_{t=1}^T \bE_t \bE_t^\top \widehat{\bR}
\right)
\boldsymbol{\Lambda}_R^{-1},
\end{aligned}
\end{equation}
Similarly,
\begin{equation}
\begin{aligned}
&\hspace{1.28em}\widehat{\bC} - \bC \bH_C\\
&=
\frac{1}{p_1^{\delta_1} p_2^{\delta_2} T}
\left(
\sum_{t=1}^T \bX_t^\top \bX_t \widehat{\bC}
-
\sum_{t=1}^T \bF_t^\top \bR'\bR\bF_t \bC^\top \widehat{\bC}
\right)
\boldsymbol{\Lambda}_C^{-1}\\
&=
\left(
\frac{1}{p_1^{\delta_1} p_2^{\delta_2} T} \sum_{t=1}^T \bC \bF_t^\top \bR^\top \bE_t \widehat{\bC}
+
\frac{1}{p_1^{\delta_1} p_2^{\delta_2} T} \sum_{t=1}^T \bE_t^\top \bR \bF_t \bC^\top \widehat{\bC}
+
\frac{1}{p_1^{\delta_1} p_2^{\delta_2} T} \sum_{t=1}^T \bE_t^\top \bE_t \widehat{\bC}
\right)
\boldsymbol{\Lambda}_C^{-1}.
\end{aligned}
\end{equation}
We begin by introducing several technical lemmas that are instrumental for deriving the asymptotic properties of the \(\alpha\)-PCA estimators. While analogous lemmas have been provided in \cite{ChenFan_JASA_2023}, the present lemmas generalize these results to accommodate the weak-loading setting.
\subsubsection{Technical Lemmas}
\begin{lemma}\label{lemma:lln for alphaPCA}
Under Assumptions~\ref{ass:weak loading}--\ref{ass:Factor-noise interaction for alpha-PCA}, we have
\[
\frac{1}{p_2^{\delta_2} \, T} \sum_{t=1}^T \bF_t \bC' \bC \bF_t'\xrightarrow[]{p}\bSigma_{1}
\quad \text{and} \quad
\frac{1}{p_1^{\delta_1} \, T} \sum_{t=1}^T \bF_t' \bR' \bR \bF_t\xrightarrow[]{p} \bSigma_{2}
\]
where \(\bSigma_1\) and \(\bSigma_2\) are given in Assumption~\ref{ass:factor}.
\end{lemma}

\begin{proof}
Since $\{\bF_t\}$ is strictly stationary and $\alpha$-mixing by Assumption~\ref{ass:mixing-noise}, and has finite moments by Assumption~\ref{ass:factor}, a law of large numbers for $\alpha$-mixing sequences applies. In particular, by Proposition 2.8 of \cite{FanYao_2003} and Assumption~\ref{ass:weak loading},
\[
\frac{1}{p_2^{\delta_2} \, T} \sum_{t=1}^T \bF_t \bC' \bC \bF_t'\xrightarrow[]{p}\bSigma_{1}
\]
The remaining results follow similarly by applying the same argument.
\end{proof}

\begin{lemma}\label{lem:row_noise_bound for alphaPCA}
Under Assumption~\ref{ass:weak loading}--\ref{ass:Factor-noise interaction for alpha-PCA}, there exists a constant $C>0$ such that, for all $i=1,\dots,p_1$ and $j=1,\dots,p_2$,
\[
\mathbb{E}\left\|
\frac{1}{\sqrt{p_1^{\delta_1} p_2 T}}
\sum_{t=1}^T \sum_{l=1}^{p_1} \sum_{h=1}^{p_2}
\bR_{l\cdot}
\left(
\be_{t,lh} \be_{t,ih} - \mathbb{E}[\be_{t,lh} \be_{t,ih}]
\right)
\right\|^2
\le C,
\]
and
\[
\mathbb{E}\left\|
\frac{1}{\sqrt{p_1 p_2^{\delta_2} T}}
\sum_{t=1}^T \sum_{l=1}^{p_1} \sum_{h=1}^{p_2}
\bC_{h\cdot}
\left(
\be_{t,lh} \be_{t,lj} - \mathbb{E}[\be_{t,lh} \be_{t,lj}]
\right)
\right\|^2
\le C.
\]
\end{lemma}

\begin{proof}
For simplicity, we take \(k_1=1\). We establish the first bound only, as the second follows by a symmetric argument.
\[
\begin{aligned}
&\hspace{1.28em}\mathbb{E}\left\|
\frac{1}{\sqrt{p_1^{\delta_1} p_2 T}}
\sum_{t=1}^T \sum_{l=1}^{p_1} \sum_{h=1}^{p_2}
\bR_{l\cdot}
\left(
\be_{t,lh} \be_{t,ih} - \mathbb{E}[\be_{t,lh} \be_{t,ih}]
\right)
\right\|^2\\
&=\mathbb{E}\paren{
\frac{1}{\sqrt{p_1^{\delta_1} p_2 T}}
\sum_{t=1}^T \sum_{l=1}^{p_1} \sum_{h=1}^{p_2}
\bR_{l\cdot}
\left(
\be_{t,lh} \be_{t,ih} - \mathbb{E}[\be_{t,lh} \be_{t,ih}]
\right)
}^2\\
&\le \frac{1}{p_1^{\delta_1}p_2T}\sum_{s,t}\sum_{l_1,l_2}\sum_{h_1,h_2}\abs{\bR_{l_1\cdot } \bR_{l_2\cdot}}\abs{\mathrm{Cov\paren{\be_{t,l_1h_1}\be_{t,ih_1},\be_{s,l_2h_2}\be_{s,ih_2}}}}\\
\le C
\end{aligned}
\]
The last inequality follows from Assumption~\ref{ass:weak loading} and Assumption~\ref{ass:noise-dependence}.
\end{proof}
\begin{lemma}\label{lemma: FCe for alphaPCA}
Under Assumption~\ref{ass:weak loading}--\ref{ass:Factor-noise interaction for alpha-PCA}, there exists a constant $C>0$ such that, for all $i=1,\dots,p_1$ and $j=1,\dots,p_2$,
\[
\mathbb{E} \Bigg\| \frac{1}{\sqrt{T \, p_2^{\delta_2}}} \sum_{t=1}^T \mathbf{F}_t \mathbf{C}' \mathbf{e}_{t,i\cdot} \Bigg\|^2 \le C
\quad and \quad
\mathbb{E} \Bigg\| \frac{1}{\sqrt{T \, p_1^{\delta_1}}} \sum_{t=1}^T \mathbf{F}_t' \mathbf{R}' \mathbf{e}_{t,\cdot{j}} \Bigg\|^2 \le C
\]
\end{lemma}
\begin{proof}
We only prove the first result; the second follows analogously.

Set
\(
\mathbf{x}_t := \frac{1}{\sqrt{p_2^{\delta_2}}} \mathbf{C}' \mathbf{e}_{t,i\cdot}.
\)
Then
\(
\frac{1}{\sqrt{T p_2^{\delta_2}}} \sum_{t=1}^T \mathbf{F}_t \mathbf{C}' \mathbf{e}_{t,i\cdot}
=
\frac{1}{\sqrt{T}} \sum_{t=1}^T \mathbf{F}_t \mathbf{x}_t.
\)
For each fixed $l \in \{1,\dots,k_1\}$, the $l$-th entry of $\mathbf{F}_t \mathbf{x}_t$ is given by
$
\mathbf{f}_{t,l\cdot}' \mathbf{x}_t
$.
Since $k_1$ is fixed, it suffices to show that for each $l$,
\(
\mathbb{E}\left|
\frac{1}{\sqrt{T}} \sum_{t=1}^T \mathbf{f}_{t,\ell\cdot}' \mathbf{x}_t
\right|^2 \le C.
\)
By Assumption~\ref{ass:mixing-noise}, we have $\mathbb{E}[\mathbf{f}_{t,\ell\cdot}' \mathbf{x}_t] = 0$,
hence
\(
\mathbb{E}\left|
\frac{1}{\sqrt{T}} \sum_{t=1}^T \mathbf{f}_{t,\ell\cdot}' \mathbf{x}_t
\right|^2
=
\var\left(
\frac{1}{\sqrt{T}} \sum_{t=1}^T \mathbf{f}_{t,\ell\cdot}' \mathbf{x}_t
\right).
\)
By Hölder's inequality and Assumption~\ref{ass:Factor-noise interaction for alpha-PCA}, we have
\(
\mathbb{E}| \mathbf{f}_{t,\ell\cdot}' \mathbf{x}_t|^\gamma < \infty.
\)
Together with Assumption~\ref{ass:mixing-noise},
the covariance inequality for $\alpha$-mixing sequences implies
\(
\sum_{h=1}^\infty |\gamma(h)| < \infty,
\)
where $\gamma(h)=\mathrm{Cov}(\mathbf{f}_{t,l\cdot}' \mathbf{x}_t,\mathbf{f}_{t+h,l\cdot}' \mathbf{x}_{t+h})$. Moreover, Theorem 2.20 of \cite{FanYao_2003} implies
\(
\frac{1}{T} \var\left(\sum_{t=1}^T \mathbf{f}_{t,\ell\cdot}' \mathbf{x}_t\right)
\to
\gamma(0) + 2\sum_{h=1}^\infty \gamma(h).
\)
Therefore,
\(
\mathbb{E}\left|
\frac{1}{\sqrt{T}} \sum_{t=1}^T \mathbf{f}_{t,\ell\cdot}' \mathbf{x}_t
\right|^2=\mathrm{Var}\left(
\frac{1}{\sqrt{T}} \sum_{t=1}^T \mathbf{f}_{t,\ell\cdot}' \mathbf{x}_t
\right)
= O\paren{1},
\)
which yields
\(
\mathbb{E}\left\|
\frac{1}{\sqrt{T p_2^{\delta_2}}}
\sum_{t=1}^T \mathbf{F}_t \mathbf{C}' \mathbf{e}_{t,i\cdot}
\right\|^2 = O\paren{1}.
\)
\end{proof}
\begin{corollary}\label{corollary: FCe for alphaPCA}
Under Assumptions~\ref{ass:weak loading}–\ref{ass:Factor-noise interaction for alpha-PCA}, we have
\[
\left\|
\sum_{t=1}^T \mathbf{F}_t \mathbf{C}' \mathbf{E}_t'
\right\|_F^2
= O_p\!\left(p_1 p_2^{\delta_2} T\right).
\]
\end{corollary}
\begin{proof}
\[
\sum_{t=1}^T \mathbf{F}_t \mathbf{C}' \mathbf{E}_t'
=
\sum_{t=1}^T \sum_{j=1}^{p_2} \mathbf{F}_t \mathbf{C}_{j\cdot} \mathbf{e}_{t,\cdot j}'.
\]
Using the identity $\mathrm{vec}(\mathbf{A}\mathbf{X}\mathbf{B}') = \paren{\mathbf{B}\otimes \mathbf{A}}\vect\paren{\mathbf{X}}$, we obtain
\[
\vect(\mathbf{F}_t \mathbf{C}_{j\cdot} \mathbf{e}_{t,\cdot j}')
=
(\mathbf{e}_{t,\cdot j} \otimes \mathbf{F}_t) \mathbf{C}_{j\cdot}.
\]

Therefore,
\[
\mathrm{vec}\!\left(\sum_{t=1}^T \mathbf{F}_t \mathbf{C}' \mathbf{E}_t'\right)
=
\sum_{t=1}^T \sum_{j=1}^{p_2} (\mathbf{e}_{t,\cdot j} \otimes \mathbf{F}_t) \mathbf{C}_{j\cdot}.
\]
\[
\begin{aligned}
\left\|
\sum_{t=1}^T \mathbf{F}_t \mathbf{C}' \mathbf{E}_t'
\right\|_F^2
&=
\left\|
\sum_{t=1}^T \sum_{j=1}^{p_2}
(\mathbf{e}_{t,\cdot j} \otimes \mathbf{F}_t)\, \mathbf{C}_{j\cdot}
\right\|_F^2 \\
&=
\sum_{i=1}^{p_1}
\left\|
\sum_{t=1}^T \sum_{j=1}^{p_2}
e_{t,ij}\, \mathbf{F}_t \mathbf{C}_{j\cdot}
\right\|_F^2 \\
&=
T p_2^{\delta_2} \sum_{i=1}^{p_1}
\left\|
\frac{1}{\sqrt{T p_2^{\delta_2}}}
\sum_{t=1}^T \mathbf{F}_t \mathbf{C}' \mathbf{e}_{t,i\cdot}
\right\|_F^2 \\
&=
O_p\!\left(T p_1 p_2^{\delta_2}\right).
\end{aligned}
\]
where the final bound follows from Lemma~\ref{lemma: FCe for alphaPCA} and Markov's inequality.
\end{proof}
\begin{lemma}\label{lemma: EtEt'-entry for alphaPCA}
\medskip

(a) For any $i,l \in [p_1]$ and $j,h \in [p_2]$,
\[
\mathbb{E}\left(
\frac{1}{\sqrt{p_2 T}} \sum_{t=1}^T \sum_{j=1}^{p_2} 
\bigl(e_{t,ij} e_{t,lj} - \mathbb{E}[e_{t,ij} e_{t,lj}]\bigr)
\right)^2
= O(1),
\]

\[
\mathbb{E}\left(
\frac{1}{\sqrt{p_1 T}} \sum_{t=1}^T \sum_{i=1}^{p_1} 
\bigl(e_{t,ij} e_{t,ih} - \mathbb{E}[e_{t,ij} e_{t,ih}]\bigr)
\right)^2
= O(1),
\]

\[
\mathbb{E}\left\|
\frac{1}{p_1 \sqrt{p_2 T}} \sum_{t=1}^T 
\bigl(\mathbf{E}_t \mathbf{E}_t^\top - \mathbb{E}[\mathbf{E}_t \mathbf{E}_t^\top]\bigr)
\right\|_F^2
= O(1),
\]

\[
\mathbb{E}\left\|
\frac{1}{p_2 \sqrt{p_1 T}} \sum_{t=1}^T 
\bigl(\mathbf{E}_t^\top \mathbf{E}_t - \mathbb{E}[\mathbf{E}_t^\top \mathbf{E}_t]\bigr)
\right\|_F^2
= O(1).
\]

\medskip

(b) For all $1 \le t \le T$,
\[
\mathbb{E}\left\|
\frac{1}{\sqrt{p_1^{\delta_1} p_2^{\delta_2}}} \mathbf{R}' \mathbf{E}_t \mathbf{C}
\right\|^2
= O(1).
\]

\end{lemma}
\begin{proof}
(a) By symmetry, it suffices to prove the first and third equations.

\[
\begin{aligned}
\mathbb{E}\Bigg[
\frac{1}{\sqrt{p_2 T}}
\sum_{t=1}^T \sum_{j=1}^{p_2}
\left(
e_{t,ij} e_{t,lj}
- \mathbb{E}[e_{t,ij} e_{t,lj}]
\right)
\Bigg]^2
&= \frac{1}{p_2 T}
\sum_{t=1}^T \sum_{j=1}^{p_2}
\cov\!\left(
e_{t,ij} e_{t,lj},\,
e_{t,ij} e_{t,lj}
\right) \\
&\quad + \frac{1}{p_2 T}
\sum_{t=1}^T \sum_{j=1}^{p_2}
\sum_{s\ne t}
\cov\!\left(
e_{t,ij} e_{t,lj},\,
e_{s,ij} e_{s,lj}
\right) \\
&\quad + \frac{1}{p_2 T}
\sum_{s=1}^T \sum_{t=1}^T \sum_{j=1}^{p_2}
\sum_{h\ne j}
\cov\!\left(
e_{t,ij} e_{t,lj},\,
e_{s,ih} e_{s,lh}
\right) \\
&:= I_1 + I_2 + I_3.
\end{aligned}
\]
We first bound $I_1$. By the Cauchy--Schwarz inequality,
\[
\left|
\cov\!\left(
e_{t,ij} e_{t,lj},\,
e_{t,ij} e_{t,lj}
\right)
\right|
\le
\mathbb{E}\left[e_{t,ij}^2 e_{t,lj}^2\right]
\le
\mathbb{E}\left[e_{t,ij}^4\right]^{1/2}
\mathbb{E}\left[e_{t,lj}^4\right]^{1/2}
\le C.
\]
Therefore, $I_1 = O(1)$.
Next, we consider $I_2$. By Davydov’s inequality (Corollary 16.2.4 in \cite{AthreyaLahiri_2006}) together with Assumption~\ref{ass:mixing-noise}, 
there exists a constant $C>0$ such that, for $s\ne t$,
\[
\left|
\cov\!\left(
e_{t,ij} e_{t,lj},\,
e_{s,ij} e_{s,lj}
\right)
\right|
\le
C\,\alpha(|s-t|)^{1-2/\gamma}.
\]
Hence,
\[
\sum_{s\ne t}
\left|
\cov\!\left(
e_{t,ij} e_{t,lj},\,
e_{s,ij} e_{s,lj}
\right)
\right|
\le
\sum_{s\ne t} C\,\alpha(|s-t|)^{1-2/\gamma}
< \infty,
\]
which implies $I_2 = O\paren{1}$.
Finally, we bound $I_3$. By Assumption~\ref{ass:noise-dependence},
\[
\sum_{s=1}^T
\sum_{h\ne j}
\cov\!\left(
e_{t,ij} e_{t,lj},\,
e_{s,ih} e_{s,lh}
\right)
\le
\sum_{m \in [p_1]} \sum_{s \in [T]} \sum_{\substack{h \in [p_2],\, h \neq j}}
\left|
\cov\!\left(
e_{t,ij} e_{t,lj},\,
e_{s,ih} e_{s,mh}
\right)
\right|
\le C,
\]
which yields $I_3 = O\paren{1}$.
Combining the above bounds, we obtain
\[
\mathbb{E}\Bigg[
\frac{1}{\sqrt{p_2 T}}
\sum_{t=1}^T \sum_{j=1}^{p_2}
\left(
e_{t,ij} e_{t,lj}
- \mathbb{E}[e_{t,ij} e_{t,lj}]
\right)
\Bigg]^2
= O\paren{1}.
\]
Next, we consider the matrix term. By expanding the Frobenius norm,
\[
\mathbb{E}\left\|
\frac{1}{p_1 \sqrt{p_2 T}} \sum_{t=1}^T 
\left(\mathbf{E}_t \mathbf{E}_t^\top - \mathbb{E}[\mathbf{E}_t \mathbf{E}_t^\top]\right)
\right\|_F^2
=
\frac{1}{p_1^2 p_2 T}
\sum_{i=1}^{p_1} \sum_{l=1}^{p_1}
\mathbb{E}\left[
\left(
\sum_{t=1}^T \sum_{j=1}^{p_2}
\left(
e_{t,ij} e_{t,lj}
- \mathbb{E}[e_{t,ij} e_{t,lj}]
\right)
\right)^2
\right]
= O\paren{1}.
\]
(b) Finally, we prove part (b). By expanding the product,
\[
\mathbb{E}\left\|
\frac{1}{\sqrt{p_1^{\delta_1} p_2^{\delta_2}}} \mathbf{R}' \mathbf{E}_t \mathbf{C}
\right\|^2
=
\mathbb{E}\left\|
\frac{1}{\sqrt{p_1^{\delta_1} p_2^{\delta_2}}}
\sum_{i=1}^{p_1} \sum_{j=1}^{p_2}
\mathbf{R}_{i\cdot}\, e_{t,ij}\, \mathbf{C}_{j\cdot}'
\right\|^2
= O\paren{1},
\]
where the last bound follows directly from Assumption~\ref{ass:Factor-noise interaction for alpha-PCA}.
This completes the proof.
\end{proof}
\begin{lemma}\label{lemma: RECF for alphaPCA}
Under Assumptions~\ref{ass:weak loading}–\ref{ass:Factor-noise interaction for alpha-PCA}, we have
\[
\mathbb{E}\left\|
\frac{1}{\sqrt{p_1^{\delta_1} p_2^{\delta_2} T}}
\sum_{t=1}^T \mathbf{R}^\top \mathbf{E}_t \mathbf{C} \mathbf{F}_t^\top
\right\|^2
= O\paren{1},
\]
and similarly,
\[
\mathbb{E}\left\|
\frac{1}{\sqrt{p_1^{\delta_1} p_2^{\delta_2} T}}
\sum_{t=1}^T \mathbf{F}_t^\top \mathbf{R}^\top \mathbf{E}_t \mathbf{C}
\right\|^2
= O\paren{1}.
\]
\end{lemma}
\begin{proof}
By symmetry, it suffices to prove the first equation.
Analogous arguments as in the proof of Lemma~\ref{lemma: FCe for alphaPCA}, define 
\(
\mathbf{U}_t := \frac{1}{\sqrt{p_1^{\delta_1} p_2^{\delta_2} }}
 \mathbf{R}^\top \mathbf{E}_t \mathbf{C}=\frac{1}{\sqrt{p_1^{\delta_1} p_2^{\delta_2} }}\sum_{i=1}^{p_1}\sum_{j=1}^{p_2}\mathbf{R}_{i\cdot}e_{t,ij}\mathbf{C}_{j\cdot}'.
\)
Then
\(
\frac{1}{\sqrt{p_1^{\delta_1} p_2^{\delta_2} T}}
\sum_{t=1}^T \mathbf{R}^\top \mathbf{E}_t \mathbf{C} \mathbf{F}_t^\top
=
\frac{1}{\sqrt{T}}\sum_{t=1}^T\mathbf{U}_t\mathbf{F}_t^\top.
\)
Since $k_1$ is fixed, it suffices that, for each $l$,
\(
\mathbb{E}\left\|
\frac{1}{\sqrt{T}} \sum_{t=1}^T \mathbf{U}_t\mathbf{f}_{t,l\cdot}
\right\|^2 = O\paren{1}.
\)
By Assumption~\ref{ass:mixing-noise}, we have $\mathbb{E}[\mathbf{U}_t\mathbf{f}_{t,l\cdot}] = 0$.
It follows that, by Assumption~\ref{ass:Factor-noise interaction for alpha-PCA}, we have
\(
\E\norm{ \mathbf{U}_t\mathbf{f}_{t,l\cdot}}^{\gamma}<\infty.
\)
Hence,
\(
\E\abs{\mathbf{u}_{t,i\cdot}'\mathbf{f}_{t,l\cdot}}^{\gamma}\le\E\norm{\mathbf{U}_t\mathbf{f}_{t,l\cdot}}^{\gamma}<\infty
\).
By strict stationarity and $\alpha$-mixing in Assumption~\ref{ass:mixing-noise},
Theorem 2.20 of \cite{FanYao_2003} implies
\(
\var\left(
\frac{1}{\sqrt{T}} \sum_{t=1}^T \mathbf{u}_{t,i\cdot}'\mathbf{f}_{t,l\cdot}
\right)
\to\gamma\paren{0}+2\sum_{h=1}^{\infty}\gamma\paren{h}
\),
where $\gamma\paren{h}=\mathrm{Cov}\paren{\mathbf{u}_{t,i\cdot}'\mathbf{f}_{t,l\cdot},\mathbf{u}_{t+h,i\cdot}'\mathbf{f}_{t+h,l\cdot}}$.
Thus,
\[
\E\norm{\frac{1}{\sqrt{T}} \sum_{t=1}^T \mathbf{U}_t\mathbf{f}_{t,l\cdot}}^2\le\E\sum_{i=1}^{k_1}\abs{\frac{1}{\sqrt{T}} \sum_{t=1}^T\mathbf{u}_{t,i\cdot}'\mathbf{f}_{t,l\cdot}}^2=\sum_{i=1}^{k_1}{\var\paren{\frac{1}{\sqrt{T}} \sum_{t=1}^T \mathbf{u}_{t,i\cdot}'\mathbf{f}_{t,l\cdot}}}=O\paren{1}
\]
This yields
\[
\mathbb{E}\left\|
\frac{1}{\sqrt{p_1^{\delta_1} p_2^{\delta_2} T}}
\sum_{t=1}^T \mathbf{R}^\top \mathbf{E}_t \mathbf{C} \mathbf{F}_t^\top
\right\|^2
= O\paren{1}.
\]
\end{proof}
\begin{lemma}\label{lemma:FCEEC for alphaPCA}
Under Assumptions~\ref{ass:weak loading}–\ref{ass:Factor-noise interaction for alpha-PCA}, we have
\[
\norm{\sum_{s=1}^T \mathbf{F}_s \mathbf{C}' \mathbf{E}_s' \mathbf{E}_t \mathbf{C}}
=
O_p\paren{
T\sqrt{p_1p_2^{3\delta_2}}+p_1p_2^{\delta_2}
}.
\]
\end{lemma}
\begin{proof}
\[
\begin{aligned}
\norm{\sum_{s=1}^T \mathbf{F}_s \mathbf{C}' \mathbf{E}_s' \mathbf{E}_t \mathbf{C}}
&\le
\norm{
\sum_{s=1}^T \sum_{j=1}^{p_2}
\mathbf{F}_s \mathbf{C}_{j\cdot} \mathbf{e}_{s,\cdot j}'
\sum_{h=1}^{p_2} \mathbf{e}_{t,\cdot h} \mathbf{C}_{h\cdot}'
} \\
&=
\norm{
\sum_{s=1}^T \sum_{j=1}^{p_2}\sum_{h=1}^{p_2}\sum_{i=1}^{p_1}
\mathbf{F}_s \mathbf{C}_{j\cdot} e_{s,ij} e_{t,ih} \mathbf{C}_{h\cdot}'
}\\
&\le
\sum_{s=1}^T
\norm{
\sum_{j=1}^{p_2}\sum_{h=1}^{p_2}\sum_{i=1}^{p_1}
\mathbf{F}_s \mathbf{C}_{j\cdot}
\paren{e_{s,ij}e_{t,ih}-\E e_{s,ij}e_{t,ih}}
\mathbf{C}_{h\cdot}'
}  \\
&\quad+
\norm{
\sum_{s=1}^T \sum_{j=1}^{p_2}\sum_{h=1}^{p_2}\sum_{i=1}^{p_1}
\mathbf{F}_s \mathbf{C}_{j\cdot}
\E e_{s,ij}e_{t,ih}
\mathbf{C}_{h\cdot}'
} \\
&:= I+II.
\end{aligned}
\]

By Assumption~\ref{ass:noise-dependence},
\[
\begin{aligned}
\E II
&\le
\sum_{s=1}^T \sum_{j,h=1}^{p_2}\sum_{i=1}^{p_1}
\E\norm{\mathbf{F}_s}\norm{\mathbf{C}_{j\cdot}}\norm{\mathbf{C}_{h\cdot}}
\abs{\E e_{s,ij}e_{t,ih}} \\
&\le
C p_1p_2^{\delta_2}.
\end{aligned}
\]

For simplicity, we assume \(k_1=k_2=1\),
\[
\E\norm{
\sum_{j=1}^{p_2}\sum_{h=1}^{p_2}\sum_{i=1}^{p_1}
\mathbf{C}_{j\cdot}
\paren{e_{s,ij}e_{t,ih}-\E e_{s,ij}e_{t,ih}}
\mathbf{C}_{h\cdot}'
}^2
\]
\[
\begin{aligned}
&\le
\sum_{j_1,j_2=1}^{p_2}
\sum_{h_1,h_2=1}^{p_2}
\sum_{i_1,i_2=1}^{p_1}
\abs{
\mathbf{C}_{j_1\cdot}\mathbf{C}_{j_2\cdot}
\mathbf{C}_{h_1\cdot}'\mathbf{C}_{h_2\cdot}'
} \\
&\quad\times
\abs{
\mathrm{Cov}\paren{
e_{s,i_1j_1}e_{t,i_1h_1},
e_{s,i_2j_2}e_{t,i_2h_2}
}
}.
\end{aligned}
\]

By Assumption~\ref{ass:alpha_spatial_noise},
\[
\le
\sum_{j_1,j_2\in \mathcal{I}_C}
\sum_{h_1\in \mathcal{I}_C}
\sum_{i_1=1}^{p_1} C
=
O\paren{p_1p_2^{3\delta_2}}.
\]

\[
\begin{aligned}
&\E\norm{
\sum_{j=1}^{p_2}\sum_{h=1}^{p_2}\sum_{i=1}^{p_1}
\mathbf{F}_s \mathbf{C}_{j\cdot}
\paren{e_{s,ij}e_{t,ih}-\E e_{s,ij}e_{t,ih}}
\mathbf{C}_{h\cdot}'
} \\
&\le
\sqrt{\E\norm{\mathbf{F}_s}^2}
\sqrt{
\E\norm{
\sum_{j=1}^{p_2}\sum_{h=1}^{p_2}\sum_{i=1}^{p_1}
\mathbf{C}_{j\cdot}
\paren{e_{s,ij}e_{t,ih}-\E e_{s,ij}e_{t,ih}}
\mathbf{C}_{h\cdot}'
}^2
} \\
&\le
C\sqrt{p_1p_2^{3\delta_2}}.
\end{aligned}
\]
Therefore, by Markov's inequality,
\[
I
=
O_p\left(T\sqrt{p_1p_2^{3\delta_2}}\right).
\]
This completes the proof.
\end{proof}
{\bf Proof of Theorem~\ref{thm:alpha-pca-estimation}\textup{(I)}.}
We begin by presenting an auxiliary lemma and a proposition.
\begin{lemma}\label{lemma: EtEt' for alphaPCA}
Under Assumptions~\ref{ass:weak loading}–\ref{ass:Factor-noise interaction for alpha-PCA}, we have
\[
\left\| \sum_{t=1}^T \mathbf{E}_t \mathbf{E}_t' \right\|_F^2
=
O_p\left(p_1^2 p_2 T\right)
+
O_p\left(p_1 p_2^2 T^2\right).
\]
\end{lemma}

\begin{proof}
By definition,
\[
\begin{aligned}
\mathbb{E}\left\| \sum_{t=1}^T \mathbf{E}_t \mathbf{E}_t' \right\|_F^2
&=
\mathbb{E}\sum_{i,j}
\left(
\sum_{t=1}^T \mathbf{e}_{t,i\cdot}' \mathbf{e}_{t,j\cdot}
\right)^2  \\
&=
\mathbb{E}\sum_{i,j}
\left(
\sum_{t,k} e_{t,ik} e_{t,jk}
\right)^2 .
\end{aligned}
\]

Write
\[
\sum_{t,k} e_{t,ik} e_{t,jk}
=
\sum_{t,k}
\bigl(
e_{t,ik} e_{t,jk}
-
\mathbb{E}(e_{t,ik} e_{t,jk})
\bigr)
+
\sum_{t,k}
\mathbb{E}(e_{t,ik} e_{t,jk}).
\]
Expanding the square and taking expectations, the cross term vanishes. Hence,
\[
\begin{aligned}
\mathbb{E}\left\| \sum_{t=1}^T \mathbf{E}_t \mathbf{E}_t' \right\|_F^2
&=
\mathbb{E}\sum_{i,j}
\left(
\sum_{t,k}
\bigl(
e_{t,ik} e_{t,jk}
-
\mathbb{E}(e_{t,ik} e_{t,jk})
\bigr)
\right)^2  \\
&\quad+
\mathbb{E}\sum_{i,j}
\left(
\sum_{t,k}
\mathbb{E}(e_{t,ik} e_{t,jk})
\right)^2  \\
&=
O\left(p_1^2 p_2 T\right)
+
O\left(p_1 p_2^2 T^2\right),
\end{aligned}
\]
where the last equality follows from Assumption~\ref{ass:alpha_spatial_noise}
and Lemma~\ref{lemma: EtEt'-entry for alphaPCA}. By Markov's inequality,
\[
\left\| \sum_{t=1}^T \mathbf{E}_t \mathbf{E}_t' \right\|_F^2
=
O_p\left(p_1^2 p_2 T\right)
+
O_p\left(p_1 p_2^2 T^2\right).
\]
This completes the proof.
\end{proof}
\begin{proposition}\label{pro:eigenvalues for alphaPCA}
Suppose Assumptions~\ref{ass:weak loading}–\ref{ass:Factor-noise interaction for alpha-PCA} and \eqref{cond:C1} hold.
Then
\[
\lambda_i(\widehat{\bM}_R)
=
\lambda_i\Bigg(
\frac{1}{p_1^{\delta_1} p_2^{\delta_2} T} \sum_{t=1}^T \bR \bF_t \bC' \bC \bF_t' \bR'
\Bigg)
+ o_p(1),
\]
and for $i\le k_1$,
\[
\lambda_i(\widehat{\bM}_R)
=
\lambda_i(\bSigma_{1}) + o_p(1).
\]
Consequently,
\[
\|\bLambda_R\| = O_p(1), \qquad \|\bLambda_R^{-1}\| = O_p(1),
\]
and analogous bounds hold for \(\bLambda_C\).
\end{proposition}
\begin{proof}
Write
\[
\begin{aligned}
\widehat{\bM}_R - \frac{1}{p_1^{\delta_1} p_2^{\delta_2} T} \sum_{t=1}^T \bR \bF_t \bC' \bC \bF_t' \bR'
=&
\frac{1}{p_1^{\delta_1} p_2^{\delta_2} T} \sum_{t=1}^T \bR \bF_t \bC' \bE_t'
+
\frac{1}{p_1^{\delta_1} p_2^{\delta_2} T} \sum_{t=1}^T \bE_t \bC \bF_t' \bR'
+
\frac{1}{p_1^{\delta_1} p_2^{\delta_2} T} \sum_{t=1}^T \bE_t \bE_t'\\
=&:II + III + IV.
\end{aligned}
\]
Under Corollary~\ref{corollary: FCe for alphaPCA}, 
\[
\|II\| \le \frac{1}{p_1^{\delta_1} p_2^{\delta_2} T} \Big\| \sum_{t=1}^T \bF_t \bC' \bE_t' \Big\| \|\bR\|
= O_p\Bigg(\sqrt{\frac{p_1^{1-\delta_1}}{T p_2^{\delta_2}}}\Bigg),
\]
and similarly,
\[
\|III\| = O_p\Bigg(\sqrt{\frac{p_1^{1-\delta_1}}{T p_2^{\delta_2}}}\Bigg).
\]
Under Lemma~\ref{lemma: EtEt' for alphaPCA},
\[
\|IV\| \le \frac{1}{p_1^{\delta_1} p_2^{\delta_2} T} \Big( O_p(p_1 \sqrt{p_2 T}) + O_p(\sqrt{p_1} p_2 T) \Big)
= O_p\Bigg(\frac{p_1^{1-\delta_1}}{p_2^{\delta_2 - \frac12}\sqrt{T}}\Bigg) + O_p\Bigg(\frac{p_2^{1-\delta_2}}{p_1^{\delta_1 - \frac12}}\Bigg).
\]
By Weyl's inequality,
\[
|\lambda_i(\widehat{\bM}_R) - \lambda_i(\frac{1}{p_1^{\delta_1} p_2^{\delta_2} T} \sum_{t=1}^T \bR \bF_t \bC' \bC \bF_t' \bR')| \lesssim \|II\| + \|III\| + \|IV\|=o_p\paren{1},
\]
where the last equation follows from the assumptions on growth rates. Moreover, For $i\le k_1$,
\[
\lambda_i\paren{\frac{1}{p_1^{\delta_1}}\bR\bSigma_{1}\bR'}=\lambda_i\paren{\frac{1}{p_1^{\delta_1}}\bSigma_{1}^{\frac{1}{2}}\bR'\bR\bSigma_{1}^{\frac{1}{2}}}\xrightarrow{p}\lambda_i\paren{\bSigma_{1}}
\]
Thus, for $i\le k_1$,
\[
\lambda_i(\widehat{\bM}_R) \xrightarrow[]{p} \lambda_i(\bSigma_{1}).
\]
\end{proof}
Now for Theorem~\ref{thm:alpha-pca-estimation}\textup{(I)}, we write
\[
\begin{aligned}
\frac{1}{p_1^{\delta_1}} \|\widehat{\bR} - \bR \bH_R\|_F^2
&= \frac{1}{p_1^{\delta_1}} \| (II+III+IV) \widehat{\bR} \bLambda_R^{-1} \|_F^2\\
&\lesssim \frac{1}{p_1^{\delta_1}} (\|II\|_F^2 + \|III\|_F^2 + \|IV\|_F^2) \|\widehat{\bR}\|_F^2\\
&\lesssim (\|II\|_F^2 + \|III\|_F^2 + \|IV\|_F^2)\\
&= O_p\Bigg(\frac{p_1^{2-2\delta_1}}{T p_2^{2\delta_2-1}}\Bigg) + O_p\Bigg(\frac{p_2^{2-2\delta_2}}{p_1^{2\delta_1-1}}\Bigg).
\end{aligned}
\]
It remains to establish the properties of the rotation matrices. For this purpose, we use the following lemma.
\begin{lemma}\label{lemma:HRHC for alphaPCA}
Suppose Assumptions~\ref{ass:weak loading}–\ref{ass:Factor-noise interaction for alpha-PCA} and \eqref{cond:C1} hold. Then
\(
\mathbf{H}_R=O_p\paren{1}\;\text{and}\;\mathbf{H}_C=O_p\paren{1},
\)
\(
\mathbf{H}_R'\mathbf{H}_R\xrightarrow[]{p}\mathbf{I}_{k_1}\;\text{and}\;\mathbf{H}_C'\mathbf{H}_C\xrightarrow[]{p}\mathbf{I}_{k_2}.
\)
Consequently,
\(
\mathbf{H}_R^{-1}=O_p\paren{1}\;\text{and}\;\mathbf{H}_C^{-1}=O_p\paren{1}.
\)
\end{lemma}

\begin{proof}
\[
\begin{aligned}
\norm{\mathbf{H}_R}
&=
\norm{
\frac{1}{{Tp_1^{\delta_1}p_2^{\delta_2}}}
\sum_{t=1}^T
\mathbf{F}_t\mathbf{C}'\mathbf{C}\mathbf{F}_t'
\mathbf{R}'\widehat{\mathbf{R}}\mathbf{\Lambda}_R^{-1}
} \\
&\le
\frac{1}{{Tp_1^{\delta_1}p_2^{\delta_2}}}
\norm{\sum_t\mathbf{F}_t\mathbf{C}'\mathbf{C}\mathbf{F}_t'}
\norm{\mathbf{R}'}
\norm{\widehat{\mathbf{R}}}
\norm{\mathbf{\Lambda}_R^{-1}} \\
&=
O_p\paren{1}.
\end{aligned}
\]
Since
$
\frac{1}{p_1^{\delta_1}}\widehat{\mathbf{R}}'\widehat{\mathbf{R}}=\mathbf{I}_{k_1}
$,
we have
\[
\frac{1}{p_1^{\delta_1}}
\paren{\widehat{\mathbf{R}}-\mathbf{R}\mathbf{H}_R+\mathbf{R}\mathbf{H}_R}'
\paren{\widehat{\mathbf{R}}-\mathbf{R}\mathbf{H}_R+\mathbf{R}\mathbf{H}_R}
=
\mathbf{I}_{k_1}.
\]
Thus, by the loading estimation bound established above and Assumption~\ref{ass:weak loading},
\[
o_p\paren{1}
+
\frac{1}{p_1^{\delta_1}}
\mathbf{H}_R'\mathbf{R}'\mathbf{R}\mathbf{H}_R
=
\mathbf{I}_{k_1}.
\]
Hence,
\[
\mathbf{H}_R'\mathbf{H}_R
=
\frac{1}{p_1^{\delta_1}}
\mathbf{H}_R'\mathbf{R}'\mathbf{R}\mathbf{H}_R
+
o_p\paren{1}
=
\mathbf{I}_{k_1}
+
o_p\paren{1}.
\]
The analogous argument applies to $\mathbf{H}_C$.
\end{proof} This completes the proof of
Theorem~\ref{thm:alpha-pca-estimation}\textup{(I)}.

{\bf Proof of Theorem~\ref{thm:alpha-pca-estimation}\textup{(II)}.}
Also, we first introduce several auxiliary lemmas and propositions.
\begin{lemma}\label{lemma:R'(R^-RHR) for alphaPCA}
Suppose Assumptions~\ref{ass:weak loading}–\ref{ass:Factor-noise interaction for alpha-PCA} and \eqref{cond:C1} hold. Then
\[
\frac{1}{p_1^{\delta_1}}
\norm{\mathbf{R}'\paren{\widehat{\mathbf{R}}-\mathbf{R}\mathbf{H}_R}}
=
O_p\paren{
\frac{p_2^{1-\delta_2}}{p_1^{\delta_1}}
+
\sqrt{\frac{1}{Tp_2^{2\delta_2-1}p_1^{2\delta_1-1}}}
+
\sqrt{\frac{p_1^{3-3\delta_1}}{T^2p_2^{3\delta_2-1}}}
+
\sqrt{\frac{1}{Tp_1^{3\delta_1-2}p_2^{3\delta_2-2}}}
}
\]
Similarly,
\[
\frac{1}{p_2^{\delta_2}}
\norm{\mathbf{C}'\paren{\widehat{\mathbf{C}}-\mathbf{C}\mathbf{H}_C}}
=
O_p\paren{
\frac{p_1^{1-\delta_1}}{p_2^{\delta_2}}
+
\sqrt{
\frac{1}{
T p_1^{2\delta_1-1}p_2^{2\delta_2-1}
}
}
+
\sqrt{
\frac{p_2^{3-3\delta_2}}{
T^2p_1^{3\delta_1-1}
}
}
+
\sqrt{
\frac{1}{
T p_2^{3\delta_2-2}p_1^{3\delta_1-2}
}
}
}.
\]
\end{lemma}

\begin{proof}
It suffices, by symmetry, to establish the first bound.
\[
\begin{aligned}
\norm{\mathbf{R}'\paren{\widehat{\mathbf{R}}-\mathbf{R}\mathbf{H}_R}}
&=
\norm{\mathbf{R}'\paren{II+III+IV}\widehat{\mathbf{R}}\mathbf{\Lambda}_R^{-1}} \\
&\le
\norm{\mathbf{R}'\paren{II+III+IV}\widehat{\mathbf{R}}}
\norm{\mathbf{\Lambda}_R^{-1}} \\
&\le
\norm{\mathbf{R}'II\widehat{\mathbf{R}}}\norm{\mathbf{\Lambda}_R^{-1}}
+
\norm{\mathbf{R}'III\widehat{\mathbf{R}}}\norm{\mathbf{\Lambda}_R^{-1}}
+
\norm{\mathbf{R}'IV\widehat{\mathbf{R}}}\norm{\mathbf{\Lambda}_R^{-1}}.
\end{aligned}
\]

For the first term,
\[
\begin{aligned}
\norm{\mathbf{R}'II\widehat{\mathbf{R}}}
&=
\norm{
\mathbf{R}'
\frac{1}{{Tp_1^{\delta_1}p_2^{\delta_2}}}
\sum_t
\mathbf{R}\mathbf{F}_t\mathbf{C}'\mathbf{E}_t'
\widehat{\mathbf{R}}
} \\
&\le
\norm{
\frac{1}{{Tp_1^{\delta_1}p_2^{\delta_2}}}
\sum_t
\mathbf{R}'\mathbf{R}\mathbf{F}_t\mathbf{C}'\mathbf{E}_t'
\paren{\widehat{\mathbf{R}}-\mathbf{R}\mathbf{H}_R}
} \\
&\quad+
\norm{
\frac{1}{{Tp_1^{\delta_1}p_2^{\delta_2}}}
\sum_t
\mathbf{R}'\mathbf{R}\mathbf{F}_t\mathbf{C}'\mathbf{E}_t'
\paren{\mathbf{R}\mathbf{H}_R}
} \\
&:= II_1+II_2.
\end{aligned}
\]

For \(II_1\),
\[
\begin{aligned}
\norm{II_1}
&\le
\frac{1}{{Tp_1^{\delta_1}p_2^{\delta_2}}}
\norm{\mathbf{R}}^2
\norm{\sum_t\mathbf{F}_t\mathbf{C}'\mathbf{E}_t'}
\norm{\widehat{\mathbf{R}}-\mathbf{R}\mathbf{H}_R} \\
&\le
\frac{1}{{Tp_1^{\delta_1}p_2^{\delta_2}}}
p_1^{\delta_1}
O_p\paren{\sqrt{Tp_1p_2^{\delta_2}}}
\norm{\widehat{\mathbf{R}}-\mathbf{R}\mathbf{H}_R} \\
&=
O_p\paren{\sqrt{\frac{p_1}{Tp_2^{\delta_2}}}}
\norm{\widehat{\mathbf{R}}-\mathbf{R}\mathbf{H}_R}.
\end{aligned}
\]

For \(II_2\),
\[
\begin{aligned}
\norm{II_2}
&\le
\frac{1}{{Tp_1^{\delta_1}p_2^{\delta_2}}}
\norm{\sum_t\mathbf{F}_t\mathbf{C}'\mathbf{E}_t'\mathbf{R}}
\norm{\mathbf{R}}^2 \\
&=
\frac{1}{{Tp_1^{\delta_1}p_2^{\delta_2}}}
O_p\paren{\sqrt{{Tp_1^{\delta_1}p_2^{\delta_2}}}}
p_1^{\delta_1} \\
&=
O_p\paren{\sqrt{\frac{p_1^{\delta_1}}{Tp_2^{\delta_2}}}}.
\end{aligned}
\]

For the second term,
\[
\begin{aligned}
\norm{\mathbf{R}'III\widehat{\mathbf{R}}}
&=
\norm{
\frac{1}{{Tp_1^{\delta_1}p_2^{\delta_2}}}
\sum_t
\mathbf{R}'\mathbf{E}_t\mathbf{C}\mathbf{F}_t'\mathbf{R}'\widehat{\mathbf{R}}
} \\
&\le
\frac{1}{{Tp_1^{\delta_1}p_2^{\delta_2}}}
\norm{\mathbf{R}}
\norm{\widehat{\mathbf{R}}}
\norm{\sum_t\mathbf{R}'\mathbf{E}_t\mathbf{C}\mathbf{F}_t'} \\
&=
\frac{1}{{Tp_1^{\delta_1}p_2^{\delta_2}}}
p_1^{\delta_1}
O_p\paren{\sqrt{{Tp_1^{\delta_1}p_2^{\delta_2}}}} \\
&=
O_p\paren{\sqrt{\frac{p_1^{\delta_1}}{Tp_2^{\delta_2}}}}.
\end{aligned}
\]

For the third term,
\[
\begin{aligned}
\norm{\mathbf{R}'IV\widehat{\mathbf{R}}}
&=
\norm{
\frac{1}{{Tp_1^{\delta_1}p_2^{\delta_2}}}
\sum_{t=1}^T
\mathbf{R}'\mathbf{E}_t\mathbf{E}_t'\widehat{\mathbf{R}}
} \\
&\le
\frac{1}{{Tp_1^{\delta_1}p_2^{\delta_2}}}
\norm{
\sum_{t=1}^T
\mathbf{R}'\paren{\mathbf{E}_t\mathbf{E}_t'-\E \mathbf{E}_t\mathbf{E}_t'}
\paren{\widehat{\mathbf{R}}-\mathbf{R}\mathbf{H}_R}
} \\
&\quad+
\frac{1}{{Tp_1^{\delta_1}p_2^{\delta_2}}}
\norm{
\sum_{t=1}^T
\mathbf{R}'\paren{\mathbf{E}_t\mathbf{E}_t'-\E \mathbf{E}_t\mathbf{E}_t'}
\paren{\mathbf{R}\mathbf{H}_R}
} \\
&\quad+
\frac{1}{{Tp_1^{\delta_1}p_2^{\delta_2}}}
\norm{
\sum_{t=1}^T
\mathbf{R}'\paren{\E \mathbf{E}_t\mathbf{E}_t'}
\paren{\widehat{\mathbf{R}}-\mathbf{R}\mathbf{H}_R}
} \\
&\quad+
\frac{1}{{Tp_1^{\delta_1}p_2^{\delta_2}}}
\norm{
\sum_{t=1}^T
\mathbf{R}'\paren{\E \mathbf{E}_t\mathbf{E}_t'}
\paren{\mathbf{R}\mathbf{H}_R}
} \\
&:=IV_1+IV_2+IV_3+IV_4.
\end{aligned}
\]

For \(IV_1\),
\[
\begin{aligned}
IV_1
&=
\frac{1}{{Tp_1^{\delta_1}p_2^{\delta_2}}}
\norm{
\sum_{t=1}^T
\mathbf{R}'\paren{\mathbf{E}_t\mathbf{E}_t'-\E \mathbf{E}_t\mathbf{E}_t'}
\paren{\widehat{\mathbf{R}}-\mathbf{R}\mathbf{H}_R}
} \\
&\le
\frac{1}{{Tp_1^{\delta_1}p_2^{\delta_2}}}
\norm{
\sum_{t=1}^T
\sum_{i,l=1}^{{p_1}}
\mathbf{R}_{l\cdot}
\sum_{j=1}^{p_2}
\paren{e_{t,lj}e_{t,ij}-\E e_{t,lj}e_{t,ij}}
\paren{\widehat{\mathbf{R}}_{i\cdot}-\mathbf{R}_{i\cdot}\mathbf{H}_R}'
} \\
&\le
\frac{1}{{Tp_1^{\delta_1}p_2^{\delta_2}}}
\paren{
\sum_{i=1}^{p_1}
\norm{
\sum_{t=1}^T
\sum_{l=1}^{{p_1}}
\mathbf{R}_{l\cdot}
\sum_{j=1}^{p_2}
\paren{e_{t,lj}e_{t,ij}-\E e_{t,lj}e_{t,ij}}
}^2
}^{\frac{1}{2}} \\
&\quad\times
\paren{
\sum_{i=1}^{p_1}
\norm{\widehat{\mathbf{R}}_{i\cdot}-\mathbf{R}_{i\cdot}\mathbf{H}_R}^2
}^{\frac{1}{2}} \\
&\le
O_p\paren{\frac{1}{{Tp_1^{\delta_1}p_2^{\delta_2}}}
\sqrt{p_1p_1^{\delta_1}p_2T}}
\norm{\widehat{\mathbf{R}}-\mathbf{R}\mathbf{H}_R}_F \\
&=
{\sqrt{\frac{p_1^{1-\delta_1}}{Tp_2^{2\delta_2-1}}}}
\norm{\widehat{\mathbf{R}}-\mathbf{R}\mathbf{H}_R}_F.
\end{aligned}
\]

For \(IV_2\),
\[
\begin{aligned}
IV_2
&=
\frac{1}{{Tp_1^{\delta_1}p_2^{\delta_2}}}
\norm{
\sum_{t=1}^T
\mathbf{R}'\paren{\mathbf{E}_t\mathbf{E}_t'-\E \mathbf{E}_t\mathbf{E}_t'}
\mathbf{R}\mathbf{H}_R
} \\
&=
\frac{1}{{Tp_1^{\delta_1}p_2^{\delta_2}}}
\norm{
\sum_{t=1}^T
\sum_{i,l=1}^{{p_1}}
\mathbf{R}_{l\cdot}
\sum_{j=1}^{p_2}
\paren{e_{t,lj}e_{t,ij}-\E e_{t,lj}e_{t,ij}}
\paren{\mathbf{R}_{i\cdot}\mathbf{H}_R}
} \\
&\le
\frac{1}{{Tp_1^{\delta_1}p_2^{\delta_2}}}
\paren{
\sum_{i=1}^{p_1}
\norm{
\sum_{t=1}^T
\sum_{l=1}^{{p_1}}
\mathbf{R}_{l\cdot}
\sum_{j=1}^{p_2}
\paren{e_{t,lj}e_{t,ij}-\E e_{t,lj}e_{t,ij}}
}^2
}^{\frac{1}{2}} \\
&\quad\times
\paren{
\sum_{i=1}^{p_1}
\norm{\mathbf{R}_{i\cdot}\mathbf{H}_R}^2
}^{\frac{1}{2}} \\
&\le O_p\paren{
\frac{1}{{Tp_1^{\delta_1}p_2^{\delta_2}}}
\sqrt{p_1p_1^{\delta_1}p_2T}
\sqrt{p_1^{\delta_1}}} \\
&=
O_p\paren{\sqrt{\frac{p_1}{Tp_2^{2\delta_2-1}}}}.
\end{aligned}
\]

For \(IV_3\),
\[
\begin{aligned}
IV_3
&=
\frac{1}{{Tp_1^{\delta_1}p_2^{\delta_2}}}
\norm{
\sum_{t=1}^T
\mathbf{R}'\paren{\E \mathbf{E}_t\mathbf{E}_t'}
\paren{\widehat{\mathbf{R}}-\mathbf{R}\mathbf{H}_R}
} \\
&\le
\frac{1}{{Tp_1^{\delta_1}p_2^{\delta_2}}}
\norm{\mathbf{R}}
\norm{\sum_{t=1}^T\E \mathbf{E}_t\mathbf{E}_t'}
\norm{\widehat{\mathbf{R}}-\mathbf{R}\mathbf{H}_R} \\
&=
O_p\paren{\frac{1}{{Tp_1^{\delta_1}p_2^{\delta_2}}}
p_1^{\frac{\delta_1}{2}}
p_2T}
\norm{\widehat{\mathbf{R}}-\mathbf{R}\mathbf{H}_R} \\
&=
O_p\paren{\frac{p_2^{1-\delta_2}}{p_1^{\frac{\delta_1}{2}}}}
\norm{\widehat{\mathbf{R}}-\mathbf{R}\mathbf{H}_R}.
\end{aligned}
\]

For \(IV_4\),
\[
\begin{aligned}
IV_4
&=
\frac{1}{{Tp_1^{\delta_1}p_2^{\delta_2}}}
\norm{
\sum_{t=1}^T
\mathbf{R}'\E\paren{\mathbf{E}_t\mathbf{E}_t'}\mathbf{R}\mathbf{H}_R
} \\
&\le
\frac{1}{{Tp_1^{\delta_1}p_2^{\delta_2}}}
\norm{\mathbf{R}}^2
\norm{\sum_{t=1}^T\E \mathbf{E}_t\mathbf{E}_t'} \\
&=
O_p\paren{\frac{1}{{Tp_1^{\delta_1}p_2^{\delta_2}}}
p_1^{\delta_1}Tp_2} \\
&=
O_p\paren{p_2^{1-\delta_2}}.
\end{aligned}
\]

Thus,
\[
\begin{aligned}
\frac{1}{p_1^{\delta_1}}
\norm{\mathbf{R}'\paren{\widehat{\mathbf{R}}-\mathbf{R}\mathbf{H}_R}}
&\le
\frac{1}{p_1^{\delta_1}}
\paren{
\norm{\mathbf{R}'II\widehat{\mathbf{R}}}
+
\norm{\mathbf{R}'III\widehat{\mathbf{R}}}
+
\norm{\mathbf{R}'IV\widehat{\mathbf{R}}}
} \\
&\le
\frac{1}{p_1^{\delta_1}}
\paren{
O_p\paren{\sqrt{\frac{p_1}{Tp_2^{\delta_2}}}}
\norm{\widehat{\mathbf{R}}-\mathbf{R}\mathbf{H}_R}
+O_p\paren{
\sqrt{\frac{p_1^{\delta_1}}{Tp_2^{\delta_2}}}
+
\sqrt{\frac{p_1^{\delta_1}}{Tp_2^{\delta_2}}}}
} \\
&\quad+
\frac{1}{p_1^{\delta_1}}
\paren{
O_p\paren{\sqrt{\frac{p_1^{1-\delta_1}}{Tp_2^{2\delta_2-1}}}}
\norm{\widehat{\mathbf{R}}-\mathbf{R}\mathbf{H}_R}_F
+
O_p\paren{\sqrt{\frac{p_1}{Tp_2^{2\delta_2-1}}}}
} \\
&\quad+
\frac{1}{p_1^{\delta_1}}
\paren{O_p\paren{
\frac{p_2^{1-\delta_2}}{p_1^{\frac{\delta_1}{2}}}}
\norm{\widehat{\mathbf{R}}-\mathbf{R}\mathbf{H}_R}
+
O_p\paren{p_2^{1-\delta_2}}
} \\
&=
\frac{1}{p_1^{\delta_1}}
O_p\paren{
p_2^{1-\delta_2}
+
\sqrt{\frac{p_1}{Tp_2^{2\delta_2-1}}}
+
\sqrt{\frac{p_1}{Tp_2^{\delta_2}}}
\sqrt{
p_1^{\delta_1}
\paren{
\frac{p_1^{2-2\delta_1}}{Tp_2^{2\delta_2-1}}
+
\frac{p_2^{2-2\delta_2}}{p_1^{2\delta_1-1}}
}
}
} \\
&=
O_p\paren{
\frac{p_2^{1-\delta_2}}{p_1^{\delta_1}}
+
\sqrt{\frac{1}{Tp_2^{2\delta_2-1}p_1^{2\delta_1-1}}}
+
\sqrt{\frac{p_1^{3-3\delta_1}}{T^2p_2^{3\delta_2-1}}}
+
\sqrt{\frac{1}{Tp_1^{3\delta_1-2}p_2^{3\delta_2-2}}}
}.
\end{aligned}
\]
\end{proof}
\begin{lemma}\label{lem:rdiff-E-C for alphaPCA}
Suppose Assumptions~\ref{ass:weak loading}–\ref{ass:Factor-noise interaction for alpha-PCA} and \eqref{cond:C1} hold. Then
\[
\norm{
\frac{1}{p_1^{\delta_1}p_2^{\delta_2}}
\left(\widehat{\mathbf{R}}-\mathbf{R}\mathbf{H}_R\right)^\top
\mathbf{E}_t\mathbf{C}
}
=
O_p\paren{
\sqrt{\frac{1}{p_1^{2\delta_1-1}p_2^{\delta_2}}}
+\frac{\sqrt{p_1^{5-5\delta_1}}}{Tp_2^{2\delta_2-1}}
+\sqrt{\frac{1}{Tp_1^{5\delta_1-4}p_2^{4\delta_2-3}}}  \\
+\sqrt{\frac{p_2^{4-4\delta_2}}{p_1^{6\delta_1-3}}}
+\sqrt{\frac{p_2^{2-2\delta_2}}{p_1^{3\delta_1-1}}}
}.
\]
Similarly,
\[
\norm{
\frac{1}{p_1^{\delta_1}p_2^{\delta_2}}
\mathbf{R}^\top
\mathbf{E}_t
\paren{\widehat{\mathbf{C}}-\mathbf{C}\mathbf{H}_C}
}
=
O_p\paren{
\sqrt{\frac{1}{p_1^{\delta_1}p_2^{2\delta_2-1}}}
+\frac{\sqrt{p_2^{5-5\delta_2}}}{T p_1^{2\delta_1-1}}  
+
\sqrt{
\frac{1}{
T p_2^{5\delta_2-4}p_1^{4\delta_1-3}
}
}
+
\sqrt{
\frac{p_1^{4-4\delta_1}}{p_2^{6\delta_2-3}}
}
+
\sqrt{
\frac{p_1^{2-2\delta_1}}{p_2^{3\delta_2-1}}
}
}.
\]
\end{lemma}

\begin{proof}
It suffices, by symmetry, to establish the first bound.
\[
\begin{aligned}
\norm{
\frac{1}{p_1^{\delta_1}p_2^{\delta_2}}
\left(\widehat{\mathbf{R}}-\mathbf{R}\mathbf{H}_R\right)^\top
\mathbf{E}_t\mathbf{C}
}
&=
\norm{
\frac{1}{p_1^{\delta_1}p_2^{\delta_2}}
\mathbf{\Lambda}_R^{-1}
\widehat{\mathbf{R}}'
\paren{II+III+IV}'
\mathbf{E}_t\mathbf{C}
}  \\
&\le
\norm{
\frac{1}{p_1^{\delta_1}p_2^{\delta_2}}
\mathbf{\Lambda}_R^{-1}
\widehat{\mathbf{R}}'
II'
\mathbf{E}_t\mathbf{C}
}
+
\norm{
\frac{1}{p_1^{\delta_1}p_2^{\delta_2}}
\mathbf{\Lambda}_R^{-1}
\widehat{\mathbf{R}}'
III'
\mathbf{E}_t\mathbf{C}
}\\
&+
\norm{
\frac{1}{p_1^{\delta_1}p_2^{\delta_2}}
\mathbf{\Lambda}_R^{-1}
\widehat{\mathbf{R}}'
IV'
\mathbf{E}_t\mathbf{C}
}.
\end{aligned}
\]

For the first term,
\[
\begin{aligned}
\norm{
\frac{1}{p_1^{\delta_1}p_2^{\delta_2}}
\mathbf{\Lambda}_R^{-1}
\widehat{\mathbf{R}}'
II'
\mathbf{E}_t\mathbf{C}
} 
&=
\norm{
\frac{1}{p_1^{\delta_1}p_2^{\delta_2}}
\mathbf{\Lambda}_R^{-1}
\widehat{\mathbf{R}}'
\frac{1}{{Tp_1^{\delta_1}p_2^{\delta_2}}}
\sum_{s=1}^T
\mathbf{R}\mathbf{F}_s\mathbf{C}'\mathbf{E}_s'\mathbf{E}_t\mathbf{C}
} \\
&\le
\norm{
\frac{1}{p_1^{\delta_1}p_2^{\delta_2}}
\paren{\widehat{\mathbf{R}}-\mathbf{R}\mathbf{H}_R}'
\frac{1}{{Tp_1^{\delta_1}p_2^{\delta_2}}}
\sum_{s=1}^T
\mathbf{R}\mathbf{F}_s\mathbf{C}'\mathbf{E}_s'\mathbf{E}_t\mathbf{C}
} \\
&+
\norm{
\frac{1}{p_1^{\delta_1}p_2^{\delta_2}}
\paren{\mathbf{R}\mathbf{H}_R}'
\frac{1}{{Tp_1^{\delta_1}p_2^{\delta_2}}}
\sum_{s=1}^T
\mathbf{R}\mathbf{F}_s\mathbf{C}'\mathbf{E}_s'\mathbf{E}_t\mathbf{C}
} \\
&:= II_1+II_2.
\end{aligned}
\]

For \(II_1\),
\[
\begin{aligned}
II_1
&\le
\frac{1}{p_1^{\delta_1}p_2^{\delta_2}}
\norm{\widehat{\mathbf{R}}-\mathbf{R}\mathbf{H}_R}
\frac{1}{{Tp_1^{\delta_1}p_2^{\delta_2}}}
\norm{\mathbf{R}}
\norm{\sum_{s=1}^T\mathbf{F}_s\mathbf{C}'\mathbf{E}_s'}
\norm{\mathbf{E}_t\mathbf{C}} \\
&\le
\frac{1}{p_1^{\delta_1}p_2^{\delta_2}}
\norm{\widehat{\mathbf{R}}-\mathbf{R}\mathbf{H}_R}
\frac{1}{{Tp_1^{\delta_1}p_2^{\delta_2}}}
p_1^{\frac{\delta_1}{2}}
\sqrt{Tp_1p_2^{\delta_2}}
\sqrt{p_1}p_2^{\delta_2} \\
&=
O_p\paren{
\frac{1}{\sqrt{T}p_1^{\frac{3}{2}\delta_1-1}p_2^{\frac{\delta_2}{2}}}
}
\norm{\widehat{\mathbf{R}}-\mathbf{R}\mathbf{H}_R} \\
&=
O_p\paren{
\sqrt{\frac{1}{Tp_1^{3\delta_1-2}p_2^{\delta_2}}}
}
\norm{\widehat{\mathbf{R}}-\mathbf{R}\mathbf{H}_R}.
\end{aligned}
\]

For \(II_2\),
\[
\begin{aligned}
II_2
&\le
\frac{1}{p_1^{\delta_1}p_2^{\delta_2}}
\frac{1}{{Tp_1^{\delta_1}p_2^{\delta_2}}}
\norm{\mathbf{R}}^2
\norm{\sum_{s=1}^T\mathbf{F}_s\mathbf{C}'\mathbf{E}_s'\mathbf{E}_t\mathbf{C}} \\
&\le
\frac{1}{Tp_1^{\delta_1}p_2^{2\delta_2}}
\norm{\sum_{s=1}^T\mathbf{F}_s\mathbf{C}'\mathbf{E}_s'\mathbf{E}_t\mathbf{C}} \\
&=
\frac{1}{Tp_1^{\delta_1}p_2^{2\delta_2}}
O_p\paren{
T\sqrt{p_1p_2^{3\delta_2}}+p_1p_2^{\delta_2}
} \\
&=
O_p\paren{
\frac{1}{\sqrt{p_1^{2\delta_1-1}p_2^{\delta_2}}}
+
\frac{p_1^{1-\delta_1}}{Tp_2^{\delta_2}}
}.
\end{aligned}
\]

For the second term,
\[
\begin{aligned}
\norm{
\frac{1}{p_1^{\delta_1}p_2^{\delta_2}}
\mathbf{\Lambda}_R^{-1}
\widehat{\mathbf{R}}'
III'
\mathbf{E}_t\mathbf{C}
}
&=
\norm{
\frac{1}{p_1^{\delta_1}p_2^{\delta_2}}
\mathbf{\Lambda}_R^{-1}
\widehat{\mathbf{R}}'
\frac{1}{{Tp_1^{\delta_1}p_2^{\delta_2}}}
\sum_{s=1}^T
\mathbf{E}_s\mathbf{C}\mathbf{F}_s'\mathbf{R}'\mathbf{E}_t\mathbf{C}
} \\
&\le
\norm{
\frac{1}{p_1^{\delta_1}p_2^{\delta_2}}
\paren{\widehat{\mathbf{R}}-\mathbf{R}\mathbf{H}_R}'
\frac{1}{{Tp_1^{\delta_1}p_2^{\delta_2}}}
\sum_{s=1}^T
\mathbf{E}_s\mathbf{C}\mathbf{F}_s'\mathbf{R}'\mathbf{E}_t\mathbf{C}
} \\
&+
\norm{
\frac{1}{p_1^{\delta_1}p_2^{\delta_2}}
\paren{\mathbf{R}\mathbf{H}_R}'
\frac{1}{{Tp_1^{\delta_1}p_2^{\delta_2}}}
\sum_{s=1}^T
\mathbf{E}_s\mathbf{C}\mathbf{F}_s'\mathbf{R}'\mathbf{E}_t\mathbf{C}
} \\
&:= III_1+III_2.
\end{aligned}
\]

For \(III_1\),
\[
\begin{aligned}
III_1
&\le
\frac{1}{p_1^{\delta_1}p_2^{\delta_2}}
\norm{\widehat{\mathbf{R}}-\mathbf{R}\mathbf{H}_R}
\frac{1}{{Tp_1^{\delta_1}p_2^{\delta_2}}}
\norm{\sum_{s=1}^T\mathbf{E}_s\mathbf{C}\mathbf{F}_s'}
\norm{\mathbf{R}'\mathbf{E}_t\mathbf{C}} \\
&=
\frac{1}{p_1^{\delta_1}p_2^{\delta_2}}
\norm{\widehat{\mathbf{R}}-\mathbf{R}\mathbf{H}_R}
\frac{1}{{Tp_1^{\delta_1}p_2^{\delta_2}}}
\sqrt{Tp_1p_2^{\delta_2}}
p_1^{\delta_1}p_2^{\delta_2} \\
&=
O_p\paren{
\sqrt{\frac{1}{Tp_1^{2\delta_1-1}p_2^{\delta_2}}}
}
\norm{\widehat{\mathbf{R}}-\mathbf{R}\mathbf{H}_R}.
\end{aligned}
\]

For \(III_2\),
\[
\begin{aligned}
III_2
&\le
\frac{1}{p_1^{\delta_1}p_2^{\delta_2}}
\frac{1}{{Tp_1^{\delta_1}p_2^{\delta_2}}}
\norm{\sum_{s=1}^T\mathbf{R}'\mathbf{E}_s\mathbf{C}\mathbf{F}_s'}
\norm{\mathbf{R}'\mathbf{E}_t\mathbf{C}} \\
&=
O_p\paren{\frac{1}{p_1^{\delta_1}p_2^{\delta_2}}
\frac{1}{{Tp_1^{\delta_1}p_2^{\delta_2}}}
\sqrt{{Tp_1^{\delta_1}p_2^{\delta_2}}}
{p_1^{\delta_1}p_2^{\delta_2}}} \\
&=
O_p\paren{
\sqrt{\frac{1}{Tp_1^{\delta_1}p_2^{\delta_2}}}
}.
\end{aligned}
\]

For the third term,
\[
\begin{aligned}
\norm{
\frac{1}{p_1^{\delta_1}p_2^{\delta_2}}
\mathbf{\Lambda}_R^{-1}
\widehat{\mathbf{R}}'
IV'
\mathbf{E}_t\mathbf{C}
}
&=
\norm{
\frac{1}{p_1^{\delta_1}p_2^{\delta_2}}
\mathbf{\Lambda}_R^{-1}
\widehat{\mathbf{R}}'
\frac{1}{{Tp_1^{\delta_1}p_2^{\delta_2}}}
\sum_{s=1}^T
\mathbf{E}_s\mathbf{E}_s'\mathbf{E}_t\mathbf{C}
}\\
&\le
\norm{
\frac{1}{p_1^{\delta_1}p_2^{\delta_2}}
\paren{\widehat{\mathbf{R}}-\mathbf{R}\mathbf{H}_R}'
\frac{1}{{Tp_1^{\delta_1}p_2^{\delta_2}}}
\sum_{s=1}^T
\paren{\mathbf{E}_s\mathbf{E}_s'-\E \mathbf{E}_s\mathbf{E}_s'}
\mathbf{E}_t\mathbf{C}
}\\
&+
\norm{
\frac{1}{p_1^{\delta_1}p_2^{\delta_2}}
\paren{\widehat{\mathbf{R}}-\mathbf{R}\mathbf{H}_R}'
\frac{1}{{Tp_1^{\delta_1}p_2^{\delta_2}}}
\sum_{s=1}^T
\paren{\E \mathbf{E}_s\mathbf{E}_s'}
\mathbf{E}_t\mathbf{C}
}\\
&+
\norm{
\frac{1}{p_1^{\delta_1}p_2^{\delta_2}}
\paren{\mathbf{R}\mathbf{H}_R}'
\frac{1}{{Tp_1^{\delta_1}p_2^{\delta_2}}}
\sum_{s=1}^T
\paren{\mathbf{E}_s\mathbf{E}_s'-\E \mathbf{E}_s\mathbf{E}_s'}
\mathbf{E}_t\mathbf{C}
} \\
&+
\norm{
\frac{1}{p_1^{\delta_1}p_2^{\delta_2}}
\paren{\mathbf{R}\mathbf{H}_R}'
\frac{1}{{Tp_1^{\delta_1}p_2^{\delta_2}}}
\sum_{s=1}^T
\paren{\E \mathbf{E}_s\mathbf{E}_s'}
\mathbf{E}_t\mathbf{C}
} \\
&:= IV_1+IV_2+IV_3+IV_4.
\end{aligned}
\]

For \(IV_1\),
\[
\begin{aligned}
IV_1
&\le
\frac{1}{p_1^{\delta_1}p_2^{\delta_2}}
\frac{1}{{Tp_1^{\delta_1}p_2^{\delta_2}}}
\norm{\widehat{\mathbf{R}}-\mathbf{R}\mathbf{H}_R}
\norm{\sum_s\paren{\mathbf{E}_s\mathbf{E}_s'-\E \mathbf{E}_s\mathbf{E}_s'}}
\norm{\mathbf{E}_t\mathbf{C}}_F \\
&\le
\frac{1}{p_1^{\delta_1}p_2^{\delta_2}}
\frac{1}{{Tp_1^{\delta_1}p_2^{\delta_2}}}
\norm{\widehat{\mathbf{R}}-\mathbf{R}\mathbf{H}_R}
\norm{\sum_s\paren{\mathbf{E}_s\mathbf{E}_s'-\E \mathbf{E}_s\mathbf{E}_s'}}
\paren{\sum_i\norm{\sum_je_{t,ij}\mathbf{C}_{j\cdot}'}^2}^{\frac{1}{2}} \\
&\le
\frac{1}{p_1^{\delta_1}p_2^{\delta_2}}
\frac{1}{{Tp_1^{\delta_1}p_2^{\delta_2}}}
\norm{\widehat{\mathbf{R}}-\mathbf{R}\mathbf{H}_R}
\norm{\sum_s\paren{\mathbf{E}_s\mathbf{E}_s'-\E \mathbf{E}_s\mathbf{E}_s'}}
\paren{\sum_i{\paren{\sum_j\norm{e_{t,ij}\mathbf{C}_{j\cdot}'}}}^2}^{\frac{1}{2}} \\
&\le
\frac{1}{p_1^{\delta_1}p_2^{\delta_2}}
\frac{1}{{Tp_1^{\delta_1}p_2^{\delta_2}}}
\norm{\widehat{\mathbf{R}}-\mathbf{R}\mathbf{H}_R}_F
O_p\paren{p_1\sqrt{p_2T}
\sqrt{p_1}p_2^{\delta_2}} \\
&=
O_p\paren{
\sqrt{\frac{1}{Tp_1^{4\delta_1-3}p_2^{2\delta_2-1}}}
}
\norm{\widehat{\mathbf{R}}-\mathbf{R}\mathbf{H}_R}_F.
\end{aligned}
\]

For \(IV_2\),
\[
\begin{aligned}
IV_2
&\le
\frac{1}{p_1^{\delta_1}p_2^{\delta_2}}
\frac{1}{{Tp_1^{\delta_1}p_2^{\delta_2}}}
\norm{\widehat{\mathbf{R}}-\mathbf{R}\mathbf{H}_R}
\norm{\sum_s\E \mathbf{E}_s\mathbf{E}_s'}
\norm{\mathbf{E}_t\mathbf{C}} \\
&\le
\frac{1}{p_1^{\delta_1}p_2^{\delta_2}}
\frac{1}{{Tp_1^{\delta_1}p_2^{\delta_2}}}
\norm{\widehat{\mathbf{R}}-\mathbf{R}\mathbf{H}_R}
O_p\paren{p_2T\sqrt{p_1}p_2^{\delta_2}} \\
&=
O_p\paren{
\frac{p_2^{1-\delta_2}}{p_1^{2\delta_1-\frac{1}{2}}}
}
\norm{\widehat{\mathbf{R}}-\mathbf{R}\mathbf{H}_R}.
\end{aligned}
\]

For \(IV_3\),
\[
\begin{aligned}
IV_3
&\le
\frac{1}{p_1^{\delta_1}p_2^{\delta_2}}
\frac{1}{{Tp_1^{\delta_1}p_2^{\delta_2}}}
\norm{\mathbf{H}_R}
\norm{\sum_s\mathbf{R}\paren{\mathbf{E}_s\mathbf{E}_s'-\E \mathbf{E}_s\mathbf{E}_s'}}
\norm{\mathbf{E}_t\mathbf{C}} \\
&\le
\frac{1}{p_1^{\delta_1}p_2^{\delta_2}}
\frac{1}{{Tp_1^{\delta_1}p_2^{\delta_2}}}
\norm{\mathbf{H}_R}
\paren{
\sum_i
\norm{
\sum_s\sum_{l,j}
\mathbf{R}_{l\cdot}
\paren{e_{s,lj}e_{s,ij}-\E e_{s,lj}e_{s,ij}}
}^2
}^{\frac{1}{2}}
\norm{\mathbf{E}_t\mathbf{C}} \\
&\le
\frac{1}{p_1^{\delta_1}p_2^{\delta_2}}
\frac{1}{{Tp_1^{\delta_1}p_2^{\delta_2}}}O_p\paren{
\sqrt{p_1Tp_1^{\delta_1}p_2}
\sqrt{p_1}p_2^{\delta_2}} \\
&=
O_p\paren{
\sqrt{
\frac{1}{Tp_1^{3\delta_1-2}p_2^{2\delta_2-1}}
}
}.
\end{aligned}
\]

For \(IV_4\),
\[
\begin{aligned}
IV_4
&\le
\frac{1}{p_1^{\delta_1}p_2^{\delta_2}}
\frac{1}{{Tp_1^{\delta_1}p_2^{\delta_2}}}
\norm{\mathbf{R}\mathbf{H}_R}
\norm{\sum_s\E \mathbf{E}_s\mathbf{E}_s'}
\norm{\mathbf{E}_t\mathbf{C}} \\
&\le
\frac{1}{p_1^{\delta_1}p_2^{\delta_2}}
\frac{1}{{Tp_1^{\delta_1}p_2^{\delta_2}}}
O_p\paren{p_1^{\frac{\delta_1}{2}}
{p_2T}
\sqrt{p_1}p_2^{\delta_2}} \\
&=
O_p\paren{
\sqrt{
\frac{p_2^{2-2\delta_2}}{p_1^{3\delta_1-1}}
}
}.
\end{aligned}
\]
Combining the preceding estimates, we conclude that
\[
\begin{aligned}
\left\|
\frac{1}{p_1^{\delta_1}p_2^{\delta_2}}
\left(\widehat{\mathbf{R}}-\mathbf{R}\mathbf{H}_R\right)^\top
\mathbf{E}_t\mathbf{C}
\right\|
&\le
II_1+II_2+III_1+III_2+IV_1+IV_2+IV_3+IV_4 \\
&\le
O_p\paren{
\sqrt{\frac{1}{Tp_1^{3\delta_1-2}p_2^{\delta_2}}}
}
\norm{\widehat{\mathbf{R}}-\mathbf{R}\mathbf{H}_R}
+
O_p\paren{
\frac{1}{\sqrt{p_1^{2\delta_1-1}p_2^{\delta_2}}}
}
+
O_p\paren{
\frac{p_1^{1-\delta_1}}{Tp_2^{\delta_2}}
} \\
&+
O_p\paren{
\sqrt{\frac{1}{Tp_1^{2\delta_1-1}p_2^{\delta_2}}}
}
\norm{\widehat{\mathbf{R}}-\mathbf{R}\mathbf{H}_R}
+
O_p\paren{
\sqrt{\frac{1}{Tp_1^{\delta_1}p_2^{\delta_2}}}
} \\
&+
O_p\paren{
\sqrt{\frac{1}{Tp_1^{4\delta_1-3}p_2^{2\delta_2-1}}}
}
\norm{\widehat{\mathbf{R}}-\mathbf{R}\mathbf{H}_R}
+
O_p\paren{
\frac{p_2^{1-\delta_2}}{p_1^{2\delta_1-\frac{1}{2}}}
}
\norm{\widehat{\mathbf{R}}-\mathbf{R}\mathbf{H}_R} \\
&+
O_p\paren{
\sqrt{\frac{1}{Tp_1^{3\delta_1-2}p_2^{2\delta_2-1}}}
}
+
O_p\paren{
\sqrt{\frac{p_2^{2-2\delta_2}}{p_1^{3\delta_1-1}}}
}.
\end{aligned}
\]
Using the bound for \(\norm{\widehat{\mathbf{R}}-\mathbf{R}\mathbf{H}_R}\), the right-hand side becomes
\[
\begin{aligned}
&O_p\paren{
\frac{1}{\sqrt{p_1^{2\delta_1-1}p_2^{\delta_2}}}
}
+
O_p\paren{
\frac{p_1^{1-\delta_1}}{Tp_2^{\delta_2}}
}
+
O_p\paren{
\sqrt{\frac{1}{Tp_1^{\delta_1}p_2^{\delta_2}}}
} +
\sqrt{\frac{p_1^{3-3\delta_1}}{Tp_2^{2\delta_2-1}}}
\sqrt{
O_p\paren{\frac{p_1^{2-2\delta_1}}{Tp_2^{2\delta_2-1}}}
+
O_p\paren{\frac{p_2^{2-2\delta_2}}{p_1^{2\delta_1-1}}}
} \\
&+
\frac{p_2^{1-\delta_2}}{p_1^{2\delta_1-1}}
\sqrt{
O_p\paren{\frac{p_1^{2-2\delta_1}}{Tp_2^{2\delta_2-1}}}
+
O_p\paren{\frac{p_2^{2-2\delta_2}}{p_1^{2\delta_1-1}}}
}+
O_p\paren{
\sqrt{\frac{1}{Tp_1^{3\delta_1-2}p_2^{2\delta_2-1}}}
}
+
O_p\paren{
\sqrt{\frac{p_2^{2-2\delta_2}}{p_1^{3\delta_1-1}}}
}.
\end{aligned}
\]
Therefore,
\[
\begin{aligned}
\left\|
\frac{1}{p_1^{\delta_1}p_2^{\delta_2}}
\left(\widehat{\mathbf{R}}-\mathbf{R}\mathbf{H}_R\right)^\top
\mathbf{E}_t\mathbf{C}
\right\|=
O_p\paren{
\frac{1}{\sqrt{p_1^{2\delta_1-1}p_2^{\delta_2}}}
+
\frac{\sqrt{p_1^{5-5\delta_1}}}{Tp_2^{2\delta_2-1}}
+
\sqrt{\frac{1}{Tp_1^{5\delta_1-4}p_2^{4\delta_2-3}}}
+
\sqrt{\frac{p_2^{4-4\delta_2}}{p_1^{6\delta_1-3}}}
+
\sqrt{\frac{p_2^{2-2\delta_2}}{p_1^{3\delta_1-1}}}
}.
\end{aligned}
\]
This proves the desired bound.
\end{proof}
\begin{lemma}\label{lem:rdiff-E-cdiff for alphaPCA}
Suppose Assumptions~\ref{ass:weak loading}–\ref{ass:Factor-noise interaction for alpha-PCA} and \eqref{cond:C1} hold. Then
\[
\norm{
\frac{1}{p_1^{\delta_1}p_2^{\delta_2}}
\paren{\widehat{\mathbf{R}}-\mathbf{R}\mathbf{H}_R}'
\mathbf{E}_t
\paren{\widehat{\mathbf{C}}-\mathbf{C}\mathbf{H}_C}
}
=
O_p\paren{
\sqrt{\frac{1}{p_1^{5\delta_1-4}p_2^{5\delta_2-4}}}
+\sqrt{\frac{p_2^{5-5\delta_2}}{Tp_1^{5\delta_1-3}}}
+\sqrt{\frac{p_1^{5-5\delta_1}}{Tp_2^{5\delta_2-3}}}
}.
\]
\end{lemma}

\begin{proof}
\[
\begin{aligned}
\norm{
\frac{1}{p_1^{\delta_1}p_2^{\delta_2}}
\paren{\widehat{\mathbf{R}}-\mathbf{R}\mathbf{H}_R}'
\mathbf{E}_t
\paren{\widehat{\mathbf{C}}-\mathbf{C}\mathbf{H}_C}
}
&\le
\frac{1}{p_1^{\delta_1}p_2^{\delta_2}}
\sqrt{p_1^{\delta_1}}
O_p\paren{
\sqrt{\frac{p_1^{2-2\delta_1}}{Tp_2^{2\delta_2-1}}}
+
\sqrt{\frac{p_2^{2-2\delta_2}}{p_1^{2\delta_1-1}}}
}
\sqrt{p_1p_2} \\
&\times
O_p\paren{
\sqrt{\frac{p_1^{2-2\delta_1}}{p_2^{2\delta_2-1}}}
+
\sqrt{\frac{p_2^{2-2\delta_2}}{Tp_1^{2\delta_1-1}}}
} \\
&=
\sqrt{p_1^{1-\delta_1}p_2^{1-\delta_2}}
O_p\paren{
\frac{1}{\sqrt{p_1^{4\delta_1-3}p_2^{4\delta_2-3}}}
+
\frac{p_2^{2-2\delta_2}}{\sqrt{T}p_1^{2\delta_1-1}}
+
\frac{p_1^{2-2\delta_1}}{\sqrt{T}p_2^{2\delta_2-1}}
} \\
&=
O_p\paren{
\sqrt{\frac{1}{p_1^{5\delta_1-4}p_2^{5\delta_2-4}}}
+
\sqrt{\frac{p_2^{5-5\delta_2}}{Tp_1^{5\delta_1-3}}}
+
\sqrt{\frac{p_1^{5-5\delta_1}}{Tp_2^{5\delta_2-3}}}
}.
\end{aligned}
\]
\end{proof}
Now we prove Theorem~\ref{thm:alpha-pca-estimation}\textup{(II)}. Recall that
\[
\widehat{\mathbf F}_t=\frac{1}{p_1^{\delta_1}p_2^{\delta_2}}\widehat{\mathbf R}'\mathbf X_t\widehat{\mathbf C}=\frac{1}{p_1^{\delta_1}p_2^{\delta_2}}\widehat{\mathbf R}'\mathbf R\mathbf F_t\mathbf C'\widehat{\mathbf C}+\frac{1}{p_1^{\delta_1}p_2^{\delta_2}}\widehat{\mathbf R}'\mathbf E_t\widehat{\mathbf C}
\].
Thus,
\begin{equation}
\begin{aligned}
\widehat{\mathbf F}_t
- \mathbf H_R^{-1} \mathbf F_t \mathbf H_C^{-1\top}
&=
\frac{1}{p_1^{\delta_1}p_2^{\delta_2}}
\widehat{\mathbf R}^\top
\left(\mathbf R-\widehat{\mathbf R} \mathbf H_R^{-1}\right)
\mathbf F_t
\left(\mathbf C-\widehat{\mathbf C} \mathbf H_C^{-1}\right)^\top
\widehat{\mathbf C}  \\
&\quad
+\frac{1}{p_1^{\delta_1}}
\widehat{\mathbf R}^\top
\left(\mathbf R-\widehat{\mathbf R} \mathbf H_R^{-1}\right)
\mathbf F_t \mathbf H_C^{-1\top}  \\
&\quad
+\frac{1}{p_2^{\delta_2}}
\mathbf H_R^{-1}\mathbf F_t
\left(\mathbf C-\widehat{\mathbf C} \mathbf H_C^{-1}\right)^\top
\widehat{\mathbf C}  \\
&\quad
+\frac{1}{p_1^{\delta_1}p_2^{\delta_2}}
\left(\widehat{\mathbf R}-\mathbf R\mathbf H_R\right)^\top
\mathbf E_t
\left(\widehat{\mathbf C}-\mathbf C\mathbf H_C\right)  \\
&\quad
+\frac{1}{p_1^{\delta_1}p_2^{\delta_2}}
\left(\widehat{\mathbf R}-\mathbf R\mathbf H_R\right)^\top
\mathbf E_t{\mathbf C}\bH_C  \\
&\quad
+\frac{1}{p_1^{\delta_1}p_2^{\delta_2}}
\bH_R^\top{\mathbf R}^\top \mathbf E_t
\left(\widehat{\mathbf C}-\mathbf C\mathbf H_C\right)  \\
&\quad
+\frac{1}{p_1^{\delta_1}p_2^{\delta_2}}
\mathbf H_R^\top \mathbf R^\top \mathbf E_t \mathbf C \mathbf H_C  \\
&=
\sum_{i=1}^7 I_i .
\end{aligned}
\end{equation}
Since
\[
p_1^{-\delta_1}\|\widehat{\mathbf R}-\mathbf R \mathbf H_R\|_F^2=o_p(1),
\qquad
p_2^{-\delta_2}\|\widehat{\mathbf C}-\mathbf C \mathbf H_C\|_F^2=o_p(1),
\]
the term \(I_1\) is negligible relative to \(I_2\) and \(I_3\). We next bound
\(I_2,\ldots,I_7\). 
\[
\begin{aligned}
\norm{I_2}
&=
\norm{\frac{1}{p_1^{\delta_1}}
\left(\widehat{\mathbf R}-\mathbf R\mathbf H_R\right)^\top
\left(\mathbf R-\widehat{\mathbf R} \mathbf H_R^{-1}\right)
\mathbf F_t \mathbf H_C^{-1\top}
+
\frac{1}{p_1^{\delta_1}}
\mathbf H_R^\top \mathbf R^\top
\left(\mathbf R-\widehat{\mathbf R} \mathbf H_R^{-1}\right)
\mathbf F_t \mathbf H_C^{-1\top}}\\
&\le\frac{1}{p_1^{\delta_1}}\norm{\widehat{\mathbf R}-\mathbf R \mathbf H_R}_F^2\norm{\mathbf F_t\mathbf H_C^{-1}}+\frac{1}{p_1^{\delta_1}}\norm{\mathbf R'\paren{\widehat{\mathbf R}-\mathbf R \mathbf H_R}}\norm{\mathbf F_t\mathbf H_C^{-1}}\\
&\le O_p\paren{\frac{p_1^{2-2\delta_1}}{Tp_2^{2\delta_2-1}}}+O_p\paren{\frac{p_2^{2-2\delta_2}}{p_1^{2\delta_1-1}}}\\
&+O_p\paren{\frac{p_2^{1-\delta_2}}{p_1^{\delta_1}}+\sqrt{\frac{1}{Tp_2^{2\delta_2-1}p_1^{2\delta_1-1}}}+\sqrt{\frac{p_1^{3-3\delta_1}}{T^2p_2^{3\delta_2-1}}}+\sqrt{\frac{1}{Tp_1^{3\delta_1-2}p_2^{3\delta_2-2}}}}\\
&=O_p\paren{\frac{p_1^{2-2\delta_1}}{Tp_2^{2\delta_2-1}}+\frac{p_2^{2-2\delta_2}}{p_1^{2\delta_1-1}}}
\end{aligned}
\]
By analogous arguments,
\[
\norm{I_3}=O_p\paren{\frac{p_2^{2-2\delta_2}}{Tp_1^{2\delta_1-1}}+\frac{p_1^{2-2\delta_1}}{p_2^{2\delta_2-1}}}
\]
By Lemma~\ref{lem:rdiff-E-cdiff for alphaPCA},
\[
\norm{I_4}=O_p\paren{\sqrt{\frac{1}{p_1^{5\delta_1-4}p_2^{5\delta_2-4}}}+\sqrt{\frac{p_2^{5-5\delta_2}}{Tp_1^{5\delta_1-3}}}+\sqrt{\frac{p_1^{5-5\delta_1}}{Tp_2^{5\delta_2-3}}}}
\]
For $I_5$,
\[
\begin{aligned}
\norm{I_5}
&=\norm{\frac{1}{p_1^{\delta_1}p_2^{\delta_2}}
\left(\widehat{\mathbf R}-\mathbf R\mathbf H_R\right)^\top
\mathbf E_t\mathbf C\mathbf H_C}\\
&=O_p\paren{\sqrt{\frac{1}{p_1^{2\delta_1-1}p_2^{\delta_2}}}+\frac{\sqrt{p_1^{5-5\delta_1}}}{Tp_2^{2\delta_2-1}}+\sqrt{\frac{1}{Tp_1^{5\delta_1-4}p_2^{4\delta_2-3}}}+\sqrt{\frac{p_2^{4-4\delta_2}}{p_1^{6\delta_1-3}}}+
\sqrt{\frac{p_2^{2-2\delta_2}}{p_1^{3\delta_1-1}}}}
\\
\end{aligned}
\]
Similarly,
\[
\norm{I_6}=
O_p\paren{
\sqrt{\frac{1}{p_1^{\delta_1}p_2^{{2\delta_2-1}}}}
+
\frac{\sqrt{p_2^{5-5\delta_2}}}{Tp_1^{2\delta_1-1}}
+
\sqrt{\frac{1}{Tp_2^{5\delta_2-4}p_1^{4\delta_1-3}}}
+
\sqrt{\frac{p_1^{4-4\delta_1}}{p_2^{6\delta_2-3}}}
+
\sqrt{\frac{p_1^{2-2\delta_1}}{p_2^{3\delta_2-1}}}
}
\]
By Lemma~\ref{lemma: EtEt'-entry for alphaPCA},
\[
\norm{I_7}\le\frac{1}{p_1^{\delta_1}p_2^{\delta_2}}O_p\paren{\sqrt{p_1^{\delta_1}p_2^{\delta_2}}}=O_p\paren{\frac{1}{\sqrt{p_1^{\delta_1}p_2^{\delta_2}}}}
\]
Finally,
\[
\begin{aligned}
\norm{\widehat{\mathbf F}_t
- \mathbf H_R^{-1} \mathbf F_t \mathbf H_C^{-1\top}}
&\le\norm{I_1}+...+\norm{I_7}\\
&\le O_p\paren{\frac{p_1^{2-2\delta_1}}{Tp_2^{2\delta_2-1}}+\frac{p_2^{2-2\delta_2}}{p_1^{2\delta_1-1}}}+O_p\paren{\frac{p_2^{2-2\delta_2}}{Tp_1^{2\delta_1-1}}+\frac{p_1^{2-2\delta_1}}{p_2^{2\delta_2-1}}}\\
&+O_p\paren{\sqrt{\frac{1}{p_1^{5\delta_1-4}p_2^{5\delta_2-4}}}+\sqrt{\frac{p_2^{5-5\delta_2}}{Tp_1^{5\delta_1-3}}}+\sqrt{\frac{p_1^{5-5\delta_1}}{Tp_2^{5\delta_2-3}}}}\\
&+O_p\paren{\sqrt{\frac{1}{p_1^{2\delta_1-1}p_2^{\delta_2}}}+\frac{\sqrt{p_1^{5-5\delta_1}}}{Tp_2^{2\delta_2-1}}+\sqrt{\frac{1}{Tp_1^{5\delta_1-4}p_2^{4\delta_2-3}}}+\sqrt{\frac{p_2^{4-4\delta_2}}{p_1^{6\delta_1-3}}}+
\sqrt{\frac{p_2^{2-2\delta_2}}{p_1^{3\delta_1-1}}}}\\
&+O_p\paren{
\sqrt{\frac{1}{p_1^{\delta_1}p_2^{{2\delta_2-1}}}}
+
\frac{\sqrt{p_2^{5-5\delta_2}}}{Tp_1^{2\delta_1-1}}
+
\sqrt{\frac{1}{Tp_2^{5\delta_2-4}p_1^{4\delta_1-3}}}
+
\sqrt{\frac{p_1^{4-4\delta_1}}{p_2^{6\delta_2-3}}}
+
\sqrt{\frac{p_1^{2-2\delta_1}}{p_2^{3\delta_2-1}}}
}
\\
&
+O_p\paren{\frac{1}{\sqrt{p_1^{\delta_1}p_2^{\delta_2}}}}\\
&
=O_p\paren{\frac{p_1^{2-2\delta_1}}{p_2^{2\delta_2-1}}+\frac{p_2^{2-2\delta_2}}{p_1^{2\delta_1-1}}+\sqrt{\frac{1}{p_1^{5\delta_1-4}p_2^{5\delta_2-4}}}}
\end{aligned}
\]
\setcounter{subsubsection}{0}
\renewcommand{\thesubsubsection}{\thesubsection.\arabic{subsubsection}}
\subsection{\texorpdfstring{Projected Estimation}{projected}}\label{sec:projected}
Recall from Section~\ref{sec:theory} that the initial estimators are denoted by 
\(\check{\bR},\check{\bC}\), and the projected estimators by 
\(\widetilde{\bR},\widetilde{\bC},\widetilde{\bF}_t\). For the initial estimators, $\check{\bH}_R,\check{\bLambda}_R$ and $\check{\bH}_C,\check{\bLambda}_C$ are defined in the same manner as $\widehat{\bH}_R,\widehat{\bLambda}_R$ and $\widehat{\bH}_C,\widehat{\bLambda}_C$. Section~\ref{sec:projected} 
is organized into two subsections: one for the analysis of the initial estimators, 
and the other for the projected estimators. For notational simplicity, throughout this subsection we present the proofs for the case \(k_1=k_2=1\). The arguments for general fixed \(k_1\) and \(k_2\) follow by applying the same bounds componentwise, since the factor dimensions remain fixed as \(T,p_1,p_2\to+\infty.\)
\subsubsection{Initial Estimators}
We begin with several auxiliary lemmas and propositions.
\begin{lemma}
Suppose Assumptions~\ref{ass:weak loading}--\ref{ass:noise-dependence}, Assumption~\ref{ass:projected_spatial_noise}--\ref{ass:Factor-noise interaction for projected} and \eqref{cond:C1} hold. Then
\[
    \sum_{t=1}^T \E\norm{\mathbf{E}_t^T\mathbf{R}}_F^2 = O\paren{Tp_1^{\delta_1}p_2},\;
    \sum_{t=1}^T \E\norm{\mathbf{E}_t\mathbf{C}}_F^2 = O\paren{Tp_1p_2^{\delta_2}}.
\]
\end{lemma}
\begin{proof}
By symmetry, it suffices to establish the first bound.
    \[
    \begin{aligned}
    \sum_{t=1}^T \E\norm{\mathbf{E}_t^T\mathbf{R}}_F^2&=\sum_{t,j}\E\norm{\mathbf{e}_{t,\cdot j}'\mathbf{R}}_F^2=\sum_{t,j}\E\paren{\sum_i e_{t,ij}\mathbf{R}_i}^2\\
    &=\sum_{t,j}\sum_{i_1,i_2}\E \mathbf{R}_{i_1}\mathbf{R}_{i_2}e_{t,i_1j}e_{t,i_2j}\\
    &=\sum_{t,j}\sum_{i_1 \in \mathcal I_R,i_2 \in \mathcal I_R}\E \mathbf{R}_{i_1}\mathbf{R}_{i_2}e_{t,i_1j}e_{t,i_2j}\\    
    &\le C\sum_{t,j}\sum_{i_1 \in \mathcal I_R,i_2 \in \mathcal I_R}\abs{\E e_{t,i_1j}e_{t,i_2j}}\\
    &\le CTp_1^{\delta_1}p_2
     \end{aligned}
    \]
    The last inequality follows from Assumption~\ref{ass:projected_spatial_noise}.
\end{proof}
\begin{lemma}
Suppose Assumptions~\ref{ass:weak loading}--\ref{ass:noise-dependence}, Assumption~\ref{ass:projected_spatial_noise}--\ref{ass:Factor-noise interaction for projected} and \eqref{cond:C1} hold. Then
    \[
\E\norm{\sum_{t}\mathbf{F}_t\mathbf{C}'\mathbf{E}_t'}_F^2=O\paren{Tp_1p_2^{\delta_2}},\;
\E\norm{\sum_{t}\mathbf{F}_t'\mathbf{R}'\mathbf{E}_t}_F^2=O\paren{Tp_1^{\delta_1}p_2},
    \]
and
    \[
\E\norm{\sum_{t}\mathbf{F}_t\mathbf{C}'\mathbf{E}_t'\mathbf{R}}_F^2=O\paren{Tp_1^{\delta_1}p_2^{\delta_2}}
,\;\E\norm{\sum_{t}\mathbf{F}_t'\mathbf{R}'\mathbf{E}_t\mathbf{C}}_F^2=O\paren{Tp_1^{\delta_1}p_2^{\delta_2}}.
    \]
\end{lemma}
\begin{proof}
By symmetry, it suffices to establish the first and third bounds, 
as the remaining bounds follow analogously.
\[    
 \begin{aligned}   \E\norm{\sum_{t}\mathbf{F}_t\mathbf{C}'\mathbf{E}_t'}_F^2
    &\le C\norm{\bC}^2\sum_{i=1}^{p_1}\E\norm{\sum_t\bF_t\frac{\bC'}{\norm{\bC}}{e_{t,i\cdot}}}_F^2\\
    &\le C \norm{\bC}^2T\sum_{i}\E\norm{\sum_{t}\frac{1}{\sqrt{T}}\mathbf{F}_t\frac{\bC'}{\norm{\bC}}e_{t,i\cdot}}_F^2\\
    &\le CTp_1p_2^{\delta_2}
    \end{aligned}
\]
The last inequality follows from Assumption~\ref{ass:weak loading} and Assumption~\ref{ass:Factor-noise interaction for projected}.
\[
\begin{aligned}
\E\norm{\sum_{t}\mathbf{F}_t\mathbf{C}'\mathbf{E}_t'\mathbf{R}}_F^2&=\norm{\bC}^2\norm{\bR}^2\E\norm{\sum_t\bF_t\frac{\bC'}{\norm{\bC}}\bE_t'\frac{\bR}{\norm{\bR}}}_F^2\\
    &\le T\norm{\bC}^2\norm{\bR}^2\E\norm{\frac{1}{\sqrt{T}}\sum_t\bF_t\frac{\bC'}{\norm{\bC}}\bE_t'\frac{\bR}{\norm{\bR}}}_F^2\\
    &\le CTp_1^{\delta_1}p_2^{\delta_2}
\end{aligned}
\]
The last inequality also follows from Assumption~\ref{ass:weak loading} and Assumption~\ref{ass:Factor-noise interaction for projected}.
\end{proof}
\begin{lemma}
Suppose Assumptions~\ref{ass:weak loading}--\ref{ass:noise-dependence}, Assumption~\ref{ass:projected_spatial_noise}--\ref{ass:Factor-noise interaction for projected} and \eqref{cond:C1} hold. Then
\[
    \E\norm{\sum_{t}\mathbf{E}_t\mathbf{e}_{t,i\cdot }}^2=O\paren{Tp_1p_2+T^2p_2^2}
,\;\E\norm{\sum_{t}\mathbf{R}'\mathbf{E}_t\mathbf{e}_{t,i\cdot }}^2=O\paren{Tp_1^{\delta_1}p_2+T^2p_2^2}
\]
\end{lemma}
\begin{proof}
On one hand, using Assumption~\ref{ass:projected_spatial_noise}, we have
\[
\begin{aligned}
\mathbb{E}\left\|
\sum_{t=1}^T \mathbf{E}_t \mathbf{e}_{t,i\cdot }
\right\|^2
&=
\sum_{i_1}
\mathbb{E}
\left(
\sum_t\sum_j
\left(
e_{t,ij}e_{t,i_1j}
-
\mathbb{E}e_{t,ij}e_{t,i_1j}
\right)
\right)^2  \\
&\quad+
\sum_{i_1}
\left(
\sum_t\sum_j
\mathbb{E}e_{t,ij}e_{t,i_1j}
\right)^2 \\
&\le
\sum_{i_1,t,j_1}\sum_{s,j_2}
\cov
\left(
e_{t,ij_1}e_{t,i_1j_1},
e_{s,ij_2}e_{s,i_1j_2}
\right) \\
&\quad+
\sum_{i_1,t,s,j_1,j_2}
\left|
\mathbb{E}e_{t,ij_1}e_{t,i_1j_1}
\right|
\left|
\mathbb{E}e_{s,ij_2}e_{s,i_1j_2}
\right| \\
&\le
cTp_1p_2
+
c\sum_{t,s,j_1,j_2}
\left(
\sum_{i_1}
\left|
\mathbb{E}e_{t,ij_1}e_{t,i_1j_1}
\right|
\right) \\
&=
O\left(Tp_1p_2+T^2p_2^2\right).
\end{aligned}
\]
On the other hand, assume $k_1=1$ so that
\[
\begin{aligned}
\mathbb{E}
\left\|
\sum_{t=1}^T \mathbf{R}^\top \mathbf{E}_t \mathbf{e}_{t,i\cdot }
\right\|^2
&=
\mathbb{E}
\left(
\sum_t\sum_{i_1}\sum_j
\left(
\mathbf{R}_{i_1}e_{t,ij}e_{t,i_1j}
-
\mathbb{E}\mathbf{R}_{i_1}e_{t,ij}e_{t,i_1j}
\right)
\right)^2 \\
&\quad+
\left(
\sum_t\sum_{i_1}\sum_j
\mathbb{E}\mathbf{R}_{i_1}e_{t,ij}e_{t,i_1j}
\right)^2 \\
&\le
\sum_{t,i_1\in \mathcal I_R,j_1}\sum_{s,i_2\in \mathcal I_R,j_2}
|\mathbf{R}_{i_1}\mathbf{R}_{i_2}|
\left|
\cov
\left(
e_{t,ij_1}e_{t,i_1j_1},
e_{s,ij_2}e_{s,i_2j_2}
\right)
\right| \\
&\quad+
\left(
\sum_t\sum_{i_1\in \mathcal I_R}\sum_j
|\mathbf{R}_{i_1}|
\left|
\mathbb{E}e_{t,ij}e_{t,i_1j}
\right|
\right)^2 \\
&\le
cTp_1^{\delta_1}p_2+cT^2p_2^2\\
&=
O\left(Tp_1^{\delta_1}p_2+T^2p_2^2\right),
\end{aligned}
\]
which concludes the lemma.
\end{proof}
\begin{proposition}
Suppose Assumptions~\ref{ass:weak loading}--\ref{ass:noise-dependence}, Assumption~\ref{ass:projected_spatial_noise}--\ref{ass:Factor-noise interaction for projected} and \eqref{cond:C1} hold. Then
\[
\lambda_j\paren{\widehat{\mathbf{M}}_R}=
\begin{cases}
\lambda_j\paren{\boldsymbol{\Sigma}_1}
+
O_p\paren{
\frac{p_1^{1-\delta_1}}{\sqrt{T}p_2^{\delta_2-\frac{1}{2}}}
}
+
O_p\paren{
\frac{p_2^{1-\delta_2}}{p_1^{\delta_1}}
}
+o_p(1),
& \text{if } j\le k_1.\\
O_p\paren{
\frac{p_1^{1-\delta_1}}{\sqrt{T}p_2^{\delta_2-\frac{1}{2}}}
}
+
O_p\paren{
\frac{p_2^{1-\delta_2}}{p_1^{\delta_1}}
},
& \text{if } j> k_1
\end{cases}
\]
consequently,
\[
\lambda_j\paren{\widehat{\mathbf{M}}_R} = 
\begin{cases}
    \lambda_j\paren{\boldsymbol{\Sigma}_1}+o_p(1), &\text{if },j\le k_1\\
    o_p(1),&\text{if }j> k_1
\end{cases}
\]
\end{proposition}
\begin{proof}
Let
\begin{equation}
\begin{aligned}
\widehat{\mathbf{M}}_R &= \frac{1}{Tp_1^{\delta_1}p_2^{\delta_2}}\sum_t\paren{\mathbf{R}\mathbf{F}_t\mathbf{C}'\mathbf{C}\mathbf{F}_t'\mathbf{R}'+\mathbf{R}\mathbf{F}_t\mathbf{C}'\mathbf{E}_t'+\mathbf{E}_t\mathbf{C}\mathbf{F}_t'\mathbf{R}'+\mathbf{E}_t\mathbf{E}_t'}\\
&=I+II+III+IV
\end{aligned}
\end{equation}
where I is the dominant term,
\[
I\xrightarrow{p}\frac{1}{p_1^{\delta_1}} \mathbf{R}\boldsymbol{\Sigma}_1 \mathbf{R}' \Rightarrow \lambda_j\paren{I}\xrightarrow{p}\lambda_j\paren{\frac{1}{p_1^{\delta_1}} \mathbf{R}\boldsymbol{\Sigma}_1 \mathbf{R}'}.
\]
For $j \le k_1$,
\(
\lambda_j\paren{\frac{1}{p_1^{\delta_1}} \mathbf{R}\boldsymbol{\Sigma}_1 \mathbf{R}'}=\lambda_j\paren{\frac{1}{p_1^{\delta_1}}\boldsymbol{\Sigma}_1^{\frac{1}{2}}\mathbf{R}'\mathbf{R}\boldsymbol{\Sigma}_1^{\frac{1}{2}}}\xrightarrow{p}\lambda_j\paren{\boldsymbol{\Sigma}_1}.
\)
Thus, for $j\le k_1$
\(
\lambda_j\paren{I}=\lambda_j\paren{\boldsymbol{\Sigma}_1}+o_p(1).
\)
For $j>k_1$,
\(
\lambda_j\paren{I} = 0.
\)
For $II$,
\[
\begin{aligned}
\norm{II}&=\norm{\frac{1}{{Tp_1^{\delta_1}p_2^{\delta_2}}}\sum_{t}\mathbf{R}\mathbf{F}_t\mathbf{C}'\mathbf{E}_t'}\lesssim\norm{\frac{1}{Tp_1^{\frac{\delta_1}{2}}p_2^{\delta_2}}\sum_{t}\mathbf{F}_t\mathbf{C}'\mathbf{E}_t'}\\
&\lesssim\frac{1}{Tp_1^{\frac{\delta_1}{2}}p_2^{\delta_2}}O_p\paren{\sqrt{Tp_1p_2^{\delta_2}}}=O_p\paren{\sqrt{\frac{p_1^{1-\delta_1}}{Tp_2^{\delta_2}}}}.
\end{aligned}
\]
Similarly,
\(
\norm{III} =O_p\paren{\sqrt{\frac{p_1^{1-\delta_1}}{Tp_2^{\delta_2}}}}.
\)
For $IV$,
\[
\norm{IV}\le\norm{IV-\frac{1}{{Tp_1^{\delta_1}p_2^{\delta_2}}}\sum_t\E \paren{\mathbf{E}_t\mathbf{E}_t'}}+\norm{\frac{1}{{Tp_1^{\delta_1}p_2^{\delta_2}}}\sum_t\E \paren{\mathbf{E}_t\mathbf{E}_t'}}
\]
By Assumption~\ref{ass:noise-dependence},Assumption~\ref{ass:projected_spatial_noise},
\[
\begin{aligned}
\E\norm{IV-\frac{1}{{Tp_1^{\delta_1}p_2^{\delta_2}}}\sum_t\E \paren{\mathbf{E}_t\mathbf{E}_t'}}_F^2&=\sum_{i,l}\E\paren{\frac{1}{{Tp_1^{\delta_1}p_2^{\delta_2}}}\sum_{t,j}e_{t,ij}e_{t,lj}-\E e_{t,ij}e_{t,lj}}^2\\
&=\paren{\frac{1}{{Tp_1^{\delta_1}p_2^{\delta_2}}}}^2\sum_{i,l}\sum_{t,j_1,s,j_2}\cov\paren{e_{t,ij_1}e_{t,lj_1},e_{s,ij_2}e_{s,lj_2}}\\
&\le \frac{1}{T^2p_1^{2\delta_1}p_2^{2\delta_2}}Cp_1^2Tp_2=C\frac{p_1^{2-2\delta_1}}{Tp_2^{2\delta_2-1}}
\end{aligned}
\]
By Assumption~\ref{ass:projected_spatial_noise},
\[
\sum_{i_1}\abs{\sum_{t,j}\E e_{t,ij}e_{t,i_1j}}\le\sum_{t,j}\sum_{i_1}\abs{\E e_{t,ij}e_{t,i_1j}}\le CTp_2
\]
Thus,
\[
\norm{\frac{1}{{Tp_1^{\delta_1}p_2^{\delta_2}}}\sum_t\E \paren{\mathbf{E}_t\mathbf{E}_t'}}\le\norm{\frac{1}{{Tp_1^{\delta_1}p_2^{\delta_2}}}\sum_t\E \paren{\mathbf{E}_t\mathbf{E}_t'}}_1\le\frac{1}{{Tp_1^{\delta_1}p_2^{\delta_2}}}CTp_2\le C\frac{p_2^{1-\delta_2}}{p_1^{\delta_1}}.
\]
Consequently,
\[
\norm{IV}\le O_p\paren{\frac{p_1^{1-\delta_1}}{\sqrt{T}p_2^{\delta_2-\frac{1}{2}}}+\frac{p_2^{1-\delta_2}}{p_1^{\delta_1}}}
\]
Combining the above results, we have
\[
\begin{aligned}
\norm{\widehat{\mathbf{M}}_R-I}&\le\norm{II}+\norm{III}+\norm{IV}\\
&\le O_p\paren{\sqrt{\frac{p_1^{1-\delta_1}}{Tp_2^{\delta_2}}}+\frac{p_1^{1-\delta_1}}{\sqrt{T}p_2^{\delta_2-\frac{1}{2}}}+\frac{p_2^{1-\delta_2}}{p_1^{\delta_1}}}\\
&=O_p\paren{\frac{p_1^{1-\delta_1}}{\sqrt{T}p_2^{\delta_2-\frac{1}{2}}}+\frac{p_2^{1-\delta_2}}{p_1^{\delta_1}}}
\end{aligned}
\]
By Weyl's theorem, the desired result follows. This completes the proof of the proposition.
\end{proof}
\begin{lemma}
Suppose Assumptions~\ref{ass:weak loading}--\ref{ass:noise-dependence}, Assumption~\ref{ass:projected_spatial_noise}--\ref{ass:Factor-noise interaction for projected} and \eqref{cond:C1} hold. Then
\[
\frac{1}{p_1^{\delta_1}}\norm{II\check{\mathbf{R}}}_F^2\le\frac{1}{Tp_1^{2\delta_1-1}p_2^{\delta_2}}\norm{\check{\mathbf{R}}-\mathbf{R}\check{\mathbf{H}}_R}_F^2+O_p\paren{\frac{1}{{Tp_1^{\delta_1}p_2^{\delta_2}}}},\;
\frac{1}{p_1^{\delta_1}}\norm{III\check{\mathbf{R}}}_F^2\le O_p\paren{\frac{p_1^{1-\delta_1}}{Tp_2^{\delta_2}}}
\]
\[
\begin{aligned}
\frac{1}{p_1^{\delta_1}}\norm{IV\check{\mathbf{R}}}_F^2&\le\paren{\frac{1}{Tp_1^{3\delta_1-2}p_2^{2\delta_2-1}}+\frac{p_2^{2-2\delta_2}}{p_1^{3\delta_1-1}}}\norm{\check{\mathbf{R}}-\mathbf{R}\check{\mathbf{H}}_R}_F^2+O_p\paren{\frac{1}{Tp_1^{2\delta_1-1}p_2^{2\delta_2-1}}+\frac{p_2^{2-2\delta_2}}{p_1^{3\delta_1-1}}}
\end{aligned}
\]
\end{lemma}
\begin{proof}
\[
\begin{aligned}
    \frac{1}{p_1^{\delta_1}}\norm{II\check{\mathbf{R}}}_F^2&=\frac{1}{p_1^{\delta_1}}\norm{\frac{1}{{Tp_1^{\delta_1}p_2^{\delta_2}}}\sum_{t}\mathbf{R}\mathbf{F}_t\mathbf{C}'\mathbf{E}_t' \check{\mathbf{R}}}_F^2
\lesssim \norm{\frac{1}{{Tp_1^{\delta_1}p_2^{\delta_2}}}\sum_{t}\mathbf{F}_t\mathbf{C}'\mathbf{E}_t'\check{\mathbf{R}}}_F^2\\
&\lesssim \norm{\frac{1}{{Tp_1^{\delta_1}p_2^{\delta_2}}}\sum_{t}\mathbf{F}_t\mathbf{C}'\mathbf{E}_t'\paren{\check{\mathbf{R}}-\mathbf{R}\check{\mathbf{H}}_R}}_F^2
+\norm{\frac{1}{{Tp_1^{\delta_1}p_2^{\delta_2}}}\sum_{t}\mathbf{F}_t\mathbf{C}'\mathbf{E}_t'\mathbf{R}\check{\mathbf{H}}_R}_F^2\\
&\lesssim \frac{1}{T^2p_1^{2\delta_1}p_2^{2\delta_2}}O_p\paren{Tp_1p_2^{\delta_2}}\norm{\check{\mathbf{R}}-\mathbf{R}\check{\mathbf{H}}_R}_F^2+\frac{1}{T^2p_1^{2\delta_1}p_2^{2\delta_2}}O_p\paren{Tp_1^{\delta_1}p_2^{\delta_2}}\\
&\lesssim \frac{1}{Tp_1^{2\delta_1-1}p_2^{\delta_2}}\norm{\check{\mathbf{R}}-\mathbf{R}\check{\mathbf{H}}_R}_F^2+O_p\paren{\frac{1}{{Tp_1^{\delta_1}p_2^{\delta_2}}}}
\end{aligned}
\]
\[
\begin{aligned}
\frac{1}{p_1^{\delta_1}}\norm{III\check{\mathbf{R}}}_F^2&=\frac{1}{p_1^{\delta_1}}\norm{\frac{1}{{Tp_1^{\delta_1}p_2^{\delta_2}}}\sum_{t}\mathbf{E}_t\mathbf{C}\mathbf{F}_t'\mathbf{R}'\check{\mathbf{R}}}_F^2\le p_1^{\delta_1}\norm{\frac{1}{{Tp_1^{\delta_1}p_2^{\delta_2}}}\sum_t\mathbf{E}_t\mathbf{C}\mathbf{F}_t'}_F^2\\
&\le\frac{p_1^{\delta_1}}{T^2p_1^{2\delta_1}p_2^{2\delta_2}}O_p\paren{Tp_1p_2^{\delta_2}}\le O_p\paren{\frac{p_1^{1-\delta_1}}{Tp_2^{\delta_2}}}
\end{aligned}
\]
\[
\begin{aligned}
\frac{1}{p_1^{\delta_1}}\norm{IV\check{\mathbf{R}}}_F^2&=\frac{1}{p_1^{\delta_1}}\norm{\frac{1}{{Tp_1^{\delta_1}p_2^{\delta_2}}}\sum_{t}\mathbf{E}_t\mathbf{E}_t'\check{\mathbf{R}}}_F^2
=\frac{1}{p_1^{\delta_1}}\frac{1}{{T^2p_1^{2\delta_1}p_2^{2\delta_2}}}\sum_{i}\norm{\sum_{t}\check{\mathbf{R}}'\mathbf{E}_t\mathbf{e}_{t,i\cdot }}_F^2\\
&\lesssim\frac{1}{p_1^{\delta_1}}\frac{1}{{T^2p_1^{2\delta_1}p_2^{2\delta_2}}}\sum_{i}\norm{\paren{\check{\mathbf{R}}-\mathbf{R}\check{\mathbf{H}}_R}'}_F^2\norm{\sum_{t}\mathbf{E}_t\mathbf{e}_{t,i\cdot }}_F^2\\
&+\frac{1}{p_1^{\delta_1}}\frac{1}{{T^2p_1^{2\delta_1}p_2^{2\delta_2}}}\sum_{i}\norm{\sum_{t}\mathbf{R}'\mathbf{E}_t\mathbf{e}_{t,i\cdot }}_F^2
\\
&\lesssim \frac{1}{p_1^{\delta_1}}\frac{1}{{T^2p_1^{2\delta_1}p_2^{2\delta_2}}}p_1O_p\paren{Tp_1p_2+T^2p_2^2}\norm{\paren{\check{\mathbf{R}}-\mathbf{R}\check{\mathbf{H}}_R}'}_F^2\\
&+\frac{1}{p_1^{\delta_1}}\frac{1}{{T^2p_1^{2\delta_1}p_2^{2\delta_2}}}p_1O_p\paren{Tp_1^{\delta_1}p_2+T^2p_2^2}\\
&\lesssim\paren{\frac{1}{Tp_1^{3\delta_1-2}p_2^{2\delta_2-1}}+\frac{p_2^{2-2\delta_2}}{p_1^{3\delta_1-1}}}\norm{\check{\mathbf{R}}-\mathbf{R}\check{\mathbf{H}}_R}_F^2\\
&+O_p\paren{\frac{1}{Tp_1^{2\delta_1-1}p_2^{2\delta_2-1}}+\frac{p_2^{2-2\delta_2}}{p_1^{3\delta_1-1}}}
\end{aligned}
\]
This completes the proof.
\end{proof} 
\begin{proposition}
Suppose Assumptions~\ref{ass:weak loading}--\ref{ass:noise-dependence}, Assumption~\ref{ass:projected_spatial_noise}--\ref{ass:Factor-noise interaction for projected} and \eqref{cond:C1} hold. Then there exists $\check{\bH}_R\in\mathbb{R}^{k_1\times k_1},\check{\bH}_C\in\mathbb{R}^{k_2\times k_2}$ satisfying $\check{\bH}_R^\top\check{\bH}_R\xrightarrow[]{p}\bI_{k_1},\;\check{\bH}_C^\top\check{\bH}_C\xrightarrow[]{p}\bI_{k_2}$ such that
\[
    \frac{1}{p_1^{\delta_1}}\norm{\check{\mathbf{R}}-\mathbf{R}\check{\mathbf{H}}_R}_F^2=O_p\paren{w_1},\;
\frac{1}{p_2^{\delta_2}}\norm{\check{\mathbf{C}}-\mathbf{C}\check{\mathbf{H}}_C}_F^2=O_p\paren{m_1},
\]
where $w_1:=\frac{p_1^{1-\delta_1}}{Tp_2^{\delta_2}}+\frac{1}{Tp_1^{2\delta_1-1}p_2^{2\delta_2-1}}+\frac{p_2^{2-2\delta_2}}{p_1^{3\delta_1-1}},$ $m_1:=\frac{p_2^{1-\delta_2}}{Tp_1^{\delta_1}}+\frac{1}{Tp_1^{2\delta_1-1}p_2^{2\delta_2-1}}+\frac{p_1^{2-2\delta_1}}{p_2^{3\delta_2-1}}$.
\end{proposition}
\begin{proof}
By symmetry, it suffices to establish the first bound.
    \[
    \begin{aligned}
    \frac{1}{p_1^{\delta_1}}\norm{\check{\mathbf{R}}-\mathbf{R}\check{\mathbf{H}}_R}_F^2&=\frac{1}{p_1^{\delta_1}}\norm{\paren{II+III+IV}\check{\mathbf{R}}\check{\boldsymbol{\Lambda}}_R^{-1}}_F^2\\
    &\le \frac{1}{Tp_1^{2\delta_1-1}p_2^{\delta_2}}\norm{\check{\mathbf{R}}-\mathbf{R}\check{\mathbf{H}}_R}_F^2+O_p\paren{\frac{1}{{Tp_1^{\delta_1}p_2^{\delta_2}}}}+O_p\paren{\frac{p_1^{1-\delta_1}}{Tp_2^{\delta_2}}}\\
    &+\paren{\frac{1}{Tp_1^{3\delta_1-2}p_2^{2\delta_2-1}}+\frac{p_2^{2-2\delta_2}}{p_1^{3\delta_1-1}}}\norm{\check{\mathbf{R}}-\mathbf{R}\check{\mathbf{H}}_R}_F^2+O_p\paren{\frac{1}{Tp_1^{2\delta_1-1}p_2^{2\delta_2-1}}+\frac{p_2^{2-2\delta_2}}{p_1^{3\delta_1-1}}}\\
    &=O_p\paren{\frac{p_1^{1-\delta_1}}{Tp_2^{\delta_2}}}+O_p\paren{\frac{1}{Tp_1^{2\delta_1-1}p_2^{2\delta_2-1}}+\frac{p_2^{2-2\delta_2}}{p_1^{3\delta_1-1}}}\\
    &+\paren{\frac{1}{Tp_1^{2\delta_1-1}p_2^{\delta_2}}+\frac{1}{Tp_1^{3\delta_1-2}p_2^{2\delta_2-1}}+\frac{p_2^{2-2\delta_2}}{p_1^{3\delta_1-1}}}\norm{\check{\mathbf{R}}-\mathbf{R}\check{\mathbf{H}}_R}_F^2
    \end{aligned}
    \]
    Thus,
    \[
    \frac{1}{p_1^{\delta_1}}\norm{\check{\mathbf{R}}-\mathbf{R}\check{\mathbf{H}}_R}_F^2=O_p\paren{\frac{p_1^{1-\delta_1}}{Tp_2^{\delta_2}}}+O_p\paren{\frac{1}{Tp_1^{2\delta_1-1}p_2^{2\delta_2-1}}+\frac{p_2^{2-2\delta_2}}{p_1^{3\delta_1-1}}}
    \]
Then we show that
$\check{\mathbf{H}}_R^{\top}\check{\mathbf{H}}_R
\xrightarrow{p} \mathbf I_{k_1}$.
By Cauchy--Schwarz inequality,
\[
\left\|
\frac{1}{p_1^{\delta_1}}\mathbf{R}^{\top}
\left(\check{\mathbf{R}}-\mathbf{R}\check{\mathbf{H}}_R\right)
\right\|_F^2
\le
\frac{\|\mathbf{R}\|_F^2}{p_1^{\delta_1}}
\frac{\|\check{\mathbf{R}}-\mathbf{R}\check{\mathbf{H}}_R\|_F^2}{p_1^{\delta_1}}
=o_p(1),
\]
and
\[
\left\|
\frac{1}{p_1^{\delta_1}}\check{\mathbf{R}}^{\top}
\left(\check{\mathbf{R}}-\mathbf{R}\check{\mathbf{H}}_R\right)
\right\|_F^2
=o_p(1).
\]
Note that $p_1^{-\delta_1}\check{\mathbf{R}}^{\top}\check{\mathbf{R}}
=\mathbf I_{k_1}$ and
$p_1^{-\delta_1}\mathbf{R}^{\top}\mathbf{R} \rightarrow \mathbf I_{k_1}$, then
\[
\mathbf I_{k_1}
=
\frac{1}{p_1^{\delta_1}}\check{\mathbf{R}}^{\top}\mathbf{R}\check{\mathbf{H}}_R
+\mathbf o_p(1)
=
\check{\mathbf{H}}_R^{\top}\check{\mathbf{H}}_R
+\mathbf o_p(1),
\]
Similarly,
\[
\mathbf I_{k_2}
=
\frac{1}{p_2^{\delta_2}}\check{\mathbf{C}}^{\top}\mathbf{C}\check{\mathbf{H}}_C
+\mathbf o_p(1)
=
\check{\mathbf{H}}_C^{\top}\check{\mathbf{H}}_C
+\mathbf o_p(1).
\]
Consequently,
\[
\norm{\check{\mathbf{H}}_R^{-1}}=\frac{1}{\sqrt{\lambda_{\min}\paren{\check{\mathbf{H}}_R^\top\check{\mathbf{H}}_R}}}=O_p\paren{1},\;\norm{\check{\mathbf{H}}_C^{-1}}=\frac{1}{\sqrt{\lambda_{\min}\paren{\check{\mathbf{H}}_C^\top\check{\mathbf{H}}_C}}}=O_p\paren{1}
\]
\end{proof}
\begin{lemma}
Suppose Assumptions~\ref{ass:weak loading}--\ref{ass:noise-dependence}, Assumption~\ref{ass:projected_spatial_noise}--\ref{ass:Factor-noise interaction for projected} and \eqref{cond:C1} hold. Then
    \[
    \begin{aligned}
    \norm{\sum_{s}\mathbf{E}_s'\paren{\check{\mathbf{R}}-\mathbf{R}\check{\mathbf{H}}_R}\mathbf{F}_s}_F^2
&=O_p\left(\frac{p_1^2}{p_2^{2\delta_2-1}}+\frac{p_1^{4-2\delta_1}}{Tp_2^{3\delta_2-2}}+\frac{p_1^{4-3\delta_1}}{Tp_2^{4\delta_2-3}}+p_1^{4-4\delta_1}p_2^{4-4\delta_2}\right.\\
&\quad\left.+p_1^{3-2\delta_1}{p_2^{3-3\delta_2}}+\frac{Tp_2^{5-4\delta_2}}{p_1^{4\delta_1-3}}+Tp_1^{2-2\delta_1}p_2^{3-2\delta_2}\right)\\
&\quad:=O_p\paren{w_2}T^2p_1^{2\delta_1}p_2^{\delta_2}
\end{aligned}
\]
where $w_2:={\frac{p_1^{2-2\delta_1}}{T^2p_2^{3\delta_2-1}}+\frac{p_1^{4-4\delta_1}}{T^3p_2^{4\delta_2-2}}+\frac{1}{T^2p_1^{4\delta_1-3}p_2^{4\delta_2-3}}+\frac{p_2^{3-3\delta_2}}{Tp_1^{4\delta_1-2}}}=o\paren{1}$.
\end{lemma}
\begin{proof}
    \[
\sum_{s}\mathbf{E}_s'\paren{\check{\mathbf{R}}-\mathbf{R}\check{\mathbf{H}}_R}\mathbf{F}_s=\sum_{s}\mathbf{E}_s'\paren{II+III+IV}\check{\mathbf{R}}\check{\boldsymbol{\Lambda}}_R^{-1}\mathbf{F}_s
    \]
    \[\begin{aligned}\norm{\sum_s\mathbf{E}_s'II\check{\mathbf{R}}\check{\boldsymbol{\Lambda}}_R^{-1}\mathbf{F}_s}_F^2&=\norm{\frac{1}{{Tp_1^{\delta_1}p_2^{\delta_2}}}\sum_s\mathbf{E}_s'\mathbf{R}\sum_t\mathbf{F}_t\mathbf{C}'\mathbf{E}_t'\check{\mathbf{R}}\mathbf{F}_s}_F^2\\
    &\lesssim\frac{1}{{T^2p_1^{2\delta_1}p_2^{2\delta_2}}}\norm{\sum_s\mathbf{E}_s'\mathbf{R}\mathbf{F}_s}_F^2\norm{\sum_t\mathbf{F}_t\mathbf{C}'\mathbf{E}_t'\check{\mathbf{R}}}_F^2\\
&\lesssim
\frac{1}{{T^2p_1^{2\delta_1}p_2^{2\delta_2}}}
\norm{\sum_s \mathbf{E}_s'\mathbf{R}\mathbf{F}_s}_F^2
\Bigg(
\norm{\sum_t \mathbf{F}_t\mathbf{C}'\mathbf{E}_t'\paren{\check{\mathbf{R}}-\mathbf{R}\check{\mathbf{H}}_R}}_F^2 \\
&
+\norm{\sum_t \mathbf{F}_t\mathbf{C}'\mathbf{E}_t'\mathbf{R}\check{\mathbf{H}}_R}_F^2
\Bigg)\\
&\lesssim\frac{1}{{T^2p_1^{2\delta_1}p_2^{2\delta_2}}}Tp_1^{\delta_1}p_2\paren{Tp_1p_2^{\delta_2}\norm{\check{\mathbf{R}}-\mathbf{R}\check{\mathbf{H}}_R}_F^2+O_p\paren{{Tp_1^{\delta_1}p_2^{\delta_2}}}}\\
&=p_1^{1-\delta_1}p_2^{1-\delta_2}p_1^{\delta_1}O_p\paren{\frac{p_1^{1-\delta_1}}{Tp_2^{\delta_2}}+\frac{1}{Tp_1^{2\delta_1-1}p_2^{2\delta_2-1}}+\frac{p_2^{2-2\delta_2}}{p_1^{3\delta_1-1}}}+O_p\paren{p_2^{1-\delta_2}}\\
&=O_p\paren{\frac{p_1^{2-\delta_1}}{Tp_2^{2\delta_2-1}}+\frac{p_1^{2-2\delta_1}}{Tp_2^{3\delta_2-2}}+\frac{p_2^{3-3\delta_2}}{p_1^{3\delta_1-2}}}+O_p\paren{p_2^{1-\delta_2}}
\end{aligned}
\]
\[
\begin{aligned}
\E\norm{\sum_{s}\mathbf{E}_s'III\check{\mathbf{R}}\check{\boldsymbol{\Lambda}}_R^{-1}\mathbf{F}_s}_F^2&=\E\norm{\frac{1}{{Tp_1^{\delta_1}p_2^{\delta_2}}}\sum_{t,s}\mathbf{E}_s'\mathbf{E}_t\mathbf{C}\mathbf{F}_t'\mathbf{R}'\check{\mathbf{R}}\mathbf{F}_s}_F^2\\
&\le\E\norm{\frac{1}{Tp_2^{\delta_2}}\sum_{t,s}\mathbf{E}_s'\mathbf{E}_t\mathbf{C}\mathbf{F}_t'\mathbf{F}_s}_F^2\\
&=\E\norm{\frac{1}{Tp_2^{\delta_2}}\sum_{t,s}\mathbf{F}_t\mathbf{F}_s\sum_j\mathbf{E}_s'\mathbf{e}_{t,\cdot j} \bC_j}_F^2
\end{aligned}
\]
\[
\begin{aligned}
\E\sum_{j_1}\norm{\frac{1}{Tp_2^{\delta_2}}\sum_{t,s}\mathbf{F}_t\mathbf{F}_s\sum_je_{s,\cdot j_1}'\mathbf{e}_{t,\cdot j} \bC_j}_F^2
&\le\E\sum_{j_1}\norm{\frac{1}{Tp_2^{\delta_2}}\sum_{t,s}\mathbf{F}_t\mathbf{F}_s\sum_{i,j}e_{s,i j_1}e_{t,ij} \bC_j}_F^2\\
&=\E\sum_{j_1}\norm{\frac{1}{Tp_2^{\delta_2}}\sum_{i,j}\sqrt{T}\bar{\boldsymbol{\zeta}}_{ij_1}\sqrt{T}\bar{\boldsymbol{\zeta}}_{ij}\mathbf{C}_j}_F^2\\
&=
\E\sum_{j_1}\norm{\frac{1}{p_2^{\delta_2}}\sum_{i,j}\bar{\boldsymbol{\zeta}}_{ij_1}\bar{\boldsymbol{\zeta}}_{ij}\mathbf{C}_j}_F^2\\
&=\E\sum_{j_1}\norm{\frac{1}{p_2^{\delta_2}}\sum_{i,j}\paren{\bar{\boldsymbol{\zeta}}_{ij_1}\bar{\boldsymbol{\zeta}}_{ij}-\E\bar{\boldsymbol{\zeta}}_{ij_1}\bar{\boldsymbol{\zeta}}_{ij}+\E\bar{\boldsymbol{\zeta}}_{ij_1}\bar{\boldsymbol{\zeta}}_{ij}}\mathbf{C}_j}_F^2\\
&=\frac{1}{p_2^{2\delta_2}}\sum_{j_1}\sum_{i_1,i_2,j,h}\mathbf{C}_j\mathbf{C}_h\cov\paren{\bar{\boldsymbol{\zeta}}_{i_1j_1}\bar{\boldsymbol{\zeta}}_{i_1j},\bar{\boldsymbol{\zeta}}_{i_2j_1}\bar{\boldsymbol{\zeta}}_{i_2h}}\\
&+\frac{1}{p_2^{2\delta_2}}\sum_{j_1}\norm{\sum_{i,j}\mathbf{C}_j\E\bar{\boldsymbol{\zeta}}_{ij_1}\bar{\boldsymbol{\zeta}}_{ij}}_F^2\\
&=\frac{1}{p_2^{2\delta_2}}p_2\sum_{i_1,i_2,j\in \mathcal I_C,h\in \mathcal I_C}\mathbf{C}_j\mathbf{C}_h\cov\paren{\bar{\boldsymbol{\zeta}}_{i_1j_1}\bar{\boldsymbol{\zeta}}_{i_1j},\bar{\boldsymbol{\zeta}}_{i_2j_1}\bar{\boldsymbol{\zeta}}_{i_2h}}\\
&+\frac{1}{p_2^{2\delta_2}}p_2\norm{\sum_{i,j}\abs{\mathbf{C}_j}\abs{\E\bar{\boldsymbol{\zeta}}_{ij_1}\bar{\boldsymbol{\zeta}}_{ij}}}_F^2\\
&\le\frac{1}{p_2^{2\delta_2}}p_2p_1p_2^{\delta_2}+\frac{1}{p_2^{2\delta_2}}p_2p_1^2\le p_1p_2^{1-\delta_2}+\frac{p_1^2}{p_2^{2\delta_2-1}}
\end{aligned}
\]
\[
\begin{aligned}
\norm{\sum_s\bE_s'IV\check{\bR}\check{\bLambda}_R^{-1}\bF_s}_F^2
&\lesssim\norm{\sum_s\mathbf{E}_s'IV\check{\mathbf{R}}\mathbf{F}_s}_F^2=\norm{\sum_s\mathbf{E}_s'\sum_t\frac{1}{{Tp_1^{\delta_1}p_2^{\delta_2}}}\mathbf{E}_t\mathbf{E}_t'\check{\mathbf{R}}\mathbf{F}_s}_F^2\\
&=\norm{Vec\paren{\sum_s\mathbf{E}_s'\sum_t\frac{1}{{Tp_1^{\delta_1}p_2^{\delta_2}}}\mathbf{E}_t\mathbf{E}_t'\check{\mathbf{R}}\mathbf{F}_s}}_F^2=\norm{\sum_sVec\paren{\mathbf{E}_s'\sum_t\frac{1}{{Tp_1^{\delta_1}p_2^{\delta_2}}}\mathbf{E}_t\mathbf{E}_t'\check{\mathbf{R}}\mathbf{F}_s}}_F^2\\
&=\norm{\sum_s\mathbf{F}_s'\otimes \mathbf{E}_s'Vec\paren{\sum_t\frac{1}{{Tp_1^{\delta_1}p_2^{\delta_2}}}\mathbf{E}_t\mathbf{E}_t'\check{\mathbf{R}}}}_F^2\le\norm{\sum_s\mathbf{F}_s'\otimes \mathbf{E}_s'}_F^2\norm{\sum_t\frac{1}{{Tp_1^{\delta_1}p_2^{\delta_2}}}\mathbf{E}_t\mathbf{E}_t'\check{\mathbf{R}}}_F^2\\
&\le\sum_{i,j}T\norm{\sum_s\frac{1}{\sqrt{T}}\mathbf{F}_s'\otimes e_{s,ij}}_F^2\norm{\frac{1}{{Tp_1^{\delta_1}p_2^{\delta_2}}}\sum_t\mathbf{E}_t\mathbf{E}_t'\check{\mathbf{R}}}_F^2\le Tp_1p_2\frac{1}{{T^2p_1^{2\delta_1}p_2^{2\delta_2}}}\norm{\sum_t\check{\mathbf{R}}'\mathbf{E}_t\mathbf{E}_t'}_F^2\\
&\lesssim\frac{1}{Tp_1^{2\delta_1-1}p_2^{2\delta_2-1}}\sum_i\norm{\sum_{t}\paren{\check{\mathbf{R}}-\mathbf{R}\check{\mathbf{H}}_R}'\mathbf{E}_t\mathbf{e}_{t,i\cdot }}_F^2\\
&+\frac{1}{Tp_1^{2\delta_1-1}p_2^{2\delta_2-1}}\sum_i\norm{\sum_{t}\paren{\mathbf{R}\check{\mathbf{H}}_R}'\mathbf{E}_t\mathbf{e}_{t,i\cdot }}_F^2\\
&=\frac{1}{Tp_1^{2\delta_1-1}p_2^{2\delta_2-1}}p_1O_p\paren{Tp_1p_2+T^2p_2^2}\norm{\check{\mathbf{R}}-\mathbf{R}\check{\mathbf{H}}_R}_F^2\\
&+\frac{1}{Tp_1^{2\delta_1-1}p_2^{2\delta_2-1}}p_1O_p\paren{Tp_1^{\delta_1}p_2+T^2p_2^2}\\
&= O_p\paren{p_1^{3-2\delta_1}p_2^{2-2\delta_2}+Tp_1^{2-2\delta_1}p_2^{3-2\delta_2}}\norm{\check{\mathbf{R}}-\mathbf{R}\check{\mathbf{H}}_R}_F^2+O_p\paren{p_1^{2-\delta_1}p_2^{2-2\delta_2}+Tp_1^{2-2\delta_1}p_2^{3-2\delta_2}}
\end{aligned}
\]
Combining the above results, we have
\[
\begin{aligned}
\norm{\sum_{s}\mathbf{E}_s'\paren{\check{\mathbf{R}}-\mathbf{R}\check{\mathbf{H}}_R}\mathbf{F}_s}_F^2
&=O_p\paren{\frac{p_1^{2-\delta_1}}{Tp_2^{2\delta_2-1}}+\frac{p_1^{2-2\delta_1}}{Tp_2^{3\delta_2-2}}+\frac{p_2^{3-3\delta_2}}{p_1^{3\delta_1-2}}}+O_p\paren{p_2^{1-\delta_2}}\\
&+O_p\paren{p_1p_2^{1-\delta_2}+\frac{p_1^2}{p_2^{2\delta_2-1}}}\\
&+ O_p\paren{p_1^{3-2\delta_1}p_2^{2-2\delta_2}+Tp_1^{2-2\delta_1}p_2^{3-2\delta_2}}\norm{\check{\mathbf{R}}-\mathbf{R}\check{\mathbf{H}}_R}_F^2\\
&+O_p\paren{p_1^{2-\delta_1}p_2^{2-2\delta_2}+Tp_1^{2-2\delta_1}p_2^{3-2\delta_2}}\\
&=O_p\paren{\frac{p_1^{2-\delta_1}}{Tp_2^{2\delta_2-1}}+\frac{p_1^{2-2\delta_1}}{Tp_2^{3\delta_2-2}}+\frac{p_2^{3-3\delta_2}}{p_1^{3\delta_1-2}}}+O_p\paren{p_2^{1-\delta_2}}\\
&+O_p\paren{p_1p_2^{1-\delta_2}+\frac{p_1^2}{p_2^{2\delta_2-1}}}\\
&+ O_p\paren{p_1^{3-\delta_1}p_2^{2-2\delta_2}+Tp_1^{2-\delta_1}p_2^{3-2\delta_2}}\times\\
&\bracket{O_p\paren{\frac{p_1^{1-\delta_1}}{Tp_2^{\delta_2}}}+O_p\paren{\frac{1}{Tp_1^{2\delta_1-1}p_2^{2\delta_2-1}}+\frac{p_2^{2-2\delta_2}}{p_1^{3\delta_1-1}}}}\\
&+O_p\paren{p_1^{2-\delta_1}p_2^{2-2\delta_2}+Tp_1^{2-2\delta_1}p_2^{3-2\delta_2}}\\
&=O_p\left(\frac{p_1^2}{p_2^{2\delta_2-1}}+\frac{p_1^{4-2\delta_1}}{Tp_2^{3\delta_2-2}}+\frac{p_1^{4-3\delta_1}}{Tp_2^{4\delta_2-3}}+p_1^{4-4\delta_1}p_2^{4-4\delta_2}\right.\\
&\left.+p_1^{3-2\delta_1}{p_2^{3-3\delta_2}}+\frac{Tp_2^{5-4\delta_2}}{p_1^{4\delta_1-3}}+Tp_1^{2-2\delta_1}p_2^{3-2\delta_2}\right)\\
&=O_p\paren{\frac{p_1^2}{p_2^{2\delta_2-1}}+\frac{p_1^{4-2\delta_1}}{Tp_2^{3\delta_2-2}}+p_1^{3-2\delta_1}{p_2^{3-3\delta_2}}
+Tp_1^{2-2\delta_1}p_2^{3-2\delta_2}}.
\end{aligned}
\]
The remaining terms are absorbed under \eqref{cond:C1}.
Set
\[
w_2:=\frac{1}{T^2p_1^{2\delta_1}p_2^{\delta_2}}\paren{\frac{p_1^2}{p_2^{2\delta_2-1}}+\frac{p_1^{4-2\delta_1}}{Tp_2^{3\delta_2-2}}+p_1^{3-2\delta_1}{p_2^{3-3\delta_2}}
+Tp_1^{2-2\delta_1}p_2^{3-2\delta_2}}
\]
\[
=\paren{\frac{p_1^{2-2\delta_1}}{T^2p_2^{3\delta_2-1}}+\frac{p_1^{4-4\delta_1}}{T^3p_2^{4\delta_2-2}}+\frac{1}{T^2p_1^{4\delta_1-3}p_2^{4\delta_2-3}}+\frac{p_2^{3-3\delta_2}}{Tp_1^{4\delta_1-2}}}
\]
This completes the proof.
\end{proof}
\subsubsection{Projected Estimators}
    \begin{equation}
    \begin{aligned}
    \widetilde{\mathbf{M}_C}&=\frac{1}{Tp_1^{2\delta_1}p_2^{\delta_2}}\sum_t\mathbf{X}_t'\check{\mathbf{R}}\check{\mathbf{R}}'\mathbf{X}_t\\
    &=\frac{1}{Tp_1^{2\delta_1}p_2^{\delta_2}}\sum_t\paren{\mathbf{C}\mathbf{F}_t'\mathbf{R}'+\mathbf{E}_t'}\check{\mathbf{R}}\check{\mathbf{R}}'\paren{\mathbf{R}\mathbf{F}_t\mathbf{C}'+\mathbf{E}_t}\\
    &=\frac{1}{Tp_1^{2\delta_1}p_2^{\delta_2}}\sum_t\left(\mathbf{C}\mathbf{F}_t'\mathbf{R}'\check{\mathbf{R}}\check{\mathbf{R}}'\mathbf{R}\mathbf{F}_t\mathbf{C}'+\mathbf{C}\mathbf{F}_t'\mathbf{R}'\check{\mathbf{R}}\check{\mathbf{R}}'\mathbf{E}_t\right.\\
&\left.+\mathbf{E}_t'\check{\mathbf{R}}\check{\mathbf{R}}'\mathbf{R}\mathbf{F}_t\mathbf{C}'+\mathbf{E}_t'\check{\mathbf{R}}\check{\mathbf{R}}'\mathbf{E}_t\right)\\
&:=I+II+III+IV
\end{aligned}
    \end{equation}
Define
\begin{equation}
\widetilde{\mathbf{H}}_C
=
\frac{1}{T p_1^{2\delta_1} p_2^{\delta_2}}
\sum_{t=1}^T
\mathbf{F}_t^{\top}\mathbf{R}^{\top}\check{\mathbf{R}}
\check{\mathbf{R}}^{\top}\mathbf{R}
\mathbf{F}_t
\mathbf{C}^{\top}\widetilde{\mathbf{C}}
\widetilde{\boldsymbol{\Lambda}}_C^{-1},
\end{equation}
then by the definition of $\widetilde{\mathbf{C}}$ we have
$\widetilde{\mathbf{C}}\widetilde{\boldsymbol{\Lambda}}_C
=
\widetilde{\mathbf{M}}_C\widetilde{\mathbf{C}}$, and
\begin{equation}
\widetilde{\mathbf{C}}
-
\mathbf{C}\widetilde{\mathbf{H}}_C
=
(II+III+IV)
\widetilde{\mathbf{C}}
\widetilde{\boldsymbol{\Lambda}}_C^{-1}.
\end{equation}
An analogous discussion applies to $\widetilde{\bM}_R$, $\widetilde{\bR}$, $\widetilde{\bLambda}_R$ and $\widetilde{\bH}_R$.
We begin by presenting several auxiliary lemmas and propositions, after which we provide the proofs of Theorem~\ref{thm:projected-estimation}\textup{(I)} and \textup{(II)}.
\begin{proposition}\label{pro:eigenvalues_for_projection}
Suppose Assumptions~\ref{ass:weak loading}--\ref{ass:noise-dependence}, Assumption~\ref{ass:projected_spatial_noise}--\ref{ass:Factor-noise interaction for projected} and \eqref{cond:C1} hold. Then
    \[
\lambda_j\paren{\widetilde{\mathbf{M}_C}} =
    \begin{cases}
\lambda_j\paren{\boldsymbol{\Sigma}_2}+o_p\paren{1},&\text {for }j\le k_2\\
        o_p(1),&\text{for }j>k_2
    \end{cases}
    \]
    Symmetrically, 
\[
\lambda_j\paren{\widetilde{\mathbf{M}_R}} =
    \begin{cases}
        \lambda_j\paren{\boldsymbol{\Sigma}_1}+o_p\paren{1},&\text {for }j\le k_1\\
        o_p(1),&\text{for }j>k_1
    \end{cases}
    \]
\end{proposition}

\begin{proof}
Without loss of generality, we set \(k_1=k_2=1\).
\[
\begin{aligned}
I
&=
\frac{1}{T p_1^{2\delta_1} p_2^{\delta_2}}
\mathbf{C}\left(
\sum_{t=1}^T
\mathbf{F}_t^{\top}\mathbf{R}^{\top}\check{\mathbf{R}}\check{\mathbf{R}}^{\top}\mathbf{R}\mathbf{F}_t
\right)\mathbf{C}^{\top} \\
&=
\frac{1}{T p_1^{2\delta_1} p_2^{\delta_2}}
\mathbf{C}\left(
\sum_{t=1}^T
\mathbf{F}_t^{\top}\mathbf{R}^{\top}
(\mathbf{R}\check{\mathbf{H}}_R+\check{\mathbf{R}}-\mathbf{R}\check{\mathbf{H}}_R)
(\mathbf{R}\check{\mathbf{H}}_R+\check{\mathbf{R}}-\mathbf{R}\check{\mathbf{H}}_R)^{\top}
\mathbf{R}\mathbf{F}_t
\right)\mathbf{C}^{\top} \\
&=
\frac{1}{T p_1^{2\delta_1} p_2^{\delta_2}}
\mathbf{C}\left(
\sum_{t=1}^T
\mathbf{F}_t^{\top}\mathbf{R}^{\top}\mathbf{R}\check{\mathbf{H}}_R
\check{\mathbf{H}}_R^{\top}\mathbf{R}^{\top}\mathbf{R}\mathbf{F}_t
\right)\mathbf{C}^{\top} \\
&\quad+
\left(
\frac{1}{T p_2^{\delta_2}}
\mathbf{C}\sum_{t=1}^T \mathbf{F}_t^{\top}\mathbf{F}_t \mathbf{C}^{\top}
\right) \\
&\quad\times
\left(
\frac{1}{p_1^{2\delta_1}}\mathbf{R}^{\top}\mathbf{R}\check{\mathbf{H}}_R
(\check{\mathbf{R}}-\mathbf{R}\check{\mathbf{H}}_R)^{\top}\mathbf{R}
+
\frac{1}{p_1^{2\delta_1}}\mathbf{R}^{\top}
(\check{\mathbf{R}}-\mathbf{R}\check{\mathbf{H}}_R)\check{\mathbf{H}}_R^{\top}\mathbf{R}^{\top}\mathbf{R}
\right.\\
&\qquad\qquad\left.
+
\frac{1}{p_1^{2\delta_1}}\mathbf{R}^{\top}
(\check{\mathbf{R}}-\mathbf{R}\check{\mathbf{H}}_R)
(\check{\mathbf{R}}-\mathbf{R}\check{\mathbf{H}}_R)^{\top}\mathbf{R}
\right).\\
\end{aligned}
\]
Thus,
\[\begin{aligned}
{I}&\xrightarrow{p}\frac{1}{Tp_2^{\delta_2}}\sum_t\mathbf{C}\mathbf{F}_t'\mathbf{F}_t\mathbf{C}'+\frac{1}{Tp_2^{\delta_2}}\sum_t\mathbf{C}\mathbf{F}_t'\mathbf{F}_t\mathbf{C}'\times o_p(1)\\
&\xrightarrow{p}\frac{1}{p_2^{\delta_2}}\mathbf{C}\boldsymbol{\Sigma}_2\mathbf{C}'+\frac{1}{p_2^{\delta_2}}\mathbf{C}\boldsymbol{\Sigma}_2\mathbf{C}'\times o_p(1)\\
&\xrightarrow{p}\frac{1}{p_2^{\delta_2}}\mathbf{C}\boldsymbol{\Sigma}_2\mathbf{C}'+ o_p(1).
\end{aligned}
\]
For $j\le k_2$,
\[
\begin{aligned}
\lambda_j\paren{I}&=\lambda_j\paren{\frac{1}{p_2^{\delta_2}}\mathbf{C}\boldsymbol{\Sigma}_2\mathbf{C}'}+o_p(1)\\
&=\lambda_j\paren{\frac{1}{p_2^{\delta_2}}\boldsymbol{\Sigma}_2^{\frac{1}{2}}\mathbf{C}'\mathbf{C}\boldsymbol{\Sigma}_2^{\frac{1}{2}}}+o_p(1)\\
&=\lambda_j\paren{\boldsymbol{\Sigma}_2}+o_p(1)
\end{aligned}
\]
For $j>k_2$,
\(
\lambda_j\paren{I}=0
\).
\[
\begin{aligned}
\norm{II}&=\norm{\frac{1}{Tp_1^{2\delta_1}p_2^{\delta_2}}
\sum_t\mathbf{C}\mathbf{F}_t'\mathbf{R}'\check{\mathbf{R}}\check{\mathbf{R}}'\mathbf{E}_t}\lesssim \norm{\frac{1}{{Tp_1^{\delta_1}p_2^{\delta_2}}}\sum_t\mathbf{C}\mathbf{F}_t'\check{\mathbf{R}}'\mathbf{E}_t}\\
&\le\norm{\frac{1}{{Tp_1^{\delta_1}p_2^{\delta_2}}}\sum_t\mathbf{C}\mathbf{F}_t'\paren{\check{\mathbf{R}}-\mathbf{R}\check{\mathbf{H}}_R}'\mathbf{E}_t}+\norm{\frac{1}{{Tp_1^{\delta_1}p_2^{\delta_2}}}\sum_t\mathbf{C}\mathbf{F}_t'\check{\mathbf{H}}_R'\mathbf{R}'\mathbf{E}_t}\\
&\le\frac{1}{p_2^{\frac{\delta_2}{2}}}\norm{\mathbf{C}}\frac{1}{Tp_1^{\delta_1}p_2^{\frac{\delta_2}{2}}}\norm{\sum_t\mathbf{F}_t'\paren{\check{\mathbf{R}}-\mathbf{R}\check{\mathbf{H}}_R}'\mathbf{E}_t}+\frac{1}{{Tp_1^{\delta_1}p_2^{\delta_2}}}\norm{\mathbf{C}}\norm{\sum_t\mathbf{F}_t'\mathbf{R}'\mathbf{E}_t}\\
&= O_p\paren{\sqrt{w_2}}+\frac{1}{{Tp_1^{\delta_1}p_2^{\delta_2}}}p_2^{\frac{\delta_2}{2}}O_p\paren{\sqrt{Tp_1^{\delta_1}p_2}}\\
&=O_p\paren{\sqrt{w_2}}+O_p\paren{\sqrt{\frac{p_2^{1-\delta_2}}{Tp_1^{\delta_1}}}}=o_p(1)
\end{aligned}
\]
Similarly,
\[
\begin{aligned}
\norm{III}&=\norm{\frac{1}{Tp_1^{2\delta_1}p_2^{\delta_2}}
\sum_t\mathbf{E}_t'\check{\mathbf{R}}\check{\mathbf{R}}'\mathbf{R}\mathbf{F}_t\mathbf{C}'}\lesssim\norm{\frac{1}{Tp_1^{\delta_1}p_2^{\delta_2}}
\sum_t\mathbf{E}_t'\check{\mathbf{R}}\mathbf{F}_t\mathbf{C}'}\\
&= O_p\paren{\sqrt{w_2}}+O_p\paren{\sqrt{\frac{p_2^{1-\delta_2}}{Tp_1^{\delta_1}}}}=o_p(1)
\end{aligned}
\]
\[
\begin{aligned}
\norm{IV}
&=\norm{\frac{1}{Tp_1^{2\delta_1}p_2^{\delta_2}}
\sum_t\mathbf{E}_t'\check{\mathbf{R}}\check{\mathbf{R}}'\mathbf{E}_t}\le\frac{1}{Tp_1^{2\delta_1}p_2^{\delta_2}}\sum_t\norm{\mathbf{E}_t'\check{\mathbf{R}}}_F^2\\
&\lesssim\frac{1}{Tp_1^{2\delta_1}p_2^{\delta_2}}\sum_t\norm{\mathbf{E}_t'\mathbf{R}}_F^2+\frac{1}{Tp_1^{2\delta_1}p_2^{\delta_2}}\sum_t\norm{\mathbf{E}_t'\paren{\check{\mathbf{R}}-\mathbf{R}\check{\mathbf{H}}_R}}_F^2\\
&=\frac{1}{Tp_1^{2\delta_1}p_2^{\delta_2}}O_p\paren{Tp_1^{\delta_1}p_2}+\frac{1}{Tp_1^{2\delta_1}p_2^{\delta_2}}O_p\paren{Tp_1p_2}p_1^{\delta_1}w_1\\
&=O_p\paren{\frac{p_2^{1-\delta_2}}{p_1^{\delta_1}}}+p_1^{1-\delta_1}p_2^{1-\delta_2}\times\paren{O_p\paren{\frac{p_1^{1-\delta_1}}{Tp_2^{\delta_2}}}+O_p\paren{\frac{1}{Tp_1^{2\delta_1-1}p_2^{2\delta_2-1}}+\frac{p_2^{2-2\delta_2}}{p_1^{3\delta_1-1}}}}\\
&=O_p\paren{\frac{p_2^{1-\delta_2}}{p_1^{\delta_1}}}+O_p\paren{\frac{p_1^{2-2\delta_1}}{Tp_2^{2\delta_2-1}}+\frac{1}{Tp_1^{3\delta_1-2}p_2^{3\delta_2-2}}+\frac{p_2^{3-3\delta_2}}{p_1^{4\delta_1-2}}}=o_p(1)
\end{aligned}
\]
Thus,
\[
    \lambda_j\paren{\widetilde{\mathbf{M}_C}} =
    \begin{cases}
        \lambda_j\paren{\boldsymbol{\Sigma}_2}+o_p\paren{1},&\text {for }j\le k_2\\
        o_p(1),&\text{for }j>k_2
    \end{cases}
    \]
By symmetry, the same arguments apply to the eigenvalues of $\widetilde{\bM}_R$.
\end{proof}
\begin{remark}
The eigenvalue-ratio method for determining the number of factors relies on
the asymptotic separation between the factor eigenvalues and the remaining
eigenvalues. Under Condition~\eqref{cond:C1},
Proposition~\ref{pro:eigenvalues_for_projection} shows that the eigenvalues
beyond the true factor number are asymptotically negligible. When Condition~\eqref{cond:C1} is not satisfied, however, this conclusion is no longer guaranteed, and the usual consistency argument for the eigenvalue-ratio method may not apply. By removing the rows and columns without factor signals, the proposed screening procedure recovers a submatrix with a pervasive loading structure relative to the retained dimensions, thereby facilitating more reliable determination of the factor numbers as stated in Corollary~\ref{thm:screened estimators}. This provides an additional justification for the proposed screening procedure.
\end{remark}

\begin{lemma}
Suppose Assumptions~\ref{ass:weak loading}--\ref{ass:noise-dependence}, Assumption~\ref{ass:projected_spatial_noise}--\ref{ass:Factor-noise interaction for projected} and \eqref{cond:C1} hold. Then
\[
\frac{1}{p_2^{\delta_2}} \norm{II\widetilde{\mathbf{C}}}_F^2=O_p\paren{w_2}+O_p\paren{\frac{p_2^{1-\delta_2}}{Tp_1^{\delta_1}}}
\]
\[
\frac{1}{p_2^{\delta_2}} \norm{III\widetilde{\mathbf{C}}}_F^2=O_p\paren{w_2}+O_p\paren{\frac{p_2^{1-\delta_2}}{Tp_1^{\delta_1}}}
\]
\end{lemma}
\begin{proof}
\[
\begin{aligned}
\frac{1}{p_2^{\delta_2}} \norm{II\widetilde{\mathbf{C}}}_F^2&=\frac{1}{p_2^{\delta_2}}\norm{\frac{1}{Tp_1^{2\delta_1}p_2^{\delta_2}}\sum_t\mathbf{C}\mathbf{F}_t'\mathbf{R}'\check{\mathbf{R}}\check{\mathbf{R}}'\mathbf{E}_t\widetilde{\mathbf{C}}}_F^2\\
&\le \norm{\frac{1}{Tp_1^{2\delta_1}p_2^{\delta_2}}\sum_t\mathbf{C}\mathbf{F}_t'\mathbf{R}'\check{\mathbf{R}}\check{\mathbf{R}}'\mathbf{E}_t}_F^2\\&=\norm{II}_F^2\\
&\le O_p\paren{w_2}+O_p\paren{\frac{p_2^{1-\delta_2}}{Tp_1^{\delta_1}}}
\end{aligned}
\]
\[
\begin{aligned}
\frac{1}{p_2^{\delta_2}}\norm{III\widetilde{\bC}}_F^2&=\frac{1}{p_2^{\delta_2}}\norm{\frac{1}{Tp_1^{2\delta_1}p_2^{\delta_2}}\sum_t\mathbf{E}_t'\check{\mathbf{R}}\check{\mathbf{R}}'\mathbf{R}\mathbf{F}_t\mathbf{C}'\widetilde{\mathbf{C}}}_F^2\\
&\le\norm{\frac{1}{Tp_1^{2\delta_1}p_2^{\delta_2}}\sum_t\mathbf{E}_t'\check{\mathbf{R}}\check{\mathbf{R}}'\mathbf{R}\mathbf{F}_t\mathbf{C}'}_F^2\\&=\norm{III}_F^2\\
&\le O_p\paren{w_2}+O_p\paren{\frac{p_2^{1-\delta_2}}{Tp_1^{\delta_1}}}
\end{aligned}
\]
\end{proof}
\begin{lemma}
Suppose Assumptions~\ref{ass:weak loading}--\ref{ass:noise-dependence}, Assumption~\ref{ass:projected_spatial_noise}--\ref{ass:Factor-noise interaction for projected} and \eqref{cond:C1} hold. Then
\[
\begin{aligned}
    \frac{1}{p_2^{\delta_2}}\norm{IV\widetilde{\mathbf{C}}}_F^2
&=O_p\paren{\frac{1}{Tp_1^{\delta_1}p_2^{2\delta_2-1}}+\frac{1}{p_1^{2\delta_1}p_2^{3\delta_2-1}}}+o_p(1)\frac{\norm{\widetilde{\mathbf{C}}-\mathbf{C}\widetilde{\mathbf{H}}_C}_F^2}{p_2^{\delta_2}}\\
&\quad+\paren{\frac{p_1^{2-2\delta_1}}{Tp_2^{2\delta_2-1}}+O_p\paren{\frac{p_1^{2-2\delta_1}}{p_2^{3\delta_2-1}}}w_1^2}
\end{aligned}
\]
\end{lemma}
\begin{proof}
    \[
    \begin{aligned}
IV\widetilde{\mathbf{C}}
&=
\frac{1}{T p_1^{2\delta_1} p_2^{\delta_2}}
\sum_{t=1}^T
\mathbf{E}_t^\top \check{\mathbf{R}} \check{\mathbf{R}}^\top \mathbf{E}_t \widetilde{\mathbf{C}}\\
&=
\frac{1}{T p_1^{2\delta_1} p_2^{\delta_2}}
\left(
\sum_{t=1}^T
\mathbf{E}_t^\top \mathbf{R} \check{\mathbf{H}}_R \check{\mathbf{H}}_R^\top \mathbf{R}^\top \mathbf{E}_t \mathbf{C} \widetilde{\mathbf{H}}_C
+
\sum_{t=1}^T
\mathbf{E}_t^\top \mathbf{R} \check{\mathbf{H}}_R \check{\mathbf{H}}_R^\top \mathbf{R}^\top \mathbf{E}_t
(\widetilde{\mathbf{C}} - \mathbf{C} \widetilde{\mathbf{H}}_C)
\right.\\
&\left.
\qquad
+
\sum_{t=1}^T
\mathbf{E}_t^\top (\check{\mathbf{R}}-\mathbf{R}\check{\mathbf{H}}_R)\check{\mathbf{H}}_R^\top \mathbf{R}^\top \mathbf{E}_t \widetilde{\mathbf{C}}
+
\sum_{t=1}^T
\mathbf{E}_t^\top \check{\mathbf{R}} (\check{\mathbf{R}}-\mathbf{R}\check{\mathbf{H}}_R)^\top \mathbf{E}_t \widetilde{\mathbf{C}}
\right)\\
&:=  V +  VI + VII + VIII.
\end{aligned}
\]
\[
\begin{aligned}
\mathbb{E}\left\|
\sum_{t=1}^T \mathbf{E}_t^\top \mathbf{R} \mathbf{R}^\top \mathbf{E}_t \mathbf{C}
\right\|_F^2
&=\sum_j\E\norm{\sum_t\mathbf{e}_{t,\cdot j}'\mathbf{R} \sum_i\mathbf{R}_i\mathbf{e}_{t,i\cdot }' \mathbf{C}}_F^2\\
&=\sum_j\E\norm{\sum_t\mathbf{R}'\mathbf{e}_{t,\cdot j} \sum_{i\in \mathcal{I}_R}\mathbf{R}_i\mathbf{e}_{t,i\cdot }' \mathbf{C}}_F^2\\
&\le\sum_{i\in \mathcal{I}_R,j}\E\norm{\mathbf{R}}_F^2\norm{\sum_t\mathbf{R}'\mathbf{e}_{t,\cdot j}\mathbf{e}_{t,i\cdot }'\mathbf{C}}_F^2
\end{aligned}
\]
\[
\begin{aligned}
\mathbb{E}\left\|
\sum_t \mathbf{R}^\top \mathbf{e}_{t,\cdot j} \mathbf{e}_{t,i\cdot }^\top \mathbf{C}
\right\|^2
&\le
\mathbb{E}\left\|
\sum_t
\left(
\mathbf{R}^\top \mathbf{e}_{t,\cdot j} \mathbf{e}_{t,i\cdot }^\top \mathbf{C}
-
\mathbb{E}\mathbf{R}^\top \mathbf{e}_{t,\cdot j} \mathbf{e}_{t,i\cdot }^\top \mathbf{C}
\right)
\right\|^2+
\left\|
\sum_t
\mathbb{E}\mathbf{R}^\top \mathbf{e}_{t,\cdot j} \mathbf{e}_{t,i\cdot }^\top \mathbf{C}
\right\|^2\\
&\le
\mathbb{E}\left\|
\sum_{t,l,h}
\left(
\mathbf{R}_l e_{t,l j} e_{t,ih} \mathbf{C}_h
-
\mathbb{E}\mathbf{R}_l e_{t,l j} e_{t,ih} \mathbf{C}_h
\right)
\right\|^2
+\left\|
\sum_{t,i_1,j_1}
\mathbb{E}\mathbf{R}_{i_1} e_{t,i_1 j} e_{t,ij_1} \mathbf{C}_{j_1}
\right\|^2\\
&\le
\sum_{t,s}\sum_{i_1,j_1}\sum_{i_2,j_2}\abs{\mathbf{R}_{i_1}\mathbf{R}_{i_2}\mathbf{C}_{j_1}\mathbf{C}_{j_2}}
\left|
\cov
\left(
e_{t,i_1 j}e_{t,i j_1},
e_{s,i_2 j}e_{s,i j_2}
\right)
\right|\\
&+
\left(
\sum_t \sum_{i_1}\sum_{j_1}\abs{\mathbf{R}_{i_1}\mathbf{C}_{j_1}}
\left |
\mathbb{E}e_{t,i_1 j}e_{t,i j_1}
\right|
\right)^2\\
&=
\sum_{t,s}\sum_{i_1\in \mathcal{I}_R,j_1\in \mathcal{I}_C}\sum_{i_2\in \mathcal{I}_R,j_2\in \mathcal{I}_C}\abs{\mathbf{R}_{i_1}\mathbf{R}_{i_2}\mathbf{C}_{j_1}\mathbf{C}_{j_2}}
\left|
\cov
\left(
e_{t,i_1 j}e_{t,i j_1},
e_{s,i_2 j}e_{s,i j_2}
\right)
\right|\\
&+
\left(
\sum_t \sum_{i_1\in \mathcal{I}_R}\sum_{j_1 \in \mathcal{I}_C}\abs{\mathbf{R}_{i_1}\mathbf{C}_{j_1}}
\left |
\mathbb{E}e_{t,i_1 j}e_{t,i j_1}
\right|
\right)^2\\
&\le
O\left(
T p_1^{\delta_1} p_2^{\delta_2}
+
T^2
\right).
\end{aligned}
\]
Thus,
\[
\norm{V}_F^2\le\frac{1}{T^2p_1^{4\delta_1}p_2^{2\delta_2}}\times O_p\paren{Tp_1^{\delta_1}p_2^{\delta_2}+T^2}\times p_1^{2\delta_1}p_2=O_p\paren{\frac{p_2^{1-\delta_2}}{Tp_1^{\delta_1}}+\frac{1}{p_1^{2\delta_1}p_2^{2\delta_2-1}}}
\]
\[
\begin{aligned}
\|VI\|_F^2
&\le
\frac{c}{T^2 p_1^{4\delta_1} p_2^{2\delta_2}}
\left(
\sum_t \|\mathbf{E}_t^\top \mathbf{R}\|_F^2
\right)
\left(
\sum_t \|\mathbf{E}_t^\top \mathbf{R}\|_F^2
\right)
\|\widetilde{\mathbf{C}} - \mathbf{C} \widetilde{\mathbf{H}}_C\|_F^2\\
&\le\frac{c}{T^2 p_1^{4\delta_1} p_2^{2\delta_2}}O_p\paren{T^2p_1^{2\delta_1}p_2^2}\|\widetilde{\mathbf{C}} - \mathbf{C} \widetilde{\mathbf{H}}_C\|_F^2=\frac{cp_2^{2-2\delta_2}}{p_1^{2\delta_1}}\norm{\widetilde{\mathbf{C}}-\mathbf{C}\widetilde{\mathbf{H}}_C}_F^2
\end{aligned}
\]
\[
\begin{aligned}
\|VII\|_F^2
&\le
\frac{c}{T^2 p_1^{4\delta_1} p_2^{2\delta_2}}
\left(
\left\|
\sum_t
\mathbf{E}_t^\top(\check{\mathbf{R}}-\mathbf{R}\check{\mathbf{H}}_R)\mathbf{R}^\top \mathbf{E}_t \mathbf{C}
\right\|_F^2
\right.\\
&\left.
+
\sum_t
\|\mathbf{E}_t^\top(\check{\mathbf{R}}-\mathbf{R}\check{\mathbf{H}}_R)\|_F^2
\sum_t
\|\mathbf{R}^\top \mathbf{E}_t\|_F^2
\|\widetilde{\mathbf{C}} - \mathbf{C}\widetilde{\mathbf{H}}_C\|_F^2
\right)\\
&\le
\frac{1}{T^2 p_1^{4\delta_1} p_2^{2\delta_2}}
\left\|
\sum_t
\mathbf{E}_t^\top(\check{\mathbf{R}}-\mathbf{R}\check{\mathbf{H}}_R)\mathbf{R}^\top \mathbf{E}_t \mathbf{C}
\right\|_F^2\\
&+\frac{1}{T^2 p_1^{4\delta_1} p_2^{2\delta_2}}O_p\paren{Tp_1p_2p_1^{\delta_1}w_1Tp_1^{\delta_1}p_2}
\|\widetilde{\mathbf{C}} - \mathbf{C}\widetilde{\mathbf{H}}_C\|_F^2\\
&=\frac{1}{T^2 p_1^{4\delta_1} p_2^{2\delta_2}}
\left\|
\sum_t
\mathbf{E}_t^\top(\check{\mathbf{R}}-\mathbf{R}\check{\mathbf{H}}_R)\mathbf{R}^\top \mathbf{E}_t \mathbf{C}
\right\|_F^2+\frac{p_2^{2-2\delta_2}}{p_1^{2\delta_1-1}}O_p\paren{w_1}\norm{\widetilde{\mathbf{C}}-\mathbf{C}\widetilde{\mathbf{H}}_C}_F^2
\end{aligned}
\]
\[
\begin{aligned}
\left\|
\sum_t \mathbf{E}_t^\top(\check{\mathbf{R}}-\mathbf{R}\check{\mathbf{H}}_R)\mathbf{R}^\top \mathbf{E}_t \mathbf{C}
\right\|_F^2&=\sum_j\norm{\sum_t\mathbf{e}_{t,\cdot j}'\paren{\check{\mathbf{R}}-\mathbf{R}\check{\mathbf{H}}_R}\mathbf{R}'\mathbf{E}_t\mathbf{C}}\\
&=\sum_j\norm{\sum_{t,i}e_{t,ij}\paren{\check{\mathbf{R}}_i-\mathbf{R}_i\check{\mathbf{H}}_R}\mathbf{R}'\mathbf{E}_t\mathbf{C}}_F^2\\
&\le\norm{\check{\mathbf{R}}-\mathbf{R}\check{\mathbf{H}}_R}_F^2\sum_{i,j}\norm{\sum_te_{t,ij}\mathbf{R}'\mathbf{E}_t\mathbf{C}}_F^2
\end{aligned}
\]
\[
\begin{aligned}
\mathbb{E}
\left\|
\sum_t e_{t,ij} \mathbf{R}^\top \mathbf{E}_t \mathbf{C}
\right\|_F^2
&=\E\norm{\sum_te_{t,ij}\sum_{l,h}\mathbf{R}_le_{t,lh}\mathbf{C}_h}_F^2
=\E\norm{\sum_{t,l,h}e_{t,ij}\mathbf{R}_le_{t,lh}\mathbf{C}_h}_F^2\\
&=\E\norm{\sum_{t,l,h}\paren{e_{t,ij}\mathbf{R}_le_{t,lh}\mathbf{C}_h-\E e_{t,ij}\mathbf{R}_le_{t,lh}\mathbf{C}_h}+\sum_t\E e_{t,ij}\mathbf{R}'\mathbf{E}_t\mathbf{C}}_F^2\\
&\le
\sum_{t,s}\sum_{i_1,j_1}\sum_{i_2,j_2}
\abs{\mathbf{R}_{i_1}\mathbf{R}_{i_2}\mathbf{C}_{j_1}\mathbf{C}_{j_2}}\left|
\cov
\left(
e_{t,ij}e_{t,i_1j_1},
e_{s,ij}e_{s,i_2j_2}
\right)
\right|\\
&+
\left\|
\sum_t
\mathbb{E}e_{t,ij}\mathbf{R}^\top \mathbf{E}_t \mathbf{C}
\right\|_F^2\\
&=\sum_{t,s}\sum_{i_1\in \mathcal{I}_R,j_1\in \mathcal{I}_C}\sum_{i_2\in \mathcal{I}_R,j_2\in \mathcal{I}_C}
\abs{\mathbf{R}_{i_1}\mathbf{R}_{i_2}\mathbf{C}_{j_1}\mathbf{C}_{j_2}}\left|
\cov
\left(
e_{t,ij}e_{t,i_1j_1},
e_{s,ij}e_{s,i_2j_2}
\right)
\right|\\
&+
\left\|
\sum_t
\mathbb{E}e_{t,ij}\mathbf{R}^\top \mathbf{E}_t \mathbf{C}
\right\|_F^2\\
&\le
T p_1^{\delta_1}p_2^{\delta_2}
+
\left(
\sum_t\sum_{i_1,j_1}\abs{\mathbf{R}_{i_1}\mathbf{C}_{j_1}}
\left|
\mathbb{E}e_{t,ij}e_{t,i_1j_1}
\right|
\right)^2\\
&=
O\left(
T p_1^{\delta_1}p_2^{\delta_2}
+
T^2
\right).
\end{aligned}
\]
Hence,
\[
\begin{aligned}
\|VII\|_F^2
&=\frac{1}{T^2 p_1^{4\delta_1} p_2^{2\delta_2}}
\left\|
\sum_t
\mathbf{E}_t^\top(\check{\mathbf{R}}-\mathbf{R}\check{\mathbf{H}}_R)\mathbf{R}^\top \mathbf{E}_t \mathbf{C}
\right\|_F^2+\frac{p_2^{2-2\delta_2}}{p_1^{2\delta_1-1}}O_p\paren{w_1}\norm{\widetilde{\mathbf{C}}-\mathbf{C}\widetilde{\mathbf{H}}_C}_F^2\\
&\le\frac{1}{T^2 p_1^{4\delta_1} p_2^{2\delta_2}}\norm{\check{\mathbf{R}}-\mathbf{R}\check{\mathbf{H}}_R}_F^2\sum_{i,j}\norm{\sum_te_{t,ij}\mathbf{R}'\mathbf{E}_t\mathbf{C}}_F^2+\frac{p_2^{2-2\delta_2}}{p_1^{2\delta_1-1}}O_p\paren{w_1}\norm{\widetilde{\mathbf{C}}-\mathbf{C}\widetilde{\mathbf{H}}_C}_F^2\\
&\le\frac{1}{T^2 p_1^{4\delta_1} p_2^{2\delta_2}}p_1^{\delta_1}w_1p_1p_2
O_p\left(
T p_1^{\delta_1}p_2^{\delta_2}
+
T^2
\right)+\frac{p_2^{2-2\delta_2}}{p_1^{2\delta_1-1}}w_1\norm{\widetilde{\mathbf{C}}-\mathbf{C}\widetilde{\mathbf{H}}_C}_F^2\\
&=O_p\paren{\frac{p_2^{1-\delta_2}}{Tp_1^{2\delta_1-1}}w_1+\frac{1}{p_1^{3\delta_1-1}p_2^{2\delta_2-1}}w_1}+\frac{p_2^{2-2\delta_2}}{p_1^{2\delta_1-1}}O_p\paren{w_1}\norm{\widetilde{\mathbf{C}}-\mathbf{C}\widetilde{\mathbf{H}}_C}_F^2
\end{aligned}
\]
\[
\begin{aligned}
\norm{VIII}_F^2&=\norm{\frac{1}{Tp_1^{2\delta_1}p_2^{\delta_2}}\sum_{t=1}^T
\mathbf{E}_t^\top \check{\mathbf{R}} (\check{\mathbf{R}}-\mathbf{R}\check{\mathbf{H}}_R)^\top \mathbf{E}_t \widetilde{\mathbf{C}}}_F^2\\
&\le
\frac{c}{T^2 p_1^{4\delta_1}p_2^{2\delta_2}}
\left(
\sum_t \|\mathbf{E}_t^\top \check{\mathbf{R}}\|_F^2
\right)
\times
\sum_t
\|(\check{\mathbf{R}}-\mathbf{R}\check{\mathbf{H}}_R)^\top \mathbf{E}_t\|_F^2
\|\widetilde{\mathbf{C}}-\mathbf{C}\widetilde{\mathbf{H}}_C\|_F^2\\
&+
\frac{c}{T^2 p_1^{4\delta_1}p_2^{2\delta_2}}
\left\|
\sum_t
\mathbf{E}_t^\top \check{\mathbf{R}}(\check{\mathbf{R}}-\mathbf{R}\check{\mathbf{H}}_R)^\top \mathbf{E}_t \mathbf{C}
\right\|_F^2\\
&\le\frac{c}{T^2 p_1^{4\delta_1}p_2^{2\delta_2}}\paren{\sum_t\norm{\mathbf{E}_t'\paren{\check{\mathbf{R}}-\mathbf{R}\check{\mathbf{H}}_R}}_F^2+\sum_t\norm{\mathbf{E}_t'\mathbf{R}\check{\mathbf{H}}_R}_F^2}\\
&\times\sum_t\norm{\mathbf{E}_t}_F^2\norm{\check{\mathbf{R}}-\mathbf{R}\check{\mathbf{H}}_R}_F^2\norm{\widetilde{\mathbf{C}}-\mathbf{C}\widetilde{\mathbf{H}}_C}_F^2\\
&+\frac{c}{T^2 p_1^{4\delta_1}p_2^{2\delta_2}}
\left\|
\sum_t
\mathbf{E}_t^\top \check{\mathbf{R}}(\check{\mathbf{R}}-\mathbf{R}\check{\mathbf{H}}_R)^\top \mathbf{E}_t \mathbf{C}
\right\|_F^2\\
&\le
\frac{c}{T^2 p_1^{4\delta_1}p_2^{2\delta_2}}O_p\paren{\paren{Tp_1p_2p_1^{\delta_1}w_1+Tp_1^{\delta_1}p_2}Tp_1p_2p_1^{\delta_1}w_1}\times
\|\widetilde{\mathbf{C}}-\mathbf{C}\widetilde{\mathbf{H}}_C\|_F^2\\
&+
\frac{1}{T^2 p_1^{4\delta_1}p_2^{2\delta_2}}
\left\|
\sum_t
\mathbf{E}_t^\top \check{\mathbf{R}}(\check{\mathbf{R}}-\mathbf{R}\check{\mathbf{H}}_R)^\top \mathbf{E}_t \mathbf{C}
\right\|_F^2\\
&=O_p\paren{p_1^{2-2\delta_1}p_2^{2-2\delta_2}w_1^2+\frac{p_2^{2-2\delta_2}}{p_1^{2\delta_1-1}}w_1}\norm{\widetilde{\mathbf{C}}-\mathbf{C}\widetilde{\mathbf{H}}_C}_F^2\\
&+
\frac{1}{T^2 p_1^{4\delta_1}p_2^{2\delta_2}}
\left\|
\sum_t
\mathbf{E}_t^\top \check{\mathbf{R}}(\check{\mathbf{R}}-\mathbf{R}\check{\mathbf{H}}_R)^\top \mathbf{E}_t \mathbf{C}
\right\|_F^2
\end{aligned}
\]
where under \eqref{cond:C1}, \(
p_1^{2-2\delta_1}p_2^{2-2\delta_2}w_1^2=o(1)
\).
\[
\begin{aligned}
\frac{1}{T^2p_1^{4\delta_1}p_2^{2\delta_2}}
\left\|
\sum_t
\mathbf{E}_t^\top \check{\mathbf{R}}(\check{\mathbf{R}}-\mathbf{R}\check{\mathbf{H}}_R)^\top \mathbf{E}_t \mathbf{C}
\right\|_F^2
&=\frac{1}{T^2p_1^{4\delta_1}p_2^{2\delta_2}}\sum_j\norm{\sum_t\mathbf{e}_{t,\cdot j}'\check{\mathbf{R}}\sum_i\paren{\check{\mathbf{R}}_i-\mathbf{R}_i\check{\mathbf{H}}_R} \mathbf{e}_{t,i\cdot }'\mathbf{C}}_F^2\\
&=\frac{1}{T^2p_1^{4\delta_1}p_2^{2\delta_2}}\sum_j\norm{\sum_{t,i}\paren{\check{\mathbf{R}}_i-\mathbf{R}_i\check{\mathbf{H}}_R}\check{\mathbf{R}}'\mathbf{e}_{t,\cdot j}\mathbf{e}_{t,i\cdot }'\mathbf{C}}_F^2\\
&\le
\frac{\|\check{\mathbf{R}}-\mathbf{R}\check{\mathbf{H}}_R\|_F^2}
{T^2p_1^{4\delta_1}p_2^{2\delta_2}}
\sum_{i=1}^{p_1}\sum_{j=1}^{p_2}
\left\|
\sum_t
\mathbf{C}^\top \mathbf{e}_{t,i\cdot }\mathbf{e}_{t,\cdot j}^\top \check{\mathbf{R}}
\right\|_F^2\\
&\le
\frac{O_p\paren{w_1}}{T^2p_1^{3\delta_1}p_2^{2\delta_2}}
\sum_{i=1}^{p_1}\sum_{j=1}^{p_2}
\left\|
\sum_t
\mathbf{C}^\top \mathbf{e}_{t,i\cdot }\mathbf{e}_{t,\cdot j}^\top \mathbf{R}
\right\|_F^2\\
&+
\frac{O_p\paren{w_1^2}}{T^2p_1^{2\delta_1}p_2^{2\delta_2}}
\sum_{i=1}^{p_1}\sum_{j=1}^{p_2}
\left\|
\sum_t
\mathbf{C}^\top \mathbf{e}_{t,i\cdot }\mathbf{e}_{t,\cdot j}^\top
\right\|_F^2
\end{aligned}
\]
\[
\mathbb{E}
\left\|
\sum_t
\mathbf{C}^\top \mathbf{e}_{t,i\cdot }\mathbf{e}_{t,\cdot j}^\top \mathbf{R}
\right\|_F^2
=
O\left(
Tp_1^{\delta_1}p_2^{\delta_2}
+
T^2
\right)
\]
\[
\begin{aligned}
\mathbb{E}
\left\|
\sum_t
\mathbf{C}^\top \mathbf{e}_{t,i\cdot }\mathbf{e}_{t,\cdot j}^\top
\right\|_F^2
&\le
\sum_{i_1=1}^{p_1}
\mathbb{E}
\left(
\sum_t\sum_{j_1}
\mathbf{C}_{j_1}e_{t,ij_1}e_{t,i_1j}
\right)^2\\
&\le
\sum_{i_1}\sum_{t,j_1}\sum_{s,j_2}\abs{\mathbf{C}_{j_1}\mathbf{C}_{j_2}}
\left|
\cov
\left(
e_{t,ij_1}e_{t,i_1j},
e_{s,ij_2}e_{s,i_1j}
\right)
\right|\\
&+
\sum_{i_1}
\left(
\sum_t\sum_{j_1}\abs{\mathbf{C}_{j_1}}
\left|
\mathbb{E}e_{t,ij_1}e_{t,i_1j}
\right|
\right)^2\\
&\le
\sum_{i_1}\sum_{t,j_1\in \mathcal{I}_C}\sum_{s,j_2\in \mathcal{I}_C}\abs{\mathbf{C}_{j_1}\mathbf{C}_{j_2}}
\left|
\cov
\left(
e_{t,ij_1}e_{t,i_1j},
e_{s,ij_2}e_{s,i_1j}
\right)
\right|\\
&+
\sum_{i_1}
\left(
\sum_t\sum_{j_1\in \mathcal{I}_C}\abs{\mathbf{C}_{j_1}}
\left|
\mathbb{E}e_{t,ij_1}e_{t,i_1j}
\right|
\right)^2\\
&\le
O\left(
Tp_1p_2^{\delta_2}
+
T^2p_1
\right)
\qquad
\end{aligned}
\]
Consequently,
\[
\begin{aligned}
\norm{VIII}_F^2&=O_p\paren{p_1^{2-2\delta_1}p_2^{2-2\delta_2}w_1^2+\frac{p_2^{2-2\delta_2}}{p_1^{2\delta_1-1}}w_1}\norm{\widetilde{\mathbf{C}}-\mathbf{C}\widetilde{\mathbf{H}}_C}_F^2\\
&+
\frac{1}{T^2 p_1^{4\delta_1}p_2^{2\delta_2}}
\left\|
\sum_t
\mathbf{E}_t^\top \check{\mathbf{R}}(\check{\mathbf{R}}-\mathbf{R}\check{\mathbf{H}}_R)^\top \mathbf{E}_t \mathbf{C}
\right\|_F^2\\
&=o_p(1)\norm{\widetilde{\mathbf{C}}-\mathbf{C}\widetilde{\mathbf{H}}_C}_F^2+
\frac{1}{T^2 p_1^{4\delta_1}p_2^{2\delta_2}}
\left\|
\sum_t
\mathbf{E}_t^\top \check{\mathbf{R}}(\check{\mathbf{R}}-\mathbf{R}\check{\mathbf{H}}_R)^\top \mathbf{E}_t \mathbf{C}
\right\|_F^2\\
&\le o_p(1)\norm{\widetilde{\mathbf{C}}-\mathbf{C}\widetilde{\mathbf{H}}_C}_F^2+
\frac{O_p\paren{w_1}}{T^2p_1^{3\delta_1}p_2^{2\delta_2}}
\sum_{i=1}^{p_1}\sum_{j=1}^{p_2}
\left\|
\sum_t
\mathbf{C}^\top \mathbf{e}_{t,i\cdot }\mathbf{e}_{t,\cdot j}^\top \mathbf{R}
\right\|_F^2\\
&+
\frac{O_p\paren{w_1^2}}{T^2p_1^{2\delta_1}p_2^{2\delta_2}}
\sum_{i=1}^{p_1}\sum_{j=1}^{p_2}
\left\|
\sum_t
\mathbf{C}^\top \mathbf{e}_{t,i\cdot }\mathbf{e}_{t,\cdot j}^\top
\right\|_F^2\\
&\le o_p(1)\norm{\widetilde{\mathbf{C}}-\mathbf{C}\widetilde{\mathbf{H}}_C}_F^2+
\frac{w_1}{T^2p_1^{3\delta_1}p_2^{2\delta_2}}p_1p_2O_p\paren{Tp_1^{\delta_1}p_2^{\delta_2}+T^2}\\
&+
\frac{w_1^2}{T^2p_1^{2\delta_1}p_2^{2\delta_2}}p_1p_2O_p\paren{Tp_1p_2^{\delta_2}+T^2p_1}\\
&=o_p(1)\norm{\widetilde{\mathbf{C}}-\mathbf{C}\widetilde{\mathbf{H}}_C}_F^2+O_p\paren{\paren{\frac{p_2^{1-\delta_2}}{Tp_1^{2\delta_1-1}}+\frac{1}{p_1^{3\delta_1-1}p_2^{2\delta_2-1}}}w_1}\\
&+O_p\paren{\paren{\frac{p_1^{2-2\delta_1}p_2^{1-\delta_2}}{T}+\frac{p_1^{2-2\delta_1}}{p_2^{2\delta_2-1}}}w_1^2}
\end{aligned}
\]
Thus,
\[
\begin{aligned}
\frac{1}{p_2^{\delta_2}}\norm{IV\widetilde{\mathbf{C}}}_F^2
&\lesssim\frac{1}{p_2^{\delta_2}}\paren{\norm{V}_F^2+\norm{VI}_F^2+\norm{VII}_F^2+\norm{VIII}_F^2}\\
&= O_p\paren{\frac{1}{p_2^{\delta_2}}\paren{\frac{p_2^{1-\delta_2}}{Tp_1^{\delta_1}}+\frac{1}{p_1^{2\delta_1}p_2^{2\delta_2-1}}+\frac{p_2^{2-2\delta_2}}{p_1^{2\delta_1}}\norm{\widetilde{\mathbf{C}}-\mathbf{C}\widetilde{\mathbf{H}}_C}_F^2}}\\
&+O_p\paren{\frac{1}{p_2^{\delta_2}}\paren{\frac{p_2^{1-\delta_2}}{Tp_1^{2\delta_1-1}}w_1+\frac{1}{p_1^{3\delta_1-1}p_2^{2\delta_2-1}}w_1+\frac{p_2^{2-2\delta_2}}{p_1^{2\delta_1-1}}w_1\norm{\widetilde{\mathbf{C}}-\mathbf{C}\widetilde{\mathbf{H}}_C}_F^2}}\\
&+O_p\paren{\frac{1}{p_2^{\delta_2}}\paren{o_p(1)\norm{\widetilde{\mathbf{C}}-\mathbf{C}\widetilde{\mathbf{H}}_C}_F^2+\paren{\frac{p_2^{1-\delta_2}}{Tp_1^{2\delta_1-1}}+\frac{1}{p_1^{3\delta_1-1}p_2^{2\delta_2-1}}}w_1}}\\
&+O_p\paren{\frac{1}{p_2^{\delta_2}}\paren{\frac{p_1^{2-2\delta_1}p_2^{1-\delta_2}}{T}+\frac{p_1^{2-2\delta_1}}{p_2^{2\delta_2-1}}}w_1^2}\\
&=O_p\paren{\frac{1}{Tp_1^{\delta_1}p_2^{2\delta_2-1}}+\frac{1}{p_1^{2\delta_1}p_2^{3\delta_2-1}}}+o_p(1)\frac{\norm{\widetilde{\mathbf{C}}-\mathbf{C}\widetilde{\mathbf{H}}_C}_F^2}{p_2^{\delta_2}}\\
&+O_p\paren{\paren{\frac{1}{Tp_1^{2\delta_1-1}p_2^{2\delta_2-1}}+\frac{1}{p_1^{3\delta_1-1}p_2^{3\delta_2-1}}}w_1}+o_p(1)\frac{\norm{\widetilde{\mathbf{C}}-\mathbf{C}\widetilde{\mathbf{H}}_C}_F^2}{p_2^{\delta_2}}\\
&+o_p(1)\frac{\norm{\widetilde{\mathbf{C}}-\mathbf{C}\widetilde{\mathbf{H}}_C}_F^2}{p_2^{\delta_2}}+O_p\paren{\paren{\frac{1}{Tp_1^{2\delta_1-1}p_2^{2\delta_2-1}}+\frac{1}{p_1^{3\delta_1-1}p_2^{3\delta_2-1}}}w_1}\\
&+O_p\paren{\paren{\frac{p_1^{2-2\delta_1}}{Tp_2^{2\delta_2-1}}+\frac{p_1^{2-2\delta_1}}{p_2^{3\delta_2-1}}}w_1^2}\\
&=O_p\paren{\frac{1}{Tp_1^{\delta_1}p_2^{2\delta_2-1}}+\frac{1}{p_1^{2\delta_1}p_2^{3\delta_2-1}}}+o_p(1)\frac{\norm{\widetilde{\mathbf{C}}-\mathbf{C}\widetilde{\mathbf{H}}_C}_F^2}{p_2^{\delta_2}}\\
&+O_p\paren{\paren{\frac{1}{Tp_1^{2\delta_1-1}p_2^{2\delta_2-1}}+\frac{1}{p_1^{3\delta_1-1}p_2^{3\delta_2-1}}}w_1+\paren{\frac{p_1^{2-2\delta_1}}{Tp_2^{2\delta_2-1}}+\frac{p_1^{2-2\delta_1}}{p_2^{3\delta_2-1}}}w_1^2}\\
&=O_p\paren{\frac{1}{Tp_1^{\delta_1}p_2^{2\delta_2-1}}+\frac{1}{p_1^{2\delta_1}p_2^{3\delta_2-1}}}+o_p(1)\frac{\norm{\widetilde{\mathbf{C}}-\mathbf{C}\widetilde{\mathbf{H}}_C}_F^2}{p_2^{\delta_2}}\\
&+O_p\paren{\paren{\frac{p_1^{1-\delta_1}w_1}{Tp_1^{\delta_1}p_2^{2\delta_2-1}}+\frac{p_1^{1-\delta_1}w_1}{p_1^{2\delta_1}p_2^{3\delta_2-1}}}+\paren{\frac{p_1^{2-2\delta_1}}{Tp_2^{2\delta_2-1}}+\frac{p_1^{2-2\delta_1}}{p_2^{3\delta_2-1}}}w_1^2}\\
&=O_p\paren{\frac{1}{Tp_1^{\delta_1}p_2^{2\delta_2-1}}+\frac{1}{p_1^{2\delta_1}p_2^{3\delta_2-1}}}+o_p(1)\frac{\norm{\widetilde{\mathbf{C}}-\mathbf{C}\widetilde{\mathbf{H}}_C}_F^2}{p_2^{\delta_2}}\\
&+O_p\paren{\paren{\frac{p_1^{2-2\delta_1}}{Tp_2^{2\delta_2-1}}+\frac{p_1^{2-2\delta_1}}{p_2^{3\delta_2-1}}}w_1^2}.
\end{aligned}
\]
The last equality follows because \({p_1^{1-\delta_1}w_1\le p_1^{1-\delta_1}p_2^{1-\delta_2}w_1=\sqrt{p_1^{2-2\delta_1}p_2^{2-2\delta_2}w_1^2}=o(1)}\).
\end{proof}
{\noindent\bf Proof of Theorem~\ref{thm:projected-estimation}\textup{(I)}.}
It suffices to establish the second assertion since the first follows from the same argument.
\[
\begin{aligned}
\frac{1}{p_2^{\delta_2}}\norm{\widetilde{\mathbf{C}}-\mathbf{C}\widetilde{\mathbf{H}}_C}_F^2&=\frac{1}{p_2^{\delta_2}}\norm{\paren{II+III+IV}\widetilde{\mathbf{C}}\widetilde{\boldsymbol{\Lambda}}_C^{-1}}_F^2\\
&\lesssim\frac{1}{p_2^{\delta_2}}\paren{\norm{II\widetilde{\mathbf{C}}}_F^2+\norm{III\widetilde{\mathbf{C}}}_F^2+\norm{IV\widetilde{\mathbf{C}}}_F^2}\\
&= O_p\paren{w_2}+O_p\paren{\frac{p_2^{1-\delta_2}}{Tp_1^{\delta_1}}}+O_p\paren{\frac{1}{Tp_1^{\delta_1}p_2^{2\delta_2-1}}+\frac{1}{p_1^{2\delta_1}p_2^{3\delta_2-1}}}\\
&+o_p(1)\frac{\norm{\widetilde{\mathbf{C}}-\mathbf{C}\widetilde{\mathbf{H}}_C}_F^2}{p_2^{\delta_2}}+\paren{\frac{p_1^{2-2\delta_1}}{Tp_2^{2\delta_2-1}}+\frac{p_1^{2-2\delta_1}}{p_2^{3\delta_2-1}}}O_p\paren{w_1^2}\\
&=O_p\paren{w_2}+O_p\paren{\frac{p_2^{1-\delta_2}}{Tp_1^{\delta_1}}}+O_p\paren{\frac{1}{p_1^{2\delta_1}p_2^{3\delta_2-1}}}+\paren{\frac{p_1^{2-2\delta_1}}{Tp_2^{2\delta_2-1}}+\frac{p_1^{2-2\delta_1}}{p_2^{3\delta_2-1}}}O_p\paren{w_1^2}\\
&=O_p\paren{\frac{p_1^{2-2\delta_1}}{T^2p_2^{3\delta_2-1}}+\frac{p_1^{4-4\delta_1}}{T^3p_2^{4\delta_2-2}}+\frac{1}{T^2p_1^{4\delta_1-3}p_2^{4\delta_2-3}}+\frac{p_2^{3-3\delta_2}}{Tp_1^{4\delta_1-2}}}\\
&+O_p\paren{\frac{p_2^{1-\delta_2}}{Tp_1^{\delta_1}}}+O_p\paren{\frac{1}{p_1^{2\delta_1}p_2^{3\delta_2-1}}}+\paren{\frac{p_1^{2-2\delta_1}}{Tp_2^{2\delta_2-1}}+\frac{p_1^{2-2\delta_1}}{p_2^{3\delta_2-1}}}\times\paren{\frac{p_1^{2-2\delta_1}}{T^2p_2^{2\delta_2}}+\frac{p_2^{4-4\delta_2}}{p_1^{6\delta_1-2}}}\\
&=O_p\paren{\frac{p_1^{2-2\delta_1}}{T^2p_2^{3\delta_2-1}}+\frac{p_1^{4-4\delta_1}}{T^3p_2^{4\delta_2-2}}+\frac{1}{T^2p_1^{4\delta_1-3}p_2^{4\delta_2-3}}+\frac{p_2^{3-3\delta_2}}{Tp_1^{4\delta_1-2}}}\\
&+O_p\paren{\frac{p_2^{1-\delta_2}}{Tp_1^{\delta_1}}}+O_p\paren{\frac{1}{p_1^{2\delta_1}p_2^{3\delta_2-1}}}+O_p\paren{\frac{p_1^{4-4\delta_1}}{T^3p_2^{4\delta_2-1}}+\frac{1}{Tp_2^{6\delta_2-5}p_1^{8\delta_1-4}}}\\
&+O_p\paren{\frac{p_1^{4-4\delta_1}}{T^2p_2^{5\delta_2-1}}+\frac{1}{p_1^{8\delta_1-4}p_2^{7\delta_2-5}}}\\
&=O_p\paren{\frac{p_1^{2-2\delta_1}}{T^2p_2^{3\delta_2-1}}+\frac{p_1^{4-4\delta_1}}{T^3p_2^{4\delta_2-2}}+\frac{1}{T^2p_1^{4\delta_1-3}p_2^{4\delta_2-3}}+\frac{p_2^{3-3\delta_2}}{Tp_1^{4\delta_1-2}}}\\
&+O_p\paren{\frac{p_2^{1-\delta_2}}{Tp_1^{\delta_1}}+\frac{1}{p_1^{2\delta_1}p_2^{3\delta_2-1}}+
\frac{1}{p_1^{8\delta_1-4}p_2^{7\delta_2-5}}}\qquad \text{$\paren{by\;\eqref{cond:C1}}$}.
\end{aligned}
\]
The result
\(
\widetilde{\mathbf{H}}_C^\top \widetilde{\mathbf{H}}_C \xrightarrow{p} \mathbf{I}_{k_2}
\)
follows from the same argument used to prove
\(
\check{\mathbf{H}}_C^\top \check{\mathbf{H}}_C \xrightarrow{p} \mathbf{I}_{k_2}.
\)\\
{\bf Proof of Theorem~\ref{thm:projected-estimation}\textup{(II)}.}
\begin{lemma}
Suppose Assumptions~\ref{ass:weak loading}--\ref{ass:noise-dependence}, Assumption~\ref{ass:projected_spatial_noise}--\ref{ass:Factor-noise interaction for projected} and \eqref{cond:C1},\eqref{cond:c3} hold. Then
\[
\frac{1}{p_1^{\delta_1}}
\mathbf{R}^\top(\widetilde{\mathbf{R}}-\mathbf{R}\widetilde{\mathbf{H}}_R)
=
O_p\paren{\frac{1}{\sqrt{{Tp_1^{\delta_1}p_2^{\delta_2}}}}+\frac{1}{p_1^{\delta_1}p_2^{\delta_2}}+\frac{p_1^{1-\delta_1}}{Tp_2^{\delta_2}}+\frac{p_2^{1-\delta_2}}{Tp_1^{\delta_1}}},
\]
\[
\frac{1}{p_2^{\delta_2}}
\mathbf{C}^\top(\widetilde{\mathbf{C}}-\mathbf{C}\widetilde{\mathbf{H}}_C)
=
O_p\paren{\frac{1}{\sqrt{{Tp_1^{\delta_1}p_2^{\delta_2}}}}+\frac{1}{p_1^{\delta_1}p_2^{\delta_2}}+\frac{p_1^{1-\delta_1}}{Tp_2^{\delta_2}}+\frac{p_2^{1-\delta_2}}{Tp_1^{\delta_1}}}
\]
\end{lemma}
\begin{proof}
By symmetry, it suffices to establish the second bound.
\[
\begin{aligned}
\frac{1}{p_2^{\delta_2}}
\mathbf{C}^\top(\widetilde{\mathbf{C}}-\mathbf{C}\widetilde{\mathbf{H}}_C)
&=
\frac{1}{p_2^{\delta_2}}
\mathbf{C}^\top
(II+III+IV)
\widetilde{\mathbf{C}}\widetilde{\boldsymbol{\Lambda}}_C^{-1}.
\end{aligned}
\]
\[
\begin{aligned}
\left\|
\frac{1}{p_2^{\delta_2}}
\mathbf{C}^\top III\widetilde{\mathbf{C}}
\right\|
&=
\left\|
\frac{1}{T p_1^{2\delta_1}p_2^{2\delta_2}}
\sum_t
\mathbf{C}^\top \mathbf{E}_t^\top \check{\mathbf{R}} \check{\mathbf{R}}^\top \mathbf{R} \mathbf{F}_t \mathbf{C}^\top \widetilde{\mathbf{C}}
\right\|\\
&\lesssim
\left\|
\frac{1}{T p_1^{\delta_1}p_2^{\delta_2}}
\sum_t
\mathbf{C}^\top \mathbf{E}_t^\top \check{\mathbf{R}} \mathbf{F}_t\
\right\|\le\left\|
\frac{1}{T p_1^{\delta_1}p_2^{\delta_2}}
\sum_t
\mathbf{F}_t\mathbf{C}^\top \mathbf{E}_t^\top \check{\mathbf{R}} \
\right\|\\
&\lesssim\frac{1}{T p_1^{\delta_1}p_2^{\delta_2}}\paren{\left\|
\sum_t
\mathbf{F}_t\mathbf{C}^\top \mathbf{E}_t^\top \paren{\check{\mathbf{R}}-\mathbf{R}\check{\mathbf{H}}_R}\
\right\|+\left\|
\sum_t
\mathbf{F}_t\mathbf{C}^\top \mathbf{E}_t^\top \mathbf{R}\check{\mathbf{H}}_R\
\right\|}\\
&=\frac{1}{T p_1^{\delta_1}p_2^{\delta_2}}O_p\paren{\sqrt{Tp_1p_2^{\delta_2}}\sqrt{p_1^{\delta_1}w_1}+\sqrt{{Tp_1^{\delta_1}p_2^{\delta_2}}}}\\
&=\sqrt{\frac{p_1^{1-\delta_1}}{Tp_2^{\delta_2}}\paren{O_p\paren{\frac{p_1^{1-\delta_1}}{Tp_2^{\delta_2}}}+O_p\paren{\frac{p_2^{2-2\delta_2}}{p_1^{3\delta_1-1}}}}}+O_p\paren{\sqrt{\frac{1}{{Tp_1^{\delta_1}p_2^{\delta_2}}}}}\\
&=
O_p\paren{\sqrt{\frac{p_1^{2-2\delta_1}}{T^2p_2^{2\delta_2}}}+\sqrt{\frac{1}{Tp_1^{4\delta_1-2}p_2^{3\delta_2-2}}}+\sqrt{\frac{1}{{Tp_1^{\delta_1}p_2^{\delta_2}}}}}\\
&=O_p\paren{\frac{p_1^{1-\delta_1}}{Tp_2^{\delta_2}}+\sqrt{\frac{1}{{Tp_1^{\delta_1}p_2^{\delta_2}}}}}\\
\end{aligned}
\]
\[
\begin{aligned}
\left\|
\frac{1}{p_2^{\delta_2}}
\mathbf{C}^\top II\widetilde{\mathbf{C}}
\right\|
&=
\left\|
\frac{1}{T p_1^{2\delta_1}p_2^{2\delta_2}}
\sum_t
\mathbf{C}^\top \mathbf{C} \mathbf{F}_t^\top \mathbf{R}^\top \check{\mathbf{R}}\check{\mathbf{R}}^\top \mathbf{E}_t\widetilde{\mathbf{C}}
\right\|\\
&\lesssim
\left\|
\frac{1}{T p_1^{\delta_1}p_2^{\delta_2}}
\sum_t
\mathbf{F}_t^\top \check{\mathbf{R}}^\top \mathbf{E}_t\widetilde{\mathbf{C}}
\right\|\\
&\le
\left\|
\frac{1}{T p_1^{\delta_1}p_2^{\delta_2}}
\sum_t
\mathbf{F}_t^\top
(\check{\mathbf{R}}-\mathbf{R}\check{\mathbf{H}}_R)^\top
\mathbf{E}_t
(\widetilde{\mathbf{C}}-\mathbf{C}\widetilde{\mathbf{H}}_C)
\right\|\\
&+
\left\|
\frac{1}{T p_1^{\delta_1}p_2^{\delta_2}}
\sum_t
\mathbf{F}_t^\top
\check{\mathbf{H}}_R^\top \mathbf{R}^\top
\mathbf{E}_t
(\widetilde{\mathbf{C}}-\mathbf{C}\widetilde{\mathbf{H}}_C)
\right\|\\
&+
\left\|
\frac{1}{T p_1^{\delta_1}p_2^{\delta_2}}
\sum_t
\mathbf{F}_t^\top
(\check{\mathbf{R}}-\mathbf{R}\check{\mathbf{H}}_R)^\top
\mathbf{E}_t
\mathbf{C}\widetilde{\mathbf{H}}_C
\right\|\\
&+
\left\|
\frac{1}{T p_1^{\delta_1}p_2^{\delta_2}}
\sum_t
\mathbf{F}_t^\top
\check{\mathbf{H}}_R^\top \mathbf{R}^\top
\mathbf{E}_t
\mathbf{C}\widetilde{\mathbf{H}}_C
\right\|\\
&\le\frac{1}{T p_1^{\delta_1}p_2^{\delta_2}}\norm{\sum_t\mathbf{E}_t'\otimes \mathbf{F}_t'}_F\norm{\check{\mathbf{R}}-\mathbf{R}\check{\mathbf{H}}_R}_F\norm{\widetilde{\mathbf{C}}-\mathbf{C}\widetilde{\mathbf{H}}_C}_F\\
&+\frac{1}{T p_1^{\delta_1}p_2^{\delta_2}}\norm{\sum_t\mathbf{F}_t'\mathbf{R}'\mathbf{E}_t}\norm{\widetilde{\mathbf{C}}-\mathbf{C}\widetilde{\mathbf{H}}_C}+\frac{1}{T p_1^{\delta_1}p_2^{\delta_2}}\norm{\sum_t\mathbf{E}_t\mathbf{C}\mathbf{F}_t'}\norm{\check{\mathbf{R}}-\mathbf{R}\check{\mathbf{H}}_R}\\
&+\frac{1}{T p_1^{\delta_1}p_2^{\delta_2}}\norm{\sum_t\mathbf{F}_t'\mathbf{R}'\mathbf{E}_t\mathbf{C}}\\
&= O_p\paren{\frac{1}{T p_1^{\delta_1}p_2^{\delta_2}}\sqrt{Tp_1p_2p_1^{\delta_1}w_1p_2^{\delta_2}\widetilde m_1}}+O_p\paren{\frac{1}{T p_1^{\delta_1}p_2^{\delta_2}}\sqrt{Tp_1^{\delta_1}p_2p_2^{\delta_2}\widetilde m_1}}\\
&+\frac{1}{T p_1^{\delta_1}p_2^{\delta_2}}O_p\paren{\sqrt{Tp_1p_2^{\delta_2}p_1^{\delta_1}w_1}}+O_p\paren{\frac{1}{T p_1^{\delta_1}p_2^{\delta_2}}\sqrt{{Tp_1^{\delta_1}p_2^{\delta_2}}}}\\
&=O_p\paren{\frac{1}{\sqrt{{Tp_1^{\delta_1}p_2^{\delta_2}}}}\paren{\sqrt{p_1w_1}+1}\paren{\sqrt{p_2\widetilde m_1}+1}}
\end{aligned}
\]
where,
\[
p_1w_1=\frac{p_1^{2-\delta_1}}{Tp_2^{\delta_2}}+\frac{p_2^{2-2\delta_2}}{p_1^{3\delta_1-2}}=\frac{p_1^{2-\delta_1}}{Tp_2^{\delta_2}}+o(1),\;
p_2\widetilde m_1=\paren{\frac{p_2^{2-\delta_2}}{Tp_1^{\delta_1}}}+o(1)
\]
Thus,
\[
\begin{aligned}
\left\|
\frac{1}{p_2^{\delta_2}}
\mathbf{C}^\top II\widetilde{\mathbf{C}}
\right\|
&=O_p\paren{\frac{1}{\sqrt{{Tp_1^{\delta_1}p_2^{\delta_2}}}}\paren{\sqrt{p_1w_1}+1}\paren{\sqrt{p_2\widetilde m_1}+1}}\\
&=O_p\paren{\frac{1}{\sqrt{{Tp_1^{\delta_1}p_2^{\delta_2}}}}\paren{\sqrt{\frac{p_1^{2-\delta_1}}{Tp_2^{\delta_2}}}+1}\paren{\sqrt{\frac{p_2^{2-\delta_2}}{Tp_1^{\delta_1}}}+1}}\\
&=O_p\paren{\paren{\frac{p_1^{1-\delta_1}}{Tp_2^{\delta_2}}+\frac{1}{\sqrt{{Tp_1^{\delta_1}p_2^{\delta_2}}}}}\paren{\sqrt{\frac{p_2^{2-\delta_2}}{Tp_1^{\delta_1}}}+1}}\\
&=O_p\paren{{\frac{p_1^{1-\delta_1}}{Tp_2^{\delta_2}}+\frac{1}{\sqrt{{Tp_1^{\delta_1}p_2^{\delta_2}}}}}+\frac{p_2^{1-\delta_2}}{Tp_1^{\delta_1}}+\frac{1}{\sqrt{T^3p_1^{3\delta_1-2}p_2^{3\delta_2-2}}}}\\
&=O_p\paren{{\frac{p_1^{1-\delta_1}}{Tp_2^{\delta_2}}+\frac{1}{\sqrt{{Tp_1^{\delta_1}p_2^{\delta_2}}}}}+\frac{p_2^{1-\delta_2}}{Tp_1^{\delta_1}}}
\end{aligned}
\]
\[
\begin{aligned}
\left\|
\frac{1}{p_2^{\delta_2}}
\mathbf{C}^\top IV\widetilde{\mathbf{C}}
\right\|
&=
\left\|
\frac{1}{T p_1^{2\delta_1}p_2^{2\delta_2}}
\sum_t
\mathbf{C}^\top \mathbf{E}_t^\top \check{\mathbf{R}} \check{\mathbf{R}}^\top \mathbf{E}_t\widetilde{\mathbf{C}}
\right\|\\
&\le
\frac{1}{Tp_1^{2\delta_1}p_2^{2\delta_2}}
\sqrt{
\sum_t
\|\mathbf{C}^\top \mathbf{E}_t^\top \check{\mathbf{R}}\|_F^2
\times
\sum_t
\|\check{\mathbf{R}}^\top \mathbf{E}_t\widetilde{\mathbf{C}}\|_F^2
}\\
&=
O_p\left(
\frac{1}{Tp_1^{2\delta_1}p_2^{2\delta_2}}
\sqrt{\sum_t\|\mathbf{C}^\top \mathbf{E}_t^\top \paren{\check{\mathbf{R}}-\mathbf{R}\check{\mathbf{H}}_R}\|_F^2+\sum_t\|\mathbf{C}^\top \mathbf{E}_t^\top \mathbf{R}\check{\mathbf{H}}_R\|_F^2
}\times
\right.\\
&\left.\sqrt{
\begin{aligned}
&\left(\sum_t\norm{\paren{\check{\mathbf{R}}-\mathbf{R}\check{\mathbf{H}}_R}'\mathbf{E}_t\paren{\widetilde{\mathbf{C}}-\mathbf{C}\widetilde{\mathbf{H}}_C}}_F^2+\sum_t\norm{\paren{\check{\mathbf{R}}-\mathbf{R}\check{\mathbf{H}}_R}'\mathbf{E}_t\mathbf{C}\widetilde{\mathbf{H}}_C}_F^2\right.\\
&\left.+\sum_t\norm{\paren{\mathbf{R}\check{\mathbf{H}}_R}'\mathbf{E}_t\paren{\widetilde{\mathbf{C}}-\mathbf{C}\widetilde{\mathbf{H}}_C}}_F^2+\sum_t\norm{\paren{\mathbf{R}\check{\mathbf{H}}_R}'\mathbf{E}_t\mathbf{C}\widetilde{\mathbf{H}}_C}_F^2\right)
\end{aligned}}
\right)
\\
&\le\frac{1}{Tp_1^{2\delta_1}p_2^{2\delta_2}}O_p\paren{\sqrt{Tp_1p_2^{\delta_2}p_1^{\delta_1}w_1+Tp_1^{\delta_1}p_2^{\delta_2}}}\\
&\times O_p\paren{\sqrt{Tp_1p_2p_1^{\delta_1}w_1p_2^{\delta_2}\widetilde m_1+Tp_1p_2^{\delta_2}p_1^{\delta_1}w_1+Tp_1^{\delta_1}p_2p_2^{\delta_2}\widetilde m_1+Tp_1^{\delta_1}p_2^{\delta_2}}}\\
&=\frac{1}{p_1^{\delta_1}p_2^{\delta_2}}O_p\paren{\sqrt{p_1w_1+1}\sqrt{\paren{p_1w_1+1}\paren{p_2\widetilde m_1+1}}}\\
&=\frac{1}{p_1^{\delta_1}p_2^{\delta_2}}O_p\paren{\paren{\sqrt{\frac{p_1^{2-\delta_1}}{Tp_2^{\delta_2}}}+1}\paren{\sqrt{\frac{p_1^{2-\delta_1}}{Tp_2^{\delta_2}}}+1}\paren{\sqrt{\frac{p_2^{2-\delta_2}}{Tp_1^{\delta_1}}}+1}}\\
&=\frac{1}{p_1^{\delta_1}p_2^{\delta_2}}O_p\paren{\paren{\frac{p_1^{2-\delta_1}}{Tp_2^{\delta_2}}+1}\paren{\sqrt{\frac{p_2^{2-\delta_2}}{Tp_1^{\delta_1}}}+1}}\\
&=\frac{1}{p_1^{\delta_1}p_2^{\delta_2}}O_p\paren{1+\frac{p_1^{2-\delta_1}}{Tp_2^{\delta_2}}+\sqrt{\frac{p_2^{2-\delta_2}}{Tp_1^{\delta_1}}}+\frac{1}{\sqrt{T^3p_1^{3\delta_1-4}p_2^{3\delta_2-2}}}}\\
&=O_p\paren{\frac{1}{p_1^{\delta_1}p_2^{\delta_2}}+\frac{p_1^{1-\delta_1}}{p_2^{\delta_2}}\frac{p_1^{1-\delta_1}}{Tp_2^{\delta_2}}+\frac{1}{\sqrt{Tp_1^{3\delta_1}p_2^{3\delta_2-2}}}+\frac{p_1}{p_1^{\delta_1}p_2^{\delta_2}}\frac{1}{\sqrt{T^3p_1^{3\delta_1-2}p_2^{3\delta_2-2}}}}\\
&=O_p\paren{\frac{1}{p_1^{\delta_1}p_2^{\delta_2}}}+o_p(1)\frac{p_1^{1-\delta_1}}{Tp_2^{\delta_2}}+o_p\paren{1}\frac{1}{\sqrt{Tp_1^{\delta_1}p_2^{\delta_2}}}+o_p(1)\frac{1}{\sqrt{T^3p_1^{3\delta_1-2}p_2^{3\delta_2-2}}}
\end{aligned}
\]
Combining the above results yields
\[
\frac{1}{p_2^{\delta_2}}
\mathbf{C}^\top(\widetilde{\mathbf{C}}-\mathbf{C}\widetilde{\mathbf{H}}_C)
=O_p\paren{\frac{1}{\sqrt{{Tp_1^{\delta_1}p_2^{\delta_2}}}}+\frac{1}{p_1^{\delta_1}p_2^{\delta_2}}+\frac{p_1^{1-\delta_1}}{Tp_2^{\delta_2}}+\frac{p_2^{1-\delta_2}}{Tp_1^{\delta_1}}}
\]
\end{proof}
\noindent Now we prove Theorem~\ref{thm:projected-estimation}\textup{(II)}.
    \[\begin{aligned}
\widetilde{\mathbf{F}}_t
&=
\frac{1}{p_1^{\delta_1}p_2^{\delta_2}}
\widetilde{\mathbf{R}}^\top \mathbf{X}_t \widetilde{\mathbf{C}}
=
\frac{1}{p_1^{\delta_1}p_2^{\delta_2}}
\widetilde{\mathbf{R}}^\top \mathbf{R} \mathbf{F}_t \mathbf{C}^\top \widetilde{\mathbf{C}}
+
\frac{1}{p_1^{\delta_1}p_2^{\delta_2}}
\widetilde{\mathbf{R}}^\top \mathbf{E}_t \widetilde{\mathbf{C}}\\
&=
\frac{1}{p_1^{\delta_1}p_2^{\delta_2}}
\widetilde{\mathbf{R}}^\top
\left(
\mathbf{R}-\widetilde{\mathbf{R}}\widetilde{\mathbf{H}}_R^{-1}
+\widetilde{\mathbf{R}}\widetilde{\mathbf{H}}_R^{-1}
\right)
\mathbf{F}_t
\left(
\mathbf{C}-\widetilde{\mathbf{C}}\widetilde{\mathbf{H}}_C^{-1}
+\widetilde{\mathbf{C}}\widetilde{\mathbf{H}}_C^{-1}
\right)^\top
\widetilde{\mathbf{C}}\\
&+
\frac{1}{p_1^{\delta_1}p_2^{\delta_2}}
\left(
\widetilde{\mathbf{R}}-\mathbf{R}\widetilde{\mathbf{H}}_R+\mathbf{R}\widetilde{\mathbf{H}}_R
\right)^\top
\mathbf{E}_t
\left(
\widetilde{\mathbf{C}}-\mathbf{C}\widetilde{\mathbf{H}}_C+\mathbf{C}\widetilde{\mathbf{H}}_C
\right).
\end{aligned}
\]
Note that
$\widetilde{\mathbf{R}}^\top \widetilde{\mathbf{R}}=p_1^{\delta_1}\mathbf{I}_{k_1}$
and
$\widetilde{\mathbf{C}}^\top \widetilde{\mathbf{C}}=p_2^{\delta_2}\mathbf{I}_{k_2}$.
Then
\[
\begin{aligned}
\widetilde{\mathbf{F}}_t-\widetilde{\mathbf{H}}_R^{-1}\mathbf{F}_t\paren{\widetilde{\mathbf{H}}_C^{-1}}^\top
&=
\frac{1}{p_1^{\delta_1}p_2^{\delta_2}}
\widetilde{\mathbf{R}}^\top
\left(
\mathbf{R}-\widetilde{\mathbf{R}}\widetilde{\mathbf{H}}_R^{-1}
\right)
\mathbf{F}_t
\left(
\mathbf{C}-\widetilde{\mathbf{C}}\widetilde{\mathbf{H}}_C^{-1}
\right)^\top
\widetilde{\mathbf{C}}\\
&+
\frac{1}{p_1^{\delta_1}}
\widetilde{\mathbf{R}}^\top
\left(
\mathbf{R}-\widetilde{\mathbf{R}}\widetilde{\mathbf{H}}_R^{-1}
\right)
\mathbf{F}_t
\left(
\widetilde{\mathbf{H}}_C^{-1}
\right)^\top
+
\frac{1}{p_2^{\delta_2}}
\widetilde{\mathbf{H}}_R^{-1}
\mathbf{F}_t
\left(
\mathbf{C}-\widetilde{\mathbf{C}}\widetilde{\mathbf{H}}_C^{-1}
\right)^\top
\widetilde{\mathbf{C}}\\
&+
\frac{1}{p_1^{\delta_1}p_2^{\delta_2}}
\left(
\widetilde{\mathbf{R}}-\mathbf{R}\widetilde{\mathbf{H}}_R
\right)^\top
\mathbf{E}_t
\left(
\widetilde{\mathbf{C}}-\mathbf{C}\widetilde{\mathbf{H}}_C
\right)\\
&+
\frac{1}{p_1^{\delta_1}p_2^{\delta_2}}
\left(
\mathbf{R}\widetilde{\mathbf{H}}_R
\right)^\top
\mathbf{E}_t
\left(
\widetilde{\mathbf{C}}-\mathbf{C}\widetilde{\mathbf{H}}_C
\right)
+
\frac{1}{p_1^{\delta_1}p_2^{\delta_2}}
\left(
\widetilde{\mathbf{R}}-\mathbf{R}\widetilde{\mathbf{H}}_R
\right)^\top
\mathbf{E}_t \mathbf{C}\widetilde{\mathbf{H}}_C\\
&+
\frac{1}{p_1^{\delta_1}p_2^{\delta_2}}
\left(
\mathbf{R}\widetilde{\mathbf{H}}_R
\right)^\top
\mathbf{E}_t \mathbf{C}\widetilde{\mathbf{H}}_C \\
&=\sum_{i=1}^{7} I_i
\end{aligned}
\]
\[
\begin{aligned}
\norm{I_1}
&=\norm{\frac{1}{p_1^{\delta_1}p_2^{\delta_2}}
\widetilde{\mathbf{R}}^\top
\left(
\mathbf{R}-\widetilde{\mathbf{R}}\widetilde{\mathbf{H}}_R^{-1}
\right)
\mathbf{F}_t
\left(
\mathbf{C}-\widetilde{\mathbf{C}}\widetilde{\mathbf{H}}_C^{-1}
\right)^\top
\widetilde{\mathbf{C}}}\\
&\le\norm{\frac{1}{p_1^{\delta_1}}
\widetilde{\mathbf{R}}^\top
\left(
\mathbf{R}-\widetilde{\mathbf{R}}\widetilde{\mathbf{H}}_R^{-1}
\right)
\mathbf{F}_t}\frac{1}{p_2^{\frac{\delta_2}{2}}}\norm{\widetilde{\mathbf{C}}}\frac{1}{p_2^{\frac{\delta_2}{2}}}\norm{\mathbf{C}-\widetilde{\mathbf{C}}\widetilde{\mathbf{H}}_C^{-1}}\\
&= o_p\paren{1}\norm{I_2}
\end{aligned}
\]
Similarly, $I_1$ is bounded by $I_3$. For $I_2$,
\[
\begin{aligned}
\norm{I_2}
&=\norm{\frac{1}{p_1^{\delta_1}}
\widetilde{\mathbf{R}}^\top
\left(
\mathbf{R}-\widetilde{\mathbf{R}}\widetilde{\mathbf{H}}_R^{-1}
\right)
\mathbf{F}_t
\left(
\widetilde{\mathbf{H}}_C^{-1}
\right)^\top}\\
&\lesssim\frac{1}{p_1^{\delta_1}}\norm{\widetilde{\mathbf{R}}^\top
\left(
\mathbf{R}-\widetilde{\mathbf{R}}\widetilde{\mathbf{H}}_R^{-1}
\right)}\\
&\le\frac{1}{p_1^{\delta_1}}\norm{\paren{\widetilde{\mathbf{R}}-\mathbf{R}\widetilde{\mathbf{H}}_R}^\top
\left(
\mathbf{R}-\widetilde{\mathbf{R}}\widetilde{\mathbf{H}}_R^{-1}
\right)}+\frac{1}{p_1^{\delta_1}}\norm{\widetilde{\mathbf{H}}_R^\top\bR^\top\left(
\mathbf{R}-\widetilde{\mathbf{R}}\widetilde{\mathbf{H}}_R^{-1}
\right)}\\
&=O_p\paren{
\frac{1}{\sqrt{{Tp_1^{\delta_1}p_2^{\delta_2}}}}
+
\frac{1}{p_1^{\delta_1}p_2^{\delta_2}}
+
\frac{p_1^{1-\delta_1}}{Tp_2^{\delta_2}}
+
\frac{p_2^{1-\delta_2}}{Tp_1^{\delta_1}}}
\end{aligned}
\]
Since the argument for $\norm{I_2}$ is analogous to that for $\norm{I_3}$, we present only the proof of the latter below.
\[
\begin{aligned}
\norm{I_3}
&=\norm{\frac{1}{p_2^{\delta_2}}
\widetilde{\mathbf{H}}_R^{-1}
\mathbf{F}_t
\left(
\mathbf{C}-\widetilde{\mathbf{C}}\widetilde{\mathbf{H}}_C^{-1}
\right)^\top
\widetilde{\mathbf{C}}}\lesssim\norm{\frac{1}{p_2^{\delta_2}}
\left(
\mathbf{C}-\widetilde{\mathbf{C}}\widetilde{\mathbf{H}}_C^{-1}
\right)^\top
\widetilde{\mathbf{C}}}\\
&\le\norm{\frac{1}{p_2^{\delta_2}}\left(
\mathbf{C}-\widetilde{\mathbf{C}}\widetilde{\mathbf{H}}_C^{-1}
\right)^\top\paren{
\widetilde{\mathbf{C}}-\mathbf{C}\widetilde{\mathbf{H}}_C}}+\norm{\frac{1}{p_2^{\delta_2}}\left(
\mathbf{C}-\widetilde{\mathbf{C}}\widetilde{\mathbf{H}}_C^{-1}
\right)^\top \mathbf{C}\widetilde{\mathbf{H}}_C}\\
&=O_p\paren{\widetilde m_1}+O_p\paren{\frac{1}{\sqrt{{Tp_1^{\delta_1}p_2^{\delta_2}}}}+\frac{1}{p_1^{\delta_1}p_2^{\delta_2}}+\frac{p_1^{1-\delta_1}}{Tp_2^{\delta_2}}+\frac{p_2^{1-\delta_2}}{Tp_1^{\delta_1}}}
\end{aligned}
\]
where $\widetilde{m}_1$ is the convergence rate of $\frac{1}{p_2^{\delta_2}}\norm{\widetilde{\bC}-\bC\widetilde{\bH}_C}_F^2$. Specifically,
\[
\begin{aligned}
\widetilde m_1
&=
{\frac{p_1^{2-2\delta_1}}{T^2p_2^{3\delta_2-1}}+\frac{p_1^{4-4\delta_1}}{T^3p_2^{4\delta_2-2}}+\frac{1}{T^2p_1^{4\delta_1-3}p_2^{4\delta_2-3}}}
+{\frac{p_2^{1-\delta_2}}{Tp_1^{\delta_1}}+\frac{1}{p_1^{2\delta_1}p_2^{3\delta_2-1}}+\frac{1}{p_1^{8\delta_1-4}p_2^{7\delta_2-5}}}+\frac{p_2^{3-3\delta_2}}{T p_1^{4\delta_1-2}}\\
&=
{
\frac{p_1^{2-2\delta_1}}{T^2p_2^{3\delta_2-1}}
+
\frac{p_2^{1-\delta_2}}{Tp_1^{\delta_1}}
+
\frac{1}{p_1^{2\delta_1}p_2^{3\delta_2-1}}
}.
\end{aligned}
\]
The last equality follows from \eqref{cond:C1} and \eqref{cond:c3}.
Thus,
\[
\begin{aligned}
\norm{I_3}
& =O_p\paren{
\frac{p_1^{2-2\delta_1}}{T^2p_2^{3\delta_2-1}}
+
\frac{p_2^{1-\delta_2}}{Tp_1^{\delta_1}}
+
\frac{1}{p_1^{2\delta_1}p_2^{3\delta_2-1}}
}
+
O_p\paren{
\frac{1}{\sqrt{{Tp_1^{\delta_1}p_2^{\delta_2}}}}
+
\frac{1}{p_1^{\delta_1}p_2^{\delta_2}}
+
\frac{p_1^{1-\delta_1}}{Tp_2^{\delta_2}}
+
\frac{p_2^{1-\delta_2}}{Tp_1^{\delta_1}}
} \\
&=
O_p\paren{
\frac{1}{\sqrt{{Tp_1^{\delta_1}p_2^{\delta_2}}}}
+
\frac{1}{p_1^{\delta_1}p_2^{\delta_2}}
+
\frac{p_1^{1-\delta_1}}{Tp_2^{\delta_2}}
+
\frac{p_2^{1-\delta_2}}{Tp_1^{\delta_1}}
}.
\end{aligned}
\]
\[
\begin{aligned}
\norm{I_4}
&=\norm{\frac{1}{p_1^{\delta_1}p_2^{\delta_2}}
\left(
\widetilde{\mathbf{R}}-\mathbf{R}\widetilde{\mathbf{H}}_R
\right)^\top
\mathbf{E}_t
\left(
\widetilde{\mathbf{C}}-\mathbf{C}\widetilde{\mathbf{H}}_C
\right)}\\
&=\frac{1}{p_1^{\delta_1}p_2^{\delta_2}}O_p\paren{p_1^{\frac{\delta_1}{2}}\sqrt{\widetilde w_1}\sqrt{p_1p_2}p_2^{\frac{\delta_2}{2}}\sqrt{\widetilde m_1}}\\
&=O_p\paren{\sqrt{p_1^{1-\delta_1}p_2^{1-\delta_2}}\sqrt{\widetilde w_1\widetilde m_1}}
\end{aligned}
\]
\[
\begin{aligned}
\norm{I_5}
&=\norm{\frac{1}{p_1^{\delta_1}p_2^{\delta_2}}
\left(
\mathbf{R}\widetilde{\mathbf{H}}_R
\right)^\top
\mathbf{E}_t
\left(
\widetilde{\mathbf{C}}-\mathbf{C}\widetilde{\mathbf{H}}_C
\right)}\\
&= O_p\paren{\frac{1}{p_1^{\delta_1}p_2^{\delta_2}}\sqrt{p_1^{\delta_1}p_2}\sqrt{p_2^{\delta_2}\widetilde m_1}}\\
&=O_p\paren{\sqrt{\frac{p_2^{1-\delta_2}}{p_1^{\delta_1}}\widetilde m_1}}
\end{aligned}
\]
\[
\begin{aligned}
\norm{I_6}
&=\norm{\frac{1}{p_1^{\delta_1}p_2^{\delta_2}}
\left(
\widetilde{\mathbf{R}}-\mathbf{R}\widetilde{\mathbf{H}}_R
\right)^\top
\mathbf{E}_t \mathbf{C}\widetilde{\mathbf{H}}_C}\\
&= O_p\paren{\frac{1}{p_1^{\delta_1}p_2^{\delta_2}}\sqrt{p_1^{\delta_1}\widetilde w_1}\sqrt{p_1p_2^{\delta_2}}}\\
&=O_p\paren{\sqrt{\frac{p_1^{1-\delta_1}}{p_2^{\delta_2}}\widetilde w_1}}
\end{aligned}
\]
\[
\begin{aligned}
\E\norm{\mathbf{R}'\mathbf{E}_t\mathbf{C}}_F^2
&=\E\norm{\sum_{i,j}\mathbf{R}_ie_{t,ij}\mathbf{C}_j}_F^2=\E\norm{\sum_{i\in \mathcal{I}_R,j\in \mathcal{I}_C}\mathbf{R}_ie_{t,ij}\mathbf{C}_j}_F^2\\
&=\E\sum_{i_1\in \mathcal{I}_R,j_1\in \mathcal{I}_C,i_2\in \mathcal{I}_R,j_2\in \mathcal{I}_C}\mathbf{R}_{i_1}\mathbf{R}_{i_2}\mathbf{C}_{j_1}\mathbf{C}_{j_2}e_{t,i_1j_1}e_{t,i_2j_2}\\
&\le\sum_{i_1\in \mathcal{I}_R,j_1\in \mathcal{I}_C,i_2\in \mathcal{I}_R,j_2\in \mathcal{I}_C}\abs{\mathbf{R}_{i_1}\mathbf{R}_{i_2}\mathbf{C}_{j_1}\mathbf{C}_{j_2}}\abs{\E e_{t,i_1j_1}e_{t,i_2j_2}}\\
&\le C
\sum_{(i_1,j_1)\in\mathcal I_R\times\mathcal I_C}
\sum_{i_2,j_2}
|\E(e_{t,i_1j_1}e_{t,i_2j_2})|\\
&=O\paren{p_1^{\delta_1}p_2^{\delta_2}}
\end{aligned}
\]
Thus,
\[
\begin{aligned}
\norm{I_7}
&=\norm{\frac{1}{p_1^{\delta_1}p_2^{\delta_2}}
\left(
\mathbf{R}\widetilde{\mathbf{H}}_R
\right)^\top
\mathbf{E}_t \mathbf{C}\widetilde{\mathbf{H}}_C }\\
&=\frac{1}{p_1^{\delta_1}p_2^{\delta_2}}O_p\paren{\sqrt{p_1^{\delta_1}p_2^{\delta_2}}}\\
&=O_p\paren{\sqrt{\frac{1}{p_1^{\delta_1}p_2^{\delta_2}}}}
\end{aligned}
\]
\[
\begin{aligned}
\norm{I_4}+\norm{I_5}+\norm{I_6}+\norm{I_7}
&=\frac{1}{\sqrt{p_1^{\delta_1}p_2^{\delta_2}}}O_p\paren{\paren{\sqrt{p_1\widetilde w_1}+1}\paren{\sqrt{p_2\widetilde m_1}+1}}\\
&=\frac{1}{\sqrt{p_1^{\delta_1}p_2^{\delta_2}}}O_p\paren{\paren{\sqrt{\frac{p_1^{2-\delta_1}}{Tp_2^{\delta_2}}}+1}\paren{\sqrt{\frac{p_2^{2-\delta_2}}{Tp_1^{\delta_1}}}+1}}\\
&=\frac{1}{\sqrt{p_1^{\delta_1}p_2^{\delta_2}}}O_p\paren{1+\frac{p_1^{1-\delta_1}p_2^{1-\delta_2}}{T}+\sqrt{\frac{p_1^{2-\delta_1}}{Tp_2^{\delta_2}}}+\sqrt{\frac{p_2^{2-\delta_2}}{Tp_1^{\delta_1}}}}\\
&=O_p\paren{\frac{1}{\sqrt{p_1^{\delta_1}p_2^{\delta_2}}}+o(1)\frac{1}{\sqrt{p_1^{\delta_1}p_2^{\delta_2}}}+\frac{p_1^{1-\delta_1}}{\sqrt{T}p_2^{\delta_2}}+\frac{p_2^{1-\delta_2}}{\sqrt{T}p_1^{\delta_1}}}
\end{aligned}
\]
Hence,
 \[
    \norm{\widetilde{\mathbf{F}}_t-\widetilde{\mathbf{H}}_R^{-1}\mathbf{F}_t\paren{\widetilde{\mathbf{H}}_C^{-1}}^\top}
=O_p\paren{\frac{1}{\sqrt{p_1^{\delta_1}p_2^{\delta_2}}}+\frac{p_1^{1-\delta_1}}{\sqrt{T}p_2^{\delta_2}}+\frac{p_2^{1-\delta_2}}{\sqrt{T}p_1^{\delta_1}}}
    \]
\subsection{Screening Consistency and Post-Screening Estimation}\label{sec:screening-proof}
In this subsection, we establish the selection consistency of both the
covariance-based and rank-based screening procedures, and then prove the
subsequent post-screening estimation results. 

We first present several auxiliary lemmas, some of which will also be used in
the analysis of the matrix-factor-augmented regression.
\begin{lemma}\label{lem:op-transfer}
Let \(X_n\) be a sequence of random variables and let \(a_n,b_n>0\) be
deterministic sequences. If
\(
X_n=O_p(a_n)
\;\text{and}\;
a_n=o(b_n),
\)
then
\(
X_n=o_p(b_n).
\)
\end{lemma}
\begin{proof}
Since \(X_n=O_p(a_n)\), for every \(\varepsilon>0\) there exists
\(C_\varepsilon<\infty\) such that
\(
\limsup_{n\to\infty}
\Pr\left(
\left|\frac{X_n}{a_n}\right|>C_\varepsilon
\right)
<\varepsilon.
\)
Moreover, \(a_n=o(b_n)\). Hence, for any fixed \(\eta>0\),
\(
\left|\frac{a_n}{b_n}\right|
<
\frac{\eta}{C_\varepsilon}
\)
for all sufficiently large \(n\). Therefore,
\(
\Pr\left(
\left|\frac{X_n}{b_n}\right|>\eta
\right)
\le
\Pr\left(
\left|\frac{X_n}{a_n}\right|>C_\varepsilon
\right)
\)
for all sufficiently large \(n\). Taking the upper limit and then letting
\(\varepsilon\downarrow0\) gives
\(
X_n=o_p(b_n).
\)
\end{proof}
\begin{lemma}\label{lem:thresholding}
Let \(I\subset[p]\) be an active index set, and let
\(\{\widehat q_i:i\in[p]\}\) be screening statistics. Suppose that, for some
deterministic sequences \(r_n\) and \(b_n\),
\(
\sup_{i\notin I}\widehat q_i=O_p(r_n),
\)
and
\(
\inf_{i\in I}\widehat q_i
\ge
C b_n-o_p(b_n)
\)
for some constant \(C>0\). Define
\(
\widehat I
:=
\{i\in[p]:\widehat q_i>\tau_n\},
\)
where
\(
r_n=o(\tau_n)
\;\text{and}\;
\tau_n=o(b_n).
\)
Then
\(
\Pr(\widehat I=I)\to1.
\)
\end{lemma}
\begin{proof}
First,
\(
\Pr(\widehat I\not\subset I)
=
\Pr\left(
\sup_{i\notin I}\widehat q_i>\tau_n
\right).
\)
Since \(\sup_{i\notin I}\widehat q_i=O_p(r_n)\) and \(r_n=o(\tau_n)\),
Lemma~\ref{lem:op-transfer} implies
\(
\sup_{i\notin I}\widehat q_i=o_p(\tau_n).
\)
Hence
\(
\Pr(\widehat I\not\subset I)\to0.
\)
Next,
\(
\Pr(I\not\subset\widehat I)
=
\Pr\left(
\inf_{i\in I}\widehat q_i\le\tau_n
\right).
\)
By assumption,
\(
\inf_{i\in I}\widehat q_i
\ge
b_n(c-o_p(1)).
\)
Since \(\tau_n=o(b_n)\), we have
\(
\frac{\tau_n}{b_n}\to0.
\)
Thus,
\[
\begin{aligned}
\Pr\left(
\inf_{i\in I}\widehat q_i\le\tau_n
\right)
&\le
\Pr\left(
c-o_p(1)\le\frac{\tau_n}{b_n}
\right)\\
&=\Pr\paren{
\{c-o_p(1)\le\frac{\tau_n}{b_n}\}\cap\{o_p\paren{1}\le \frac{c}{2}\}}\\
&+\Pr\paren{
\{c-o_p(1)\le\frac{\tau_n}{b_n}\}\cap\{o_p\paren{1}>\frac{c}{2}\}}\\
&\le\Pr\paren{
\frac{c}{2}\le\frac{\tau_n}{b_n}}+\Pr\paren{
o_p\paren{1}>\frac{c}{2}}\to0.
\end{aligned}
\]
Therefore,
\(
\Pr(I\not\subset\widehat I)\to0.
\)
Combining the two bounds,
\(
\Pr(\widehat I=I)
\ge
1-\Pr(\widehat I\not\subset I)
-\Pr(I\not\subset\widehat I)
\to1.
\)
\end{proof}
\begin{lemma}\label{lem:high-probability-transfer}
Let $\mathcal E_n$ be a sequence of events such that
$\Pr(\mathcal E_n)\to1$, and let $X_n$ and $Y_n$ be random elements
satisfying \(
X_n=Y_n
\;\text{on }\mathcal E_n.
\)
Then, for any deterministic sequence $a_n>0$:

\begin{enumerate}
    \item if $\|Y_n\|=O_p(a_n)$, then
    $\|X_n\|=O_p(a_n)$;

    \item if $\|Y_n\|=o_p(a_n)$, then
    $\|X_n\|=o_p(a_n)$.
\end{enumerate}
Moreover, if $\Pr(Y_n\in A_n)\to1$ for some sequence of sets $A_n$,
then
\[
\Pr(X_n\in A_n)\to1.
\]
\end{lemma}

\begin{proof}
For any $M>0$,
\[
\begin{aligned}
\Pr\{\|X_n\|>Ma_n\}
&=\Pr\{\{\|X_n\|>Ma_n\}\cap\mathcal{E}_n\}+\Pr\{\{\|X_n\|>Ma_n\}\cap\mathcal{E}_n^c\}\\
&\le
\Pr\{\|Y_n\|>Ma_n\}
+
\Pr(\mathcal E_n^c).
\end{aligned}
\]
The first two conclusions follow immediately from
$\Pr(\mathcal E_n^c)\to0$. Similarly,
\[
\begin{aligned}
\Pr(X_n\notin A_n)
&=\Pr\paren{\{X_n\notin A_n\}\cap\mathcal{E}_n}+\Pr\paren{\{X_n\notin A_n\}\cap\mathcal{E}_n^c}\\
&\le
\Pr(Y_n\notin A_n)
+
\Pr(\mathcal E_n^c),
\end{aligned}
\]
which proves the last assertion.
\end{proof}
\begin{lemma}
\label{lem:distribution convergence}
Let \(\ba_n\) and \(\bb_n\) be random vectors in
\(\mathbb R^d\), and let \(\{\mathcal A_n\}\) be a sequence of events
such that
\(
\Pr(\mathcal A_n)\to1.
\)
Suppose that
\(
\ba_n=\bb_n
\;\text{on }\mathcal A_n.
\)
If
\(
\ba_n
\xrightarrow{d}
\mathcal N(\bzero,\bSigma_a),
\)
then
\(
\bb_n
\xrightarrow{d}
\mathcal N(\bzero,\bSigma_a).
\)
\end{lemma}
\begin{proof}
For every \(\varepsilon>0\),
\[
\Pr\left(
\|\bb_n-\ba_n\|>\varepsilon
\right)
\le
\Pr(\mathcal A_n^c)
\to0.
\]
Hence,
\(
\bb_n-\ba_n=o_p(1).
\)
The result then follows from Slutsky's theorem.
\end{proof}
\noindent{\bf Proof of Theorem~\ref{thm:screening consistency}.}
By the model equations,
\(
X_{t,ij}
=
\bR_{i\cdot}^\top \bF_t \bC_{j\cdot}+e_{t,ij},
\;
y_{t+h}
=
\balpha^\top \bF_t\bbeta+\epsilon_{t+h},
\)
we have 
\[
\begin{aligned}
\E(y_{t+h}X_{t,ij})
&=
\E\left[
(\balpha^\top \bF_t\bbeta+\epsilon_{t+h})
(\bR_{i\cdot}^\top \bF_t \bC_{j\cdot}+e_{t,ij})
\right] \\
&=
\E\left[
(\balpha^\top \bF_t\bbeta)
(\bR_{i\cdot}^\top \bF_t \bC_{j\cdot})
\right],
\end{aligned}
\]
where the last equality follows from Assumptions~\ref{ass:mixing-noise} and \ref{ass:epsilon-t}.
Then \(\E Z_{t,ij}=0\). By the union bound,
\[
\begin{aligned}
\Pr\left(
\sup_{i\in[p_1],\,j\in[p_2]}
|\frac1T\sum_{t=1}^T Z_{t,ij}|
>\lambda
\right) 
&\le
\sum_{i=1}^{p_1}\sum_{j=1}^{p_2}
\Pr\left(
\left|
\frac1T\sum_{t=1}^T Z_{t,ij}
\right|
>\lambda
\right) \\
&\le
\sum_{i=1}^{p_1}\sum_{j=1}^{p_2}
\Pr\left(
\sup_{1\le m\le T}
\left|
\sum_{t=1}^m Z_{t,ij}
\right|
>
T\lambda
\right).
\end{aligned}
\]

Applying the Bernstein-type inequality of \cite{MerlevedePeligradRio_PTRF_2011} to the centered sequence \(\{Z_{t,ij}\}_{t=1}^T\), by
Assumptions~\ref{ass:screening} and \ref{ass:v-bound}, yields, for
\(T\lambda\ge C(\log T)^\eta\),
\[
\Pr\left(
\sup_{1\le m\le T}
\left|
\sum_{t=1}^m Z_{t,ij}
\right|
>
T\lambda
\right)
\le
(T+1)\exp\left(-\frac{(T\lambda)^{\gamma_s}}{C_1}\right)
+
\exp\left(
-\frac{(T\lambda)^2}{C_2+C_2TV}
\right),
\]
where \(C,C_1,C_2,\eta\) are positive constants independent of
\(i,j,p_1,p_2,T\). Consequently,
\[
\begin{aligned}
&\Pr\left(
\sup_{i,j}
|\frac1T\sum_{t=1}^T Z_{t,ij}|
>\lambda
\right)\le
p_1p_2(T+1)\exp\left(-\frac{(T\lambda)^{\gamma_s}}{C_1}\right)
+
p_1p_2
\exp\left(
-\frac{(T\lambda)^2}{C_2+C_2TV}
\right).
\end{aligned}
\]

Let
\(r_{T,p_1,p_2}
:=
\sqrt{\frac{\log(p_1p_2)}{T}}.
\)
We show that the preceding probability converges to zero when
\[
\lambda=a_{T,p_1,p_2}r_{T,p_1,p_2},
\;
a_{T,p_1,p_2}\to+\infty.
\]
For the first term, it suffices that
\(
\frac{(T\lambda)^{\gamma_s}}{C_1}
-
\log\{p_1p_2(T+1)\}
\to+\infty.
\)
A sufficient condition is
\(
\lambda
\gg
\frac{\{\log[p_1p_2(T+1)]\}^{1/\gamma_s}}{T}.
\)
Under Condition~\ref{cond:C4}, we have
\[
\frac{
\{\log[p_1p_2(T+1)]\}^{1/\gamma_s}/T
}{
\sqrt{\log(p_1p_2)/T}
}
=
\left(
\frac{
\log(p_1p_2)+\log(T+1)
}{
[T\log(p_1p_2)]^{\gamma_s/2}
}
\right)^{1/\gamma_s}
=
O(1).
\]
Moreover,
\(
\frac{(\log T)^\eta/T}{r_{T,p_1,p_2}}
=
\frac{(\log T)^\eta}{\sqrt{T\log(p_1p_2)}}
=o(1).
\)
Hence \(T\lambda\ge C(\log T)^\eta\) for all sufficiently large \(T\)
whenever \(\lambda=a_{T,p_1,p_2}r_{T,p_1,p_2}\) and
\(a_{T,p_1,p_2}\to+\infty\).
For the second term, since \(V=O(1)\),
\[
\sqrt{
\frac{(C_2+C_2TV)\log(p_1p_2)}{T^2}
}
=
O\left(
\sqrt{\frac{\log(p_1p_2)}{T}}
\right)
=
O(r_{T,p_1,p_2}).
\]
Therefore, for every \(a_{T,p_1,p_2}\to+\infty\),
\(
\Pr\left(
\sup_{i,j}
|\frac1T\sum_{t=1}^T Z_{t,ij}|
>
a_{T,p_1,p_2}r_{T,p_1,p_2}
\right)
\to0.
\)
Then, we have
\[
\sup_{i,j}
|\frac1T\sum_{t=1}^T Z_{t,ij}|
=
O_p(r_{T,p_1,p_2}).
\]
We next derive the signal--noise separation for the row screening statistic.
For any \(i\notin \mathcal{I}_R\), \(\bR_{i\cdot}=0\). Hence,
\[
\begin{aligned}
\sup_{i\notin \mathcal{I}_R}\widehat s_i^{\paren{{R}}}
&=\sup_{i\notin \mathcal{I}_R}\frac{1}{p_2}\sum_{j=1}^{p_2}\abs{\frac{1}{T}\sum_{t=1}^{T}y_{t+h}X_{t,ij}}
=\sup_{i\notin \mathcal{I}_R}\frac{1}{p_2}\sum_{j=1}^{p_2}\abs{\frac1T\sum_{t=1}^T Z_{t,ij}}\le
\sup_{i,j}
\abs{\frac1T\sum_{t=1}^T Z_{t,ij}}
=
O_p(r_{T,p_1,p_2}).
\end{aligned}
\]

For any \(i\in \mathcal{I}_R\), by Assumption~\ref{ass:signal},
\(
\left|
\E\left[
(\balpha^\top \bF_t\bbeta)
(\bR_{i\cdot}^\top \bF_t \bC_{j\cdot})
\right]
\right|
\ge C,
\; j\in \mathcal{I}_C.
\)
Since \(|\mathcal{I}_C|\asymp p_2^{\delta_2}\), we have
\(
\frac1{p_2}\sum_{j\in \mathcal{I}_C}\abs{\frac1T\sum_{t=1}^T \E y_{t+h}X_{t,ij}}
\ge
C p_2^{\delta_2-1},
\)
for any \(i\in\mathcal{I}_R\)
Thus,
\[
\begin{aligned}
\inf_{i\in \mathcal{I}_R}\widehat s_i^{\paren{{R}}}
&\gtrsim
\inf_{i\in \mathcal{I}_R}
\frac1{p_2}
\sum_{j=1}^{p_2}\abs{\frac1T\sum_{t=1}^T y_{t+h}X_{t,ij}} \\
&\ge
\inf_{i\in \mathcal{I}_R}
\left\{
\frac1{p_2}
\sum_{j=1}^{p_2}\abs{\frac1T\sum_{t=1}^T \E y_{t+h}X_{t,ij}}
-
\frac1{p_2}
\sum_{j=1}^{p_2}\abs{\frac1T\sum_{t=1}^T Z_{t,ij}}
\right\} \\
&\ge
\inf_{i\in \mathcal{I}_R}
\left\{
\frac1{p_2}
\sum_{j\in \mathcal{I}_C}\abs{\frac1T\sum_{t=1}^T \E y_{t+h}X_{t,ij}}
-
\frac1{p_2}
\sum_{j=1}^{p_2}\abs{\frac1T\sum_{t=1}^T Z_{t,ij}}
\right\} \\
&\ge
C p_2^{\delta_2-1}
-
\sup_{i,j}|\frac{1}{T}\sum_{t=1}^T Z_{t,ij}|.
\end{aligned}
\]
Since \(p_2^{2-2\delta_2}\log\paren{p_1p_2}=o\paren{T}\),
we have
\(
r_{T,p_1,p_2}
=
\sqrt{\frac{\log(p_1p_2)}{T}}
=
o(p_2^{\delta_2-1}).
\)
Hence, by Lemma~\ref{lem:op-transfer},
\(
\sup_{i,j}|\frac{1}{T}\sum_{t=1}^T Z_{t,ij}|
=
o_p(p_2^{\delta_2-1}).
\)
Consequently,
\(
\inf_{i\in \mathcal{I}_R}\widehat s_i^{\paren{{R}}}
\ge
Cp_2^{\delta_2-1}
-
o_p(p_2^{\delta_2-1}).
\)
Choose the row screening threshold as
\(
\tau_R
=
M_{T,p_1,p_2}r_{T,p_1,p_2},
\)
where
\(
M_{T,p_1,p_2}\to+\infty
\)
and
\(
\tau_R
=
o(p_2^{\delta_2-1}).
\)
Such a sequence exists under Condition~\ref{cond:C4}. 
Applying Lemma~\ref{lem:thresholding} with
\(
\widehat q_i=\widehat s_i^{\paren{{R}}},
\;
r_n=r_{T,p_1,p_2},
\;
b_n=p_2^{\delta_2-1},
\;
I=\mathcal{I}_R,
\)
gives
\(
\Pr(\widehat {\mathcal{I}}_R=\mathcal{I}_R)\to1.
\)

Let
\(\mathcal{E}_R=\{\widehat {\mathcal{I}}_R=\mathcal{I}_R\}\). By the preceding results, we have \(\Pr\paren{\mathcal{E}_R}\to1\). Define the true-support column-screening statistic
\(
\widehat s_j^{(C,\mathrm{true})}
=
\frac{1}{|\mathcal I_R|}
\sum_{i\in\mathcal I_R}
\widehat s_{ij},
\; j\in[p_2],
\)
and let
\(
\widehat{\mathcal I}_C^{\mathrm{true}}
=
\left\{
j\in[p_2]:
\widehat s_j^{(C,\mathrm{true})}>\tau_C
\right\}.
\)
By the same Bernstein argument as above, applied only to
\((i,j)\in \mathcal{I}_R\times[p_2]\), we have
\[
\begin{aligned}
\Pr\left(
\sup_{i\in \mathcal{I}_R,j}
|\frac1T\sum_{t=1}^T Z_{t,ij}|
>\lambda
\right)
&\le
p_1^{\delta_1}p_2(T+1)\exp\left(-\frac{(T\lambda)^{\gamma_s}}{C_1}\right)+
p_1^{\delta_1}p_2
\exp\left(
-\frac{(T\lambda)^2}{C_2+C_2TV}
\right).
\end{aligned}
\]
Let
\(
r_{T,p_1,p_2}'
:=
\sqrt{\frac{\log(p_1^{\delta_1}p_2)}{T}}.
\)
We show that the preceding probability converges to zero when
\[
\lambda=a_{T,p_1,p_2}r_{T,p_1,p_2}',
\;
a_{T,p_1,p_2}\to+\infty.
\]
For the first term, it suffices that
\(
\frac{(T\lambda)^{\gamma_s}}{C_1}
-
\log\{p_1^{\delta_1}p_2(T+1)\}
\to+\infty.
\)
A sufficient condition is
\(
\lambda
\gg
\frac{\{\log[p_1^{\delta_1}p_2(T+1)]\}^{1/\gamma_s}}{T}.
\)
Under Condition~\ref{cond:C4}
we have
\[
\frac{
\{\log[p_1^{\delta_1}p_2(T+1)]\}^{1/\gamma_s}/T
}{
\sqrt{\log(p_1^{\delta_1}p_2)/T}
}
=
\left(
\frac{
\log(p_1^{\delta_1}p_2)+\log(T+1)
}{
[T\log(p_1^{\delta_1}p_2)]^{\gamma_s/2}
}
\right)^{1/\gamma_s}
=
O(1).
\]
Moreover,
\(
\frac{(\log T)^\eta/T}{r_{T,p_1,p_2}'}
=
\frac{(\log T)^\eta}{\sqrt{T\log(p_1^{\delta_1}p_2)}}
=o(1).
\)
Hence \(T\lambda\ge C(\log T)^\eta\) for all sufficiently large \(T\)
whenever \(\lambda=a_{T,p_1,p_2}r_{T,p_1,p_2}'\) and
\(a_{T,p_1,p_2}\to+\infty\).
For the second term, since \(V=O(1)\),
\[
\sqrt{
\frac{(C_2+C_2TV)\log(p_1^{\delta_1}p_2)}{T^2}
}
=
O\left(
\sqrt{\frac{\log(p_1^{\delta_1}p_2)}{T}}
\right)
=
O(r_{T,p_1,p_2}').
\]
Therefore, for every \(a_{T,p_1,p_2}\to+\infty\),
\(
\Pr\left(
\sup_{i\in \mathcal{I}_R,j}
|\frac1T\sum_{t=1}^T Z_{t,ij}|
>
a_{T,p_1,p_2}r_{T,p_1,p_2}'
\right)
\to0.
\)
Thus,
\[
\sup_{i\in \mathcal{I}_R,j}
|\frac1T\sum_{t=1}^T Z_{t,ij}|
=
O_p(r_{T,p_1,p_2}').
\]
For \(j\notin \mathcal{I}_C\), \(\bC_{j\cdot}=0\), and hence \(y_{t+h}X_{t,ij}=Z_{t,ij}\). Therefore,
\(
\sup_{j\notin \mathcal{I}_C}\widehat{s}_j^{(C,\mathrm{true})}=O_p(r_{T,p_1,p_2}').
\)
For \(j\in \mathcal{I}_C\), Assumption~\ref{ass:signal} gives
\(
\left|
\E(X_{t,ij}y_{t+h})
\right|
\ge C,
\; i\in \mathcal{I}_R.
\)
Therefore,
\[
\begin{aligned}
\inf_{j\in \mathcal{I}_C}\widehat{s}_j^{(C,\mathrm{true})}
&\gtrsim\inf_{j\in \mathcal{I}_C}\frac{1}{p_1^{\delta_1}}\sum_{i\in \mathcal{I}_R}\abs{\frac{1}{T}\sum_{t=1}^T y_{t+h}X_{t,ij}}\\
&\ge\inf_{j\in \mathcal{I}_C}\frac{1}{p_1^{\delta_1}}\sum_{i\in \mathcal{I}_R}\paren{\abs{\frac{1}{T}\sum_{t=1}^T \E y_{t+h}X_{t,ij}}-\abs{\frac{1}{T}\sum_{t=1}^T Z_{t,ij}}}\\
&\ge
C-\sup_{i\in \mathcal{I}_R,j\in[p_2]}\abs{\frac{1}{T}\sum_{t=1}^T Z_{t,ij}}\\
&=
C-O_p(r_{T,p_1,p_2}').
\end{aligned}
\]
Under Condition~\ref{cond:C4}
implying \(r_{T,p_1,p_2}'=o(1)\), we have
\(
\inf_{j\in \mathcal{I}_C}\widehat{s}_j^{\paren{C,\mathrm{true}}}
\ge
C-o_p(1).
\)
Use the threshold \(\tau_C=M_{T,p_1,p_2}r_{T,p_1,p_2}'\), where \(M_{T,p_1,p_2}\to+\infty,\) as \(\min\{T,p_1,p_2\}\to+\infty\;\text{and}\;\tau_C=o\paren{1}\).
Applying Lemma~\ref{lem:thresholding} with
\(
\widehat q_j=\widehat{s}_j^{(C,\mathrm{true})},
\;
r_n=r_{T,p_1,p_2}',
\;
b_n=1,
\;
I=\mathcal{I}_C,
\)
yields
\(
\Pr(\widehat {\mathcal{I}}_C^{\mathrm{true}}=\mathcal{I}_C)\to1.
\) Since \(\widehat{\mathcal{I}}_C=\widehat{\mathcal{I}}_C^{\mathrm{true}}\) on the event \(\mathcal{E}_R\), by Lemma~\ref{lem:high-probability-transfer}, we have \(\Pr\paren{\widehat{\mathcal{I}}_C={\mathcal{I}}_C}\to 1.\)
Finally,
\[
\begin{aligned}
\Pr(\widehat {\mathcal{I}}_R=\mathcal{I}_R,\widehat {\mathcal{I}}_C=\mathcal{I}_C)
&\ge1-
\Pr(\widehat {\mathcal{I}}_R\neq\mathcal{I}_R)-
\Pr(\widehat {\mathcal{I}}_C\neq\mathcal{I}_C) \to1.
\end{aligned}
\]
\begin{remark}
The main procedure uses average-type screening statistics because cross-sectional aggregation can stabilize sampling fluctuations. Maximum-type statistics, such as \(\max_j\widehat s_{ij}\) and \(\max_i\widehat s_{ij}\), provide a possible alternative that preserves the strongest entrywise signal and avoids averaging it over inactive coordinates, but they are more sensitive to extreme sampling fluctuations. A formal comparison therefore involves a signal--noise trade-off and requires different cross-sectional concentration arguments. We leave such a comparison for future work.
\end{remark}
We next establish a robustness extension based on the rank statistic
introduced in Section~\ref{subsec:screening}. The result is stated under
dependence conditions tailored to the U-statistic representation of
Spearman's rho. These conditions are imposed only for the rank-based
extension.
\begin{assumption}[Rank-screening dependence I]\label{ass:rank-beta}
For each \((i,j)\), let
\(
\mathbf Z_t^{(ij)}
=
(X_{t,ij},y_{t+h})^\top.
\)
The bivariate process
\(\{\mathbf Z_t^{(ij)}:t\in\mathbb Z\}\)
is strictly stationary. Let \(\beta_{ij}(n)\) denote its \(\beta\)-mixing
coefficient. There exist constants \(C_\beta>0\) and \(\delta\ge1\),
independent of \(i,j,p_1,p_2,T\), such that
\[
\sup_{i\in[p_1],\,j\in[p_2]}
\beta_{ij}(n)
\le
C_\beta n^{-\delta},
\qquad n\ge1.
\]
\end{assumption}
\begin{assumption}[Rank-screening dependence II]
\label{ass:rank-condmix}
For each \((i,j)\), let
\(
\mathbf Z_t^{(ij)}
=
(X_{t,ij},y_{t+h})^\top .
\)
For any integer \(J\) satisfying \(1\le J\le r-1\) and any
\(t_1<\cdots<t_J\), conditional on
\(
\mathbf Z_{t_1}^{(ij)},\ldots,\mathbf Z_{t_J}^{(ij)},
\)
the sequence
\(
\{\mathbf Z_t^{(ij)}\}_{t=t_J+1}^{\infty}
\)
is exponentially \(\alpha\)-mixing.

More precisely, let
\(
\mathcal F_{a}^{b,(ij)}
=
\sigma\left(
\mathbf Z_t^{(ij)}:a\le t\le b
\right).
\)
There exist constants \(C_0,C_1>0\), independent of
\(i,j,p_1,p_2,T\), such that
\[
\sup_{i\in[p_1],\,j\in[p_2]}
\alpha_{ij}\left(
n;
\mathbf Z_{t_1}^{(ij)},\ldots,\mathbf Z_{t_J}^{(ij)}
\right)
\le
C_0 e^{-C_1 n},
\qquad n\ge1,
\]
almost surely, where \(
\alpha_{ij}\left(
n;
\mathbf Z_{t_1}^{(ij)},\ldots,\mathbf Z_{t_J}^{(ij)}
\right)
=
\sup_{s\ge t_J+1}
\alpha\left(
\mathcal F_{t_J+1}^{s,(ij)},
\mathcal F_{s+n}^{\infty,(ij)};
\Pr\left(
\cdot\mid
\mathbf Z_{t_1}^{(ij)},\ldots,\mathbf Z_{t_J}^{(ij)}
\right)
\right).
\)
\end{assumption}

For the Spearman statistic considered here, the U-statistic order \(r=3\), so the condition is required only for \(J=1,2\). Assumptions~\ref{ass:rank-beta} and~\ref{ass:rank-condmix}
are the counterparts, imposed uniformly over \((i,j)\), of
Assumptions~(A1) and~(A2) in \citet{Fanghan2018}.
They are used in \citet{Fanghan2018} to derive exponential
inequalities for fixed-order U-statistics under temporal dependence.
In our setting, these conditions are applied to the bivariate processes
\(
\mathbf Z_t^{(ij)}
=
\bigl(X_{t,ij},y_{t+h}\bigr)^\top,
\)
which enter the U-statistic representation of Spearman's rho.
These conditions are sufficient for the rank-based screening theory
developed below, but are not intended to be minimal. Similar support-recovery
results may continue to hold under weaker dependence conditions with
appropriately modified concentration arguments. A systematic investigation
of such weaker conditions is beyond the scope of the present paper.

\begin{assumption}[Rank-signal separation]
\label{ass:rank-signal}
For each \((i,j)\), let \(F_{ij}\) and \(F_y\) denote the marginal
distribution functions of \(X_{t,ij}\) and \(y_{t+h}\), respectively,
and suppose that the marginal distributions are continuous. Denote the population Spearman correlation
between \(X_{t,ij}\) and \(y_{t+h}\) as 
\(
\rho_{ij}
=
12\E\left[
\left\{F_{ij}(X_{t,ij})-\frac12\right\}
\left\{F_y(y_{t+h})-\frac12\right\}
\right].
\)
There exists a constant \(c>0\) such that
\(
\min_{i\in\mathcal I_R,\;j\in\mathcal I_C}
|\rho_{ij}|
\ge c,
\)
whereas
\(
\max_{(i,j)\notin\mathcal I_R\times\mathcal I_C}
|\rho_{ij}|
=
O\left(
\sqrt{\frac{\log(p_1p_2)}{T}}
\right).
\)
\end{assumption}

Assumption~\ref{ass:rank-signal} is the rank-based analogue of
Assumption~\ref{ass:signal} for the covariance screening procedure.
Hence the first part of
Assumption~\ref{ass:rank-signal} requires the predictive rank
association over the informative block to remain uniformly separated
from zero, while the second part allows inactive entries to retain
weak residual rank dependence of at most the uniform stochastic fluctuation order.

A sufficient setting is joint Gaussianity. Suppose that
\(
\bigl(\vect(\bF_t)^\top,e_{t,ij},\epsilon_{t+h}\bigr)^\top
\)
is jointly Gaussian. Then the zero-covariance restrictions between
\(e_{t,ij}\) and \(\bF_t\), together with the independence between
\(e_{t,ij}\) and \(\epsilon_{t+h}\), imply that \(e_{t,ij}\) is
independent of \((\bF_t,\epsilon_{t+h})\). Hence the inactive
population Spearman correlations are again zero. Moreover, for an
informative entry, if the original covariance signal is uniformly
bounded away from zero and the marginal variances are uniformly
bounded above, the corresponding Pearson correlation is uniformly
bounded away from zero. For a bivariate Gaussian pair,
\(
\rho_{ij}
=
\frac{6}{\pi}
\arcsin\left(\frac{\rho_{ij}^{P}}{2}\right)
\) (\citet{croux2010influence} Eq.(5)), 
where \(\rho_{ij}^{P}\) is the population Pearson correlation.
Thus, under joint Gaussianity, the covariance-based signal condition
together with the existing orthogonality conditions provides a
sufficient condition for Assumption~\ref{ass:rank-signal}.

\noindent{\bf Proof of Corollary~\ref{cor:rank-screening}.}
We prove the row-first case; the column-first case follows symmetrically. Let
\(
r_{T,p_1,p_2}^{\mathrm{rank}}
:=
\sqrt{\frac{\log(p_1p_2)}{T}}.
\)
We first establish the uniform concentration bound
\begin{equation}
\label{eq:rank-uniform-rate}
\max_{i\in[p_1],\,j\in[p_2]}
\left|
\widehat\rho_{ij}-\rho_{ij}
\right|
=
O_p\left(
r_{T,p_1,p_2}^{\mathrm{rank}}
\right).
\end{equation}

As shown in \citet{Fanghan2018}, the usual sample Spearman correlation therefore admits the
U-statistic representation
\(
\widehat\rho_{ij}
=
\frac{T-2}{T+1}U_{ij,T}
+
\frac{3}{T+1}\widehat\tau_{ij},
\)
where \(U_{ij,T}\) is a third-order U-statistic with a uniformly
bounded kernel, whose population target is \(\rho_{ij}\), and
\(\widehat\tau_{ij}\) is the sample Kendall correlation. In particular,
\(
|\widehat\tau_{ij}|\le1
\)
and the kernel bound is uniform over \((i,j)\).

Under Assumptions~\ref{ass:rank-beta} and
\ref{ass:rank-condmix}, Theorem~2.1 of
\citet{Fanghan2018} applies uniformly over \((i,j)\). Hence there
exist constants \(C_1,C_2,C_3>0\), independent of
\(i,j,p_1,p_2,T\), such that, for every \(x>0\) and all sufficiently
large \(T\),
\begin{equation}
\Pr\left(
|U_{ij,T}-\rho_{ij}|
>
\frac{C_1}{\sqrt T}+x
\right)
\le
2\exp\left\{
-
\frac{
C_2Tx^2
}{
1+C_3x(\log T)(\log\log 4T)
}
\right\}.
\end{equation}
Since \(|\rho_{ij}|\le1\) and \(|\widehat\tau_{ij}|\le1\),
the preceding U-statistic representation gives, after modifying
constants if necessary,
\begin{equation}
\Pr\left(
|\widehat\rho_{ij}-\rho_{ij}|
>
\frac{C_1}{\sqrt T}
+\frac{C_2}{T}
+x
\right)
\le
2\exp\left\{
-
\frac{
C_3Tx^2
}{
1+C_3x(\log T)(\log\log 4T)
}
\right\}.
\end{equation}
Therefore, by the union bound,
\begin{equation}
\label{eq:rank-union-bound}
\begin{aligned}
\Pr\left(
\max_{i\in[p_1],\,j\in[p_2]}
|\widehat\rho_{ij}-\rho_{ij}|
>
\frac{C_1}{\sqrt T}
+\frac{C_2}{T}
+x
\right)
\le
2p_1p_2
\exp\left\{
-
\frac{
C_3Tx^2
}{
1+C_3x(\log T)(\log\log 4T)
}
\right\}.
\end{aligned}
\end{equation}

Let \(a_{T,p_1,p_2}\to+\infty\) be arbitrary and set
\(
x
=
\frac12
a_{T,p_1,p_2}
r_{T,p_1,p_2}^{\mathrm{rank}}.
\)
Because
\(
\log(p_1p_2)=O(T^{1-v})
\)
for some \(v\in(0,1)\),
\[
\begin{aligned}
q_T^2
&:=
\left\{
r_{T,p_1,p_2}^{\mathrm{rank}}
(\log T)(\log\log 4T)
\right\}^2
=
\frac{\log(p_1p_2)}{T}
(\log T)^2(\log\log 4T)^2
=
O\left(
\frac{
(\log T)^2(\log\log 4T)^2
}{
T^v
}
\right)
=o(1).
\end{aligned}
\]
Hence \(q_T=o(1)\). Moreover,
\(
\frac{
a_{T,p_1,p_2}^2
}{
1+C a_{T,p_1,p_2}q_T
}
\to+\infty.
\)
Since
\(
\frac1{\sqrt T}
=
o\left(
r_{T,p_1,p_2}^{\mathrm{rank}}
\right),
\;
\frac1T
=
o\left(
r_{T,p_1,p_2}^{\mathrm{rank}}
\right),
\)
for all sufficiently large \(T,p_1,p_2\),
\(
\frac{C_1}{\sqrt T}+\frac{C_2}{T}
\le
\frac12
a_{T,p_1,p_2}
r_{T,p_1,p_2}^{\mathrm{rank}}.
\)
Substituting
\(
x=
a_{T,p_1,p_2}r_{T,p_1,p_2}^{\mathrm{rank}}/2
\)
into \eqref{eq:rank-union-bound} therefore yields
\[
\begin{aligned}
&\Pr\left(
\max_{i,j}
|\widehat\rho_{ij}-\rho_{ij}|
>
a_{T,p_1,p_2}
r_{T,p_1,p_2}^{\mathrm{rank}}
\right)
\le
2p_1p_2
\exp\left\{
-
C
\log(p_1p_2)
\frac{
a_{T,p_1,p_2}^2
}{
1+C a_{T,p_1,p_2}q_T
}
\right\}
\to0 .
\end{aligned}
\]
Since \(a_{T,p_1,p_2}\to+\infty\) is arbitrary, it follows that
\(
\max_{i,j}
|\widehat\rho_{ij}-\rho_{ij}|
=
O_p\left(
\sqrt{\frac{\log(p_1p_2)}{T}}
\right),
\)
which proves \eqref{eq:rank-uniform-rate}.

Recall that
\(
\widehat s_{ij}^{\,\mathrm{rank}}
=
|\widehat\rho_{ij}|
\)
and
\(
s_{ij}^{\mathrm{rank}}
=
|\rho_{ij}|.
\) 
Then we have \(
\left|
\widehat s_{ij}^{\,\mathrm{rank}}
-
s_{ij}^{\mathrm{rank}}
\right|
\le
|\widehat\rho_{ij}-\rho_{ij}|,
\)
and hence
\begin{equation}
\label{eq:rank-score-uniform-rate}
\max_{i,j}
\left|
\widehat s_{ij}^{\,\mathrm{rank}}
-
s_{ij}^{\mathrm{rank}}
\right|
=
O_p\left(
\sqrt{\frac{\log(p_1p_2)}{T}}
\right).
\end{equation}

It remains to establish signal--noise separation. This follows by the
same argument as in the proof of
Theorem~\ref{thm:screening consistency}. Indeed, defining
\(
s_i^{(R,\mathrm{rank})}
=
\frac1{p_2}
\sum_{j=1}^{p_2}
s_{ij}^{\mathrm{rank}},
\)
Assumption~\ref{ass:rank-signal} and
\(|\mathcal I_C|\asymp p_2^{\delta_2}\) imply
\(
\sup_{i\notin\mathcal I_R}
s_i^{(R,\mathrm{rank})}
=
O\left(
r_{T,p_1,p_2}^{\mathrm{rank}}
\right),
\;
\inf_{i\in\mathcal I_R}
s_i^{(R,\mathrm{rank})}
\gtrsim
p_2^{\delta_2-1}.
\)
Together with \eqref{eq:rank-score-uniform-rate},
\(
\sup_{i\notin\mathcal I_R}
\widehat s_i^{(R,\mathrm{rank})}
=
O_p\left(
r_{T,p_1,p_2}^{\mathrm{rank}}
\right),
\)
whereas
\(
\inf_{i\in\mathcal I_R}
\widehat s_i^{(R,\mathrm{rank})}
\ge
C p_2^{\delta_2-1}
-
O_p\left(
r_{T,p_1,p_2}^{\mathrm{rank}}
\right).
\)
Since
\[
p_2^{2-2\delta_2}\log(p_1p_2)=o(T),
\]
we have
\(
r_{T,p_1,p_2}^{\mathrm{rank}}
=
o(p_2^{\delta_2-1}).
\)
The first-stage exact recovery therefore follows from the threshold
conditions on \(\tau_R^{\mathrm{rank}}\) and the same thresholding
argument used in the covariance-based proof.

On the event
\(
\widehat{\mathcal I}_R^{\,\mathrm{rank}}=\mathcal I_R,
\)
repeat the preceding concentration argument over
\(\mathcal I_R\times[p_2]\). Since
\(
|\mathcal I_R|\asymp p_1^{\delta_1},
\)
this gives
\(
\max_{i\in\mathcal I_R,\,j\in[p_2]}
|\widehat\rho_{ij}-\rho_{ij}|
=
O_p\left(
\sqrt{
\frac{\log(p_1^{\delta_1}p_2)}{T}
}
\right).
\)
The corresponding population column score is bounded away from zero
for \(j\in\mathcal I_C\), while it is of the stochastic fluctuation
order for \(j\notin\mathcal I_C\), by
Assumption~\ref{ass:rank-signal}. Hence the second-stage exact recovery
follows from the threshold conditions on
\(\tau_C^{\mathrm{rank}}\), exactly as in the proof of
Theorem~\ref{thm:screening consistency}. Therefore,
\(
\Pr\left(
\widehat{\mathcal I}_R^{\,\mathrm{rank}}=\mathcal I_R,\,
\widehat{\mathcal I}_C^{\,\mathrm{rank}}=\mathcal I_C
\right)
\to1.
\)\\

\noindent{\bf Proof of Corollary~\ref{thm:screened estimators}.}
Let
\(
d_R=\abs{\mathcal I_R},
\;
d_C=\abs{\mathcal I_C}.
\)
By assumption,
\(
d_R\sim p_1^{\delta_1},
\;
d_C\sim p_2^{\delta_2}.
\)
Define the exact-screening event
\(
\mathcal E
=
\left\{
\widehat{\mathcal I}_R=\mathcal I_R,\,
\widehat{\mathcal I}_C=\mathcal I_C
\right\}.
\)
By Theorem~\ref{thm:screening consistency},
\(
\Pr(\mathcal E)\to1.
\)

Let
\(
\widetilde{\bR}^{\mathrm{true}}
\)
and
\(
\widetilde{\bC}^{\mathrm{true}}
\)
denote the projected loading estimators obtained by applying the
projected method to the submatrix indexed by the true active sets
\(\mathcal I_R\) and \(\mathcal I_C\).
Similarly, let
\(
\widetilde{\bF}_t^{\mathrm{true}}
\)
denote the corresponding factor estimator.
The rotation matrices for the screened and true-set estimators are
defined using the same construction as in
\citet{YuHeKongZhang_JoE_2022}.
Hence, on $\mathcal E$,
\(
\widetilde{\bR}^{\mathrm{scr}}
=
\widetilde{\bR}^{\mathrm{true}},
\;
\widetilde{\bC}^{\mathrm{scr}}
=
\widetilde{\bC}^{\mathrm{true}},
\;
\widetilde{\bF}_t^{\mathrm{scr}}
=
\widetilde{\bF}_t^{\mathrm{true}},
\)
and
\(
\widetilde{\bH}_R^{\mathrm{scr}}
=
\widetilde{\bH}_R^{\mathrm{true}},
\;
\widetilde{\bH}_C^{\mathrm{scr}}
=
\widetilde{\bH}_C^{\mathrm{true}}.
\)

Under the true active sets, the resulting matrix factor model has
dimensions $d_R$ and $d_C$ and satisfies the pervasive-loading
condition. As discussed in Section~\ref{sec:theory}, the remaining
conditions required by \citet{YuHeKongZhang_JoE_2022} continue to hold for the screened process.
By Corollary 3.1 of
\citet{YuHeKongZhang_JoE_2022},
\[
\frac{1}{d_R}
\norm{
\widetilde{\bR}^{\mathrm{true}}
-
\bR(\mathcal I_R,:)
\widetilde{\bH}_R^{\mathrm{true}}
}_F^2
=
O_p\left(
\frac{1}{Td_C}
+
\frac{1}{d_R^2d_C^2}
+
\frac{1}{T^2d_R^2}
\right).
\]
Since the corresponding screened and true-set quantities coincide on
$\mathcal E$, Lemma~\ref{lem:high-probability-transfer} implies
\[
\frac{1}{d_R}
\norm{
\widetilde{\bR}^{\mathrm{scr}}
-
\bR^{\mathrm{scr}}
\widetilde{\bH}_R^{\mathrm{scr}}
}_F^2
=
O_p\left(
\frac{1}{Td_C}
+
\frac{1}{d_R^2d_C^2}
+
\frac{1}{T^2d_R^2}
\right).
\]
Using
\(
d_R\sim p_1^{\delta_1}
\)
and
\(
d_C\sim p_2^{\delta_2}
\),
we obtain
\[
\frac{1}{p_1^{\delta_1}}
\norm{
\widetilde{\bR}^{\mathrm{scr}}
-
\bR^{\mathrm{scr}}
\widetilde{\bH}_R^{\mathrm{scr}}
}_F^2
=
O_p\left(
\frac{1}{Tp_2^{\delta_2}}
+
\frac{1}{p_1^{2\delta_1}p_2^{2\delta_2}}
+
\frac{1}{T^2p_1^{2\delta_1}}
\right).
\]
Applying the column-loading result in Corollary 3.1 of
\citet{YuHeKongZhang_JoE_2022} in the same manner yields
\[
\frac{1}{p_2^{\delta_2}}
\norm{
\widetilde{\bC}^{\mathrm{scr}}
-
\bC^{\mathrm{scr}}
\widetilde{\bH}_C^{\mathrm{scr}}
}_F^2
=
O_p\left(
\frac{1}{Tp_1^{\delta_1}}
+
\frac{1}{p_1^{2\delta_1}p_2^{2\delta_2}}
+
\frac{1}{T^2p_2^{2\delta_2}}
\right).
\]
Next, by Theorem 3.5 of
\citet{YuHeKongZhang_JoE_2022}, applied to the true-set model,
\[
\norm{
\widetilde{\bF}_t^{\mathrm{true}}
-
(\widetilde{\bH}_R^{\mathrm{true}})^{-1}
\bF_t
(\widetilde{\bH}_C^{\mathrm{true}})^{-\top}
}
=
O_p\left(
\frac{1}{\sqrt{T}\min\{d_R,d_C\}}
+
\frac{1}{\sqrt{d_Rd_C}}
\right).
\]
Therefore, by
Lemma~\ref{lem:high-probability-transfer},
\[
\norm{
\widetilde{\bF}_t^{\mathrm{scr}}
-
(\widetilde{\bH}_R^{\mathrm{scr}})^{-1}
\bF_t
(\widetilde{\bH}_C^{\mathrm{scr}})^{-\top}
}
=
O_p\left(
\frac{1}{\sqrt{T}\min\{d_R,d_C\}}
+
\frac{1}{\sqrt{d_Rd_C}}
\right).
\]
Since
\(
d_R\sim p_1^{\delta_1}
\)
and
\(
d_C\sim p_2^{\delta_2}
\),
this gives
\[
\norm{
\widetilde{\bF}_t^{\mathrm{scr}}
-
(\widetilde{\bH}_R^{\mathrm{scr}})^{-1}
\bF_t
(\widetilde{\bH}_C^{\mathrm{scr}})^{-\top}
}
=
O_p\left(
\frac{1}{
\sqrt{T}\min\{p_1^{\delta_1},p_2^{\delta_2}\}
}
+
\frac{1}{
\sqrt{p_1^{\delta_1}p_2^{\delta_2}}
}
\right).
\]
It remains to establish the consistency of the factor-number
estimators. By Theorem~\ref{thm:screening consistency},
\(
\Pr(\mathcal E)\to1.
\)

For a fixed integer \(q\), let
\(
\widehat k_1^{(s)}(q)
\)
denote the row-factor-number update at iteration \(s\) when the
projected row-side matrix is constructed using \(q\) column factors.
Similarly, let
\(
\widehat k_2^{(s)}(q)
\)
denote the column-factor-number update computed using \(q\) row
factors.

For each fixed
\(
q\in\{k_2,\ldots,k_{\max}\},
\)
the fixed-\(q\) argument in the proof of Theorem~3.8 of
\citet{YuHeKongZhang_JoE_2022}, applied to the true-support submatrix
with effective dimensions
\(
|\mathcal I_R|
\)
and
\(
|\mathcal I_C|,
\)
gives
\(
\Pr\left\{
\widehat k_{1,\mathrm{true}}^{(s)}(q)\ne k_1
\right\}
\to0.
\)
On the event \(\mathcal E\), the screened and true-support updates
coincide. Hence,
\(
\Pr\left\{
\widehat k_1^{(s)}(q)\ne k_1
\right\}
\le
\Pr(\mathcal E^c)
+
\Pr\left\{
\widehat k_{1,\mathrm{true}}^{(s)}(q)\ne k_1
\right\}\to0.
\)
Similarly, for each fixed
\(
q\in\{k_1,\ldots,k_{\max}\},
\)
\(
\Pr\left\{
\widehat k_2^{(s)}(q)\ne k_2
\right\}
\to0.
\)
We now establish the result by induction.

For the first iteration,
\(
\widehat k_1^{(0)}
=
\widehat k_2^{(0)}
=
k_{\max},
\)
where \(k_{\max}\ge\max\{k_1,k_2\}\) is deterministic. Therefore,
\(
\widehat k_1^{(1)}
=
\widehat k_1^{(1)}(k_{\max}),
\)
and the preceding fixed-input result yields
\(
\Pr\left\{
\widehat k_1^{(1)}\ne k_1
\right\}
\to0.
\)

On the event
\(
\{\widehat k_1^{(1)}=k_1\},
\)
the actual column-side update coincides with the update computed using
the fixed true row factor number \(k_1\), namely,
\(
\widehat k_2^{(1)}
=
\widehat k_2^{(1)}(k_1).
\)
Consequently,
\[
\begin{aligned}
\Pr\left\{
\widehat k_2^{(1)}\ne k_2
\right\}
&\le
\Pr\left\{
\widehat k_1^{(1)}\ne k_1
\right\}
+
\Pr\left\{
\widehat k_2^{(1)}(k_1)\ne k_2
\right\}\to0.
\end{aligned}
\]
It follows from the union bound that
\[
\begin{aligned}
&\Pr\left\{
\widehat k_1^{(1)}=k_1,\,
\widehat k_2^{(1)}=k_2
\right\}\ge
1-
\Pr\left\{
\widehat k_1^{(1)}\ne k_1
\right\}
-
\Pr\left\{
\widehat k_2^{(1)}\ne k_2
\right\}
\to1.
\end{aligned}
\]
Suppose now that, for some fixed \(s\ge2\),
\(
\Pr\left\{
\widehat k_1^{(s-1)}=k_1,\,
\widehat k_2^{(s-1)}=k_2
\right\}
\to1.
\)
On the event
\(
\{\widehat k_2^{(s-1)}=k_2\},
\)
the actual row-side update coincides with
\(
\widehat k_1^{(s)}(k_2).
\)
Therefore,
\[
\begin{aligned}
\Pr\left\{
\widehat k_1^{(s)}\ne k_1
\right\}
&\le
\Pr\left\{
\widehat k_2^{(s-1)}\ne k_2
\right\}
+
\Pr\left\{
\widehat k_1^{(s)}(k_2)\ne k_1
\right\}\to0.
\end{aligned}
\]
Likewise, on the event
\(
\{\widehat k_1^{(s)}=k_1\},
\)
the actual column-side update coincides with
\(
\widehat k_2^{(s)}(k_1).
\)
It follows that
\[
\begin{aligned}
\Pr\left\{
\widehat k_2^{(s)}\ne k_2
\right\}
&\le
\Pr\left\{
\widehat k_1^{(s)}\ne k_1
\right\}
+
\Pr\left\{
\widehat k_2^{(s)}(k_1)\ne k_2
\right\}\to0.
\end{aligned}
\]
Hence,
\(
\Pr\left\{
\widehat k_1^{(s)}=k_1,\,
\widehat k_2^{(s)}=k_2
\right\}
\to1.
\)
By induction, the preceding conclusion holds for every fixed iteration \(s\ge1\).
For theoretical convenience, if the stopping criterion is met before
the maximum iteration number \(m\), define all subsequent iterates to be equal to the values at the stopping iteration. If \(m\) is fixed, the estimates returned by Algorithm~\ref{alg:factor-number} therefore
satisfy
\(
\Pr\left\{
\widehat k_1=k_1,\,
\widehat k_2=k_2
\right\}
\to1.
\)
This completes the proof.

\begin{lemma}
\label{lem:avg-screened-factor-error}
Under the conditions of Corollary~\ref{thm:screened estimators},
\[
\frac{1}{T}\sum_{t=1}^T
\left\|
\widetilde{\bF}_t^{\mathrm{scr}}
-
(\widetilde{\bH}_R^{\mathrm{scr}})^{-1}
\bF_t
\left((\widetilde{\bH}_C^{\mathrm{scr}})^{-1}\right)^\top
\right\|_F^2
=
O_p\left(
\frac{1}{T p_1^{2\delta_1}}
+
\frac{1}{T p_2^{2\delta_2}}
+
\frac{1}{p_1^{\delta_1}p_2^{\delta_2}}
\right).
\]
\end{lemma}
\begin{proof}
By Theorem~\ref{thm:screening consistency},
\(\Pr(\mathcal E)\to1\). On \(\mathcal E\), the post-screening
estimator coincides with the projected estimator applied to the true
active submatrix. 

The factor-error decomposition used in the proof of Theorem~\ref{thm:projected-estimation}\textup{(II)} applies verbatim to
the true-support model. Since this is an algebraic identity, it holds
simultaneously for all \(t=1,\ldots,T\). Specifically, on
\(\mathcal E\),
\[
\begin{aligned}
&
\widetilde{\bF}_t^{\mathrm{scr}}
-
(\widetilde{\bH}_R^{\mathrm{scr}})^{-1}
\bF_t
(\widetilde{\bH}_C^{\mathrm{scr}})^{-\top}
\\
={}&
\frac{1}{|\mathcal I_R||\mathcal I_C|}
(\widetilde{\bR}^{\mathrm{scr}})^\top
\left[
\bR^{\mathrm{scr}}
-
\widetilde{\bR}^{\mathrm{scr}}
(\widetilde{\bH}_R^{\mathrm{scr}})^{-1}
\right]
\bF_t
\left[
\bC^{\mathrm{scr}}
-
\widetilde{\bC}^{\mathrm{scr}}
(\widetilde{\bH}_C^{\mathrm{scr}})^{-1}
\right]^\top
\widetilde{\bC}^{\mathrm{scr}}
\\
&+
\frac{1}{|\mathcal I_R|}
(\widetilde{\bR}^{\mathrm{scr}})^\top
\left[
\bR^{\mathrm{scr}}
-
\widetilde{\bR}^{\mathrm{scr}}
(\widetilde{\bH}_R^{\mathrm{scr}})^{-1}
\right]
\bF_t
(\widetilde{\bH}_C^{\mathrm{scr}})^{-\top}
\\
&+
\frac{1}{|\mathcal I_C|}
(\widetilde{\bH}_R^{\mathrm{scr}})^{-1}
\bF_t
\left[
\bC^{\mathrm{scr}}
-
\widetilde{\bC}^{\mathrm{scr}}
(\widetilde{\bH}_C^{\mathrm{scr}})^{-1}
\right]^\top
\widetilde{\bC}^{\mathrm{scr}}
\\
&+
\frac{1}{|\mathcal I_R||\mathcal I_C|}
\left(
\widetilde{\bR}^{\mathrm{scr}}
-
\bR^{\mathrm{scr}}\widetilde{\bH}_R^{\mathrm{scr}}
\right)^\top
\bE_t^{\mathrm{scr}}
\left(
\widetilde{\bC}^{\mathrm{scr}}
-
\bC^{\mathrm{scr}}\widetilde{\bH}_C^{\mathrm{scr}}
\right)
\\
&+
\frac{1}{|\mathcal I_R||\mathcal I_C|}
\left(
\bR^{\mathrm{scr}}\widetilde{\bH}_R^{\mathrm{scr}}
\right)^\top
\bE_t^{\mathrm{scr}}
\left(
\widetilde{\bC}^{\mathrm{scr}}
-
\bC^{\mathrm{scr}}\widetilde{\bH}_C^{\mathrm{scr}}
\right)
\\
&+
\frac{1}{|\mathcal I_R||\mathcal I_C|}
\left(
\widetilde{\bR}^{\mathrm{scr}}
-
\bR^{\mathrm{scr}}\widetilde{\bH}_R^{\mathrm{scr}}
\right)^\top
\bE_t^{\mathrm{scr}}
\bC^{\mathrm{scr}}\widetilde{\bH}_C^{\mathrm{scr}}
\\
&+
\frac{1}{|\mathcal I_R||\mathcal I_C|}
\left(
\bR^{\mathrm{scr}}\widetilde{\bH}_R^{\mathrm{scr}}
\right)^\top
\bE_t^{\mathrm{scr}}
\bC^{\mathrm{scr}}\widetilde{\bH}_C^{\mathrm{scr}}
\\
=:{}&
\sum_{\ell=1}^7 I_{\ell t}.
\end{aligned}
\]
Consequently,
\[
\frac{1}{T}\sum_{t=1}^T
\left\|
\widetilde{\bF}_t^{\mathrm{scr}}
-
(\widetilde{\bH}_R^{\mathrm{scr}})^{-1}
\bF_t
(\widetilde{\bH}_C^{\mathrm{scr}})^{-\top}
\right\|_F^2
\le
7\sum_{\ell=1}^7
\frac{1}{T}\sum_{t=1}^T
\|I_{\ell t}\|_F^2.
\]

The first three terms involve \(\bF_t\) multiplied by quantities that do
not depend on \(t\). Hence their time averages are controlled by the
post-screening loading-estimation bounds together with
\(
\frac{1}{T}\sum_{t=1}^T\|\bF_t\|_F^2=O_p(1).
\)
For the remaining four terms, the same moment and weak-dependence
bounds used in the proof of
Theorem~\ref{thm:projected-estimation}\textup{(II)}, now averaged over
\(t\), apply. For example,
\[
\begin{aligned}
\frac{1}{T}\sum_{t=1}^T\|I_{7t}\|_F^2
&\lesssim
\frac{1}{|\mathcal I_R|^2|\mathcal I_C|^2}
\frac{1}{T}\sum_{t=1}^T
\left\|
(\bR^{\mathrm{scr}})^\top
\bE_t^{\mathrm{scr}}
\bC^{\mathrm{scr}}
\right\|_F^2
=
O_p\left(
\frac{1}{|\mathcal I_R||\mathcal I_C|}
\right).
\end{aligned}
\]
Applying the corresponding bounds to \(I_{1t},\ldots,I_{6t}\) and using
Cauchy--Schwarz for the mixed terms therefore gives
\[
\sum_{\ell=1}^7
\frac{1}{T}\sum_{t=1}^T
\|I_{\ell t}\|_F^2
=
O_p\left(
\frac{1}{T|\mathcal I_R|^2}
+
\frac{1}{T|\mathcal I_C|^2}
+
\frac{1}{|\mathcal I_R||\mathcal I_C|}
\right).
\]
Finally, since
\(
|\mathcal I_R|\sim p_1^{\delta_1},
\;
|\mathcal I_C|\sim p_2^{\delta_2},
\)
the stated result follows on \(\mathcal E\). Since \(\Pr(\mathcal E)\to1\), the same result holds unconditionally.
\end{proof}

\subsection{Factor-Augmented Matrix Regression}
{\bf Proof of Theorem~\ref{thm:regression coefficient estimator}.}
On the event \(\mathcal E\), the screened matrix
\(
\bX_t(
\widehat{\mathcal I}_R,
\widehat{\mathcal I}_C
)
\)
coincides with the fixed true-support submatrix
\(
\bX_t(\mathcal I_R,\mathcal I_C).
\)
Because the same factor-estimation, normalization, and
regression-estimation procedures are applied, the corresponding
factor estimates, rotation matrices, regression estimates, and
centered regression statistics also coincide on \(\mathcal E\).
Moreover, by Theorem~\ref{thm:screening consistency},
\(
\Pr(\mathcal E)\to1.
\)
Therefore, by Lemma~\ref{lem:distribution convergence}, it suffices
to establish the asserted asymptotic results for the
estimators constructed from
\(\bX_t(\mathcal I_R,\mathcal I_C)\).
We work with this true-support submatrix throughout the proof of
Theorem~\ref{thm:regression coefficient estimator}. To avoid introducing a separate set of superscripts throughout the
lengthy calculations below, we retain the notation
\(
\bX_t^{\mathrm{scr}},
\widetilde{\bR}^{\mathrm{scr}},
\widetilde{\bC}^{\mathrm{scr}},
\widetilde{\bF}_t^{\mathrm{scr}},
\widetilde{\bH}_R^{\mathrm{scr}},
\widetilde{\bH}_C^{\mathrm{scr}},
\widehat{\btheta},
\widehat{\balpha},
\widehat{\bbeta}
\)
for the corresponding quantities constructed from
\(\bX_t(\mathcal I_R,\mathcal I_C)\) throughout the following
derivation. This is only a local notational convention.

Consider the predictive regression \eqref{equa:factor-augmented-regression}.
\[
\begin{aligned}
y_{t+h}
&=
\balpha_0^\top \bF_t\bbeta_0+\epsilon_{t+h}
=
\left(\bbeta_0^\top\otimes \balpha_0^\top\right)\vect(\bF_t)
+\epsilon_{t+h}\\
&=
\paren{\bbeta_0^\top\otimes\balpha_0^\top}
\vect\paren{\widetilde{\bH}_R^{\mathrm{scr}}\widetilde{\bF}_t^{\mathrm{scr}}(\widetilde{\bH}_C^{\mathrm{scr}})^\top}
+
\paren{\bbeta_0^\top\otimes\balpha_0^\top}
\vect\paren{\bF_t-\widetilde{\bH}_R^{\mathrm{scr}}\widetilde{\bF}_t^{\mathrm{scr}}(\widetilde{\bH}_C^{\mathrm{scr}})^\top}
+\epsilon_{t+h}\\
&=
\paren{\bbeta_0^\top\otimes\balpha_0^\top}
\paren{\widetilde{\bH}_C^{\mathrm{scr}}\otimes\widetilde{\bH}_R^{\mathrm{scr}}}
\vect\paren{\widetilde{\bF}_t^{\mathrm{scr}}}
+
\paren{\bbeta_0^\top\otimes\balpha_0^\top}
\vect\paren{\bF_t-\widetilde{\bH}_R^{\mathrm{scr}}\widetilde{\bF}_t^{\mathrm{scr}}(\widetilde{\bH}_C^{\mathrm{scr}})^\top}
+\epsilon_{t+h}\\
&=
\paren{\paren{\bbeta_0^\top\widetilde{\bH}_C^{\mathrm{scr}}}\otimes
\paren{\balpha_0^\top\widetilde{\bH}_R^{\mathrm{scr}}}}
\vect\paren{\widetilde{\bF}_t^{\mathrm{scr}}}
+
\paren{\bbeta_0^\top\otimes\balpha_0^\top}
\vect\paren{\bF_t-\widetilde{\bH}_R^{\mathrm{scr}}\widetilde{\bF}_t^{\mathrm{scr}}(\widetilde{\bH}_C^{\mathrm{scr}})^\top}
+\epsilon_{t+h}
\end{aligned}
\]
To get the LSE estimator, write the squared error as
\begin{equation}
\mathcal{L}:=
\sum_t
\paren{y_{t+h}-\btheta^\top\vect\paren{\widetilde{\bF}_t^{\mathrm{scr}}}}^2
\end{equation}
Differentiating with respect to \(\btheta\) yields the condition
\begin{equation}
\sum_t
\paren{
y_{t+h}
-
{\btheta^\top}
\vect\paren{\widetilde{\bF}_t^{\mathrm{scr}}}
}
\vect\paren{\widetilde{\bF}_t^{\mathrm{scr}}}
=0
\end{equation}
Equivalently,
\begin{equation}
\sum_t
\vect\paren{\widetilde{\bF}_t^{\mathrm{scr}}}
\vect\paren{\widetilde{\bF}_t^{\mathrm{scr}}}^\top{\btheta}
=
\sum_t
y_{t+h}
\vect\paren{\widetilde{\bF}_t^{\mathrm{scr}}}
\end{equation}
Thus,
\[
\begin{aligned}
\widehat\btheta
&=
\left[
\sum_t
\vect\paren{\widetilde{\bF}_t^{\mathrm{scr}}}
\vect\paren{\widetilde{\bF}_t^{\mathrm{scr}}}^\top
\right]^{-1}
\sum_t
y_{t+h}
\vect\paren{\widetilde{\bF}_t^{\mathrm{scr}}}\\
&=
\left[
\sum_t
\vect\paren{\widetilde{\bF}_t^{\mathrm{scr}}}
\vect\paren{\widetilde{\bF}_t^{\mathrm{scr}}}^\top
\right]^{-1}
\sum_t
\left(
\paren{
\paren{\bbeta_0^\top\widetilde{\bH}_C^{\mathrm{scr}}}
\otimes
\paren{\balpha_0^\top\widetilde{\bH}_R^{\mathrm{scr}}}
}
\vect\paren{\widetilde{\bF}_t^{\mathrm{scr}}}
\right.\\
&\left.
+
\paren{\bbeta_0^\top\otimes\balpha_0^\top}
\vect\paren{\bF_t-\widetilde{\bH}_R^{\mathrm{scr}}\widetilde{\bF}_t^{\mathrm{scr}}(\widetilde{\bH}_C^{\mathrm{scr}})^\top}
+\epsilon_{t+h}
\right)
\vect\paren{\widetilde{\bF}_t^{\mathrm{scr}}}
\end{aligned}
\]
Hence,
\[
\begin{aligned}
&\quad\widehat\btheta
-
\paren{
\paren{(\widetilde{\bH}_C^{\mathrm{scr}})^\top\bbeta_0}
\otimes
\paren{(\widetilde{\bH}_R^{\mathrm{scr}})^\top\balpha_0}
}\\
&=
\left[
\sum_t
\vect\paren{\widetilde{\bF}_t^{\mathrm{scr}}}
\vect\paren{\widetilde{\bF}_t^{\mathrm{scr}}}^\top
\right]^{-1}
\sum_t
{
\vect\paren{\widetilde{\bF}_t^{\mathrm{scr}}}
\vect\paren{
\bF_t-\widetilde{\bH}_R^{\mathrm{scr}}\widetilde{\bF}_t^{\mathrm{scr}}(\widetilde{\bH}_C^{\mathrm{scr}})^\top
}^\top
}
\paren{\bbeta_0\otimes\balpha_0}\\
&+
\left[
\sum_t
\vect\paren{\widetilde{\bF}_t^{\mathrm{scr}}}
\vect\paren{\widetilde{\bF}_t^{\mathrm{scr}}}^\top
\right]^{-1}
\sum_t
\epsilon_{t+h}
\vect\paren{\widetilde{\bF}_t^{\mathrm{scr}}}\\
&=
\left[
\frac{1}{T}
\sum_t
\vect\paren{\widetilde{\bF}_t^{\mathrm{scr}}}
\vect\paren{\widetilde{\bF}_t^{\mathrm{scr}}}^\top
\right]^{-1}
\frac{1}{T}
\sum_t
{
\vect\paren{\widetilde{\bF}_t^{\mathrm{scr}}}
\vect\paren{
\bF_t-\widetilde{\bH}_R^{\mathrm{scr}}\widetilde{\bF}_t^{\mathrm{scr}}(\widetilde{\bH}_C^{\mathrm{scr}})^\top
}^\top
}
\paren{\bbeta_0\otimes\balpha_0}
\\
&+
\left[
\frac{1}{T}
\sum_t
\vect\paren{\widetilde{\bF}_t^{\mathrm{scr}}}
\vect\paren{\widetilde{\bF}_t^{\mathrm{scr}}}^\top
\right]^{-1}
\frac{1}{T}
\sum_t
\epsilon_{t+h}
\vect\paren{\widetilde{\bF}_t^{\mathrm{scr}}}
\\
&:=I+II.
\end{aligned}
\]
First, we show that \(\norm{\left[
\frac{1}{T}
\sum_t
\vect\paren{\widetilde{\bF}_t^{\mathrm{scr}}}
\vect\paren{\widetilde{\bF}_t^{\mathrm{scr}}}^\top
\right]^{-1}}=O_p\paren{1}.\)
\[
\begin{aligned}
&\quad\frac{1}{T}\sum_t
\vect\paren{\widetilde{\bF}_t^{\mathrm{scr}}}
\vect\paren{\widetilde{\bF}_t^{\mathrm{scr}}}^\top\\
&=
\frac{1}{T}\sum_t
\vect\paren{
\widetilde{\bF}_t^{\mathrm{scr}}
-
(\widetilde{\bH}_R^{\mathrm{scr}})^{-1}
\bF_t
\left((\widetilde{\bH}_C^{\mathrm{scr}})^{-1}\right)^\top
}
\vect\paren{
\widetilde{\bF}_t^{\mathrm{scr}}
-
(\widetilde{\bH}_R^{\mathrm{scr}})^{-1}
\bF_t
\left((\widetilde{\bH}_C^{\mathrm{scr}})^{-1}\right)^\top
}^\top\\
&+
\frac{1}{T}\sum_t
\vect\paren{
\widetilde{\bF}_t^{\mathrm{scr}}
-
(\widetilde{\bH}_R^{\mathrm{scr}})^{-1}
\bF_t
\left((\widetilde{\bH}_C^{\mathrm{scr}})^{-1}\right)^\top
}
\vect\paren{
(\widetilde{\bH}_R^{\mathrm{scr}})^{-1}
\bF_t
\left((\widetilde{\bH}_C^{\mathrm{scr}})^{-1}\right)^\top
}^\top\\
&+
\frac{1}{T}\sum_t
\vect\paren{
(\widetilde{\bH}_R^{\mathrm{scr}})^{-1}
\bF_t
\left((\widetilde{\bH}_C^{\mathrm{scr}})^{-1}\right)^\top
}
\vect\paren{
\widetilde{\bF}_t^{\mathrm{scr}}
-
(\widetilde{\bH}_R^{\mathrm{scr}})^{-1}
\bF_t
\left((\widetilde{\bH}_C^{\mathrm{scr}})^{-1}\right)^\top
}^\top\\
&+
\frac{1}{T}\sum_t
\vect\paren{
(\widetilde{\bH}_R^{\mathrm{scr}})^{-1}
\bF_t
\left((\widetilde{\bH}_C^{\mathrm{scr}})^{-1}\right)^\top
}
\vect\paren{
(\widetilde{\bH}_R^{\mathrm{scr}})^{-1}
\bF_t
\left((\widetilde{\bH}_C^{\mathrm{scr}})^{-1}\right)^\top
}^\top\\
&=
\frac{1}{T}\sum_t
\paren{(\widetilde{\bH}_C^{\mathrm{scr}})^{-1}\otimes(\widetilde{\bH}_R^{\mathrm{scr}})^{-1}}
\vect\paren{\bF_t}
\vect\paren{\bF_t}^\top
\paren{(\widetilde{\bH}_C^{\mathrm{scr}})^{-1}\otimes(\widetilde{\bH}_R^{\mathrm{scr}})^{-1}}^\top
+o_p\paren{1}.
\end{aligned}
\]
Thus,
\[
\begin{aligned}
&\quad\lambda_{\min}\paren{
\frac{1}{T}\sum_t
\vect\paren{\widetilde{\bF}_t^{\mathrm{scr}}}
\vect\paren{\widetilde{\bF}_t^{\mathrm{scr}}}^\top
}\\
&=
\lambda_{\min}\paren{
\frac{1}{T}\sum_t
\paren{(\widetilde{\bH}_C^{\mathrm{scr}})^{-1}\otimes(\widetilde{\bH}_R^{\mathrm{scr}})^{-1}}
\vect\paren{\bF_t}
\vect\paren{\bF_t}^\top
\paren{(\widetilde{\bH}_C^{\mathrm{scr}})^{-1}\otimes(\widetilde{\bH}_R^{\mathrm{scr}})^{-1}}^\top
}
+o_p(1).
\end{aligned}
\]
Since
\(
\widetilde{\bH}_C^{\mathrm{scr}}(\widetilde{\bH}_C^{\mathrm{scr}})^{\top}\xrightarrow[]{p}\bI_{k_2},
\;
\widetilde{\bH}_R^{\mathrm{scr}}(\widetilde{\bH}_R^{\mathrm{scr}})^{\top}\xrightarrow[]{p}\bI_{k_1},
\)
we have
\[
\paren{(\widetilde{\bH}_C^{\mathrm{scr}})^{-1}\otimes(\widetilde{\bH}_R^{\mathrm{scr}})^{-1}}^\top
\paren{(\widetilde{\bH}_C^{\mathrm{scr}})^{-1}\otimes(\widetilde{\bH}_R^{\mathrm{scr}})^{-1}}
=
\paren{
\paren{\widetilde{\bH}_C^{\mathrm{scr}}(\widetilde{\bH}_C^{\mathrm{scr}})^\top}^{-1}
\otimes
\paren{\widetilde{\bH}_R^{\mathrm{scr}}(\widetilde{\bH}_R^{\mathrm{scr}})^\top}^{-1}
}
\xrightarrow[]{p}
\bI_{k_1k_2}.
\]
By Rayleigh quotient characterization, for any \(\bS\succeq0\),
\[
\lambda_{\min}\paren{\bA^\top\bA}\lambda_{\min}\paren{\bS}
\le
\lambda_{\min}\paren{\bA^\top\bS\bA}
\le
\lambda_{\max}\paren{\bA^\top\bA}\lambda_{\min}\paren{\bS}.
\]
Then,
\[
\lambda_{\min}\paren{
\frac{1}{T}\sum_t
\vect\paren{\widetilde{\bF}_t^{\mathrm{scr}}}
\vect\paren{\widetilde{\bF}_t^{\mathrm{scr}}}^\top
}
=
\lambda_{\min}\paren{
\frac{1}{T}\sum_t
\vect\paren{\bF_t}
\vect\paren{\bF_t}^\top
}
+o_p(1).
\]
Hence, the smallest eigenvalue of
\(
\frac{1}{T}\sum_t
\vect\paren{\widetilde{\bF}_t^{\mathrm{scr}}}
\vect\paren{\widetilde{\bF}_t^{\mathrm{scr}}}^\top
\)
is away from zero. Thus,
\[
\norm{
\paren{
\frac{1}{T}\sum_t
\vect\paren{\widetilde{\bF}_t^{\mathrm{scr}}}
\vect\paren{\widetilde{\bF}_t^{\mathrm{scr}}}^\top
}^{-1}
}
=
O_p\paren{1}.
\]
Moreover, we show that \([\frac{1}{T}\sum_t
\vect\paren{\widetilde{\bF}_t^{\mathrm{scr}}}
\vect\paren{\widetilde{\bF}_t^{\mathrm{scr}}}^\top]^{-1}\xrightarrow[]{p}\paren{\bGamma_2\otimes\bGamma_1}^\top
\bSigma_F^{-1}
\paren{\bGamma_2\otimes\bGamma_1}.\)
Following the arguments in Section A.2 of the supplementary material of
\citet{CaiKongWuZhao_JBES_2025} and the proof of Theorem 3.4 of \cite{YuHeKongZhang_JoE_2022}, we have, by choosing column signs of \(\widetilde{\bH}_R\) and \(\widetilde{\bH}_C\), 
\(
\widetilde{\bH}_R^{\mathrm{scr}}\xrightarrow[]{p}\bGamma_1,
\;
\widetilde{\bH}_C^{\mathrm{scr}}\xrightarrow[]{p}\bGamma_2.
\)
Thus,
\[
\begin{aligned}
&\quad\frac{1}{T}\sum_t
\vect\paren{\widetilde{\bF}_t^{\mathrm{scr}}}
\vect\paren{\widetilde{\bF}_t^{\mathrm{scr}}}^\top\\
&=
\frac{1}{T}\sum_t
\paren{(\widetilde{\bH}_C^{\mathrm{scr}})^{-1}\otimes(\widetilde{\bH}_R^{\mathrm{scr}})^{-1}}
\vect\paren{\bF_t}
\vect\paren{\bF_t}^\top
\paren{(\widetilde{\bH}_C^{\mathrm{scr}})^{-1}\otimes(\widetilde{\bH}_R^{\mathrm{scr}})^{-1}}^\top
+
o_p\paren{1}.
\end{aligned}
\]
Hence,
\[
\begin{aligned}
&\quad\left[
\frac{1}{T}\sum_t
\vect\paren{\widetilde{\bF}_t^{\mathrm{scr}}}
\vect\paren{\widetilde{\bF}_t^{\mathrm{scr}}}^\top
\right]^{-1}\\
&=
\paren{\widetilde{\bH}_C^{\mathrm{scr}}\otimes\widetilde{\bH}_R^{\mathrm{scr}}}^\top
\left[
\frac{1}{T}\sum_t
\vect\paren{\bF_t}
\vect\paren{\bF_t}^\top
\right]^{-1}
\paren{\widetilde{\bH}_C^{\mathrm{scr}}\otimes\widetilde{\bH}_R^{\mathrm{scr}}}
+
o_p\paren{1}\\
&=
\paren{\bGamma_2\otimes\bGamma_1}^\top
\bSigma_F^{-1}
\paren{\bGamma_2\otimes\bGamma_1}
+
o_p\paren{1}.
\end{aligned}
\]
Next, we show that \[
\norm{
\frac{1}{T}\sum_t
\vect\paren{\widetilde{\bF}_t^{\mathrm{scr}}}
\vect\paren{
\bF_t-
\widetilde{\bH}_R^{\mathrm{scr}}
\widetilde{\bF}_t^{\mathrm{scr}}
(\widetilde{\bH}_C^{\mathrm{scr}})^\top
}^\top
}=O_p\paren{
\frac{1}{Tp_1^{\delta_1}}
+
\frac{1}{Tp_2^{\delta_2}}
+
\frac{1}{p_1^{\delta_1}p_2^{\delta_2}}
+
\sqrt{\frac{1}{Tp_1^{\delta_1}p_2^{\delta_2}}}}
\]
By decomposition \(\widetilde{\bF}_t^{\mathrm{scr}}
=
\widetilde{\bF}_t^{\mathrm{scr}}
-
(\widetilde{\bH}_R^{\mathrm{scr}})^{-1}
\bF_t
\left(
(\widetilde{\bH}_C^{\mathrm{scr}})^{-1}
\right)^{\top}
+
(\widetilde{\bH}_R^{\mathrm{scr}})^{-1}
\bF_t
\left(
(\widetilde{\bH}_C^{\mathrm{scr}})^{-1}
\right)^{\top}
\), we have
\[
\begin{aligned}
&\quad\norm{
\frac{1}{T}\sum_t
\vect\paren{\widetilde{\bF}_t^{\mathrm{scr}}}
\vect\paren{
\bF_t-
\widetilde{\bH}_R^{\mathrm{scr}}
\widetilde{\bF}_t^{\mathrm{scr}}
(\widetilde{\bH}_C^{\mathrm{scr}})^\top
}^\top
}\\
&\le
\norm{
\frac{1}{T}\sum_t
\vect\paren{
\widetilde{\bF}_t^{\mathrm{scr}}
-
(\widetilde{\bH}_R^{\mathrm{scr}})^{-1}
\bF_t
\left((\widetilde{\bH}_C^{\mathrm{scr}})^{-1}\right)^\top
}
\vect\paren{
\bF_t-
\widetilde{\bH}_R^{\mathrm{scr}}
\widetilde{\bF}_t^{\mathrm{scr}}
(\widetilde{\bH}_C^{\mathrm{scr}})^\top
}^\top
}\\
&+
\norm{
\frac{1}{T}\sum_t
\vect\paren{
(\widetilde{\bH}_R^{\mathrm{scr}})^{-1}
\bF_t
\left((\widetilde{\bH}_C^{\mathrm{scr}})^{-1}\right)^\top
}
\vect\paren{
\bF_t-
\widetilde{\bH}_R^{\mathrm{scr}}
\widetilde{\bF}_t^{\mathrm{scr}}
(\widetilde{\bH}_C^{\mathrm{scr}})^\top
}^\top
}\\
&:=I_1+I_2.
\end{aligned}
\]
Since \(k_1,k_2\) are fixed, without loss of generality, assume \(k_1=k_2=1\). Then
\[
\begin{aligned}
I_1&=
\norm{
\frac{1}{T}\sum_t
\vect\paren{
\widetilde{\bF}_t^{\mathrm{scr}}
-
(\widetilde{\bH}_R^{\mathrm{scr}})^{-1}
\bF_t
\left((\widetilde{\bH}_C^{\mathrm{scr}})^{-1}\right)^\top
}
\vect\paren{
\bF_t-
\widetilde{\bH}_R^{\mathrm{scr}}
\widetilde{\bF}_t^{\mathrm{scr}}
(\widetilde{\bH}_C^{\mathrm{scr}})^\top
}^\top
}\\
&=
\norm{
\frac{1}{T}\sum_t
\paren{
\widetilde{\bF}_t^{\mathrm{scr}}
-
(\widetilde{\bH}_R^{\mathrm{scr}})^{-1}
\bF_t
\left((\widetilde{\bH}_C^{\mathrm{scr}})^{-1}\right)^\top
}
\paren{
\bF_t-
\widetilde{\bH}_R^{\mathrm{scr}}
\widetilde{\bF}_t^{\mathrm{scr}}
(\widetilde{\bH}_C^{\mathrm{scr}})^\top
}^\top
}\\
&\le
{\frac{1}{T}\sum_t
\norm{
\widetilde{\bF}_t^{\mathrm{scr}}
-
(\widetilde{\bH}_R^{\mathrm{scr}})^{-1}
\bF_t
\left((\widetilde{\bH}_C^{\mathrm{scr}})^{-1}\right)^\top
}}^2
\norm{\widetilde{\bH}_R^{\mathrm{scr}}}
\norm{\widetilde{\bH}_C^{\mathrm{scr}}}\\
&=
O_p\paren{
\frac{1}{Tp_1^{2\delta_1}}
+
\frac{1}{Tp_2^{2\delta_2}}
+
\frac{1}{p_1^{\delta_1}p_2^{\delta_2}}
}.
\end{aligned}
\]
By Lemma A.1 and Lemma A.2 in supplementary material of \cite{CaiKongWuZhao_JBES_2025} with cases \(m=n=1\) and Corollary~\ref{thm:screened estimators}, it is not hard to show that
\[
\begin{aligned}
I_2&=
\norm{
\frac{1}{T}\sum_t
\vect\paren{
(\widetilde{\bH}_R^{\mathrm{scr}})^{-1}
\bF_t
\left((\widetilde{\bH}_C^{\mathrm{scr}})^{-1}\right)^\top
}
\vect\paren{
\bF_t-
\widetilde{\bH}_R^{\mathrm{scr}}
\widetilde{\bF}_t^{\mathrm{scr}}
(\widetilde{\bH}_C^{\mathrm{scr}})^\top
}^\top
}\\
&=
\norm{
\frac{1}{T}\sum_t
\paren{
(\widetilde{\bH}_R^{\mathrm{scr}})^{-1}
\bF_t
\left((\widetilde{\bH}_C^{\mathrm{scr}})^{-1}\right)^\top
}
\paren{
\bF_t-
\widetilde{\bH}_R^{\mathrm{scr}}
\widetilde{\bF}_t^{\mathrm{scr}}
(\widetilde{\bH}_C^{\mathrm{scr}})^\top
}^\top
}\\
&=
O_p\paren{
\frac{1}{Tp_1^{\delta_1}}
+
\frac{1}{Tp_2^{\delta_2}}
+
\frac{1}{p_1^{\delta_1}p_2^{\delta_2}}
+
\sqrt{\frac{1}{Tp_1^{\delta_1}p_2^{\delta_2}}}
}.
\end{aligned}
\]
For \(II\),
\[
\begin{aligned}
&\quad\frac{1}{T}\sum_t
\epsilon_{t+h}
\vect\paren{\widetilde{\bF}_t^{\mathrm{scr}}}\\
&=
\frac{1}{T}\sum_t
\epsilon_{t+h}
\vect\paren{
(\widetilde{\bH}_R^{\mathrm{scr}})^{-1}
\bF_t
\left((\widetilde{\bH}_C^{\mathrm{scr}})^{-1}\right)^\top
}\\
&+
\frac{1}{T}\sum_t
\epsilon_{t+h}
\vect\paren{
\widetilde{\bF}_t^{\mathrm{scr}}
-
(\widetilde{\bH}_R^{\mathrm{scr}})^{-1}
\bF_t
\left((\widetilde{\bH}_C^{\mathrm{scr}})^{-1}\right)^\top
}\\
&:=II_1+II_2.
\end{aligned}
\]
For \(II_2\), without loss of generality, we assume \(k_1=k_2=1\). Then by Lemma A.1 and Lemma A.2 in supplementary material of \cite{CaiKongWuZhao_JBES_2025} with cases \(m=n=1\) and Corollary~\ref{thm:screened estimators}, it is not hard to show that
\[
\begin{aligned}
\norm{II_2}
&=
\norm{
\frac{1}{T}\sum_t
\epsilon_{t+h}
\vect\paren{
\widetilde{\bF}_t^{\mathrm{scr}}
-
(\widetilde{\bH}_R^{\mathrm{scr}})^{-1}
\bF_t
\left((\widetilde{\bH}_C^{\mathrm{scr}})^{-1}\right)^\top
}
}\\
&=
\norm{
\frac{1}{T}\sum_t
\epsilon_{t+h}
\paren{
\widetilde{\bF}_t^{\mathrm{scr}}
-
(\widetilde{\bH}_R^{\mathrm{scr}})^{-1}
\bF_t
\left((\widetilde{\bH}_C^{\mathrm{scr}})^{-1}\right)^\top
}
}\\
&=
O_p\paren{
\frac{1}{Tp_1^{\delta_1}}
+
\frac{1}{Tp_2^{\delta_2}}
+
\frac{1}{p_1^{\delta_1}p_2^{\delta_2}}
+
\sqrt{\frac{1}{Tp_1^{\delta_1}p_2^{\delta_2}}}
}.
\end{aligned}
\]
For \(II_1\),
\[
\sqrt{T}II_1
=
\paren{(\widetilde{\bH}_C^{\mathrm{scr}})^{-1}\otimes(\widetilde{\bH}_R^{\mathrm{scr}})^{-1}}
\frac{1}{\sqrt{T}}\sum_t
\epsilon_{t+h}
\vect\paren{\bF_t}
\xrightarrow[]{d}
\mathcal{N}\paren{
\bzero,
\paren{\bGamma_2^{-1}\otimes\bGamma_1^{-1}}
\bSigma_{F\epsilon}
\paren{\bGamma_2^{-1}\otimes\bGamma_1^{-1}}^\top
}.
\]
Denote
\(
\bE^*_{p_1,p_2,T}
:=
-I
-
\left[
\frac{1}{T}\sum_t
\vect\paren{\widetilde{\bF}_t^{\mathrm{scr}}}
\vect\paren{\widetilde{\bF}_t^{\mathrm{scr}}}^\top
\right]^{-1}
II_2.
\)
Hence, if
\(
\frac{\sqrt{T}}{p_1^{\delta_1}p_2^{\delta_2}}
=o\paren{1},
\)
\[
\sqrt{T}
\paren{
\widehat\btheta
-
\paren{
\paren{(\widetilde{\bH}_C^{\mathrm{scr}})^\top\bbeta_0}
\otimes
\paren{(\widetilde{\bH}_R^{\mathrm{scr}})^\top\balpha_0}
}
+
\bE^*_{p_1,p_2,T}
}
\xrightarrow[]{d}
\mathcal{N}\paren{
\bzero,
\paren{\bGamma_2\otimes\bGamma_1}^\top
\bSigma_F^{-1}
\bSigma_{F\epsilon}
(\bSigma_F^{-1})^\top
\paren{\bGamma_2\otimes\bGamma_1}
}.
\]
where \(\bE^*_{p_1,p_2,T}\) is the error term, with
\(
\bE^*_{p_1,p_2,T}
=
O_p\paren{
\frac{1}{T\min\{p_1^{\delta_1},p_2^{\delta_2}\}}
+
\sqrt{\frac{1}{Tp_1^{\delta_1}p_2^{\delta_2}}}
+
\frac{1}{p_1^{\delta_1}p_2^{\delta_2}}
}=o_p\paren{\frac{1}{\sqrt{T}}}.
\)
Consequently, by Slutsky's theorem, 
\[
\sqrt{T}
\paren{
\widehat\btheta
-
\paren{
\paren{(\widetilde{\bH}_C^{\mathrm{scr}})^\top\bbeta_0}
\otimes
\paren{(\widetilde{\bH}_R^{\mathrm{scr}})^\top\balpha_0}
}
}
\xrightarrow[]{d}
\mathcal{N}\paren{
\bzero,
\paren{\bGamma_2\otimes\bGamma_1}^\top
\bSigma_F^{-1}
\bSigma_{F\epsilon}
(\bSigma_F^{-1})^\top
\paren{\bGamma_2\otimes\bGamma_1}
}.
\]
Next, we establish the asymptotic distribution of \(\paren{\begin{matrix}
        \widehat{\balpha}-\paren{\widetilde \bH_R^{\mathrm{scr}}}'\balpha_0\\
        \widehat{\bbeta}-\paren{\widetilde \bH_C^{\mathrm{scr}}}'\bbeta_0
    \end{matrix}}\). Define \(\bPhi=\bbeta'\otimes\balpha'\) and \(\bPhi_0=(\bbeta_0'{\widetilde \bH_C^{\mathrm{scr}}})\otimes(\balpha_0'{\widetilde \bH_R^{\mathrm{scr}}})\). We write the objective function as
\begin{equation}
\mathcal{L}\paren{\bPhi}=\sum_t\paren{y_{t+h}-\bPhi\vect\paren{{\widetilde{\bF}^{\mathrm{scr}}_t}}}^2.
\end{equation}
By an argument analogous to that used in Lemma B.1 of
\citet{CaiKongWuZhao_JBES_2025}, specialized to the scalar-response case \(m=n=1\) and with the observable-predictor
component \(\bU\bW_t\bV^\top\) omitted, we obtain the following
result.
\begin{lemma}\label{lem:sqrt T convergence rate}
Suppose the conditions of Theorem~\ref{thm:regression coefficient estimator} hold. Then, for any sequence \(c_T\to\infty\),
    \[
\Pr\paren{\inf_{\sqrt{T}\norm{\widetilde{\bPhi}-\bPhi_0}\ge c_T}\mathcal{L}\paren{\widetilde\bPhi}>\mathcal{L}\paren{\bPhi_0}}\to1
    \]
\end{lemma}
\begin{proof}
Recall that
\(
\bPhi_0
=
\paren{\bbeta_0^\top\widetilde{\bH}_C^{\mathrm{scr}}}
\otimes
\paren{\balpha_0^\top\widetilde{\bH}_R^{\mathrm{scr}}},
\)
and let
\(
\bz_t
=
\vect\paren{\widetilde{\bF}_t^{\mathrm{scr}}}.
\)
Since
\[
\bPhi_0\bz_t
=
\left[
\paren{\bbeta_0^\top\widetilde{\bH}_C^{\mathrm{scr}}}
\otimes
\paren{\balpha_0^\top\widetilde{\bH}_R^{\mathrm{scr}}}
\right]
\vect\paren{\widetilde{\bF}_t^{\mathrm{scr}}}=
\balpha_0^\top
\widetilde{\bH}_R^{\mathrm{scr}}
\widetilde{\bF}_t^{\mathrm{scr}}
\paren{\widetilde{\bH}_C^{\mathrm{scr}}}^\top
\bbeta_0,
\]
we may write
\(
y_{t+h}
=
\bPhi_0\bz_t+u_{t+h},
\)
where
\(
u_{t+h}
=
\epsilon_{t+h}
+
\balpha_0^\top
\left[
\bF_t
-
\widetilde{\bH}_R^{\mathrm{scr}}
\widetilde{\bF}_t^{\mathrm{scr}}
\paren{\widetilde{\bH}_C^{\mathrm{scr}}}^\top
\right]
\bbeta_0.
\)

For any \(\widetilde{\bPhi}\), define
\(
\bDelta
=
\widetilde{\bPhi}-\bPhi_0.
\)
Then
\(
y_{t+h}
-
\widetilde{\bPhi}\bz_t
=
u_{t+h}-\bDelta\bz_t,
\)
and hence
\[
\begin{aligned}
\mathcal{L}\paren{\widetilde{\bPhi}}
-
\mathcal{L}\paren{\bPhi_0}
&=
\sum_{t=1}^T
\left[
\paren{u_{t+h}-\bDelta\bz_t}^2-u_{t+h}^2
\right]
=
\bDelta
\left(
\sum_{t=1}^T\bz_t\bz_t^\top
\right)
\bDelta^\top
-
2\bDelta
\sum_{t=1}^T\bz_tu_{t+h}.
\end{aligned}
\]
Define
\(
\widehat{\bQ}_T
=
\frac{1}{T}
\sum_{t=1}^T
\bz_t\bz_t^\top,
\;
\widehat{\bgg}_T
=
\frac{1}{T}
\sum_{t=1}^T
\bz_tu_{t+h}.
\)
Therefore,
\begin{equation}\label{eq:local-objective-expansion}
\mathcal{L}\paren{\widetilde{\bPhi}}
-
\mathcal{L}\paren{\bPhi_0}
=
T\bDelta\widehat{\bQ}_T\bDelta^\top
-
2T\bDelta\widehat{\bgg}_T.
\end{equation}

The preceding arguments in the proof of
Theorem~\ref{thm:regression coefficient estimator} show that
the smallest eigenvalue of \(\widehat{\bQ}_T\) is bounded away
from zero with probability tending to one. Hence, there exists
a constant \(c_0>0\) such that
\(
\Pr\left\{
\lambda_{\min}\paren{\widehat{\bQ}_T}
\ge c_0
\right\}
\to1.
\) Moreover, by the definition of \(u_{t+h}\),
\[
\begin{aligned}
\widehat{\bgg}_T
&=
\frac{1}{T}
\sum_{t=1}^T
\epsilon_{t+h}
\vect\paren{\widetilde{\bF}_t^{\mathrm{scr}}}+
\frac{1}{T}
\sum_{t=1}^T
\vect\paren{\widetilde{\bF}_t^{\mathrm{scr}}}
\vect\paren{
\bF_t
-
\widetilde{\bH}_R^{\mathrm{scr}}
\widetilde{\bF}_t^{\mathrm{scr}}
\paren{\widetilde{\bH}_C^{\mathrm{scr}}}^\top
}^\top
\paren{\bbeta_0\otimes\balpha_0}.
\end{aligned}
\]
The bounds established above in the proof of
Theorem~\ref{thm:regression coefficient estimator} imply
\(
\frac{1}{T}
\sum_{t=1}^T
\epsilon_{t+h}
\vect\paren{\widetilde{\bF}_t^{\mathrm{scr}}}
=
O_p\paren{T^{-1/2}},
\)
and
\(
\norm{
\frac{1}{T}
\sum_{t=1}^T
\vect\paren{\widetilde{\bF}_t^{\mathrm{scr}}}
\vect\paren{
\bF_t
-
\widetilde{\bH}_R^{\mathrm{scr}}
\widetilde{\bF}_t^{\mathrm{scr}}
\paren{\widetilde{\bH}_C^{\mathrm{scr}}}^\top
}^\top
}
=
O_p\paren{
\frac{1}{Tp_1^{\delta_1}}
+
\frac{1}{Tp_2^{\delta_2}}
+
\frac{1}{p_1^{\delta_1}p_2^{\delta_2}}
+
\sqrt{
\frac{1}{
Tp_1^{\delta_1}p_2^{\delta_2}
}
}
}.
\)
Since
\(
\frac{\sqrt{T}}
{p_1^{\delta_1}p_2^{\delta_2}}
=o(1),
\)
and \(k_1,k_2\) are fixed, it follows that
\(
\sqrt{T}\norm{\widehat{\bgg}_T}
=
O_p(1).
\)

We first establish the result on the boundary
\(
\sqrt{T}
\norm{\widetilde{\bPhi}-\bPhi_0}
=
c_T.
\)
For any \(\varepsilon>0\), there exists a finite constant
\(M_\varepsilon>0\) such that, for all sufficiently large \(T\),
\(
\Pr\left\{
\lambda_{\min}\paren{\widehat{\bQ}_T}\ge c_0,
\;
\sqrt{T}\norm{\widehat{\bgg}_T}\le M_\varepsilon
\right\}
\ge1-\varepsilon.
\)
On this event, uniformly over all \(\widetilde{\bPhi}\) satisfying
\(
\sqrt{T}
\norm{\widetilde{\bPhi}-\bPhi_0}
=
c_T,
\)
Equation~\eqref{eq:local-objective-expansion} gives
\[
\begin{aligned}
\mathcal{L}\paren{\widetilde{\bPhi}}
-
\mathcal{L}\paren{\bPhi_0}
&\ge
T\lambda_{\min}\paren{\widehat{\bQ}_T}
\norm{\bDelta}^2
-
2T\norm{\bDelta}
\norm{\widehat{\bgg}_T}
\ge
c_0c_T^2
-
2M_\varepsilon c_T
=
c_T\paren{c_0c_T-2M_\varepsilon}.
\end{aligned}
\]
Since \(c_T\to\infty\), the right-hand side is strictly positive
for all sufficiently large \(T\). Therefore,
\[
\Pr\left\{
\inf_{\sqrt{T}\norm{\widetilde{\bPhi}-\bPhi_0}=c_T}
\mathcal{L}\paren{\widetilde{\bPhi}}
>
\mathcal{L}\paren{\bPhi_0}
\right\}
\to1.
\]

It remains to extend the boundary result to the entire exterior region. It can be directly given by the convexity of \(\mathcal{L}\paren{\bPhi}\). This completes the proof.
\end{proof}
Let \(c_T\) be any sequence satisfying \(c_T\to\infty\) and \(c_T/\sqrt{T}\to0\). Define \(\mathcal{B}=\{\balpha\in \mathbb{R}^{k_1\times1}:\norm{\balpha}_F=1,\be_1'\balpha>0\}\) and \(\mathcal{G}=\{\{\widetilde{\balpha},\widetilde{\bbeta}\}:T\norm{\widetilde{\balpha}-(\widetilde{\bH}_R^{\mathrm{scr}})'\balpha_0}_F^2+T\norm{\widetilde{\bbeta}-(\widetilde{\bH}_C^{\mathrm{scr}})'\bbeta_0}_F^2\ge c_T^2\}\). 
Since \(\bPhi=\bbeta'\otimes\balpha'\), the objective function can be rewritten as
\begin{equation}
\mathcal{L}\paren{\balpha,\bbeta}=\sum_t\paren{y_{t+h}-\paren{{\bbeta'}\otimes{\balpha'}}\vect\paren{\widetilde{\bF}_t^{\mathrm{scr}}}}^2.
\end{equation}
Following the proof of Theorem 3.1 in
\citet{CaiKongWuZhao_JBES_2025}, if \(\widetilde{\balpha}\in\mathcal{B}\) and \(\paren{\widetilde{\balpha},\widetilde{\bbeta}}\in\mathcal{G}\), then \begin{equation}\sqrt{T}\norm{\widetilde{\bbeta}'\otimes\widetilde{\balpha}'-\paren{\bbeta_0'\widetilde{\bH}_C^{\mathrm{scr}}}\otimes\paren{\balpha_0'\widetilde{\bH}_R^{\mathrm{scr}}}}_F\ge C c_T,\end{equation} where \(C\) is a positive constant. It therefore follows from
Lemma~\ref{lem:sqrt T convergence rate} that \begin{equation}\Pr\left\{\inf_{\paren{\widetilde{\balpha},\widetilde{\bbeta}}\in\mathcal{G}}\mathcal{L}\paren{\widetilde\balpha,\widetilde\bbeta}>\mathcal{L}\paren{(\widetilde{\bH}_R^{\mathrm{scr}})'\balpha_0,(\widetilde{\bH}_C^{\mathrm{scr}})'\bbeta_0}\right\}\to1,\end{equation} which implies that, for any sequence \(c_T\to\infty\) and \(c_T/\sqrt{T}\to0\), \begin{equation}\Pr\left\{T\norm{\widehat{\balpha}-(\widetilde{\bH}_R^{\mathrm{scr}})'\balpha_0}_F^2+T\norm{\widehat{\bbeta}-(\widetilde{\bH}_C^{\mathrm{scr}})'\bbeta_0}_F^2\le c_T^2\right\}\to1\end{equation} with \(\widehat{\balpha}\in\mathcal{B}\). Consequently, \begin{equation}
\widehat\balpha-(\widetilde{\bH}_R^{\mathrm{scr}})'\balpha_0=O_p\paren{\sqrt{\frac{1}{T}}},\;\widehat{\bbeta}-(\widetilde{\bH}_C^{\mathrm{scr}})'\bbeta_0=O_p\paren{\sqrt{\frac{1}{T}}}.
\end{equation}
The corresponding first-order conditions are
\begin{equation}
\sum_ty_{t+h}\widetilde{\bF}_t^{\mathrm{scr}}\widehat{\bbeta}=\sum_t\widetilde{\bF}_t^{\mathrm{scr}}\widehat{\bbeta}\widehat{\bbeta}'\paren{\widetilde{\bF}_t^{\mathrm{scr}}}^\top\widehat{\balpha}
\end{equation}
\begin{equation}
\sum_ty_{t+h}\paren{\widetilde{\bF}_t^{\mathrm{scr}}}^\top\widehat{\balpha}=\sum_t\paren{\widetilde{\bF}_t^{\mathrm{scr}}}^\top\widehat{\balpha}\widehat{\balpha}'\paren{\widetilde{\bF}_t^{\mathrm{scr}}}\widehat{\bbeta}
\end{equation}
Using arguments analogous to those in the proof of
\citet{CaiKongWuZhao_JBES_2025}, together with the corresponding
decompositions of \(\widetilde{\bF}_t^{\mathrm{scr}}\),
\(\widehat{\balpha}\), and \(\widehat{\bbeta}\), we obtain
\begin{equation}\label{eq:gradient condition 1 result}
\begin{aligned}
&\quad\sum_t(\widetilde{\bH}_R^{\mathrm{scr}})^{-1}\bF_t\bbeta_0\bbeta_0'\bF_t'\paren{(\widetilde{\bH}_R^{\mathrm{scr}})^{-1}}'\paren{\widehat{\balpha}-(\widetilde{\bH}_R^{\mathrm{scr}})'\balpha_0}\\
&+\sum_t(\widetilde{\bH}_R^{\mathrm{scr}})^{-1}\bF_t\bbeta_0\paren{\widehat{\bbeta}'-\bbeta_0'\widetilde{\bH}_C^{\mathrm{scr}}}{(\widetilde{\bH}_C^{\mathrm{scr}})^{-1}}\bF_t'\balpha_0\\
&
=\sum_t\epsilon_{t+h}(\widetilde{\bH}_R^{\mathrm{scr}})^{-1}\bF_t\bbeta_0+O_p\paren{1+\frac{1}{\min\{p_1^{\delta_1},p_2^{\delta_2}\}}+\frac{T}{p_1^{\delta_1}p_2^{\delta_2}}+\sqrt{\frac{T}{p_1^{\delta_1}p_2^{\delta_2}}}}
\end{aligned}
\end{equation}
\begin{equation}\label{eq:gradient condition 2 result}
\begin{aligned}
&\quad\sum_t\paren{\widehat{\bbeta}'-\bbeta_0'\widetilde{\bH}_C^{\mathrm{scr}}}(\widetilde{\bH}_C^{\mathrm{scr}})^{-1}\bF_t'\balpha_0\balpha_0'\bF_t\paren{(\widetilde{\bH}_C^{\mathrm{scr}})^{-1}}^\top\\
&+\sum_t\bbeta_0'\bF_t'\paren{(\widetilde{\bH}_R^{\mathrm{scr}})^{-1}}^\top\paren{\widehat\balpha-(\widetilde{\bH}_R^{\mathrm{scr}})'\balpha_0}\balpha_0'\bF_t\paren{(\widetilde{\bH}_C^{\mathrm{scr}})^{-1}}^\top\\
&=\sum_t\epsilon_{t+h}\balpha_0'\bF_t\paren{(\widetilde{\bH}_C^{\mathrm{scr}})^{-1}}^\top+O_p\paren{1+\frac{1}{\min\{p_1^{\delta_1},p_2^{\delta_2}\}}+\frac{T}{p_1^{\delta_1}p_2^{\delta_2}}+\sqrt{\frac{T}{p_1^{\delta_1}p_2^{\delta_2}}}}
\end{aligned}
\end{equation}
Vectorizing both sides of \eqref{eq:gradient condition 1 result} and \eqref{eq:gradient condition 2 result} yields 
\begin{equation}\label{eq:vector gradient condition 1 result}
    \begin{aligned}
&\quad\sum_t(\widetilde{\bH}_R^{\mathrm{scr}})^{-1}\bF_t\bbeta_0\bbeta_0'\bF_t'\paren{(\widetilde{\bH}_R^{\mathrm{scr}})^{-1}}'\paren{\widehat{\balpha}-(\widetilde{\bH}_R^{\mathrm{scr}})'\balpha_0}\\
&+\sum_t\bracket{\paren{\balpha_0'\bF_t\paren{(\widetilde{\bH}_C^{\mathrm{scr}})^{-1}}^\top}\otimes(\widetilde{\bH}_R^{\mathrm{scr}})^{-1}\bF_t\bbeta_0}\paren{\widehat{\bbeta}-(\widetilde{\bH}_C^{\mathrm{scr}})'\bbeta_0}\\
        &=\sum_t\epsilon_{t+h}(\widetilde{\bH}_R^{\mathrm{scr}})^{-1}\bF_t\bbeta_0+O_p\paren{1+\frac{1}{\min\{p_1^{\delta_1},p_2^{\delta_2}\}}+\frac{T}{p_1^{\delta_1}p_2^{\delta_2}}+\sqrt{\frac{T}{p_1^{\delta_1}p_2^{\delta_2}}}}
    \end{aligned}
\end{equation}
\begin{equation}\label{eq:vector gradient condition 2 result}
    \begin{aligned}
&\quad\sum_t(\widetilde{\bH}_C^{\mathrm{scr}})^{-1}\bF_t'\balpha_0\balpha_0'\bF_t\paren{(\widetilde{\bH}_C^{\mathrm{scr}})^{-1}}^\top\paren{\widehat{\bbeta}-(\widetilde{\bH}_C^{\mathrm{scr}})'\bbeta_0}\\
&+\sum_t\bracket{\paren{(\widetilde{\bH}_C^{\mathrm{scr}})^{-1}\bF_t'\balpha_0}\otimes\paren{\bbeta_0^{\top}\bF_t^\top\paren{(\widetilde{\bH}_R^{\mathrm{scr}})^{-1}}^\top}}\paren{\widehat{\balpha}-(\widetilde{\bH}_R^{\mathrm{scr}})'\balpha_0}\\
        &=\sum_t\epsilon_{t+h}(\widetilde{\bH}_C^{\mathrm{scr}})^{-1}\bF_t'\balpha_0+O_p\paren{1+\frac{1}{\min\{p_1^{\delta_1},p_2^{\delta_2}\}}+\frac{T}{p_1^{\delta_1}p_2^{\delta_2}}+\sqrt{\frac{T}{p_1^{\delta_1}p_2^{\delta_2}}}}.
    \end{aligned}
\end{equation}
Denote \(\bar\bK_t^{}=\bracket{\begin{matrix}
    (\widetilde{\bH}_R^{\mathrm{scr}})^{-1}\bF_t\bbeta_0\\
    (\widetilde{\bH}_C^{\mathrm{scr}})^{-1}\bF_t'\balpha_0
\end{matrix}}\), \(\bPsi=\bracket{\begin{matrix}
    \widetilde{\bH}_R^{\mathrm{scr}}& \\
     &\widetilde{\bH}_C^{\mathrm{scr}}
\end{matrix}}\) and \(\bPsi_0=\bracket{\begin{matrix}
    \bGamma_1& \\
     &\bGamma_2
\end{matrix}}\).
Then,
\eqref{eq:vector gradient condition 1 result} and
\eqref{eq:vector gradient condition 2 result} can be written jointly as
\begin{equation}\label{eq:vectorization equation of bilinear regression}
    \sum_t\bar\bK_t\bar\bK_t'\paren{\begin{matrix}
        \widehat{\balpha}-(\widetilde{\bH}_R^{\mathrm{scr}})'\balpha_0\\
        \widehat{\bbeta}-(\widetilde{\bH}_C^{\mathrm{scr}})'\bbeta_0
\end{matrix}}=\sum_t\epsilon_{t+h}\bar\bK_t+O_p\paren{1+\frac{1}{\min\{p_1^{\delta_1},p_2^{\delta_2}\}}+\frac{T}{p_1^{\delta_1}p_2^{\delta_2}}+\sqrt{\frac{T}{p_1^{\delta_1}p_2^{\delta_2}}}}
\end{equation}
\[
\frac{1}{T}\sum_t\bar\bK_t\bar\bK_t'=\bPsi_0'\E\paren{\bK_t\bK_t'}\bPsi_0+o_p\paren{1}=\bPsi_0'\bSigma_{kk}\bPsi_0+o_p\paren{1}
\]
Since \(\bPsi_0'\E\paren{\bK_t\bK_t'}\bPsi_0\cdot\bPsi_0'\paren{{\begin{matrix}
    \balpha_0\\-\bbeta_0
\end{matrix}}}=0\), the matrix \(\bPsi_0'\bSigma_{kk}\bPsi_0\) is not full-rank. Let \(\bgamma=\paren{\balpha_0'\widetilde{\bH}_R^{\mathrm{scr}},\bzero'}^\top\in\mathbb{R}^{(k_1+k_2)\times1}\), \(\bgamma_0=\paren{\balpha_0'\bGamma_1,\bzero'}^\top\). Denote \(\bTheta=\frac{1}{T}\sum_t\bar\bK_t\bar\bK_t'+\bgamma\bgamma'\) and \(\bTheta_0=\bPsi_0'\bSigma_{kk}\bPsi_0+\bgamma_0\bgamma_0'\) where the rank-one correction
\(\bgamma_0\bgamma_0^\top\) renders \(\bTheta_0\) nonsingular. And we have \(\bTheta=\bTheta_0+o_p\paren{1}\) by the preceding asymptotic results.
\begin{lemma}\label{lem:HR'HR-I}
Suppose the conditions of Corollary~\ref{thm:screened estimators} hold. In addition, assume that \(\frac{1}{\abs{\mathcal{I}_R}}\bR'\bR=\bI_{k_1},\;\frac{1}{\abs{\mathcal{I}_C}}\bC'\bC=\bI_{k_2}\). Then
\[
\norm{(\widetilde{\bH}_R^{\mathrm{scr}})'\widetilde{\bH}_R^{\mathrm{scr}}-\bI_{k_1}}=O_p\paren{\frac{1}{T\min\{p_1^{\delta_1},p_2^{\delta_2}\}}+\frac{1}{p_1^{\delta_1}p_2^{\delta_2}}+\frac{1}{\sqrt{Tp_1^{\delta_1}p_2^{\delta_2}}}}
\]
\end{lemma}
\begin{proof}
On the event \(\mathcal{E}\), we have  \(\paren{\widetilde{\bR}^{\mathrm{scr}}}^\top\widetilde{\bR}^{\mathrm{scr}}/\abs{\mathcal{I}_R}=\bI_{k_1}\)
Decomposing $\widetilde{\bR}^{\mathrm{scr}}$, we obtain
\[
\begin{aligned}
&\qquad\paren{\widetilde{\bR}^{\mathrm{scr}}-\bR^{\mathrm{scr}}\widetilde{\bH}_R^{\mathrm{scr}}}^\top\paren{\widetilde{\bR}^{\mathrm{scr}}-\bR^{\mathrm{scr}}\widetilde{\bH}_R^{\mathrm{scr}}}+\paren{\widetilde{\bR}^{\mathrm{scr}}-\bR^{\mathrm{scr}}\widetilde{\bH}_R^{\mathrm{scr}}}^\top\bR^{\mathrm{scr}}\widetilde{\bH}_R^{\mathrm{scr}}\\
&\qquad\quad+\paren{\bR^{\mathrm{scr}}\widetilde{\bH}_R^{\mathrm{scr}}}^\top\paren{\widetilde{\bR}^{\mathrm{scr}}-\bR^{\mathrm{scr}}\widetilde{\bH}_R^{\mathrm{scr}}}+\paren{\bR^{\mathrm{scr}}\widetilde{\bH}_R^{\mathrm{scr}}}^\top\bR^{\mathrm{scr}}\widetilde{\bH}_R^{\mathrm{scr}}=\abs{\mathcal{I}_R}\bI_{k_1}\\
\end{aligned}\]
Consequently,
\[
\begin{aligned}
&\norm{\paren{\widetilde{\bH}_R^{\mathrm{scr}}}^\top\widetilde{\bH}_R^{\mathrm{scr}}-\bI_{k_1}}\le\frac{1}{\abs{\mathcal{I}_R}}\norm{\widetilde{\bR}^{\mathrm{scr}}-\bR^{\mathrm{scr}}\widetilde{\bH}_R^{\mathrm{scr}}}_F^2+2\frac{1}{\abs{\mathcal{I}_R}}\norm{\paren{\widetilde{\bR}^{\mathrm{scr}}-\bR^{\mathrm{scr}}\widetilde{\bH}_R^{\mathrm{scr}}}^\top\bR^{\mathrm{scr}}\widetilde{\bH}_R^{\mathrm{scr}}}_F
\end{aligned}\]
The stated result follows from the convergence rates of the
post-screening estimators, together with Lemma E.1 in the supplementary material of \cite{YuHeKongZhang_JoE_2022}.
\end{proof}
Since \(
\left\|
\widetilde{\bH}_R^{\mathrm{scr}}
(\widetilde{\bH}_R^{\mathrm{scr}})^\top-\bI_{k_1}
\right\|_F
=
\left\|
(\widetilde{\bH}_R^{\mathrm{scr}})^\top
\widetilde{\bH}_R^{\mathrm{scr}}-\bI_{k_1}
\right\|_F,
\) by Lemma~\ref{lem:HR'HR-I}, we have
\[\balpha_0'\widetilde{\bH}_R^{\mathrm{scr}}
\paren{\widehat{\balpha}-
(\widetilde{\bH}_R^{\mathrm{scr}})'\balpha_0}
=o_p\paren{\frac{1}{\sqrt{T}}}.\]
To verify this relation, note that \(
0=\norm{\balpha_0}_F^2-\norm{\widehat\balpha}_F^2.\) Decomposing \(\widehat{\balpha},\) we have
\[\begin{aligned}
-2\left\langle
\widehat{\balpha}-
(\widetilde{\bH}_R^{\mathrm{scr}})'\balpha_0,
(\widetilde{\bH}_R^{\mathrm{scr}})'\balpha_0
\right\rangle
&=
\Tr\paren{
(\widetilde{\bH}_R^{\mathrm{scr}})'
\balpha_0\balpha_0'
\widetilde{\bH}_R^{\mathrm{scr}}
}
-\Tr\paren{\balpha_0\balpha_0'}+\left\langle
\widehat{\balpha}-
(\widetilde{\bH}_R^{\mathrm{scr}})'\balpha_0,
\widehat{\balpha}-(\widetilde{\bH}_R^{\mathrm{scr}})'\balpha_0
\right\rangle\\
&=\Tr\paren{
\paren{
\widetilde{\bH}_R^{\mathrm{scr}}
(\widetilde{\bH}_R^{\mathrm{scr}})'
-\bI_{k_1}
}
\paren{\balpha_0\balpha_0'}
}+\norm{
\widehat{\balpha}-
(\widetilde{\bH}_R^{\mathrm{scr}})'\balpha_0}_F^2
\end{aligned}
\]
Therefore,
\[
2\abs{\left\langle
\widehat{\balpha}-
(\widetilde{\bH}_R^{\mathrm{scr}})'\balpha_0,
(\widetilde{\bH}_R^{\mathrm{scr}})'\balpha_0
\right\rangle}
\le
\norm{
\widetilde{\bH}_R^{\mathrm{scr}}
(\widetilde{\bH}_R^{\mathrm{scr}})'
-\bI_{k_1}
}_F
\norm{\balpha_0\balpha_0'}+\norm{
\widehat{\balpha}-
(\widetilde{\bH}_R^{\mathrm{scr}})'\balpha_0}_F^2
\]
Since \(\frac{\sqrt{T}} {p_1^{\delta_1}p_2^{\delta_2}}=o\paren{1}\) and \(\norm{
\widehat{\balpha}-
(\widetilde{\bH}_R^{\mathrm{scr}})'\balpha_0}_F^2=O_p\paren{\frac{1}{T}}\), by
Lemma~\ref{lem:HR'HR-I}, we have the desired result.
By \eqref{eq:vectorization equation of bilinear regression} and the preceding results, we have 
\[
\sqrt{T}\bTheta\paren{\begin{matrix}
        \widehat{\balpha}-(\widetilde{\bH}_R^{\mathrm{scr}})'\balpha_0\\
        \widehat{\bbeta}-(\widetilde{\bH}_C^{\mathrm{scr}})'\bbeta_0
    \end{matrix}}+o_p\paren{1}=\frac{1}{\sqrt{T}}\sum_t\epsilon_{t+h}\bar\bK_t+O_p\paren{\sqrt{\frac{1}{T}}+\frac{1}{\sqrt{T}\min\{p_1^{\delta_1},p_2^{\delta_2}\}}+\frac{\sqrt{T}}{p_1^{\delta_1}p_2^{\delta_2}}+\sqrt{\frac{1}{p_1^{\delta_1}p_2^{\delta_2}}}}
\]
Since \(\frac{\sqrt{T}}{p_1^{\delta_1}p_2^{\delta_2}}=o\paren{1}\), by Equation~\eqref{ass:bilinear normal condition} and the
preceding results, we have
\[
\begin{aligned}
&\sqrt{T}\bTheta_0\paren{\begin{matrix}
        \widehat{\balpha}-(\widetilde{\bH}_R^{\mathrm{scr}})'\balpha_0\\
        \widehat{\bbeta}-(\widetilde{\bH}_C^{\mathrm{scr}})'\bbeta_0
    \end{matrix}}+o_p\paren{1}=\frac{1}{\sqrt{T}}\sum_t\epsilon_{t+h}\bar\bK_t+O_p\paren{\sqrt{\frac{1}{T}}+\frac{1}{\sqrt{T}\min\{p_1^{\delta_1},p_2^{\delta_2}\}}+\frac{\sqrt{T}}{p_1^{\delta_1}p_2^{\delta_2}}+\sqrt{\frac{1}{p_1^{\delta_1}p_2^{\delta_2}}}}
    \end{aligned}
\]
Thus,
\[
\begin{aligned}\sqrt{T}\bTheta_0
\begin{pmatrix}
\widehat{\balpha}
-(\widetilde{\bH}_R^{\mathrm{scr}})^\top\balpha_0\\
\widehat{\bbeta}-(\widetilde{\bH}_C^{\mathrm{scr}})^\top\bbeta_0
\end{pmatrix}
\xrightarrow{d}
\mathcal{N}\left(
\bzero,\,
\bPsi_0^\top\bSigma_{kk,\epsilon}\bPsi_0
\right).
\end{aligned}
\]
Since \(\bTheta_0\) is nonsingular, premultiplying by
\(\bTheta_0^{-1}\) and applying the continuous mapping theorem yield
\[
\sqrt{T}
\begin{pmatrix}
\widehat{\balpha}
-(\widetilde{\bH}_R^{\mathrm{scr}})^\top\balpha_0\\
\widehat{\bbeta}-(\widetilde{\bH}_C^{\mathrm{scr}})^\top\bbeta_0
\end{pmatrix}
\xrightarrow{d}
\mathcal{N}\left(
\bzero,\,
\bTheta_0^{-1}
\bPsi_0^\top\bSigma_{kk,\epsilon}\bPsi_0
\bTheta_0^{-1}
\right),
\]
which proves the desired result.
\section{Data Description}
This section provides detailed descriptions of the two matrix-valued datasets
used in the empirical analysis. The first dataset is a quarterly multicountry
macroeconomic panel constructed from the OECD database. The second dataset is a monthly U.S. state--industry employment panel constructed from the Bureau of Labor Statistics (BLS). The latter dataset is combined with U.S. manufacturing production from the Federal Reserve as the forecast target.
\subsection{Multicountry Macroeconomic Dataset}
This subsection describes the macroeconomic dataset used in the empirical
analysis. The data are obtained from the OECD database and contain nine
quarterly macroeconomic indicators for ten countries over the
period from 1988.Q4 to 2024.Q4. A similar dataset is adopted in
\cite{ChenFan_JASA_2023,ChenTsayChen_JASA_2020,YuHeKongZhang_JoE_2022},
among others. The countries are listed in
Table~\ref{tab:country_codes}, together with their codes. The countries are the
United States, Canada, New Zealand, Australia, Norway, Finland, Sweden,
France, the Netherlands, and Germany.

For each macroeconomic indicator, Table~\ref{tab:macro_indicators} reports
its short name, OECD series code, data transformation, and brief description.
In the transformation column, $\Delta x_t=x_t-x_{t-1}$ denotes the first
difference of a series, $\Delta\ln x_t=\ln x_t-\ln x_{t-1}$ denotes the
first difference of the logarithm, and
$\Delta^2\ln x_t=\Delta(\Delta\ln x_t)$ denotes the second difference of
the logarithm. The dataset contains only one missing observation, corresponding to
the three-month interest rate of the United States in 2020.Q2. We impute this observation using the average of the preceding three observations from the same calendar quarter, namely,
\(
\widehat{x}_{2020.Q2}
=
\frac{
x_{2017.Q2}+x_{2018.Q2}+x_{2019.Q2}
}{3}.
\)
This past-only imputation rule avoids the use of future information.
Since the missing observation accounts for approximately \(0.0077\%\)
of all entries in the raw dataset, the amount of imputation is
negligible relative to the overall sample size.

Before transformation, the dataset contains 145 quarterly observations.
Because the CPI series are transformed using second log differences, the
common transformed sample contains 143 quarterly observations from 1989.Q2
to 2024.Q4. At each time point, the observations are arranged as a
$10\times9$ matrix, where the rows correspond to countries and the columns
correspond to macroeconomic indicators.

\begin{table}[htbp]
    \centering
    \caption{Countries in the Macroeconomic Dataset}
    \label{tab:country_codes}
    \begin{tabular}{ll@{\hspace{1.2cm}}ll}
        \toprule
        Country & Code
        & Country & Code \\
        \midrule
        United States of America & USA
        & Finland & FIN \\

        Canada & CAN
        & Sweden & SWE \\

        New Zealand & NZL
        & France & FRA \\

        Australia & AUS
        & Netherlands & NLD \\

        Norway & NOR
        & Germany & DEU \\
        \bottomrule
    \end{tabular}
\end{table}

\begin{table}[htbp]
    \centering
    \caption{Macroeconomic Indicators and Data Transformations}
    \label{tab:macro_indicators}
    \renewcommand{\arraystretch}{1.15}
    \small

    \begin{tabularx}{\textwidth}{
        >{\raggedright\arraybackslash}p{2.0cm}
        >{\raggedright\arraybackslash}p{2.2cm}
        >{\centering\arraybackslash}p{2.2cm}
        >{\raggedright\arraybackslash}X
    }
        \toprule
        Short name & Series codes & transformation & Description \\
        \midrule

        CPI: Food
        & CPI--CP01
        & $\Delta^{2}\ln$
        & Consumer price index: Food and non-alcoholic beverages \\

        CPI: Ener
        & CPI--CP045\_0722
        & $\Delta^{2}\ln$
        & Consumer price index: Energy \\

        CPI: Tot
        & CPI--\_T
        & $\Delta^{2}\ln$
        & Consumer price index: Total \\

        IR: Long
        & IRLT
        & $\Delta$
        & Long-term interest rates, percent per annum \\

        IR: 3-Mon
        & IR3TIB
        & $\Delta$
        & Short-term interest rates, percent per annum \\

        P: TIEC
        & PRVM--BTE
        & $\Delta\ln$
        & Production volume: Industry except construction \\

        P: TM
        & PRVM--C
        & $\Delta\ln$
        & Production volume: Manufacturing \\

        IT: Ex
        & X--C
        & $\Delta\ln$
        & Exports of commodities, US dollars, calendar and seasonally adjusted \\

        IT: Im
        & M--C
        & $\Delta\ln$
        & Imports of commodities, US dollars, calendar and seasonally adjusted \\

        \bottomrule
    \end{tabularx}

    \vspace{0.15cm}

    \begin{minipage}{\textwidth}
        \footnotesize
        \raggedright
        \textit{Note:} The series codes are constructed from the relevant
        OECD code dimensions. For example, CPI--CP01 combines
        \texttt{MEASURE = CPI} and \texttt{EXPENDITURE = CP01}.
        Only dimensions that identify the economic content of a series are
        reported. OECD placeholder codes such as \texttt{\_Z}
        (\emph{not applicable}) are omitted. The CPI and production-volume
        indices use 2015 as the base period; exports and imports are expressed
        in US dollars converted using exchange rates.
    \end{minipage}
\end{table}

\begin{figure}
    \caption{Standardized transformed multinational macroeconomic series and
    their stationarity test results. Blue: stationary by both ADF and KPSS at
    5\%; red: not confirmed (81/90 confirmed; 1989-Q2--2024-Q4).}
    \label{fig:stationarity test}
    \centering
    \includegraphics[width=\linewidth]{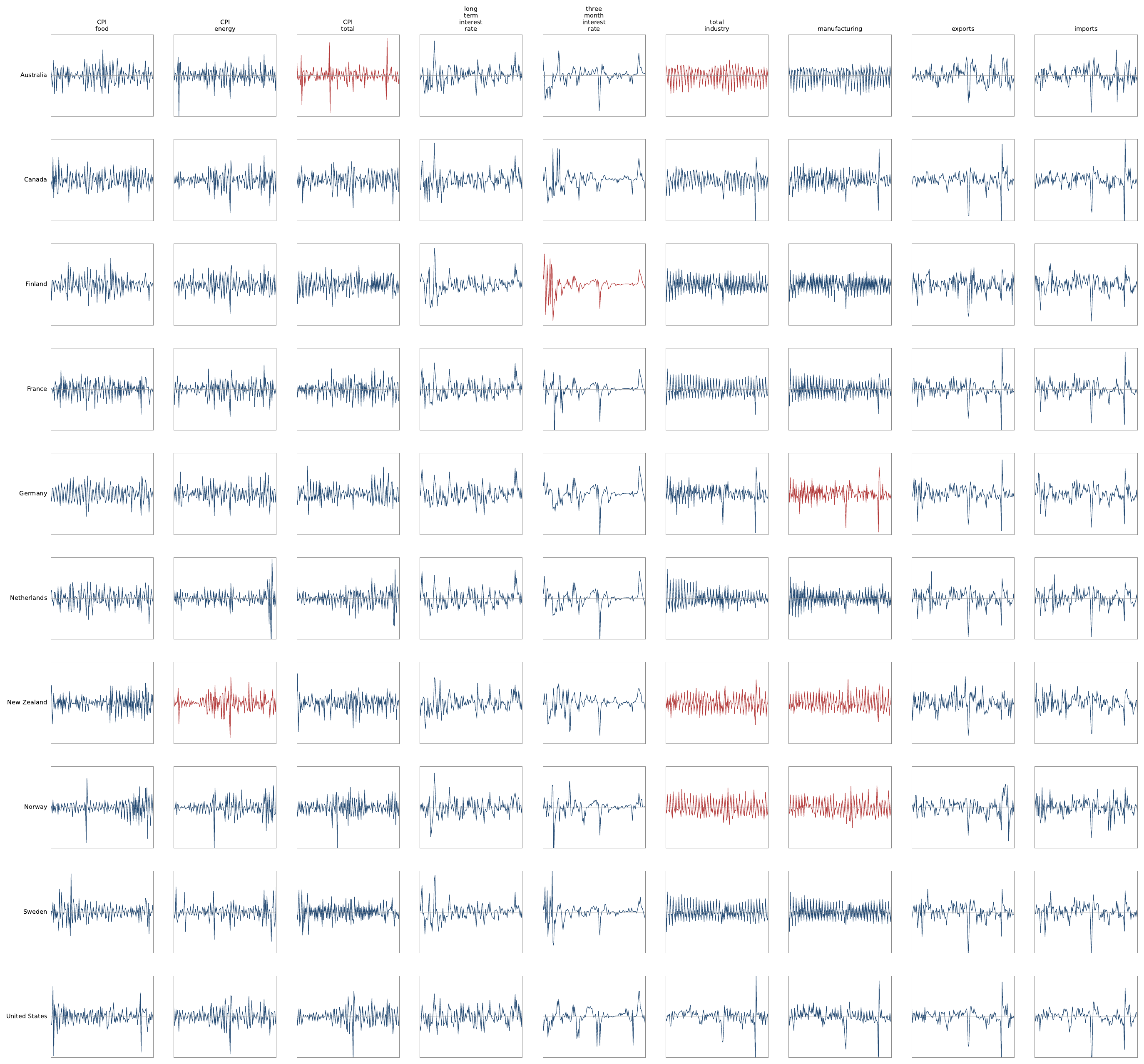}

\end{figure}

We assess the stationarity of the 90 transformed series using the
augmented Dickey--Fuller (ADF) test
\citep{SaidDickey_Biometrika_1984} and the
Kwiatkowski--Phillips--Schmidt--Shin (KPSS) test
\citep{KwiatkowskiPhillipsSchmidtShin_JoE_1992}, using a 5\%
significance level. 
The ADF test takes the presence of a unit root as its null hypothesis,
whereas the KPSS test takes level stationarity as its null hypothesis.
Accordingly, rejection of the ADF null and nonrejection of the KPSS null
are interpreted as evidence consistent with stationarity. Among the 90
transformed series, 81 (90.0\%) satisfy both criteria. For the remaining
nine series, the two tests yield different conclusions, although each
series satisfies at least one of the two criteria. For none of the 90 series does the ADF test fail to reject the unit-root
null while the KPSS test rejects the null of level stationarity. Figure~\ref{fig:stationarity test} displays the 90 multicountry
macroeconomic series after transformation and standardization.

\subsection{U.S. State--Industry Employment Dataset}

This subsection describes the employment dataset used in the empirical
analysis, which is constructed from monthly state-level employment data
provided by the Bureau of Labor Statistics (BLS) State and Area Current
Employment Statistics. The employment series are seasonally adjusted and
measured in thousands of employees. At each month \(t\), the predictor
matrix is arranged as
\(
\mathbf{X}_t\in\mathbb{R}^{41\times20},
\)
where the rows correspond to 41 U.S. states and the columns correspond to
20 industries. The states included in the analysis are reported in
Table~\ref{tab:state-codes}, while Table~\ref{tab:industry-series} reports
the industry names, BLS series codes, transformations, and descriptions.
Thus, the two dimensions of \(\mathbf{X}_t\) retain the natural
geographical and sectoral structure of U.S. employment data.

The forecast target is the seasonally adjusted U.S. manufacturing
production index from the Federal Reserve G.17 release, corresponding to
the G.17 series code \(\texttt{GMF}\) and FRED series
\(\texttt{IPMAN}\). The common level sample for the employment predictors
and manufacturing production extends from January 2003 to February 2026,
providing 278 monthly level observations.

Both the employment series and manufacturing production are transformed
using monthly natural-log first differences. Accordingly, each element of
\(\mathbf{X}_t\) represents the monthly log change in employment for a
particular state--industry pair, while the response variable is the
monthly log change in U.S. manufacturing production. We use the
state--industry employment matrix at month \(t\) to forecast manufacturing
production growth at month \(t+1\). After applying the log-difference
transformation and aligning the predictor and response series, the
forecasting sample contains 276 observations.

\begin{table}[htbp]
\centering
\caption{States in the State--Industry Employment Dataset}
\label{tab:state-codes}
\small
\begin{adjustbox}{max width=\textwidth}
\begin{tabular*}{\textwidth}{
    @{\extracolsep{\fill}}
    lc
    lc
    lc
    lc
    @{}
}
\toprule
State & Code
& State & Code
& State & Code
& State & Code \\
\midrule
Alabama     & 01 & Illinois      & 17 & Nebraska       & 31 & South Carolina & 45 \\
Alaska      & 02 & Indiana       & 18 & Nevada         & 32 & Tennessee      & 47 \\
Arkansas    & 05 & Iowa          & 19 & New Hampshire  & 33 & Texas          & 48 \\
California  & 06 & Kansas        & 20 & New Jersey     & 34 & Utah           & 49 \\
Colorado    & 08 & Kentucky      & 21 & New York       & 36 & Virginia       & 51 \\
Connecticut & 09 & Maine         & 23 & North Carolina & 37 & Washington     & 53 \\
Delaware    & 10 & Maryland      & 24 & North Dakota   & 38 & West Virginia  & 54 \\
Florida     & 12 & Massachusetts & 25 & Ohio           & 39 & Wisconsin      & 55 \\
Georgia     & 13 & Michigan      & 26 & Oklahoma       & 40 &                &    \\
Hawaii      & 15 & Minnesota     & 27 & Oregon         & 41 &                &    \\
Idaho       & 16 & Mississippi   & 28 & Pennsylvania   & 42 &                &    \\
\bottomrule
\end{tabular*}
\end{adjustbox}
\end{table}

\begin{table}[htbp]
\centering
\caption{State--Industry Employment Series and Forecast Target}
\label{tab:industry-series}
\renewcommand{\arraystretch}{1.10}
\small
\begin{adjustbox}{max width=\textwidth}
\begin{tabularx}{\textwidth}{
    >{\raggedright\arraybackslash}p{2.3cm}
    >{\centering\arraybackslash}p{2.1cm}
    >{\centering\arraybackslash}p{2.3cm}
    >{\raggedright\arraybackslash}X
}
\toprule
Short name & Series code & Transformation & Description \\
\midrule

MLC
& 15000000
& $\Delta\ln$
& Mining, logging, and construction employment \\

Mfg.: Durable
& 31000000
& $\Delta\ln$
& Durable goods manufacturing employment \\

Mfg.: Nondurable
& 32000000
& $\Delta\ln$
& Nondurable goods manufacturing employment \\

Wholesale
& 41000000
& $\Delta\ln$
& Wholesale trade employment \\

Retail
& 42000000
& $\Delta\ln$
& Retail trade employment \\

Transp./Util.
& 43000000
& $\Delta\ln$
& Transportation, warehousing, and utilities employment \\

Information
& 50000000
& $\Delta\ln$
& Information-sector employment \\

Finance
& 55520000
& $\Delta\ln$
& Finance and insurance employment \\

Real Estate
& 55530000
& $\Delta\ln$
& Real estate and rental and leasing employment \\

Prof./Tech.
& 60540000
& $\Delta\ln$
& Professional, scientific, and technical services employment \\

Management
& 60550000
& $\Delta\ln$
& Management of companies and enterprises employment \\

Admin./Waste
& 60560000
& $\Delta\ln$
& Administrative, support, waste management, and remediation services employment \\

Education
& 65610000
& $\Delta\ln$
& Private educational services employment \\

Health/Social
& 65620000
& $\Delta\ln$
& Health care and social assistance employment \\

Arts/Rec.
& 70710000
& $\Delta\ln$
& Arts, entertainment, and recreation employment \\

Accom./Food
& 70720000
& $\Delta\ln$
& Accommodation and food services employment \\

Other Services
& 80000000
& $\Delta\ln$
& Other services employment \\

Federal Gov.
& 90910000
& $\Delta\ln$
& Federal government employment \\

State Gov.
& 90920000
& $\Delta\ln$
& State government employment \\

Local Gov.
& 90930000
& $\Delta\ln$
& Local government employment \\

\midrule

IP: Manufacturing
& GMF
& $\Delta\ln$
& U.S. manufacturing production index used as the forecast target
(FRED series \texttt{IPMAN}) \\

\bottomrule
\end{tabularx}
\end{adjustbox}
\vspace{0.15cm}

\end{table}
\section{Additional Simulation and Empirical Results}
\label{sec:additional-simulation-results}
\subsection{Additional Simulation Results}
\subsubsection{Finite-Sample Performance of Data-Driven Iterative Screening}
\label{sec:iterative-screening-simulation}
To examine the finite-sample performance of the data-driven implementation, we apply Algorithm~\ref{alg:iterative-screening} to the same Gaussian baseline design used in Section~\ref{sec:screening-consistency-by-covariance}. The active proportions are \(r\in\{1/5,1/3,1/2,2/3\}\), and each configuration is based on 500 Monte Carlo replications. Unlike the threshold-based procedures, the iterative method determines the retained row and column dimensions from the largest break in the ordered screening statistics. Table~\ref{tab:iterative-ratio-grid-exact-support} reports the joint exact-recovery frequencies for both the covariance- and rank-based scores over several values of the ridge multiplier \(c\). We set
\(
p_{R,\max}=\left\lfloor 0.8p_1\right\rfloor,
\;
p_{C,\max}=\left\lfloor 0.8p_2\right\rfloor,
\;
L_{\max}=20.
\)
At iteration $\ell$, the ridge scales are updated as
\(
\nu_R^{(\ell)}
=
\sqrt{
\frac{
\log\!\left(
p_1\left|\widehat{\mathcal I}_C^{(\ell-1)}\right|
\right)
}{T}
},
\;
\nu_C^{(\ell)}
=
\sqrt{
\frac{
\log\!\left(
\left|\widehat{\mathcal I}_R^{(\ell)}\right|p_2
\right)
}{T}
}.
\)
We consider the ridge constants
\(
c\in\left\{0,10^{-4},10^{-2}\right\}.
\)
\begin{table}[!htbp]
\centering
\caption{Exact-support recovery frequencies for data-driven iterative screening.}
\label{tab:iterative-ratio-grid-exact-support}

\small
\setlength{\tabcolsep}{6pt}
\renewcommand{\arraystretch}{1.05}

\begin{tabular}{@{\hspace{0.5em}}ccc*{4}{cc}@{\hspace{0.5em}}}
\toprule
$(p_1,p_2)$ & $T$ & $c$
& \multicolumn{2}{c}{$r=1/5$}
& \multicolumn{2}{c}{$r=1/3$}
& \multicolumn{2}{c}{$r=1/2$}
& \multicolumn{2}{c}{$r=2/3$} \\
\cmidrule(lr){4-5}\cmidrule(lr){6-7}\cmidrule(lr){8-9}\cmidrule(lr){10-11}
& & & Cov. & Rank & Cov. & Rank & Cov. & Rank & Cov. & Rank \\
\midrule
$(100,100)$ & 100 & $0$ & 0.998 & 0.476 & 1.000 & 0.584 & 1.000 & 0.686 & 0.998 & 0.686 \\
 &  & $10^{-4}$ & 0.998 & 0.476 & 1.000 & 0.584 & 1.000 & 0.686 & 0.998 & 0.686 \\
 &  & $10^{-2}$ & 0.998 & 0.478 & 1.000 & 0.574 & 1.000 & 0.680 & 0.998 & 0.680 \\
 & 200 & $0$ & 1.000 & 0.924 & 1.000 & 0.944 & 1.000 & 0.972 & 1.000 & 0.978 \\
 &  & $10^{-4}$ & 1.000 & 0.924 & 1.000 & 0.944 & 1.000 & 0.972 & 1.000 & 0.978 \\
 &  & $10^{-2}$ & 1.000 & 0.924 & 1.000 & 0.940 & 1.000 & 0.970 & 1.000 & 0.976 \\
\addlinespace[5pt]
$(200,200)$ & 200 & $0$ & 1.000 & 0.966 & 1.000 & 0.974 & 1.000 & 0.962 & 1.000 & 0.984 \\
 &  & $10^{-4}$ & 1.000 & 0.966 & 1.000 & 0.974 & 1.000 & 0.962 & 1.000 & 0.984 \\
 &  & $10^{-2}$ & 1.000 & 0.966 & 1.000 & 0.974 & 1.000 & 0.962 & 1.000 & 0.982 \\
 & 400 & $0$ & 1.000 & 0.998 & 1.000 & 1.000 & 1.000 & 0.998 & 1.000 & 1.000 \\
 &  & $10^{-4}$ & 1.000 & 0.998 & 1.000 & 1.000 & 1.000 & 0.998 & 1.000 & 1.000 \\
 &  & $10^{-2}$ & 1.000 & 0.998 & 1.000 & 1.000 & 1.000 & 0.998 & 1.000 & 1.000 \\
\addlinespace[5pt]
$(300,300)$ & 300 & $0$ & 1.000 & 0.992 & 1.000 & 0.998 & 1.000 & 0.996 & 1.000 & 0.996 \\
 &  & $10^{-4}$ & 1.000 & 0.992 & 1.000 & 0.998 & 1.000 & 0.996 & 1.000 & 0.996 \\
 &  & $10^{-2}$ & 1.000 & 0.990 & 1.000 & 0.998 & 1.000 & 0.996 & 1.000 & 0.996 \\
 & 500 & $0$ & 1.000 & 1.000 & 1.000 & 1.000 & 1.000 & 1.000 & 1.000 & 1.000 \\
 &  & $10^{-4}$ & 1.000 & 1.000 & 1.000 & 1.000 & 1.000 & 1.000 & 1.000 & 1.000 \\
 &  & $10^{-2}$ & 1.000 & 1.000 & 1.000 & 1.000 & 1.000 & 1.000 & 1.000 & 1.000 \\
\bottomrule
\end{tabular}%

\end{table}

Table~\ref{tab:iterative-ratio-grid-exact-support} shows that the iterative procedure accurately recovers the informative block over a broad range of configurations. Under the Gaussian design, covariance-based screening achieves nearly perfect exact recovery even in the smaller-sample settings. Rank-based screening is less accurate in the most difficult small-sample configurations, but its recovery frequency improves rapidly with the sample size and matrix dimensions and approaches one in the larger configurations. This difference is consistent with the greater efficiency of the covariance score under the Gaussian baseline; the robustness advantage of the rank-based score under contamination is documented separately in Section~\ref{sec:robust-screening}.

The results are also essentially unchanged across the reported values of \(c\). Thus, in these experiments, the location of the dominant break in the ordered screening statistics is sufficiently pronounced that the retained dimensions are insensitive to moderate ridge adjustment. Compared with the threshold-based results in the main text, the iterative updates are particularly useful in configurations where the initial full-matrix separation is weak, since successive row and column refinements reduce cross-mode signal dilution without requiring prespecified screening thresholds. These results support the iterative procedure as a practical data-driven implementation of the sequential screening principle developed in the main paper.
\subsubsection{Finite-Sample Performance of Rank-Based Screening}
Under the Gaussian design, the corresponding informative Spearman correlations are also separated from zero, so that the same data-generating process in Section~\ref{sec:screening-consistency-by-covariance} can be used to evaluate the rank-based
screening procedure.
For rank-based screening, we use the same threshold-exponent family, with a fixed
score-scale adjustment described in the table notes. Table~\ref{tab:rank-screening-consistency} reports the corresponding
exact-selection frequencies for the rank-based procedure.
\begin{table}[htbp]
\centering
\caption{Exact-selection frequencies under rank-based screening and alternative threshold exponents.}
\label{tab:rank-screening-consistency}

\small
\setlength{\tabcolsep}{6pt}
\renewcommand{\arraystretch}{1.05}

\begin{threeparttable}
\begin{tabular}{ccc*{5}{r}}
\toprule
$(p_1,p_2)$ & $T$ & $r$
& $a=0.20$ & $a=0.35$ & ${a=0.50}$
& $a=0.65$ & $a=0.80$ \\
\midrule

$(100,100)$ & 100 & $1/5$
& 0.000 & 0.000 & 0.000 & 0.000 & 0.000 \\
& & $1/3$ & 0.060 & 0.192 & 0.246 & 0.196 & 0.016 \\
& & $1/2$ & 0.026 & 0.156 & 0.390 & 0.626 & 0.482 \\
& & $2/3$ & 0.002 & 0.062 & 0.262 & 0.552 & 0.696 \\
\addlinespace[3pt]
& 200 & $1/5$ & 0.000 & 0.000 & 0.000 & 0.002 & 0.000 \\
& & $1/3$ & 0.218 & 0.682 & 0.894 & 0.962 & 0.532 \\
& & $1/2$ & 0.068 & 0.468 & 0.828 & 0.948 & 0.944 \\
& & $2/3$ & 0.018 & 0.308 & 0.778 & 0.950 & 0.992 \\
\addlinespace[3pt]
$(200,200)$ & 200 & $1/5$
& 0.000 & 0.012 & 0.206 & 0.618 & 0.728 \\
& & $1/3$ & 0.076 & 0.494 & 0.800 & 0.920 & 0.968 \\
& & $1/2$ & 0.012 & 0.266 & 0.672 & 0.868 & 0.960 \\
& & $2/3$ & 0.002 & 0.164 & 0.564 & 0.852 & 0.968 \\
\addlinespace[3pt]
& 400 & $1/5$ & 0.478 & 0.960 & 0.994 & 0.998 & 0.986 \\
& & $1/3$ & 0.224 & 0.846 & 0.982 & 0.998 & 1.000 \\
& & $1/2$ & 0.046 & 0.706 & 0.962 & 0.992 & 0.998 \\
& & $2/3$ & 0.014 & 0.584 & 0.952 & 0.996 & 1.000 \\
\addlinespace[3pt]
$(300,300)$ & 300 & $1/5$
& 0.256 & 0.842 & 0.974 & 0.990 & 0.996 \\
& & $1/3$ & 0.076 & 0.574 & 0.874 & 0.970 & 0.996 \\
& & $1/2$ & 0.014 & 0.346 & 0.822 & 0.966 & 0.996 \\
& & $2/3$ & 0.000 & 0.290 & 0.832 & 0.972 & 0.996 \\
\addlinespace[3pt]
& 500 & $1/5$ & 0.496 & 0.968 & 0.998 & 1.000 & 1.000 \\
& & $1/3$ & 0.156 & 0.834 & 0.984 & 0.998 & 1.000 \\
& & $1/2$ & 0.022 & 0.740 & 0.984 & 1.000 & 1.000 \\
& & $2/3$ & 0.004 & 0.690 & 0.986 & 1.000 & 1.000 \\

\bottomrule
\end{tabular}

\begin{tablenotes}[flushleft]
\footnotesize
\item \textit{Notes:}
Each entry reports the frequency of exact recovery of both
the informative row set and the informative column set
over 500 Monte Carlo replications.
Here, $r=|\mathcal I_R|/p_1=|\mathcal I_C|/p_2$.
Because the normalized Spearman score has a different numerical scale
from the covariance score, the rank-based thresholds are rescaled by the
fixed constants \(\kappa_R=0.300\) and \(\kappa_C=0.275\).
These constants are held fixed across all simulation configurations and
threshold exponents.
\end{tablenotes}
\end{threeparttable}
\end{table}
\subsubsection{Fixed-Size Ranking Benchmark under Heavy Tails and Contamination}
To further distinguish failures of the screening scores from those due to threshold calibration, Table~\ref{tab:robust-fixed-size} reports a fixed-size ranking benchmark. For each replication, we rank the row screening scores and retain exactly the top \(|\mathcal I_R|\) rows. Using these retained rows, we then recompute the column scores and retain exactly the top \(|\mathcal I_C|\) columns. The reported frequency is the proportion of replications for which both retained sets coincide exactly with \(\mathcal I_R\) and \(\mathcal I_C\).

\begin{table}[htbp]
\centering
\caption{Exact-selection frequencies under fixed-size screening.}
\label{tab:robust-fixed-size}

\small
\setlength{\tabcolsep}{6pt}
\renewcommand{\arraystretch}{1.05}

\begin{threeparttable}
\begin{tabular}{@{}ccc
@{\hspace{1.5em}}cc
@{\hspace{1.5em}}cc
@{\hspace{1.5em}}cc
@{\hspace{1.5em}}cc@{}}
\toprule
&&&
\multicolumn{2}{c}{$t_3$}
& \multicolumn{2}{c}{$\epsilon=0.01$}
& \multicolumn{2}{c}{$\epsilon=0.05$}
& \multicolumn{2}{c}{$\epsilon=0.10$} \\
\cmidrule(lr){4-5}
\cmidrule(lr){6-7}
\cmidrule(lr){8-9}
\cmidrule(l){10-11}
$(p_1,p_2)$ & $T$ & $r$
& Cov. & Rank
& Cov. & Rank
& Cov. & Rank
& Cov. & Rank \\
\midrule

$(100,100)$ & 100 & $1/5$
& 0.998 & 0.506
& 0.004 & 0.520
& 0.000 & 0.314
& 0.000 & 0.176 \\

& & $1/3$
& 1.000 & 0.772
& 0.046 & 0.732
& 0.000 & 0.696
& 0.000 & 0.494 \\

& & $1/2$
& 1.000 & 0.820
& 0.198 & 0.818
& 0.000 & 0.776
& 0.000 & 0.636 \\

& & $2/3$
& 1.000 & 0.884
& 0.366 & 0.852
& 0.000 & 0.810
& 0.000 & 0.726 \\

\addlinespace[1.5pt]

& 200 & $1/5$
& 1.000 & 0.952
& 0.008 & 0.964
& 0.000 & 0.882
& 0.000 & 0.730 \\

& & $1/3$
& 1.000 & 0.998
& 0.058 & 0.996
& 0.000 & 0.984
& 0.000 & 0.942 \\

& & $1/2$
& 1.000 & 0.998
& 0.182 & 0.998
& 0.000 & 0.996
& 0.000 & 0.982 \\

& & $2/3$
& 1.000 & 0.998
& 0.412 & 0.996
& 0.002 & 0.996
& 0.000 & 0.984 \\

\addlinespace[2.5pt]

$(200,200)$ & 200 & $1/5$
& 1.000 & 0.980
& 0.000 & 0.966
& 0.000 & 0.870
& 0.000 & 0.644 \\

& & $1/3$
& 1.000 & 0.992
& 0.004 & 0.992
& 0.000 & 0.972
& 0.000 & 0.940 \\

& & $1/2$
& 1.000 & 1.000
& 0.042 & 0.996
& 0.000 & 0.988
& 0.000 & 0.970 \\

& & $2/3$
& 1.000 & 1.000
& 0.134 & 0.992
& 0.000 & 0.996
& 0.000 & 0.984 \\

\addlinespace[1.5pt]

& 400 & $1/5$
& 1.000 & 1.000
& 0.000 & 1.000
& 0.000 & 1.000
& 0.000 & 0.988 \\

& & $1/3$
& 1.000 & 1.000
& 0.002 & 1.000
& 0.000 & 1.000
& 0.000 & 1.000 \\

& & $1/2$
& 1.000 & 1.000
& 0.040 & 1.000
& 0.000 & 1.000
& 0.000 & 1.000 \\

& & $2/3$
& 1.000 & 1.000
& 0.154 & 1.000
& 0.000 & 1.000
& 0.000 & 1.000 \\

\addlinespace[2.5pt]

$(300,300)$ & 300 & $1/5$
& 1.000 & 1.000
& 0.000 & 0.998
& 0.000 & 0.994
& 0.000 & 0.918 \\

& & $1/3$
& 1.000 & 1.000
& 0.000 & 1.000
& 0.000 & 0.994
& 0.000 & 0.986 \\

& & $1/2$
& 1.000 & 1.000
& 0.004 & 1.000
& 0.000 & 1.000
& 0.000 & 1.000 \\

& & $2/3$
& 1.000 & 1.000
& 0.058 & 1.000
& 0.000 & 0.998
& 0.000 & 1.000 \\

\addlinespace[1.5pt]

& 500 & $1/5$
& 1.000 & 1.000
& 0.000 & 1.000
& 0.000 & 1.000
& 0.000 & 0.996 \\

& & $1/3$
& 1.000 & 1.000
& 0.000 & 1.000
& 0.000 & 1.000
& 0.000 & 1.000 \\

& & $1/2$
& 1.000 & 1.000
& 0.010 & 1.000
& 0.000 & 1.000
& 0.000 & 1.000 \\

& & $2/3$
& 1.000 & 1.000
& 0.048 & 1.000
& 0.000 & 1.000
& 0.000 & 1.000 \\
\bottomrule
\end{tabular}
\end{threeparttable}
\end{table}
Table~\ref{tab:robust-fixed-size} shows that this distinction is substantial. Under the \(t_3\) design, covariance-based scores retain nearly perfect ranking performance in most configurations, whereas under Cauchy contamination their fixed-size exact-recovery frequencies collapse, particularly for \(\epsilon=0.05\) and \(0.10\). This indicates that the deterioration of covariance-based screening under contamination is driven primarily by distortion of the covariance ranking itself, rather than by an unfavorable choice of threshold. In contrast, the rank-based scores preserve substantially more ranking information across the contaminated designs.
\subsubsection{Additional Results on Factor-Number Determination}
For completeness, Table~\ref{tab:factor-number-joint-selection-whole} reports the joint factor-number selection frequencies for all three ridge constants \(c\in\{0,10^{-4},10^{-2}\}\).
\begin{table}[htbp]
\centering
\begin{threeparttable}
\caption{Joint factor-number selection frequencies under sparse loading support}
\label{tab:factor-number-joint-selection-whole}
\footnotesize
\setlength{\tabcolsep}{3pt}
\renewcommand{\arraystretch}{1.1}
\begin{tabular}{@{}ccc*{12}{r}@{}}
\toprule
 & & & \multicolumn{4}{c}{$c=0$} & \multicolumn{4}{c}{$c=10^{-4}$} & \multicolumn{4}{c}{$c=10^{-2}$} \\
\cmidrule(lr){4-7}\cmidrule(lr){8-11}\cmidrule(l){12-15}
$T$ & $r$ & $\sigma_E$ & A & PS-A & P & PS-P & A & PS-A & P & PS-P & A & PS-A & P & PS-P \\
\midrule
\multicolumn{15}{@{}l}{
$p_1=p_2=100$} \\
100 & $1/5$ & 1 & 0.000 & 0.996 & 1.000 & 1.000 & 0.000 & 0.996 & 1.000 & 1.000 & 0.000 & 0.996 & 1.000 & 1.000 \\
 &  & 5 & 0.000 & 0.000 & 0.000 & 0.502 & 0.000 & 0.000 & 0.000 & 0.502 & 0.000 & 0.000 & 0.000 & 0.506 \\
\addlinespace[1.0pt]
 & $1/3$ & 1 & 0.000 & 1.000 & 1.000 & 1.000 & 0.000 & 1.000 & 1.000 & 1.000 & 0.000 & 1.000 & 1.000 & 1.000 \\
 &  & 5 & 0.000 & 0.000 & 0.000 & 0.966 & 0.000 & 0.000 & 0.000 & 0.966 & 0.000 & 0.000 & 0.000 & 0.966 \\
\addlinespace[1.0pt]
\addlinespace[1.5pt]
200 & $1/5$ & 1 & 0.000 & 1.000 & 1.000 & 1.000 & 0.000 & 1.000 & 1.000 & 1.000 & 0.000 & 1.000 & 1.000 & 1.000 \\
 &  & 5 & 0.000 & 0.008 & 0.000 & 0.952 & 0.000 & 0.008 & 0.000 & 0.952 & 0.000 & 0.008 & 0.000 & 0.952 \\
\addlinespace[1.0pt]
 & $1/3$ & 1 & 0.000 & 1.000 & 1.000 & 1.000 & 0.000 & 1.000 & 1.000 & 1.000 & 0.000 & 1.000 & 1.000 & 1.000 \\
 &  & 5 & 0.000 & 0.000 & 0.000 & 1.000 & 0.000 & 0.000 & 0.000 & 1.000 & 0.000 & 0.000 & 0.000 & 1.000 \\
\addlinespace[1.0pt]
\addlinespace[2pt]
\multicolumn{15}{@{}l}{
$p_1=p_2=200$} \\
200 & $1/5$ & 1 & 0.000 & 1.000 & 1.000 & 1.000 & 0.000 & 1.000 & 1.000 & 1.000 & 0.000 & 1.000 & 1.000 & 1.000 \\
 &  & 5 & 0.000 & 0.056 & 0.002 & 0.994 & 0.000 & 0.056 & 0.002 & 0.994 & 0.000 & 0.056 & 0.002 & 0.994 \\
\addlinespace[1.0pt]
 & $1/3$ & 1 & 0.000 & 1.000 & 1.000 & 1.000 & 0.000 & 1.000 & 1.000 & 1.000 & 0.000 & 1.000 & 1.000 & 1.000 \\
 &  & 5 & 0.000 & 0.000 & 1.000 & 1.000 & 0.000 & 0.000 & 1.000 & 1.000 & 0.000 & 0.000 & 1.000 & 1.000 \\
\addlinespace[1.0pt]
\addlinespace[1.5pt]
400 & $1/5$ & 1 & 0.000 & 1.000 & 1.000 & 1.000 & 0.000 & 1.000 & 1.000 & 1.000 & 0.000 & 1.000 & 1.000 & 1.000 \\
 &  & 5 & 0.000 & 0.050 & 0.000 & 1.000 & 0.000 & 0.050 & 0.000 & 1.000 & 0.000 & 0.050 & 0.000 & 1.000 \\
\addlinespace[1.0pt]
 & $1/3$ & 1 & 0.000 & 1.000 & 1.000 & 1.000 & 0.000 & 1.000 & 1.000 & 1.000 & 0.000 & 1.000 & 1.000 & 1.000 \\
 &  & 5 & 0.000 & 0.000 & 1.000 & 1.000 & 0.000 & 0.000 & 1.000 & 1.000 & 0.000 & 0.000 & 1.000 & 1.000 \\
\addlinespace[1.0pt]
\addlinespace[2pt]
\multicolumn{15}{@{}l}{
$p_1=p_2=300$} \\
300 & $1/5$ & 1 & 0.000 & 1.000 & 1.000 & 1.000 & 0.000 & 1.000 & 1.000 & 1.000 & 0.000 & 1.000 & 1.000 & 1.000 \\
 &  & 5 & 0.000 & 0.588 & 1.000 & 1.000 & 0.000 & 0.588 & 1.000 & 1.000 & 0.000 & 0.588 & 1.000 & 1.000 \\
\addlinespace[1.0pt]
 & $1/3$ & 1 & 0.002 & 1.000 & 1.000 & 1.000 & 0.002 & 1.000 & 1.000 & 1.000 & 0.002 & 1.000 & 1.000 & 1.000 \\
 &  & 5 & 0.000 & 0.002 & 1.000 & 1.000 & 0.000 & 0.002 & 1.000 & 1.000 & 0.000 & 0.002 & 1.000 & 1.000 \\
\addlinespace[1.0pt]
\addlinespace[1.5pt]
500 & $1/5$ & 1 & 0.000 & 1.000 & 1.000 & 1.000 & 0.000 & 1.000 & 1.000 & 1.000 & 0.000 & 1.000 & 1.000 & 1.000 \\
 &  & 5 & 0.000 & 0.518 & 1.000 & 1.000 & 0.000 & 0.518 & 1.000 & 1.000 & 0.000 & 0.518 & 1.000 & 1.000 \\
\addlinespace[1.0pt]
 & $1/3$ & 1 & 0.002 & 1.000 & 1.000 & 1.000 & 0.002 & 1.000 & 1.000 & 1.000 & 0.002 & 1.000 & 1.000 & 1.000 \\
 &  & 5 & 0.000 & 0.002 & 1.000 & 1.000 & 0.000 & 0.002 & 1.000 & 1.000 & 0.000 & 0.002 & 1.000 & 1.000 \\
\addlinespace[1.0pt]
\bottomrule
\end{tabular}
\begin{tablenotes}[flushleft]
\footnotesize
\item[] \textit{Notes:}
Each entry reports the proportion of 500 Monte Carlo replications in which both factor numbers are correctly selected, $(\widehat{k}_1,\widehat{k}_2)=(3,3).$
A and P denote $\alpha$-PCA and projected estimators, respectively, while PS-A and PS-P denote their post-screening
counterparts. Screening is row-first with $a=0.9$, and
$r=|\mathcal I_R|/p_1=|\mathcal I_C|/p_2$.
\end{tablenotes}
\end{threeparttable}
\end{table}
\subsubsection{Normal Q--Q Plots for the Regression Coefficient Estimator}
\begin{figure}[htbp]
\centering
\caption{Normal Q--Q plots for the standardized regression coefficient estimator. The reference line corresponds to the standard normal quantile line.}
\begin{subfigure}{0.31\textwidth}
    \centering
    \includegraphics[width=\linewidth]{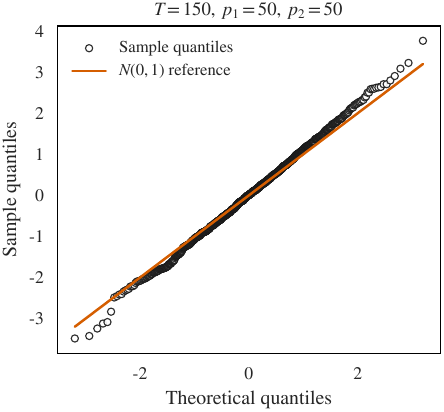}

\end{subfigure}
\hfill
\begin{subfigure}{0.31\textwidth}
    \centering
    \includegraphics[width=\linewidth]{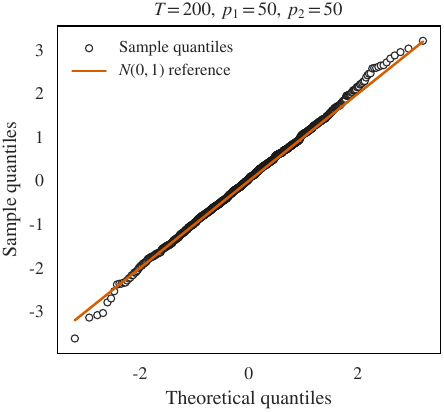}

\end{subfigure}
\hfill
\begin{subfigure}{0.31\textwidth}
    \centering
    \includegraphics[width=\linewidth]{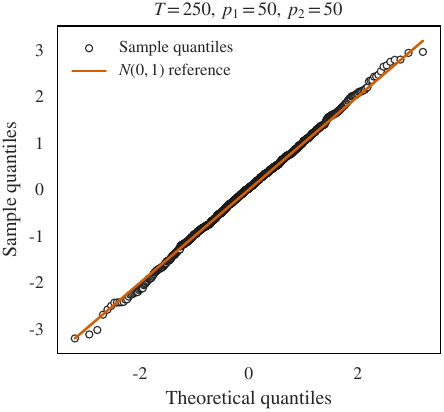}

\end{subfigure}

\label{fig:regression-coefficient-qq}
\end{figure}

Figure~\ref{fig:regression-coefficient-qq} displays the Q-Q plots of the first coordinate of
\(
\sqrt{T}\,
(\bSigma^{\mathrm{sim}})^{-1/2}
\left[
\widehat{\btheta}
-
\left\{
(\widetilde\bH_C^{\mathrm{scr}})^\top\bbeta_0
\otimes
(\widetilde\bH_R^{\mathrm{scr}})^\top\balpha_0
\right\}
\right],
\)
under different settings in Section~\ref{sec:inference-simulation}.
\subsection{Additional Empirical Results}
\subsubsection{Ordered Screening Statistics and Successive Ratios for Macroeconomic Application}
\begin{figure}[htbp]
    \caption{Ordered row and column screening statistics and their
    successive ratios for United States total CPI, computed from the
    initial training sample. The upper-left and upper-right panels display
    the ordered row screening statistics and their successive ratios,
    respectively, while the lower-left and lower-right panels report the
    corresponding quantities for the columns.}
    \label{fig:sta-distribution}
    \centering
    \includegraphics[width=0.75\linewidth]{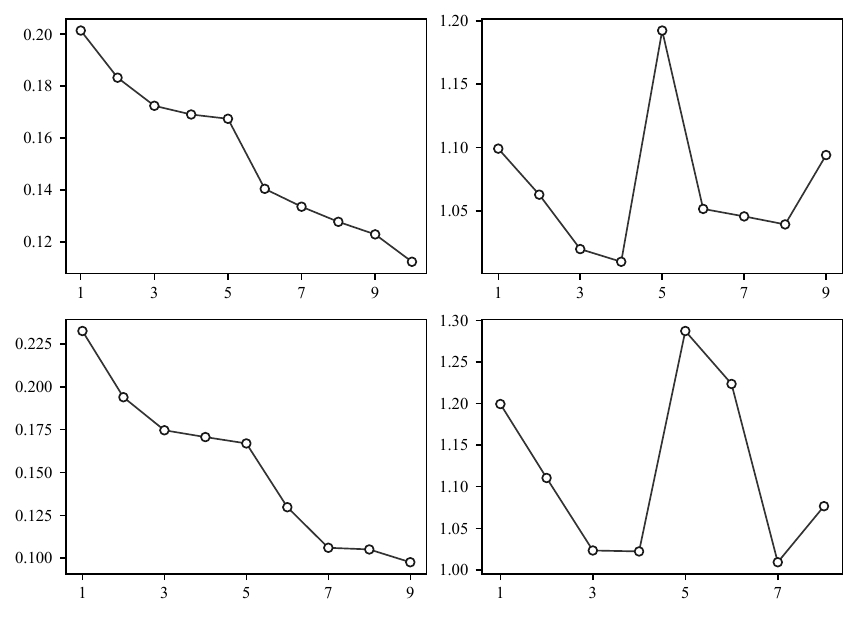}
\end{figure}
Specifically, \(\tau_R=0.165\) lies
between the fifth and sixth ordered row screening statistics, where the
most prominent peak in the corresponding successive-ratio plot is
observed. In contrast, the column threshold \(\tau_C=0.215\) is applied
at the second screening stage, after the row restriction, and is
therefore based on column statistics recomputed from the retained rows.
It is consequently not directly comparable with the initial marginal
column statistics displayed in the figure. Moreover, because the figure
is constructed using only the initial training sample, the locations and
relative prominence of the empirical separations may change as the
estimation window expands. Overall, the selected specification is
consistent with the most pronounced row-wise separation in the initial
sample.
\subsubsection{Threshold-Grid Results for Macroeconomic Application}
\label{sec:empirical-threshold-grid}
For compact presentation, Tables~\ref{tab:tau-pivot-rowfirst-screened}--\ref{tab:tau-pivot-colfirst-unscreened} report a subset of the threshold combinations evaluated in the forecasting exercise. The response MSE is minimized over the full admissible grid with spacing \(0.005\) in both directions. In each table, we retain the corresponding minimizing threshold pair and display additional threshold values at increments of \(0.04\) moving outward from the optimum, together with the zero-threshold specification. Thus, the displayed entries provide a representative view of the local and broader variation in forecasting performance without reporting the entire fine grid. In all four tables, rows index \(\tau_R\) and columns index \(\tau_C\). Each entry is reported as \linebreak[4]
\(
\left[
\begin{smallmatrix}
\mathrm{Zero\text{-}insert\ MSE} & \mathrm{Retained\ MSE}\\
y\mathrm{\ MSE}\times 10^{3} & R^2
\end{smallmatrix}
\right].
\)
A boxed entry marks a threshold pair attaining the minimum response MSE over the full admissible fine grid, rather than only over the threshold combinations displayed in the table. Dashes denote threshold pairs for which the retained matrix is treated
as inadmissible, including cases where
\(p_1^{\mathrm{scr}}\leq 1\) or \(p_2^{\mathrm{scr}}\leq 1\), or where
a retained dimension is smaller than the corresponding factor number.
\begin{table}[htbp]
\centering
\small
\setlength{\tabcolsep}{4pt}
\renewcommand{\arraystretch}{1.5}
\caption{Threshold-grid performance under row-first screening with factor numbers estimated from the screened initial training sample. Displayed row and column thresholds are selected from the full fine grid at intervals of 0.04 anchored at the full-grid minimizer, with
additional boundary values included for reference. The minimum response MSE over the full admissible grid is attained at $\paren{\tau_R,\tau_C}=\paren{0.165,0.215}$, with $y\mathrm{\ MSE}=0.0310\times10^{-3}$ and $R^2=0.490$.}
\label{tab:tau-pivot-rowfirst-screened}
\begin{adjustbox}{max width=\textwidth}
\begin{tabular}{cccccccccc}
\toprule
$\tau_R\backslash\tau_C$ & 0.000 & 0.015 & 0.055 & 0.095 & 0.135 & 0.175 & \textbf{0.215} & 0.255 & 0.260 \\
\midrule
0.000 &
$\left[\begin{smallmatrix}0.626&0.626\\0.0564&0.072\end{smallmatrix}\right]$ &
$\left[\begin{smallmatrix}0.626&0.626\\0.0564&0.072\end{smallmatrix}\right]$ &
$\left[\begin{smallmatrix}0.626&0.626\\0.0564&0.072\end{smallmatrix}\right]$ &
$\left[\begin{smallmatrix}0.646&0.613\\0.0564&0.073\end{smallmatrix}\right]$ &
$\left[\begin{smallmatrix}0.810&0.654\\0.0561&0.077\end{smallmatrix}\right]$ &
$\left[\begin{smallmatrix}0.857&0.520\\0.0458&0.247\end{smallmatrix}\right]$ &
-- &
-- &
-- \\
0.005 &
$\left[\begin{smallmatrix}0.626&0.626\\0.0564&0.072\end{smallmatrix}\right]$ &
$\left[\begin{smallmatrix}0.626&0.626\\0.0564&0.072\end{smallmatrix}\right]$ &
$\left[\begin{smallmatrix}0.626&0.626\\0.0564&0.072\end{smallmatrix}\right]$ &
$\left[\begin{smallmatrix}0.646&0.613\\0.0564&0.073\end{smallmatrix}\right]$ &
$\left[\begin{smallmatrix}0.810&0.654\\0.0561&0.077\end{smallmatrix}\right]$ &
$\left[\begin{smallmatrix}0.857&0.520\\0.0458&0.247\end{smallmatrix}\right]$ &
-- &
-- &
-- \\
0.045 &
$\left[\begin{smallmatrix}0.626&0.626\\0.0564&0.072\end{smallmatrix}\right]$ &
$\left[\begin{smallmatrix}0.626&0.626\\0.0564&0.072\end{smallmatrix}\right]$ &
$\left[\begin{smallmatrix}0.626&0.626\\0.0564&0.072\end{smallmatrix}\right]$ &
$\left[\begin{smallmatrix}0.646&0.613\\0.0564&0.073\end{smallmatrix}\right]$ &
$\left[\begin{smallmatrix}0.810&0.654\\0.0561&0.077\end{smallmatrix}\right]$ &
$\left[\begin{smallmatrix}0.857&0.520\\0.0458&0.247\end{smallmatrix}\right]$ &
-- &
-- &
-- \\
0.085 &
$\left[\begin{smallmatrix}0.626&0.626\\0.0564&0.072\end{smallmatrix}\right]$ &
$\left[\begin{smallmatrix}0.626&0.626\\0.0564&0.072\end{smallmatrix}\right]$ &
$\left[\begin{smallmatrix}0.626&0.626\\0.0564&0.072\end{smallmatrix}\right]$ &
$\left[\begin{smallmatrix}0.646&0.613\\0.0564&0.073\end{smallmatrix}\right]$ &
$\left[\begin{smallmatrix}0.810&0.654\\0.0561&0.077\end{smallmatrix}\right]$ &
$\left[\begin{smallmatrix}0.857&0.520\\0.0458&0.247\end{smallmatrix}\right]$ &
-- &
-- &
-- \\
0.125 &
$\left[\begin{smallmatrix}0.620&0.581\\0.0589&0.032\end{smallmatrix}\right]$ &
$\left[\begin{smallmatrix}0.620&0.581\\0.0589&0.032\end{smallmatrix}\right]$ &
$\left[\begin{smallmatrix}0.620&0.581\\0.0589&0.032\end{smallmatrix}\right]$ &
$\left[\begin{smallmatrix}0.630&0.563\\0.0586&0.036\end{smallmatrix}\right]$ &
$\left[\begin{smallmatrix}0.831&0.677\\0.0559&0.081\end{smallmatrix}\right]$ &
$\left[\begin{smallmatrix}0.896&0.669\\0.0567&0.067\end{smallmatrix}\right]$ &
$\left[\begin{smallmatrix}0.935&0.341\\0.0385&0.367\end{smallmatrix}\right]$ &
-- &
-- \\
\textbf{0.165} &
$\left[\begin{smallmatrix}0.826&0.573\\0.0560&0.080\end{smallmatrix}\right]$ &
$\left[\begin{smallmatrix}0.826&0.573\\0.0560&0.080\end{smallmatrix}\right]$ &
$\left[\begin{smallmatrix}0.826&0.573\\0.0560&0.080\end{smallmatrix}\right]$ &
$\left[\begin{smallmatrix}0.834&0.582\\0.0528&0.132\end{smallmatrix}\right]$ &
$\left[\begin{smallmatrix}0.932&0.760\\0.0555&0.088\end{smallmatrix}\right]$ &
$\left[\begin{smallmatrix}0.990&0.767\\0.0547&0.101\end{smallmatrix}\right]$ &
\fbox{$\left[\begin{smallmatrix}0.906&0.306\\0.0310&0.490\end{smallmatrix}\right]$} &
$\left[\begin{smallmatrix}1.042&0.317\\0.0376&0.381\end{smallmatrix}\right]$ &
$\left[\begin{smallmatrix}1.043&0.281\\0.0358&0.411\end{smallmatrix}\right]$ \\
0.180 &
$\left[\begin{smallmatrix}0.878&0.432\\0.0504&0.172\end{smallmatrix}\right]$ &
$\left[\begin{smallmatrix}0.878&0.432\\0.0504&0.172\end{smallmatrix}\right]$ &
$\left[\begin{smallmatrix}0.878&0.432\\0.0504&0.172\end{smallmatrix}\right]$ &
$\left[\begin{smallmatrix}0.889&0.446\\0.0459&0.245\end{smallmatrix}\right]$ &
$\left[\begin{smallmatrix}1.008&0.875\\0.0517&0.149\end{smallmatrix}\right]$ &
$\left[\begin{smallmatrix}1.010&0.803\\0.0537&0.118\end{smallmatrix}\right]$ &
$\left[\begin{smallmatrix}1.015&0.747\\0.0485&0.202\end{smallmatrix}\right]$ &
$\left[\begin{smallmatrix}1.034&0.786\\0.0479&0.213\end{smallmatrix}\right]$ &
$\left[\begin{smallmatrix}1.048&0.767\\0.0468&0.230\end{smallmatrix}\right]$ \\
\bottomrule
\end{tabular}%
\end{adjustbox}
\vspace{0.8cm}
\setlength{\tabcolsep}{6pt}
\renewcommand{\arraystretch}{1.5}
\caption{Threshold-grid performance under row-first screening with factor numbers estimated from the unscreened initial training sample. Displayed row and column thresholds are selected from the full fine grid at intervals of 0.04 anchored at the full-grid minimizer, with
additional boundary values included for reference. The minimum response MSE over the full admissible grid is attained at $\paren{\tau_R,\tau_C}=\paren{0.110,0.145}$, with $y\mathrm{\ MSE}=0.0445\times10^{-3}$ and $R^2=0.268$.}
\label{tab:tau-pivot-rowfirst-unscreened}
\begin{adjustbox}{max width=\textwidth}
\begin{tabular}{ccccccc}
\toprule
$\tau_R\backslash\tau_C$ & 0.000 & 0.025 & 0.065 & 0.105 & \textbf{0.145} & 0.180 \\
\midrule
0.000 &
$\left[\begin{smallmatrix}0.626&0.626\\0.0564&0.072\end{smallmatrix}\right]$ &
$\left[\begin{smallmatrix}0.626&0.626\\0.0564&0.072\end{smallmatrix}\right]$ &
$\left[\begin{smallmatrix}0.626&0.626\\0.0564&0.072\end{smallmatrix}\right]$ &
$\left[\begin{smallmatrix}0.677&0.573\\0.0546&0.102\end{smallmatrix}\right]$ &
$\left[\begin{smallmatrix}0.810&0.638\\0.0456&0.249\end{smallmatrix}\right]$ &
-- \\
0.030 &
$\left[\begin{smallmatrix}0.626&0.626\\0.0564&0.072\end{smallmatrix}\right]$ &
$\left[\begin{smallmatrix}0.626&0.626\\0.0564&0.072\end{smallmatrix}\right]$ &
$\left[\begin{smallmatrix}0.626&0.626\\0.0564&0.072\end{smallmatrix}\right]$ &
$\left[\begin{smallmatrix}0.677&0.573\\0.0546&0.102\end{smallmatrix}\right]$ &
$\left[\begin{smallmatrix}0.810&0.638\\0.0456&0.249\end{smallmatrix}\right]$ &
-- \\
0.070 &
$\left[\begin{smallmatrix}0.626&0.626\\0.0564&0.072\end{smallmatrix}\right]$ &
$\left[\begin{smallmatrix}0.626&0.626\\0.0564&0.072\end{smallmatrix}\right]$ &
$\left[\begin{smallmatrix}0.626&0.626\\0.0564&0.072\end{smallmatrix}\right]$ &
$\left[\begin{smallmatrix}0.677&0.573\\0.0546&0.102\end{smallmatrix}\right]$ &
$\left[\begin{smallmatrix}0.810&0.638\\0.0456&0.249\end{smallmatrix}\right]$ &
-- \\
\textbf{0.110} &
$\left[\begin{smallmatrix}0.622&0.628\\0.0563&0.074\end{smallmatrix}\right]$ &
$\left[\begin{smallmatrix}0.622&0.628\\0.0563&0.074\end{smallmatrix}\right]$ &
$\left[\begin{smallmatrix}0.622&0.628\\0.0563&0.074\end{smallmatrix}\right]$ &
$\left[\begin{smallmatrix}0.680&0.582\\0.0548&0.099\end{smallmatrix}\right]$ &
\fbox{$\left[\begin{smallmatrix}0.807&0.636\\0.0445&0.268\end{smallmatrix}\right]$} &
-- \\
0.150 &
$\left[\begin{smallmatrix}0.757&0.618\\0.0561&0.078\end{smallmatrix}\right]$ &
$\left[\begin{smallmatrix}0.757&0.618\\0.0561&0.078\end{smallmatrix}\right]$ &
$\left[\begin{smallmatrix}0.757&0.618\\0.0561&0.078\end{smallmatrix}\right]$ &
$\left[\begin{smallmatrix}0.771&0.626\\0.0537&0.116\end{smallmatrix}\right]$ &
-- &
-- \\
0.190 &
$\left[\begin{smallmatrix}0.916&0.403\\0.0514&0.155\end{smallmatrix}\right]$ &
$\left[\begin{smallmatrix}0.916&0.403\\0.0514&0.155\end{smallmatrix}\right]$ &
$\left[\begin{smallmatrix}0.917&0.400\\0.0497&0.183\end{smallmatrix}\right]$ &
$\left[\begin{smallmatrix}0.924&0.379\\0.0498&0.181\end{smallmatrix}\right]$ &
$\left[\begin{smallmatrix}0.945&0.345\\0.0543&0.106\end{smallmatrix}\right]$ &
-- \\
0.200 &
$\left[\begin{smallmatrix}0.932&0.301\\0.0526&0.135\end{smallmatrix}\right]$ &
$\left[\begin{smallmatrix}0.932&0.301\\0.0526&0.135\end{smallmatrix}\right]$ &
$\left[\begin{smallmatrix}0.934&0.297\\0.0520&0.146\end{smallmatrix}\right]$ &
$\left[\begin{smallmatrix}0.937&0.269\\0.0536&0.118\end{smallmatrix}\right]$ &
$\left[\begin{smallmatrix}0.962&0.220\\0.0581&0.044\end{smallmatrix}\right]$ &
$\left[\begin{smallmatrix}0.978&0.163\\0.0583&0.041\end{smallmatrix}\right]$ \\
\bottomrule
\end{tabular}%
\end{adjustbox}
\end{table}

\begin{table}[htbp]
\centering
\small
\setlength{\tabcolsep}{6pt}
\renewcommand{\arraystretch}{1.5}
\caption{Threshold-grid performance under column-first screening with factor numbers estimated from the screened initial training sample. Displayed row and column thresholds are selected from the full fine grid at intervals of 0.04 anchored at the full-grid minimizer, with
additional boundary values included for reference. The minimum response MSE over the full admissible grid is attained at $\paren{\tau_R,\tau_C}=\paren{0.195,0.190}$, with $y\mathrm{\ MSE}=0.0397\times10^{-3}$ and $R^2=0.348$.}
\label{tab:tau-pivot-colfirst-screened}
\begin{adjustbox}{max width=\textwidth}
\begin{tabular}{ccccccc}
\toprule
$\tau_R\backslash\tau_C$ & 0.000 & 0.030 & 0.070 & 0.110 & 0.150 & \textbf{0.190} \\
\midrule
0.000 &
$\left[\begin{smallmatrix}0.626&0.626\\0.0564&0.072\end{smallmatrix}\right]$ &
$\left[\begin{smallmatrix}0.626&0.626\\0.0564&0.072\end{smallmatrix}\right]$ &
$\left[\begin{smallmatrix}0.626&0.626\\0.0564&0.072\end{smallmatrix}\right]$ &
$\left[\begin{smallmatrix}0.748&0.623\\0.0561&0.078\end{smallmatrix}\right]$ &
$\left[\begin{smallmatrix}0.821&0.659\\0.0537&0.117\end{smallmatrix}\right]$ &
$\left[\begin{smallmatrix}0.890&0.368\\0.0457&0.249\end{smallmatrix}\right]$ \\
0.035 &
$\left[\begin{smallmatrix}0.626&0.626\\0.0564&0.072\end{smallmatrix}\right]$ &
$\left[\begin{smallmatrix}0.626&0.626\\0.0564&0.072\end{smallmatrix}\right]$ &
$\left[\begin{smallmatrix}0.626&0.626\\0.0564&0.072\end{smallmatrix}\right]$ &
$\left[\begin{smallmatrix}0.748&0.623\\0.0561&0.078\end{smallmatrix}\right]$ &
$\left[\begin{smallmatrix}0.821&0.659\\0.0537&0.117\end{smallmatrix}\right]$ &
$\left[\begin{smallmatrix}0.890&0.368\\0.0457&0.249\end{smallmatrix}\right]$ \\
0.075 &
$\left[\begin{smallmatrix}0.626&0.626\\0.0564&0.072\end{smallmatrix}\right]$ &
$\left[\begin{smallmatrix}0.626&0.626\\0.0564&0.072\end{smallmatrix}\right]$ &
$\left[\begin{smallmatrix}0.626&0.626\\0.0564&0.072\end{smallmatrix}\right]$ &
$\left[\begin{smallmatrix}0.748&0.623\\0.0561&0.078\end{smallmatrix}\right]$ &
$\left[\begin{smallmatrix}0.821&0.659\\0.0537&0.117\end{smallmatrix}\right]$ &
$\left[\begin{smallmatrix}0.899&0.377\\0.0427&0.298\end{smallmatrix}\right]$ \\
0.115 &
$\left[\begin{smallmatrix}0.641&0.638\\0.0560&0.078\end{smallmatrix}\right]$ &
$\left[\begin{smallmatrix}0.641&0.638\\0.0560&0.078\end{smallmatrix}\right]$ &
$\left[\begin{smallmatrix}0.641&0.638\\0.0560&0.078\end{smallmatrix}\right]$ &
$\left[\begin{smallmatrix}0.761&0.625\\0.0575&0.054\end{smallmatrix}\right]$ &
$\left[\begin{smallmatrix}0.851&0.660\\0.0532&0.126\end{smallmatrix}\right]$ &
$\left[\begin{smallmatrix}0.955&0.395\\0.0435&0.285\end{smallmatrix}\right]$ \\
0.155 &
$\left[\begin{smallmatrix}0.733&0.569\\0.0581&0.045\end{smallmatrix}\right]$ &
$\left[\begin{smallmatrix}0.733&0.569\\0.0581&0.045\end{smallmatrix}\right]$ &
$\left[\begin{smallmatrix}0.733&0.569\\0.0581&0.045\end{smallmatrix}\right]$ &
$\left[\begin{smallmatrix}0.818&0.635\\0.0553&0.091\end{smallmatrix}\right]$ &
$\left[\begin{smallmatrix}0.869&0.686\\0.0532&0.125\end{smallmatrix}\right]$ &
$\left[\begin{smallmatrix}0.971&0.445\\0.0409&0.327\end{smallmatrix}\right]$ \\
\textbf{0.195} &
-- &
-- &
-- &
$\left[\begin{smallmatrix}0.935&0.499\\0.0535&0.121\end{smallmatrix}\right]$ &
$\left[\begin{smallmatrix}0.917&0.763\\0.0520&0.145\end{smallmatrix}\right]$ &
\fbox{$\left[\begin{smallmatrix}0.984&0.480\\0.0397&0.348\end{smallmatrix}\right]$} \\
0.235 &
-- &
-- &
-- &
-- &
-- &
$\left[\begin{smallmatrix}1.033&0.566\\0.0421&0.307\end{smallmatrix}\right]$ \\
0.270 &
-- &
-- &
-- &
-- &
-- &
$\left[\begin{smallmatrix}1.086&0.874\\0.0504&0.171\end{smallmatrix}\right]$ \\
\bottomrule
\end{tabular}%
\end{adjustbox}
\vspace{0.8cm}
\setlength{\tabcolsep}{6pt}
\renewcommand{\arraystretch}{1.5}
\caption{Threshold-grid performance under column-first screening with factor numbers estimated from the unscreened initial training sample. Displayed row and column thresholds are selected from the full fine grid at intervals of 0.04 anchored at the full-grid minimizer, with
additional boundary values included for reference. The minimum response MSE over the full admissible grid is attained at $\paren{\tau_R,\tau_C}=\paren{0.110,0.145}$, with $y\mathrm{\ MSE}=0.0450\times10^{-3}$ and $R^2=0.259$. The same minimum is also attained at $\paren{0.110,0.150}$, $\paren{0.110,0.155}$, $\paren{0.110,0.160}$ in the full grid.}
\label{tab:tau-pivot-colfirst-unscreened}
\begin{tabular}{ccccccc}
\toprule
$\tau_R\backslash\tau_C$ & 0.000 & 0.025 & 0.065 & 0.105 & \textbf{0.145} & \textbf{0.160} \\
\midrule
0.000 &
$\left[\begin{smallmatrix}0.626&0.626\\0.0564&0.072\end{smallmatrix}\right]$ &
$\left[\begin{smallmatrix}0.626&0.626\\0.0564&0.072\end{smallmatrix}\right]$ &
$\left[\begin{smallmatrix}0.626&0.626\\0.0564&0.072\end{smallmatrix}\right]$ &
$\left[\begin{smallmatrix}0.677&0.573\\0.0546&0.102\end{smallmatrix}\right]$ &
$\left[\begin{smallmatrix}0.810&0.638\\0.0456&0.249\end{smallmatrix}\right]$ &
$\left[\begin{smallmatrix}0.810&0.638\\0.0456&0.249\end{smallmatrix}\right]$ \\
0.030 &
$\left[\begin{smallmatrix}0.626&0.626\\0.0564&0.072\end{smallmatrix}\right]$ &
$\left[\begin{smallmatrix}0.626&0.626\\0.0564&0.072\end{smallmatrix}\right]$ &
$\left[\begin{smallmatrix}0.626&0.626\\0.0564&0.072\end{smallmatrix}\right]$ &
$\left[\begin{smallmatrix}0.677&0.573\\0.0546&0.102\end{smallmatrix}\right]$ &
$\left[\begin{smallmatrix}0.810&0.638\\0.0456&0.249\end{smallmatrix}\right]$ &
$\left[\begin{smallmatrix}0.810&0.638\\0.0456&0.249\end{smallmatrix}\right]$ \\
0.070 &
$\left[\begin{smallmatrix}0.626&0.626\\0.0564&0.072\end{smallmatrix}\right]$ &
$\left[\begin{smallmatrix}0.626&0.626\\0.0564&0.072\end{smallmatrix}\right]$ &
$\left[\begin{smallmatrix}0.626&0.626\\0.0564&0.072\end{smallmatrix}\right]$ &
$\left[\begin{smallmatrix}0.677&0.573\\0.0546&0.102\end{smallmatrix}\right]$ &
$\left[\begin{smallmatrix}0.810&0.638\\0.0456&0.249\end{smallmatrix}\right]$ &
$\left[\begin{smallmatrix}0.810&0.638\\0.0456&0.249\end{smallmatrix}\right]$ \\
\textbf{0.110} &
$\left[\begin{smallmatrix}0.622&0.628\\0.0563&0.074\end{smallmatrix}\right]$ &
$\left[\begin{smallmatrix}0.622&0.628\\0.0563&0.074\end{smallmatrix}\right]$ &
$\left[\begin{smallmatrix}0.622&0.628\\0.0563&0.074\end{smallmatrix}\right]$ &
$\left[\begin{smallmatrix}0.672&0.569\\0.0549&0.097\end{smallmatrix}\right]$ &
\fbox{$\left[\begin{smallmatrix}0.827&0.627\\0.0450&0.259\end{smallmatrix}\right]$} &
\fbox{$\left[\begin{smallmatrix}0.827&0.627\\0.0450&0.259\end{smallmatrix}\right]$} \\
0.150 &
$\left[\begin{smallmatrix}0.757&0.618\\0.0561&0.078\end{smallmatrix}\right]$ &
$\left[\begin{smallmatrix}0.757&0.618\\0.0561&0.078\end{smallmatrix}\right]$ &
$\left[\begin{smallmatrix}0.757&0.618\\0.0561&0.078\end{smallmatrix}\right]$ &
$\left[\begin{smallmatrix}0.779&0.585\\0.0554&0.089\end{smallmatrix}\right]$ &
$\left[\begin{smallmatrix}0.848&0.640\\0.0461&0.241\end{smallmatrix}\right]$ &
$\left[\begin{smallmatrix}0.848&0.640\\0.0461&0.241\end{smallmatrix}\right]$ \\
0.190 &
$\left[\begin{smallmatrix}0.916&0.403\\0.0514&0.155\end{smallmatrix}\right]$ &
$\left[\begin{smallmatrix}0.916&0.403\\0.0514&0.155\end{smallmatrix}\right]$ &
$\left[\begin{smallmatrix}0.916&0.403\\0.0514&0.155\end{smallmatrix}\right]$ &
$\left[\begin{smallmatrix}0.932&0.460\\0.0517&0.149\end{smallmatrix}\right]$ &
$\left[\begin{smallmatrix}0.900&0.703\\0.0479&0.212\end{smallmatrix}\right]$ &
$\left[\begin{smallmatrix}0.900&0.703\\0.0479&0.212\end{smallmatrix}\right]$ \\
0.230 &
-- &
-- &
-- &
-- &
$\left[\begin{smallmatrix}0.958&0.652\\0.0514&0.155\end{smallmatrix}\right]$ &
$\left[\begin{smallmatrix}0.958&0.652\\0.0514&0.155\end{smallmatrix}\right]$ \\
0.250 &
-- &
-- &
-- &
-- &
$\left[\begin{smallmatrix}1.063&0.110\\0.0489&0.196\end{smallmatrix}\right]$ &
$\left[\begin{smallmatrix}1.063&0.110\\0.0489&0.196\end{smallmatrix}\right]$ \\
\bottomrule
\end{tabular}%
\end{table}
\subsubsection{Estimated Loading Structures for Macroeconomic Application}
Based on the initial unscreened training sample, the estimated row and column factor numbers are
\(
(\widehat{k}_1,\widehat{k}_2)=(1,5).
\)
For the representative row-first Post-screening PE specification with
\(
(\tau_R,\tau_C)=(0.165,0.215),
\)
the corresponding estimates from the screened initial training sample are
\(
(\widehat{k}_1,\widehat{k}_2)=(3,2).
\)
Tables~\ref{tab:row-loadings-united-states-cpi} and
\ref{tab:column-loadings-united-states-cpi} report the estimated row and
column loading patterns obtained using \(\alpha\)-PCA, PE, and
Post-screening PE.

\begin{table}[htbp]
\centering
\caption{Estimated row loading matrices with
United States--Total CPI as the target series.}
\label{tab:row-loadings-united-states-cpi}

\begin{adjustbox}{max width=\textwidth}
\begin{tabular}{llrrrrrrrrrr}
\toprule
Method & Factor
& AUS & CAN & FIN & FRA & DEU & NLD & NZL & NOR & SWE & USA \\
\midrule
\(\alpha\)-PCA
& 1
& -2
& -3
& -3
& \cellcolor{blue!10}-4
& \cellcolor{blue!10}-4
& \cellcolor{blue!10}-4
& -3
& -3
& -3
& -3 \\

PE
& 1
& -2
& -3
& \cellcolor{blue!10}-4
& \cellcolor{blue!10}-4
& \cellcolor{blue!10}-4
& -3
& -3
& -3
& -3
& -3 \\

\midrule
\multirow{3}{*}{Post-screening PE}
& 1
& 0
& -2
& 
& -6
& 
& 
& 
& \cellcolor{blue!10}-7
& 
& 2 \\

& 2
& 0
& -6
& 
& -2
& 
& 
& 
& 1
& 
& \cellcolor{blue!10}-8 \\

& 3
& \cellcolor{blue!10}-10
& 2
& 
& 0
& 
& 
& 
& 0
& 
& -2 \\

\bottomrule
\end{tabular}
\end{adjustbox}

\vspace{1mm}
\begin{minipage}{0.97\linewidth}
\footnotesize
\textit{Notes:}
For Post-screening PE, row-first screening with
\((\tau_R,\tau_C)=(0.165,0.215)\) retains Australia, Canada, France,
Norway, and the United States. Blank entries correspond to countries
excluded by the screening procedure. The estimated row factor number is
one for \(\alpha\)-PCA and PE and three for Post-screening PE.
Varimax rotation is therefore
applied only to the Post-screening PE row loading matrix. Each loading
vector is normalized to have unit Euclidean norm, multiplied by 10, and
rounded to the nearest integer. Shaded entries indicate the largest
displayed loading(s) in absolute value within each factor. Because the
loading vectors are normalized separately and are defined over different
sets of countries, their displayed magnitudes should not be compared
directly across methods.
\end{minipage}
\end{table}

\begin{table}[htbp]
\centering
\caption{Estimated column loading matrices with
United States--Total CPI as the target series.}
\label{tab:column-loadings-united-states-cpi}

\begin{adjustbox}{max width=\textwidth}
\begin{tabular}{llrrrrrrrrr}
\toprule
Method & Factor
& CPI:F & CPI:E & CPI:T & IR:L & IR:3M & P:TI & P:M & EX & IM \\
\midrule
\multirow{5}{*}{\(\alpha\)-PCA}
& 1
& 0
& -1
& 0
& 0
& 0
& 0
& 0
& \cellcolor{blue!10}-7
& \cellcolor{blue!10}-7 \\

& 2
& 0
& \cellcolor{blue!10}-7
& -6
& -2
& 2
& 0
& 0
& 0
& 0 \\

& 3
& 0
& -1
& 1
& -1
& 1
& \cellcolor{blue!10}7
& \cellcolor{blue!10}7
& 0
& 0 \\

& 4
& \cellcolor{blue!10}9
& -2
& 2
& -1
& 1
& 1
& -1
& 0
& 0 \\

& 5
& 0
& 0
& 0
& \cellcolor{blue!10}7
& \cellcolor{blue!10}7
& 0
& 0
& 0
& 0 \\

\midrule
\multirow{5}{*}{PE}
& 1
& -1
& 1
& 0
& -1
& 2
& 0
& 0
& \cellcolor{blue!10}7
& \cellcolor{blue!10}7 \\

& 2
& -1
& -1
& 1
& 0
& 1
& \cellcolor{blue!10}7
& \cellcolor{blue!10}7
& 0
& 0 \\

& 3
& 1
& \cellcolor{blue!10}8
& 6
& 1
& -3
& 0
& 0
& 0
& 0 \\

& 4
& 1
& 0
& 0
& \cellcolor{blue!10}-9
& -3
& 0
& 0
& 0
& 0 \\

& 5
& \cellcolor{blue!10}-9
& 1
& -2
& 1
& -4
& -1
& 1
& 0
& 0 \\

\midrule
\multirow{2}{*}{Post-screening PE}
& 1
& 
& 
& 
& 
& 
& \cellcolor{blue!10}-7
& \cellcolor{blue!10}-7
& 
& 0 \\

& 2
& 
& 
& 
& 
& 
& 0
& 0
& 
& \cellcolor{blue!10}10 \\

\bottomrule
\end{tabular}
\end{adjustbox}

\vspace{1mm}
\begin{minipage}{0.97\linewidth}
\footnotesize
\textit{Notes:}
For Post-screening PE, row-first screening with
\((\tau_R,\tau_C)=(0.165,0.215)\) retains total-industry production,
manufacturing production, and imports. Blank entries correspond to
indicators excluded by the screening procedure. The estimated column
factor number is five for \(\alpha\)-PCA and PE and two for
Post-screening PE. Varimax rotation
is applied to all three column loading matrices. Each loading vector is
normalized to have unit Euclidean norm, multiplied by 10, and rounded to
the nearest integer. Shaded entries indicate the largest displayed loading(s)
in absolute value within each factor. 
\end{minipage}
\end{table}

Tables~\ref{tab:row-loadings-united-states-cpi} and
\ref{tab:column-loadings-united-states-cpi} reveal several notable
features of the estimated loading structures. First, the row loading
patterns obtained by \(\alpha\)-PCA and PE from the unscreened predictor
matrix are broadly similar. For \(\alpha\)-PCA, the largest absolute row
loadings occur for France, Germany, and the Netherlands, whereas for PE
they occur for Finland, France, and Germany. On the column side, the two
methods also identify broadly similar groups of macroeconomic variables,
although these groups appear in different factor positions. In
particular, both loading structures contain factors primarily associated
with international trade, production, consumer prices, and interest
rates. This broad similarity is consistent with their relatively close
reconstruction and forecasting performance reported in
Table~\ref{tab:representative-method-comparison}.

The Post-screening PE row loadings exhibit a more concentrated
cross-country structure. In absolute magnitude, the first row factor is
primarily associated with Norway and France, the second places its
largest loadings on the United States and Canada, and the third is
dominated by Australia. One possible interpretation is that the
predictive variation relevant to United States total CPI is concentrated
in a small number of cross-country components, rather than in a broad
loading pattern spanning all ten countries. 

On the column side, Post-screening PE retains total-industry production,
manufacturing production, and imports. The first column factor has equal
loadings of \(-7\) on the two production indicators and zero loading on
imports, whereas the second has zero loadings on the two production
indicators and a loading of \(10\) on imports. The resulting structure
therefore separates a real-activity component from an external-trade
component, both of which may contain predictive variation for United
States total CPI beyond that captured by the unscreened factor estimates.
\subsubsection{Diebold--Mariano Tests for Two Empirical Applications}
To assess whether the forecast improvements are statistically
distinguishable, Table~\ref{tab:representative-dm-tests} and ~\ref{tab:ip-dm-tests} report pairwise
Diebold--Mariano tests \citep{DieboldMariano_JBES_1995} comparing
Post-screening PE with each competing method. The loss differential is
defined as the squared forecast loss of Post-screening PE minus that of
the benchmark, so a negative value favors Post-screening PE. The null of
equal expected forecast loss is tested against the one-sided alternative
that Post-screening PE has the smaller expected loss.

\begin{table}[!htb]
\centering
\small
\setlength{\tabcolsep}{8pt}
\renewcommand{\arraystretch}{1.15}
\caption{Pairwise Diebold--Mariano tests for United States--Total CPI.}
\label{tab:representative-dm-tests}

\begin{tabular}{lrrr}
\toprule
Benchmark
& Mean loss difference \((\times 10^{3})\)
& DM statistic
& One-sided \(p\)-value \\
\midrule

\(\alpha\)-PCA
& -0.02632 & -3.841 & \(0.0002^{***}\) \\

PE
& -0.02545 & -4.010 & \(0.0001^{***}\) \\

Vec-PCA (ER, \(r=1\))
& -0.02477 & -4.138 & \(0.00009^{***}\) \\

AR(2)
& -0.01223 & -1.306 & \(0.0996^{*}\) \\

Lasso
& -0.01294 & -2.145 & \(0.0191^{**}\) \\

\bottomrule
\end{tabular}

\begin{minipage}{0.92\textwidth}
\footnotesize
\textit{Notes:}
For benchmark \(m\), the loss differential is
\(
d_t
=
\left(y_t-\widehat y_t^{\,\mathrm{Post}}\right)^2
-
\left(y_t-\widehat y_t^{\,(m)}\right)^2,
\)
where \(\widehat y_t^{\,\mathrm{Post}}\) denotes the forecast produced
by Post-screening PE. Thus, negative mean loss differences and negative
DM statistics favor Post-screening PE. The tests use 40 one-step-ahead
forecasts and HAC lag zero. The Harvey--Leybourne--Newbold small-sample
correction is applied. The reported \(p\)-values correspond to the
one-sided alternative
\(
\mathbb E(d_t)<0.
\) \(^{***}p<0.01\), \(^{**}p<0.05\), and \(^{*}p<0.10\).
\end{minipage}
\end{table}

\begin{table}[htbp]
\centering
\small
\setlength{\tabcolsep}{8pt}
\renewcommand{\arraystretch}{1.15}
\caption{Pairwise Diebold--Mariano tests for U.S. manufacturing
production growth.}
\label{tab:ip-dm-tests}

\begin{tabular}{lrrr}
\toprule
Benchmark
& Mean loss difference \((\times10^{3})\)
& DM statistic
& One-sided \(p\)-value \\
\midrule
\(\alpha\)-PCA
& -0.00452 & -1.586 & \(0.0595^{*}\) \\

PE
& -0.00452 & -1.587 & \(0.0593^{*}\) \\

Vec-PCA (ER, \(r=1\))
& -0.00452 & -1.587 & \(0.0593^{*}\) \\

AR(2)
& -0.00641 & -1.857 & \(0.0346^{**}\) \\

Lasso
& -0.00470 & -1.602 & \(0.0576^{*}\) \\
\bottomrule
\end{tabular}

\begin{minipage}{0.92\textwidth}
\footnotesize
\textit{Notes:}
For benchmark \(m\), the loss differential is
\(
d_t
=
(y_t-\widehat y_t^{\,\mathrm{Post}})^2
-
(y_t-\widehat y_t^{\,(m)})^2.
\)
Thus, negative mean loss differences and negative DM statistics favor
Post-screening PE. The tests use 52 one-step-ahead forecasts and HAC lag
zero. The Harvey--Leybourne--Newbold small-sample correction is applied.
The reported \(p\)-values correspond to the one-sided alternative
\(\mathbb E(d_t)<0\). \(^{***}p<0.01\), \(^{**}p<0.05\), and \(^{*}p<0.10\).
\end{minipage}
\end{table}

\clearpage
\onehalfspacing
\bibliographystyle{jasa-asa}
\let\oldbibliography\thebibliography
\renewcommand{\thebibliography}[1]{%
  \oldbibliography{#1}%
  \setlength{\itemsep}{3pt}%
}
	{\footnotesize
		\bibliography{reference}
	}
	\onehalfspacing

\end{document}

%% file: preamble.tex
\usepackage[table]{xcolor}
\usepackage{tikz}
\usetikzlibrary{positioning,arrows.meta,fit,backgrounds,calc,decorations.pathreplacing}
\usepackage{pdflscape}
\usepackage{booktabs}
\usepackage{rotating}
\usepackage{amsmath}
\usepackage{makecell}
\usepackage{threeparttable,tabularx,array}
\usepackage{multirow,multicol}
\usepackage{booktabs}
\usepackage{adjustbox}
\usepackage[margin=1in]{geometry} 
\usepackage{amsmath,amsthm,amssymb, graphicx, multicol, array, bbm, comment, subcaption,  float, booktabs, xpatch, color}
\usepackage{setspace} 
\usepackage{longtable}
\usepackage{enumerate}
\usepackage{algorithm}
\usepackage{algorithmicx}
\usepackage{float}
\usepackage{algpseudocode}
\usepackage{xr-hyper}
\usepackage{hyperref}
\algrenewcommand\algorithmicrequire{\textbf{Input:}}
\algrenewcommand\algorithmicensure{\textbf{Output:}}

\def\*#1{\mathbf{#1}}
\def\+#1{\mathbb{#1}}

\newcommand{\RNum}[1]{\uppercase\expandafter{\romannumeral #1\relax}}

\usepackage[T1]{fontenc}
\usepackage[utf8]{inputenc}
\usepackage{lmodern}
\usepackage[english]{babel}
\usepackage[autostyle]{csquotes}
\usepackage{amsmath}
\usepackage{natbib}
\setcitestyle{
    open={(},
    close={)},
    aysep={},
    citesep={;}
}

\makeatletter
\newcommand*{\addFileDependency}[1]{%
    \typeout{(#1)}%
    \@addtofilelist{#1}%
    \IfFileExists{#1}{}{\typeout{No file #1.}}%
}
\makeatother
\newcommand*{\myexternaldocument}[1]{%
    \IfFileExists{xr-auto-#1.aux}{%
        \externaldocument[][nocite]{xr-auto-#1}[#1.pdf]%
    }{%
        \ifdefined\XRAUXONLY
            \typeout{XR-AUTO: peer cache absent during initial auxiliary pass.}%
        \else
            \PackageWarningNoLine{xr-auto}{Missing xr-auto-#1.aux.
            Check that the automatic latexmkrc is installed
            and the compiler is pdfLaTeX}%
        \fi
    }%
    \ifdefined\XRAUXONLY
    \else
        \addFileDependency{#1.tex}%
        \IfFileExists{xr-auto-#1.deps.tex}{\input{xr-auto-#1.deps.tex}}{}%
    \fi
}

\usepackage{enumitem}
\setlist{itemsep=.01em}
\setlist{topsep=.5em}

\newcommand{\norm}[1]{\left\lVert#1\right\rVert}

\newtheorem{innercustomgeneric}{\customgenericname}
\providecommand{\customgenericname}{}

\newtheorem{theorem}{Theorem}
\newtheorem{lemma}{Lemma}
\newtheorem{proposition}{Proposition}
\newtheorem{remark}{Remark}
\newtheorem{corollary}{Corollary}

\newtheorem{assumption}{{Assumption}}

\ifdefined\XRAUXONLY
    \setkeys{Gin}{draft}%
\fi

\def\beq{\begin{equation}}
\def\eeq{\end{equation}}
\def\beqr{\begin{eqnarray}}
\def\eeqr{\end{eqnarray}}
\def\beqrs{\begin{eqnarray*}}
\def\eeqrs{\end{eqnarray*}}
\def\bet{\begin{theorem}}
\def\eet{\end{theorem}}
\def\bel{\begin{lemma}}
\def\eel{\end{lemma}}
\def\bep{\begin{proposition}}
\def\eep{\end{proposition}}
\def\bg{\begin{figure}[tbph]\begin{center}}
\def\eg{\end{center}\end{figure}}

\def\bc{\begin{center}}
\def\ec{\end{center}}

\def\var{\mbox{var}}
\def\cov{\mbox{Cov}}

\def\argmax{\mbox{argmax}}

\newcommand{\Cov}{\textnormal{Cov}}

\newcommand{\vect}{\textnormal{vec}}

\newcommand{\bA}{{\mathbf A}}
\newcommand{\bB}{{\mathbf B}}
\newcommand{\bF}{{\mathbf F}}
\newcommand{\bE}{{\mathbf E}}
\newcommand{\bG}{{\mathbf G}}
\newcommand{\bH}{{\mathbf H}}
\newcommand{\bI}{{\mathbf I}}
\newcommand{\bK}{{\mathbf K}}
\newcommand{\bL}{{\mathbf L}}
\newcommand{\bM}{{\mathbf M}}

\newcommand{\bQ}{{\mathbf Q}}

\newcommand{\bR}{{\mathbf R}}
\newcommand{\bS}{{\mathbf S}}

\newcommand{\bU}{{\mathbf U}}
\newcommand{\bV}{{\mathbf V}}
\newcommand{\bW}{{\mathbf W}}
\newcommand{\bX}{{\mathbf X}}
\newcommand{\bY}{{\mathbf Y}}

\newcommand{\ba}{{\mathbf a}}
\newcommand{\bb}{{\mathbf b}}

\newcommand{\be}{{\mathbf e}}

 \newcommand{\bgg}{{\mathbf g}}

\newcommand{\bv}{{\mathbf v}}
\newcommand{\bw}{{\mathbf w}}
\newcommand{\bx}{{\mathbf x}}

\newcommand{\bz}{{\mathbf z}}
\newcommand{\balpha} {\boldsymbol{\alpha}}
\newcommand{\bbeta}  {\boldsymbol{\beta}}

\newcommand{\bSigma}{\boldsymbol{\Sigma}}
\newcommand{\bDelta}{\boldsymbol{\Delta}}
\newcommand{\bgamma}{\boldsymbol{\gamma}}

\newcommand{\bTheta} {\boldsymbol{\Theta}}
\newcommand{\bPhi} {\boldsymbol{\Phi}}
\newcommand{\bPsi} {\boldsymbol{\Psi}}

\newcommand{\btheta} {\boldsymbol{\theta}}

\newcommand{\bzeta} {\boldsymbol{\zeta}}
\newcommand{\bGamma} {\boldsymbol{\Gamma}}
\newcommand{\bLambda} {\boldsymbol{\Lambda}}
\newcommand{\bC}{{\mathbf C}}

\newcommand{\bzero}{{\mathbf 0}}

\newcommand{\ve}{{\varepsilon}}
\renewcommand{\epsilon}{{\ve}}

\newcommand{\Tr}{\mbox{tr}}
\newcommand{\paren}[1]{\left(#1\right)}
\newcommand{\bracket}[1]{\left[#1\right]}
\newcommand{\bracee}[1]{\left\{#1\right\}}
\newcommand{\abs}[1]{\left|#1\right|}
\newcommand{\E}{\mathbb{E}}

\newcolumntype{L}[1]{>{\raggedright\arraybackslash}p{#1}}

\usepackage[margin=10pt,labelfont=bf]{caption}

\newcommand{\simfigdir}{simulation_figures}

%% file: reference.bib
@book{AthreyaLahiri_2006,
  author    = {Athreya, Krishna B. and Lahiri, Soumendra N.},
  title     = {Measure Theory and Probability Theory},
  series    = {Springer Texts in Statistics},
  publisher = {Springer},
  address   = {New York, NY},
  year      = {2006},
  doi       = {10.1007/978-0-387-35434-7}
}

@article{Bai_Econometrica_2003,
author = {Bai, Jushan},
title = {Inferential Theory for Factor Models of Large Dimensions},
journal = {Econometrica},
volume = {71},
number = {1},
pages = {135-171},
doi = {10.1111/1468-0262.00392},
url = {https://onlinelibrary.wiley.com/doi/abs/10.1111/1468-0262.00392},
eprint = {https://onlinelibrary.wiley.com/doi/pdf/10.1111/1468-0262.00392},
year = {2003}
}

@article{BaiNg_Econometrica_2006,
author = {Bai, Jushan and Ng, Serena},
title = {Confidence Intervals for Diffusion Index Forecasts and Inference for Factor-Augmented Regressions},
journal = {Econometrica},
volume = {74},
number = {4},
pages = {1133-1150},
doi = {10.1111/j.1468-0262.2006.00696.x},
url = {https://onlinelibrary.wiley.com/doi/abs/10.1111/j.1468-0262.2006.00696.x},
eprint = {https://onlinelibrary.wiley.com/doi/pdf/10.1111/j.1468-0262.2006.00696.x},
year = {2006}
}

@article{BaiNg_JoE_2023,
title = {Approximate Factor Models with Weaker Loadings},
journal = {Journal of Econometrics},
volume = {235},
number = {2},
pages = {1893-1916},
year = {2023},
issn = {0304-4076},
doi = {10.1016/j.jeconom.2023.01.027},
url = {https://www.sciencedirect.com/science/article/pii/S030440762300060X},
author = {Jushan Bai and Serena Ng}
}

@article{CaiKongWuZhao_JBES_2025,
author = {Xiong Cai and Xinbing Kong and Xinlei Wu and Peng Zhao},
title = {Matrix-Factor-Augmented Regression},
journal = {Journal of Business \& Economic Statistics},
volume = {43},
number = {4},
pages = {1145--1157},
year = {2025},
publisher = {Taylor \& Francis},
doi = {10.1080/07350015.2025.2478986},
URL = { 10.1080/07350015.2025.2478986
},
eprint = {10.1080/07350015.2025.2478986
}
}

@article{ChenTsayChen_JASA_2020,
  title={Constrained Factor Models for High-Dimensional Matrix-Variate Time Series},
  author={Chen, Elynn Y and Tsay, Ruey S and Chen, Rong},
  journal={Journal of the American Statistical Association},
  volume={115},
  number={530},
  pages={775--793},
  year={2020},
  doi = {10.1080/01621459.2019.1584899},
  publisher={Taylor \& Francis}
}

@article{ChenFan_JASA_2023,
author = {Elynn Y. Chen and Jianqing Fan},
title = {Statistical Inference for High-Dimensional Matrix-Variate Factor Models},
journal = {Journal of the American Statistical Association},
volume = {118},
number = {542},
pages = {1038--1055},
year = {2023},
publisher = {Taylor \& Francis},
doi = {10.1080/01621459.2021.1970569},
URL = {10.1080/01621459.2021.1970569
},
eprint = {10.1080/01621459.2021.1970569
}
}

@article{ChenFanZhu_JASA_2026,
title={Factor Augmented Matrix Regression},
author={Chen, Elynn Y and Fan, Jianqing and Zhu, Xiaonan},
journal={Journal of the American Statistical Association},
year={2026},
publisher={Taylor \& Francis},
doi={10.1080/01621459.2025.2595734},
volume = {121},
number = {554},
pages = {1192--1205},
}

@book{FanYao_2003,
  author    = {Fan, Jianqing and Yao, Qiwei},
  title     = {Nonlinear Time Series: Nonparametric and Parametric Methods},
  series    = {Springer Series in Statistics},
  publisher = {Springer},
  address   = {New York, NY},
  year      = {2003},
  doi       = {10.1007/978-0-387-69395-8}
}

@article{lamyao2012,
         journal = {The Annals of Statistics},
           title = {Factor Modeling for High-Dimensional Time Series: Inference for the Number of Factors},
             doi = {10.1214/12-AOS970},
            year = {2012},
       publisher = {Institute of Mathematical Statistics},
          number = {2},
           pages = {694--726},
          volume = {40},
             url = {http://www.imstat.org/aos/},
          author = {Lam, Clifford and Yao, Qiwei},
            issn = {0090-5364}
}

@article{LamYaoBathia_Biometrika_2011,
    author = {Lam, Clifford and Yao, Qiwei and Bathia, Neil},
    title = {Estimation of Latent Factors for High-Dimensional Time Series},
    journal = {Biometrika},
    volume = {98},
    number = {4},
    pages = {901-918},
    year = {2011},
    month = {12},
    issn = {0006-3444},
    doi = {10.1093/biomet/asr048},
    url = {10.1093/biomet/asr048},
    eprint = {https://academic.oup.com/biomet/article-pdf/98/4/901/639859/asr048.pdf},
}

@article{MerlevedePeligradRio_PTRF_2011,
  title={A Bernstein Type Inequality and Moderate Deviations for Weakly Dependent Sequences},
  author={Merlevède, Florence and Peligrad, Magda and Rio, Emmanuel},
  journal={Probability Theory and Related Fields},
  volume={151},
  number={3--4},
  pages={435--474},
  year={2011},
  publisher={Springer},
  doi={10.1007/s00440-010-0304-9}
}

@article{WangLiuChen_JoE_2019,
title = {Factor Models for Matrix-Valued High-Dimensional Time Series},
journal = {Journal of Econometrics},
volume = {208},
number = {1},
pages = {231-248},
year = {2019},
issn = {0304-4076},
doi = {10.1016/j.jeconom.2018.09.013},
url = {https://www.sciencedirect.com/science/article/pii/S0304407618301787},
author = {Dong Wang and Xialu Liu and Rong Chen}
}

@article{YuHeKongZhang_JoE_2022,
title = {Projected Estimation for Large-Dimensional Matrix Factor Models},
journal = {Journal of Econometrics},
volume = {229},
number = {1},
pages = {201-217},
year = {2022},
issn = {0304-4076},
doi = {10.1016/j.jeconom.2021.04.001},
url = {https://www.sciencedirect.com/science/article/pii/S0304407621001123},
author = {Long Yu and Yong He and Xinbing Kong and Xinsheng Zhang}
}

@article{StockWatson_JBES_2002,
author = {James H Stock and Mark W Watson},
title = {Macroeconomic Forecasting Using Diffusion Indexes},
journal = {Journal of Business \& Economic Statistics},
volume = {20},
number = {2},
pages = {147--162},
year = {2002},
publisher = {Taylor \& Francis},
doi = {10.1198/073500102317351921},
URL = { 10.1198/073500102317351921},
eprint = { 10.1198/073500102317351921
}
}

@article{GaoTsay_JASA_2025,
journal={Journal of the American Statistical Association},
author={Zhaoxing Gao and Ruey S. Tsay},
title={Supervised Dynamic PCA: Linear Dynamic Forecasting with Many Predictors},
year={2025},
month={April},
pages={869-883},
volume={120},
number={550},
doi={10.1080/01621459.2024.2370592},
url={https://ideas.repec.org/a/taf/jnlasa/v120y2025i550p869-883.html},
}

@article{StockWatson_JASA_2002,
author = {James H Stock and Mark W Watson},
title = {Forecasting Using Principal Components From a Large Number of Predictors},
journal = {Journal of the American Statistical Association},
volume = {97},
number = {460},
pages = {1167--1179},
year = {2002},
publisher = {Taylor \& Francis},
doi = {10.1198/016214502388618960},
URL = {10.1198/016214502388618960},
eprint = {10.1198/016214502388618960}}

@article{ChenYangZhang_JASA_2022,
author = {Rong Chen and Dan Yang and Cun-Hui Zhang},
title = {Factor Models for High-Dimensional Tensor Time Series},
journal = {Journal of the American Statistical Association},
volume = {117},
number = {537},
pages = {94--116},
year = {2022},
publisher = {Taylor \& Francis},
doi = {10.1080/01621459.2021.1912757},
URL = {10.1080/01621459.2021.1912757
},
eprint = {10.1080/01621459.2021.1912757
}
}

@article{FanLv_JRSSB_2008,
    author = {Fan, Jianqing and Lv, Jinchi},
    title = {Sure Independence Screening for Ultrahigh Dimensional Feature Space},
    journal = {Journal of the Royal Statistical Society Series B: Statistical Methodology},
    volume = {70},
    number = {5},
    pages = {849-911},
    year = {2008},
    month = {11},
    issn = {1369-7412},
    doi = {10.1111/j.1467-9868.2008.00674.x},
    url = {10.1111/j.1467-9868.2008.00674.x},
    eprint = {https://academic.oup.com/jrsssb/article-pdf/70/5/849/49796154/jrsssb_70_5_849.pdf},
}

@article{HuangJiangLiTongZhou_MS_2022,
author = {Huang, Dashan and Jiang, Fuwei and Li, Kunpeng and Tong, Guoshi and Zhou, Guofu},
title = {Scaled PCA: A New Approach to Dimension Reduction},
journal = {Management Science},
volume = {68},
number = {3},
pages = {1678-1695},
year = {2022},
doi = {10.1287/mnsc.2021.4020},
URL = { 10.1287/mnsc.2021.4020
},
eprint = {10.1287/mnsc.2021.4020
}}

@article{Onatski_JoE_2012,
title = {Asymptotics of the Principal Components Estimator of Large Factor Models With Weakly Influential Factors},
journal = {Journal of Econometrics},
volume = {168},
number = {2},
pages = {244-258},
year = {2012},
issn = {0304-4076},
doi = {10.1016/j.jeconom.2012.01.034},
url = {https://www.sciencedirect.com/science/article/pii/S0304407612000449},
author = {Alexei Onatski}
}

@article{UematsuYamagata_JBES_2023_Estimation,
author = {Yoshimasa Uematsu and Takashi Yamagata},
title = {Estimation of Sparsity-Induced Weak Factor Models},
journal = {Journal of Business \& Economic Statistics},
volume = {41},
number = {1},
pages = {213--227},
year = {2023},
publisher = {Taylor \& Francis},
doi = {10.1080/07350015.2021.2008405},
URL = {10.1080/07350015.2021.2008405
},
eprint = {10.1080/07350015.2021.2008405
}
}

@article{ChenLam_AoS_2024,
author = {Weilin Chen and Clifford Lam},
title = {Rank and Factor Loadings Estimation in Time Series Tensor Factor Model by Pre-Averaging},
volume = {52},
journal = {The Annals of Statistics},
number = {1},
publisher = {Institute of Mathematical Statistics},
pages = {364--391},
year = {2024},
doi = {10.1214/23-AOS2350},
URL = {10.1214/23-AOS2350}
}

@article{HanChenYangZhang_AoS_2024,
author = {Yuefeng Han and Rong Chen and Dan Yang and Cun-Hui Zhang},
title = {Tensor Factor Model Estimation by Iterative Projection},
volume = {52},
journal = {The Annals of Statistics},
number = {6},
publisher = {Institute of Mathematical Statistics},
pages = {2641--2667},
year = {2024},
doi = {10.1214/24-AOS2412},
URL = {10.1214/24-AOS2412}
}

@article{HanChenZhang_EJS_2022,
  title={Rank Determination in Tensor Factor Model},
  author={Han, Yuefeng and Chen, Rong and Zhang, Cun-Hui},
  journal={Electronic Journal of Statistics},
  volume={16},
  number={1},
  pages={1726--1803},
  year={2022},
  publisher={Institute of Mathematical Statistics and Bernoulli Society},
  doi={10.1214/22-EJS1991}
}

@article{BaiNg_JoE_2008,
title = {Forecasting Economic Time Series Using Targeted Predictors},
journal = {Journal of Econometrics},
volume = {146},
number = {2},
pages = {304-317},
year = {2008},
issn = {0304-4076},
doi = {10.1016/j.jeconom.2008.08.010},
url = {https://www.sciencedirect.com/science/article/pii/S0304407608001085},
author = {Jushan Bai and Serena Ng}
}

@article{AhnHorenstein_Econometrica_2013,
author = {Ahn, Seung C. and Horenstein, Alex R.},
title = {Eigenvalue Ratio Test for the Number of Factors},
journal = {Econometrica},
volume = {81},
number = {3},
pages = {1203-1227},
doi = {10.3982/ECTA8968},
url = {https://onlinelibrary.wiley.com/doi/abs/10.3982/ECTA8968},
eprint = {https://onlinelibrary.wiley.com/doi/pdf/10.3982/ECTA8968},
year = {2013}
}

@misc{ChenLam_arXiv_2024,
  author       = {Chen, Weilin and Lam, Clifford},
  title        = {Factor Strength Estimation in Vector and Matrix Time Series Factor Models},
  year         = {2024},
  howpublished = {arXiv preprint, arXiv:2405.07294},
  doi          = {10.48550/arXiv.2405.07294}
}

@article{ChenXiaoYang_JoE_2021,
title = {Autoregressive Models for Matrix-Valued Time Series},
journal = {Journal of Econometrics},
volume = {222},
number = {1, Part B},
pages = {539-560},
year = {2021},
issn = {0304-4076},
doi = {10.1016/j.jeconom.2020.07.015},
url = {https://www.sciencedirect.com/science/article/pii/S0304407620302050},
author = {Rong Chen and Han Xiao and Dan Yang}
}

@article{SaidDickey_Biometrika_1984,
    author = {Said, Said E. and Dickey, David A.},
    title = {Testing for Unit Roots in Autoregressive-Moving Average Models of Unknown Order},
    journal = {Biometrika},
    volume = {71},
    number = {3},
    pages = {599-607},
    year = {1984},
    month = {12},
    issn = {0006-3444},
    doi = {10.1093/biomet/71.3.599},
    url = {10.1093/biomet/71.3.599},
    eprint = {https://academic.oup.com/biomet/article-pdf/71/3/599/719376/71-3-599.pdf},
}

@article{KwiatkowskiPhillipsSchmidtShin_JoE_1992,
title = {Testing the Null Hypothesis of Stationarity Against the Alternative of a Unit Root: How Sure Are We That Economic Time Series Have a Unit Root?},
journal = {Journal of Econometrics},
volume = {54},
number = {1--3},
pages = {159-178},
year = {1992},
issn = {0304-4076},
doi = {10.1016/0304-4076(92)90104-Y},
url = {https://www.sciencedirect.com/science/article/pii/030440769290104Y},
author = {Denis Kwiatkowski and Peter C.B. Phillips and Peter Schmidt and Yongcheol Shin}
}

@article{DieboldMariano_JBES_1995,
author = {Francis X. Diebold and Roberto S. Mariano},
title = {Comparing Predictive Accuracy},
journal = {Journal of Business \& Economic Statistics},
volume = {13},
number = {3},
pages = {253--263},
year = {1995},
publisher = {Taylor \& Francis},
doi = {10.1080/07350015.1995.10524599},
URL = {https://www.tandfonline.com/doi/abs/10.1080/07350015.1995.10524599
},
eprint = {https://www.tandfonline.com/doi/pdf/10.1080/07350015.1995.10524599
}
}

@article{BOIVINNG_JOE_2006,
title = {Are More Data Always Better for Factor Analysis?},
journal = {Journal of Econometrics},
volume = {132},
number = {1},
pages = {169-194},
year = {2006},
issn = {0304-4076},
doi = {10.1016/j.jeconom.2005.01.027},
url = {https://www.sciencedirect.com/science/article/pii/S030440760500045X},
author = {Jean Boivin and Serena Ng}
}

@article{BairHastiePaulTibshirani_JASA_2006,
author = {Eric Bair and Trevor Hastie and Debashis Paul and Robert Tibshirani},
title = {Prediction by Supervised Principal Components},
journal = {Journal of the American Statistical Association},
volume = {101},
number = {473},
pages = {119--137},
year = {2006},
publisher = {Taylor \& Francis},
doi = {10.1198/016214505000000628},
URL = {  10.1198/016214505000000628
},
eprint = {10.1198/016214505000000628
}
}

@article{FanXueYao_JoE_2017,
title = {Sufficient Forecasting Using Factor Models},
journal = {Journal of Econometrics},
volume = {201},
number = {2},
pages = {292-306},
year = {2017},
issn = {0304-4076},
doi = {10.1016/j.jeconom.2017.08.009},
url = {https://www.sciencedirect.com/science/article/pii/S0304407617301616},
author = {Jianqing Fan and Lingzhou Xue and Jiawei Yao}
}

@article{FanSamworthWu_JMLR_2009,
  author  = {Jianqing Fan and Richard Samworth and Yichao Wu},
  title   = {Ultrahigh Dimensional Feature Selection: Beyond the Linear Model},
  journal = {Journal of Machine Learning Research},
  year    = {2009},
  volume  = {10},
  number  = {70},
  pages   = {2013--2038},
  url     = {http://jmlr.org/papers/v10/fan09a.html}
}

@article{FanSong_AoS_2010,
author = {Jianqing Fan and Rui Song},
title = {Sure Independence Screening in Generalized Linear Models with NP-Dimensionality},
volume = {38},
journal = {The Annals of Statistics},
number = {6},
publisher = {Institute of Mathematical Statistics},
pages = {3567--3604},
year = {2010},
doi = {10.1214/10-AOS798},
URL = {10.1214/10-AOS798}
}

@article{FanFengSong_JASA_2011,
author = {Jianqing Fan and Yang Feng and Rui Song},
title = {Nonparametric Independence Screening in Sparse Ultra-High Dimensional Additive Models},
journal = {Journal of the American Statistical Association},
volume = {106},
number = {494},
pages = {544--557},
year = {2011},
publisher = {Taylor \& Francis},
doi = {10.1198/jasa.2011.tm09779},
URL = {10.1198/jasa.2011.tm09779
},
eprint = {10.1198/jasa.2011.tm09779}
}

@Article{Fanghan2018,
journal={Journal of Theoretical Probability},
author={Fang Han},
title={An Exponential Inequality for U-Statistics Under Mixing Conditions},
year={2018},
month={March},
pages={556-578},
volume={31},
number={1},
doi={10.1007/s10959-016-0722-4},
url={https://ideas.repec.org/a/spr/jotpro/v31y2018i1d10.1007_s10959-016-0722-4.html},
}

@article{croux2010influence,
  title   = {Influence Functions of the Spearman and Kendall Correlation Measures},
  author  = {Croux, Christophe and Dehon, Catherine},
  journal = {Statistical Methods \& Applications},
  volume  = {19},
  number  = {4},
  pages   = {497--515},
  year    = {2010},
  doi     = {10.1007/s10260-010-0142-z}
}
